\documentclass[epj]{svjour}
\usepackage{graphics}
\begin{document}

\title{Electric and magnetic dipole modes in high-resolution inelastic proton scattering at $0^\circ$}
\titlerunning{E1 and M1 modes in high-resolution inelastic proton scattering at $0^\circ$}
\author{Peter~von~Neumann-Cosel\inst{1} and Atsushi~Tamii\inst{2}
}                     % Do not remove
\institute{
Institut f\"{u}r Kernphysik, Technische Universit\"{a}t Darmstadt, D-64289 Darmstadt, Germany
\and
Research Center for Nuclear Physics, Osaka University, 10-1 Mihogaoka, Ibaraki 567-0047, Japan
}
\date{Received: date / Revised version: date}
\abstract{Inelastic proton scattering under extreme forward angles including $0^\circ$ and at energies of a few hundred MeV has been established as a new spectroscopic tool for the study of complete dipole strength distributions in nuclei. 
Such data allow an extraction of the electric dipole polarizability which provides important constraints for parameters of the symmetry energy, which determine the neutron-skin thickness and the equation of state (EOS) of neutron-rich matter. 
Also, new insight into the much-debated nature of the pygmy dipole resonance (PDR) is obtained.
Additionally, the isovector spin-M1 resonance can be studied in heavy nuclei, where only limited experimental information exists so far. 
Together with much improved results on the isoscalar spin-M1 strength distributions in $N = Z$ nuclei, these data shed new light on the phenomenon of quenching of the nuclear spin response.    
Using dispersion matching techniques, high energy resolution ($\Delta E/E \leq 10^{-4}$ full width at half maximum, FWHM) can be achieved in the experiments. 
In spherical-vibrational nuclei considerable fine structure is observed in the energy region of the isovector giant dipole resonance (IVGDR). 
A quantitative analysis of the fine structure with wavelet methods provides information on the role of different damping mechanisms contributing to the width of the IVGDR. 
Furthermore, level densities can be extracted from a fluctuation analysis at excitation energies well above neutron threshold, a region hardly accessible by other means. 
The combination of the gamma strength function (GSF) extracted from the E1 and M1 strength distributions with the independently derived level density permits novel tests of the Brink-Axel hypothesis underlying all calculations of statistical model reaction cross sections in astrophysical applications in the energy region of the PDR.     
 % end of PACS codes
} %end of abstract
\maketitle

\tableofcontents

\section{Introduction}
\label{sec1}

Experimental studies of nuclear reactions with the highest possible energy resolution and selectivity are a driving force of our understanding of nuclear structure.
A famous example are nuclear resonance fluorescence experiments, also called $(\gamma,\gamma^\prime)$ reactions, where the energy resolution is achieved by the use of germanium detectors and the selectivity by the dominance of dipole electromagnetic transition matrix elements \cite{kne06}.
For inelastic scattering or other direct reactions with charged particles (electrons, hadrons), high energy resolution can be achieved using magnetic spectrometers combined with a dispersion matching of the beam.
At sufficiently high velocities (corresponding to energies $>$ 100 MeV/nucleon for ions), the measured inelastic scattering cross sections are approximately proportional to the transition matrix elements.
Selectivity to the transferred angular momentum can be achieved by varying the scattering angle (i.e., the momentum transfer).    
For a target with ground state (g.s.) spin and parity $J^\pi = 0^+$ this determines the spin and often also the parity of the excited states.

In general, (p,p$^\prime$) scattering is not a very selective reaction because natural and unnatural-parity states are excited with comparable probability.
However, when going to very small angles (i.e., low momentum transfers), a strong selectivity towards dipole excitations is observed.
One the one hand,due to the strong spin-isospin part of the effective proton-nucleus interactions at small momentum transfers \cite{lov81} transitions with angular momentum transfer $\Delta L = 0$ and spin transfer $\Delta S = 1$ are favored populating $1^+$ states in targets with a $0^+$ ground state.
The excited states form the spin-flip M1 resonance representing the isospin analog of the Gamow-Teller (GT) resonances \cite{fuj11}.
On the other hand, Coulomb-excitation cross sections become very large at relativistic particle velocities and actually dominate the (p,p$^\prime$) response for scattering angles near $0^\circ$, making it an excellent probe to study the E1 strength distribution in nuclei. 
The present review discusses recent progress in using the reaction as a spectroscopic tool for electric and magnetic dipole strength in nuclei by realizing experiments with energy resolutions $\Delta E/ E \approx 10^{-4}$ at extreme forward angles including $0^\circ$.
 
Such experiments present a challenge due to the very small magnetic rigidity difference between the incident and scattered particles.
Consequently, they are known to be extremely sensitive to beam halo and background from atomic small-angle scattering in the target since the primary beam exits the spectrometer very close to the position of the focal-plane detectors. 
Historically, the technical development of (p,p$^\prime$) measurements under 0$^\circ$ at energies of $100-500$ MeV was performed in the 1980s at the Los Alamos National Laboratory \cite{mcc84,mcc85} and at the Indiana University Cyclotron (IUCF) \cite{IUCF}, but the background still limited applications to light and medium-mass nuclei.
However, at the Research Center for Nuclear Physics (RCNP) \cite{tam09} and later at the iThemba Laboratory for Accelerator Based Sciences (iThemba LABS) near Cape Town, South Africa \cite{nev11}, the method has been developed to a level allowing for essentially background-free measurements up to the heaviest stable nuclei.
While the research programs at these $0^\circ$ facilities are ongoing and also involve other topics like the study of isoscalar giant resonances with $\alpha$ scattering \cite{gar18}, investigation of the GT strength with charge-exchange reactions \cite{fuj11,fre18}, astrophysically relevant transitions \cite{lon18,ads17} or giant pairing vibrations \cite{mou11}, we focus here on the (p,p$^\prime$) experiments providing novel information on the E1 and spin-M1 resonances, thereby contributing to a variety of nuclear structure problems of current interest.

Low-energy electric dipole strength in neutron-rich nuclei, commonly termed PDR, is currently a topic of great interest \cite{sav13}. 
It occurs at energies well below the isovector giant dipole resonance (IVGDR) and exhausts a considerable fraction (up to about 10\%) of the total E1 strength in nuclei with a large neutron-to-proton ratio \cite{adr05,kli07,wie09,ros13}. 
The properties of the mode are claimed to provide insight into the formation of a neutron skin \cite{kli07,pie06,tso08,pie11,ina11} and the density dependence of the symmetry energy \cite{kli07,car10,fat12,tsa12}, although this is questioned \cite{rei10,rei13,rei14}.
Furthermore, dipole strength in the vicinity of the neutron threshold may lead to significant changes of neutron-capture rates in the astrophysical $r$-process \cite{gor04,lit09,dao12}.

A successful description of collective phenomena in nuclei can be achieved by the mean-field approach. 
The respective models can be understood as an approximate realization of a nuclear energy density functional (EDF) \cite{ben03}.
Many of the models favor an explanation of the PDR as an oscillation of a neutron skin - emerging with an increasing $N/Z$ ratio - against an approximately isospin-saturated core. 
This conclusion is based on the analysis of theoretical transition densities which differ significantly from those in the IVGDR region \cite{paa07}.
However, at least for stable nuclei with a moderate neutron excess this question is far from being settled, see, e.g., ref.~\cite{pap14}. 
Quantitative predictions of the centroid energy and strength of the PDR as well as the corresponding collectivity as a function of neutron excess differ considerably.
This is due partly to the properties of the underlying mean-field description (e.g., non-relativistic Skyrme, Gogny or relativistic interactions) but also partly results from the difficulty to separate experimentally strength belonging either to the PDR or the IVGDR.
E1 strength distributions at low excitation energies are also strongly modified when complex configurations beyond the 1 particle - 1 hole (1p1h) level are included in the models (see, e.g.,\ refs.~\cite{rye02,ton10,lit10}). 

Because of their neutron character one expects excitation of the states forming the PDR when using both isoscalar (IS) and isovector (IV) probes.
The IS response has been investigated with $\alpha$  \cite{sav13,pol92,sav06,end10} and heavy-ion \cite{bra15} scattering.
Most data on the IV response stem from $(\gamma,\gamma^\prime)$ experiments (see, e.g., refs.~\cite{sav13,bra19} and references therein) selective towards ground-state (g.s.) transitions. 
Possible branching ratios to excited states, which usually cannot be measured because of the large background, are often neglected, but statistical model calculations of the branching ratios suggest potentially large corrections of the deduced E1 strength \cite{rus08}.
Recent work using self-absorption techniques \cite{rom15} and $(\gamma,\gamma^\prime \gamma^{\prime\prime})$ coincidence experiments \cite{loe16,isa19} suggest that the decay of the PDR is non-statistical but the corrections for decay to excited states are sizable.
Here, the (p,p$^\prime$) experiments provide an important benchmark because the experimental cross sections measure the total absorption independent of the branching ratios.  

The nuclear EOS describing the energy of nuclear matter as function of its density has wide impact on nuclear physics and astrophysics \cite{lat12,heb15} as well as physics beyond the standard model~\cite{wen09,pol99}.  
The EOS of symmetric nuclear matter with equal proton and neutron densities is well constrained from the ground state properties of finite nuclei, especially in the region of saturation density $\rho_{0} \simeq 0.16$ fm$^{-3}$ \cite{dan02}.
However, the description of astrophysical systems as, e.g., neutron stars requires knowledge of the EOS for asymmetric matter \cite{Sto06,lat04,lat14,heb10} which is related to the leading isovector parameters of nuclear matter, viz.\ the symmetry energy (called $J$) and its derivative with respect to density (called $L$ or $S_{\rm V}$) \cite{mol95ra}.  
For recent overviews of experimental and theoretical studies of the symmetry energy see, e.g., refs.~\cite{roc18,epj50}. 
The observation of gravitational waves from merging neutron stars \cite{abb17} provided a wealth of new experimental constraints on the EOS of neutron-rich matter at high density \cite{fat18,mos18}, which calls for an improved description at low densities from nuclear properties.  
However, in spite of steady extension of knowledge on exotic nuclei, these isovector properties are poorly determined by fits to experimental ground-state data because the valley of nuclear stability is still extremely narrow along isotopic chains \cite{klu09,erl13,naz14}. 
Thus, one needs observables in finite nuclei specifically sensitive to isovector properties to better confine $J$ and $L$.
There are two such observables, the neutron skin thickness $r_\mathrm{skin}$ in nuclei with large neutron excess and the (static) dipole polarizability $\alpha_\mathrm{D}$.

The dipole polarizability is related to the second inverse moment of the photoabsorption cross section \cite{boh81}.
Thus, its determination requires knowledge of the complete E1 strength distribution.
Although the low-energy strength (the PDR) exhausts a small fraction of the E1 strength only, it becomes important for $\alpha_{\rm D}$ because of the inverse energy weighting. 
Most experimental information stems from the $(\gamma,\gamma^\prime)$ reaction below neutron threshold and from ($\gamma$,xn) studies \cite{ber75} above neutron threshold.
In the light of the above discussed problems to extract the full E1 strength from NRF data, results from the (p,p$^\prime$) reaction provide an important alternative method where the E1 strength in the resonance region is consistently extracted below and above threshold. 

The isovector spin-flip M1 (IVSM1) resonance is a fundamental excitation mode of nuclei \cite{hey10}.
Its properties impact on diverse fields like the description of neutral-current neutrino interactions in supernovae \cite{lan04,lan08}, $\gamma$ strength functions utilized for physics of reactor design \cite{cha11} or for modeling of reaction cross sections in large-scale nucleosynthesis network calculations \cite{loe12}, and the evolution of single-particle properties leading to new shell closures in neutron-rich nuclei \cite{ots05,ots10}.
It also contributes to the long-standing problem of quenching of the spin-isospin response in nuclei \cite{ost92}, whose understanding is, e.g., a prerequisite for reliable calculations of nuclear matrix elements needed to determine absolute neutrino masses from a positive neutrinoless double-$\beta$ decay experiment \cite{ver12}.
GT and IV spin-M1 resonances involve transitions between spin-orbit partners and the properties of single-particle states near the Fermi surface confine the strength to a certain energy region.
Quenching is then defined as the observation of a reduction of the experimental strength with respect to microscopic model predictions.
A systematic reduction of the GT strengths by a factor of two is observed for medium-mass and heavy nuclei \cite{ich06}.   

The strength distributions of the IVSM1 resonance in light and medium-mass ($fp$-shell) nuclei have been studied extensively using (p,p$^\prime$) scattering and electromagnetic probes like electron scattering and NRF.
Systematic comparison in self-conjugate $sd$-shell nuclei showed clear differences of GT and IVSM1 quenching factors with respect to shell-model calculations attributed to the different role of meson-exchange currents \cite{ric90,lue96,vnc97,hof02}.  
In medium-mass nuclei at the $N =28$ shell closure, a quenching of the IVSM1 resonance by a factor of two comparable to the GT strength in $\beta$ decay \cite{mar96} was found \cite{vnc98}.    
However, information on the IVSM1 in heavy nuclei is limited to a few magic nuclei studied with photon scattering \cite{ton10,las86,las87,las88,ala89,rus13}, and it is questionable whether the full strength has been observed, since the method is limited to the energy region below neutron threshold.

Pioneering forward-angle (p,p$^\prime$) experiments observed bumps identified as IVSM1 resonance in heavy nuclei \cite{dja82,fre90}.       
However, the classical method of extraction of the matrix elements depends on model wave functions of the initial and final states and on the description of the projectile-target interaction, leading to large systematic uncertainties. 
The new (p,p$^\prime$) results from the $0^\circ$ facilities put the extraction of the M1 cross sections on a safer ground.
Furthermore, a new method to extract the transition strengths with the so-called `unit cross section' method in analogy to charge-exchange reactions \cite{tad87,zeg07} was developed \cite{bir16}, which provides improved results with important consequences for the quenching problem.    

The results discussed here also contribute to the problem of a possible quenching of the isoscalar spin-M1 strength, which is theoretically predicted to be of comparable size \cite{bro87,tow87,ari87}.
Because of the dominance of the IV over the IS spin part of the effective proton-nucleus interaction \cite{lov81}, the latter contribution is hard to extract from (p,p$^\prime$) cross sections.
The only clean cases are self-conjugate nuclei with g.s.\ isospin $T = 0$, where the isospin selection rules demand either $T = 0$ or $T = 1$ for the final states corresponding to pure IS and IV transitions, respectively. 
Studies of $N = Z$ nuclei in the $sd$-shell reported conflicting results \cite{ana84,cra89}.
The new experiments allow for a much improved identification of IS M1 transitions and a systematic study of all $N = Z$ nuclei in the $sd$ shell provides new and unexpected insight into the quenching of spin strength.   

The information on the E1 and spin-M1 strength distributions from the (p,p$^\prime$) data also allows to extract the gamma strength function (GSF) \cite{bar72}.
The GSF describes the average $\gamma$ decay behavior of a nucleus.
Its knowledge is required for applications of statistical nuclear theory in astrophysics \cite{arn07}, reactor design \cite{cha11}, and waste transmutation \cite{sal11}.
The impact of the GSF near threshold on astrophysical reaction rates and the resulting abundances in the $r$ process have been discussed, e.g., in refs. \cite{gor04,lit09,dao12}.
Although all electromagnetic multipoles contribute, the GSF is dominated by the E1 radiation with smaller contributions from M1 strength.
Above particle threshold it is governed by the IVGDR but at lower excitation energies the situation is complex.
In nuclei with neutron excess one observes the PDR \cite{sav13} sitting on the low-energy tail of the IVGDR.
Furthermore, the spinflip M1 resonance overlaps with the energy region of the PDR \cite{hey10}

Most applications imply an environment of finite temperature, notably in stellar scenarios \cite{wie12}, and thus reactions on initially excited states become relevant.
Their contributions to the reaction rates are usually estimated applying the generalized Brink-Axel (BA) hypothesis \cite{bri55,axe62}, which states that the GSF is independent of the properties of the initial and final states and thus should be the same in $\gamma$ emission and absorption experiments.
Although historically formulated for the IVGDR, where it seems to hold approximately for not too high temperatures \cite{bbb98}, this is nowadays a commonly used assumption to calculate the low-energy E1 and M1 strength functions at finite temperature. 
Recent theoretical studies \cite{joh15,hun17} put that into question demonstrating that the strength functions of collective modes built on excited states do show an energy dependence.
However, the majority of data used to construct the GSF below particle thresholds stem from NRF experiments with the aforementioned problems to extract the full E1 strength. 
The (p,p$^\prime$) results discussed in this review simultaneously provide information on the E1 and M1 strengths and thus the full GSF can be constructed \cite{bas16}.
Comparison with GSF data extracted from the compound nucleus $\gamma$ decay in the same nuclides with the so-called Oslo method (see, e.g., ref.~\cite{lar17} and references therein) allows novel tests of the BA hypothesis \cite{mar17}.

Finally, the excellent energy resolution of $20 - 40$ keV (FWHM) achieved in these experiments 
unveils the phenomenon of fine structure of the IVGDR.
Indeed, fine structure of giant resonances has been shown to appear globally for electric and magnetic resonances across the nuclear chart \cite{vnc19a}.
There are two types of information which can be extracted from the fine structure.
Firstly, several methods have been developed aiming at a quantification of characteristic features of the fine structure \cite{aib99,aib11,lac99,lac00,she08,hei10}, of which wavelet analysis was shown to be particularly promising \cite{she08}.
By comparison with EDF calculations one is able to single out the relative importance of competing mechanisms \cite{bbb98} contributing to the total width of a giant resonance.
The present high-resolution data on the IVGDR offer the possibility of a systematic comparison with previous studies of the isoscalar giant quadrupole resonance (ISGQR) \cite{she04,she09,usm11}.  
Secondly, the magnitude of the observed cross section fluctuations depends on the level density (LD) and it is possible to extract them with a fluctuation analysis \cite{han90} in cases where a single type of excitation dominates the spectra.
The LDs extracted with this technique are quite unique in several ways: (i) In even-even nuclei they possess a given spin and parity ($J^\pi = 1^-$ in the case of the IVGDR). (ii) They are extracted directly from the data and do not depend on the indirect comparison of measured cross sections with statistical model calculations. (iii) The excitation-energy region is typically well above the neutron threshold, where hardly any LD data exist.
Such data contribute to a variety of open questions like a systematic description of the spin distribution \cite{alh15} or a possible parity dependence \cite{kal07} in certain shell regions.
Combined with the GSF results they also provide an independent test of basic assumptions underlying the Oslo method, where only the product of GSF and LD can be determined and additional assumptions are needed for a decomposition \cite{sch00}.

The paper is organized as follows.
Section \ref{sec2} provides information on the experimental techniques including a description of setups at RCNP (\ref{subsec22}) and iThemba LABS (\ref{subsec23}) and the data extraction.
Representative examples of the experimental observables are discussed in sec.~\ref{subsec26}.
Section \ref{sec3} summarizes the methods used to extract E1 and M1 cross sections and their conversion to strength distributions.
The contributions of the present work to our current understanding of the E1 response in nuclei are presented in sec.~\ref{sec4}.
These include measurements of the dipole polarizability as a measure of the neutron-skin thickness and symmetry-energy parameters (\ref{subsec41}), new information on the PDR (\ref{subsec42}) and new data on the evolution of the IVGDR with deformation (\ref{subsec43}).
New results on the spin-M1 resonance and the quenching of the spin response in nuclei are discussed in sec.~\ref{sec5}.
The fine structure analysis with wavelet techniques is presented in sec.~\ref{subsec61} and the relation of the resulting characteristic scales to different decay mechanisms of the IVGDR in sec.~\ref{subsec62}.
Section \ref{subsec63} describes the level density analysis.
Finally, the GSFs deduced from the data and their relevance for tests of the Brink-Axel hypothesis are discussed in sec.~\ref{sec7}.
A summary and an outlook on future work is given in the concluding sec.~\ref{sec8}.       

%%%%%%%%%%%%%%%%% Modified by AT - beginning %%%%%%%%%%%%%%%%%%%%%%%
\section{Experimental techniques}
\label{sec2}

After a brief history of the technical development on the proton scattering measurement at zero degrees, we focus on the general features of the two representative facilities, 
RCNP and iThemba LABS, and the differences between them.
Further detailed technical information can be found in ref.~\cite{tam09} for the RCNP and in ref.~\cite{nev11} for the iThemba LABS setup.
We also describe the polarization-transfer analysis at RCNP not discussed in ref.~\cite{tam09}.

\subsection{A brief history of $0^\circ$ proton scattering experiments}
\label{subsec21}

Realization of measurement at zero degrees and very forward angles is essential for the study of the $E1$ and spin-$M1$ excitations in nuclei with the (p,p$^\prime$) reaction, since they have the maximum cross section at $0^\circ$ due to the dominance of Coulomb excitation for $E1$ transitions and the $\Delta L=0$ angular momentum transfer of nuclear excitation of spin-$M1$ transitions. 
Technical developments to realize proton inelastic scattering measurement at zero degrees was attempted at Los Alamos National Laboratory~\cite{mcc84,mcc85} and at IUCF, Indiana University~\cite{IUCF,berg93,mer94}. 
Physics results were reported from RCNP for differential cross sections~\cite{sak95}, and for extension to polarization-transfer data~\cite{tam99,kaw02}, followed by the measurement applying dispersion matching for higher energy resolution at IUCF~\cite{fuj07}.
Further developments were achieved at RCNP on dispersion matching, better angular resolution, background subtraction, and polarization-transfer analysis~\cite{tam09}.
A technical development along the same line has been realized at iThemba LABS \cite{nev11}.
Gamma-coincidence measurements at $0^\circ$ represent the most recent progress at both RCNP~\cite{ree19,cagra,scylla} and iThemba LABS~\cite{bagel,alba}.

%--------------------------------------------------%
\subsection{Experimental setup at RCNP}
\label{subsec22}
%--------------------------------------------------%
%
%%
\subsubsection{Accelerators and beam line}
%%
%%%% --- Figure ---%%%%
\begin{figure}[t]
\begin{center}
\resizebox{0.48\textwidth}{!}{\includegraphics{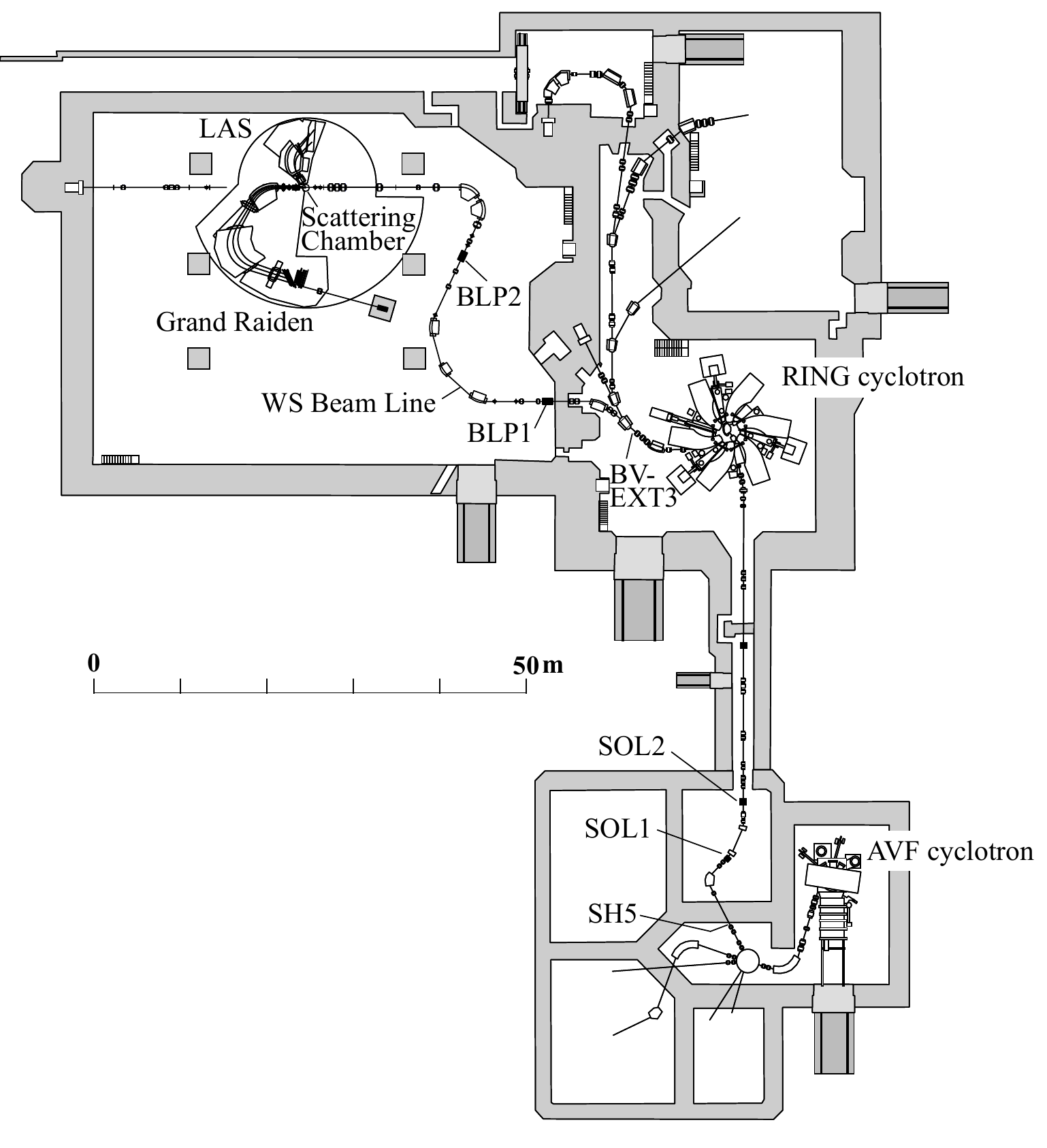}}
\caption{
%Figure221. 
Cyclotron facility of the Research Center for Nuclear Physics (RCNP), Osaka University. 
For details see text.  
Figure taken from ref.~\cite{tam09}.} 
\label{fig221}  
\end{center} 
\end{figure}
%%%%%%%%%%%%%
An overview of the cyclotron facility at RCNP is shown in Fig.~\ref{fig221}.
So far, experimental programs using the zero-degree setup have been performed at proton energies of 295 MeV and 392 MeV. 
A polarized or unpolarized proton beam at 54 (64) MeV from the injector Azimuthally Varying Field (AVF) cyclotron is accelerated to 295 (392) MeV by the RING cyclotron. 
The beam is transported through the West South (WS) beam line~\cite{wak02} to the scattering chamber of the high-resolution spectrometer Grand Raiden~\cite{fuj99}.

Acceleration of a stable, low-emittance and halo-free beam by the cyclotrons is essential for the realization of the zero-degree proton scattering measurement.
Single-turn extraction of the beam from the AVF cyclotron is not fully applicable~\cite{tam09}.
A horizontal slit (SH5) at a dispersive focus point in the beam line after the AVF cyclotron is used to improve the selection of a single turn and to define further the beam energy spread.
Other slits in the beam line between the two cyclotrons are used to cut the remaining beam halos depending on the beam profile.
Single-turn extraction of the RING cyclotron is fully established. 
The transmission efficiency of the beam in the RING cyclotron is $80-90$\%.
No collimator is used after the acceleration by the RING cyclotron to prevent the production of beam halo by slit-edge scattering.
The beam axis is carefully adjusted to the center of the beam line without any beam loss.
The beam energy spread at 295 MeV is typically  60 keV.

The beam spot size at BV-EXT3 -- the object point of the WS beam line -- is minimized to less than 1 mm (typically $0.3 - 0.5$ mm) for achieving the best resolution by applying dispersion matching~\cite{fuj97} between the WS beam line and Grand Raiden. 
A faint beam method~\cite{fuj02} is applied to optimize the matching condition.
The typical resolution after the dispersion matching is $20-30$ keV (FWHM) at 295 MeV.

When a polarized proton beam is accelerated, the polarization axis is fully controlled by the two superconducting solenoids, SOL1 and SOL2 placed in between the two cyclotrons.
It is aligned in vertical direction for injection into the RING cyclotron.
The absolute polarization and the polarization axis after the acceleration by the RING cyclotron are monitored by two beam-line polarimeters, BLP1 and BLP2, located in the WS beam line.
The BLPs measure proton scattering from a thin polyethylene (CH$_2$) foil by detecting two protons in coincidence using four pairs of plastic scintillators.
The BLPs are also used to monitor the beam current.

\subsubsection{Spectrometer}
%%

%%%% --- Figure ---%%%%
\begin{figure}
\begin{center}
\resizebox{0.48\textwidth}{!}{\includegraphics{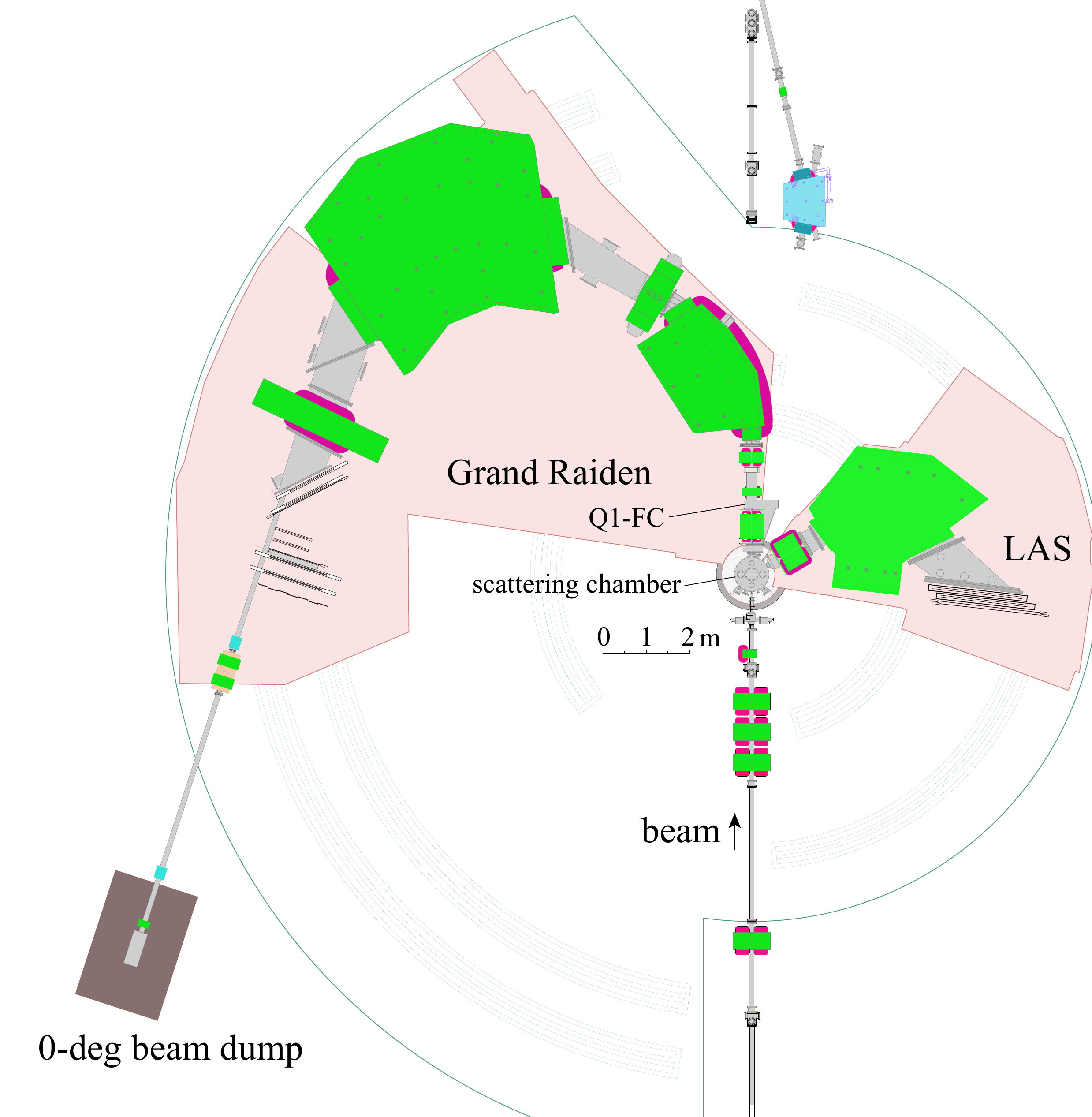}}
\caption{
%Figure222. 
Experimental setup for the proton scattering experiments at RCNP. 
For details see text. 
Figure taken from ref.~\cite{tam09}.} 
\label{fig222}  
\end{center} 
\end{figure}
%%%%%%%%%%%%%

The spectrometer Grand Raiden in the configuration for the zero-degree transmission mode is shown in Fig.~\ref{fig222}.
A proton beam bombards the target foil placed in the scattering chamber. 
Since the spectrometer is placed at $0^\circ$ to cover the inelastically scattered particles in this direction, the primary beam also enters the spectrometer.
Separation of the scattered protons from the unscattered beam is only possible after the bending magnets.
%In Fig.~\ref{fig212}, the spectrometer is in the $D_{SS}$ setup. See sec.~\ref{sec:spin} for the details.

\begin{figure}
\begin{center}
\resizebox{0.4\textwidth}{!}{\includegraphics{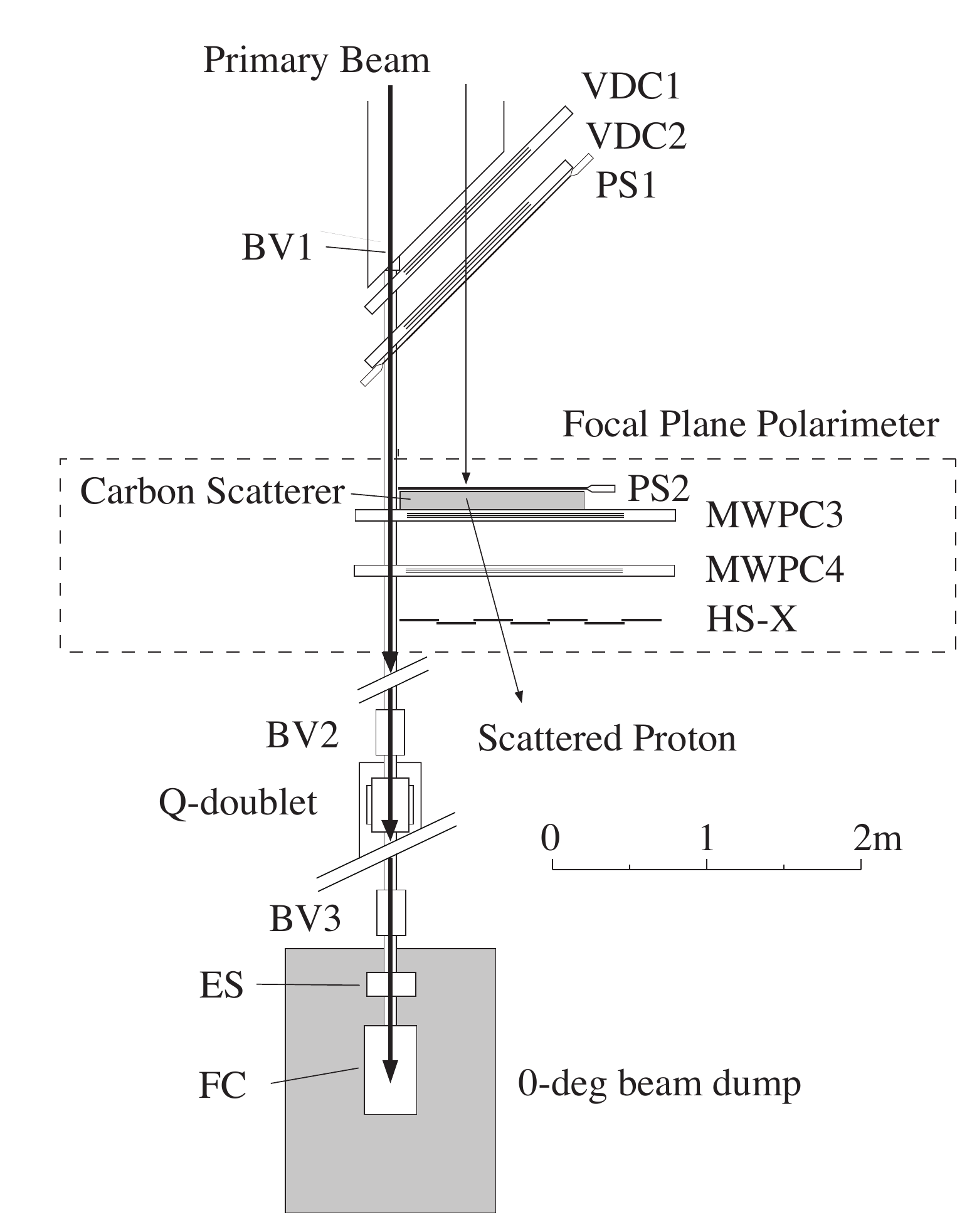}}
\caption{
%Figure223. 
Detectors placed at the focal plane of Grand Raiden spectrometer.
For details see text.
}
\label{fig223}  
\end{center} 
\end{figure}

The scattered protons are measured by the focal-plane detectors and the focal-plane polarimeter (FPP, see Fig.~\ref{fig223}).
The focal plane detectors consist of two multi-wire drift chambers of vertical drift type (VDC1 and VDC2) for tracking of the scattered protons, and a plastic scintillator (PS1) with a thickness of 3 mm.
The wire chambers are horizontally displaced from the standard central position to allow for a beam duct to be placed through the holes of the wire chambers. 
The distance between the beam center and the sensitive area of the wire chambers is fixed to 200 mm.
Accordingly, the excitation energy acceptance of GR at $0^\circ$ is fixed to $5-22$ MeV and $7-30$ MeV at proton beam energies of 295 and 392 MeV, respectively.
The FPP consists of a plastic scintillator (PS2) with a thickness of 3~mm, two multiwire proportional chambers (MWPC3 and 4), and a plastic scintillator hodoscope (HS-X).
The trigger for the data acquisition is produced by a coincidence of the PS1 and PS2 signals.
MWPC3 and 4 also have holes for the beam duct.
In the case of an unpolarized beam, PS1 is replaced by two plastic scintillators with a thickness of 10~mm to measure the energy loss and to produce trigger signals by their coincidence.
%%%% --- Figure ---%%%%
%%%%%%%%%%%%

The beam is focused by a quadrupole doublet after passing through the FPP and is stopped in a Faraday cup (FC) mounted in the beam dump.
A small steering magnet, electron sweeper (ES), is placed before the FC to remove upstream electrons.
The FC is equipped with permanent magnets to avoid the loss of electrons produced by the beam hit.
The FC is placed 12~m downstream of the focal plane and is shielded by 1 m thick concrete on the sides and 40 cm thick iron blocks on the top.
The shielding reduces the radiation from the beam stopper to a level which allows the coincidence measurement of gamma-rays from the target, as discussed in sec.~\ref{sec8}.
The charge collection of the FC is 97\% at maximum. 
The number decreases depending on the target and the beam conditions.
Thus, the total beam charge is more accurately monitored by measuring the event rate of the BLPs, which is proportional to the beam current. 
The proportionality coefficient is calibrated for each experiment by using a beam stopper in the scattering chamber.
The BLP targets are inserted into the beam line periodically, typically 10 out of every 100 seconds. 
During beam transportation, the beam profiles are monitored by using three luminescence viewers, BV1, 2 and 3.
%They are retracted from the beam axis before the proton scattering measurement is started.

Typical beam intensities at $0^\circ$ are $1-2$ nA, the average beam polarization amounts to 70\%, and typical target thicknesses are $1-4$ mg/cm$^{2}$.
The Grand Raiden acceptance is $\pm 1.2^\circ$ and $\pm3^\circ$ for the horizontal (dispersive) and vertical (non-dispersive) directions, respectively.
Measurements at finite angles are performed by placing the spectrometer, e.g., at $2.5^\circ$ or $4.0^\circ$ stopping the beam at the Q1-FC (cf.\ fig.~\ref{fig222}).
A part of the beam duct between the final quadrupole doublet and the zero-degree beam dump needs to be removed for changing the spectrometer angle.
The Q1-FC is horizontally movable for optimizing the position to stop the beam deflected from the central orbit of the spectrometer by the first quadrupole.
For spectrometer angles larger than $6^\circ$, the beam is stopped in a Faraday cup placed in the scattering chamber.
With these settings, angular distributions of the (p,p$^\prime)$ cross sections can be measured continuously from $0^\circ$ to a maximum angle of $70^\circ$.

The large acceptance spectrometer (LAS) also shown in fig.~\ref{fig222} is used to monitor the vertical position of the beam spot by measuring protons from the target produced by quasi-free scattering.
This monitoring is important for the operation of a mild off-focus mode of the spectrometer for the realization of sufficient vertical scattering angle resolution.

%%%% --- Figure ---%%%%
\begin{figure*}
\begin{center}
\resizebox{0.9\textwidth}{!}{\includegraphics{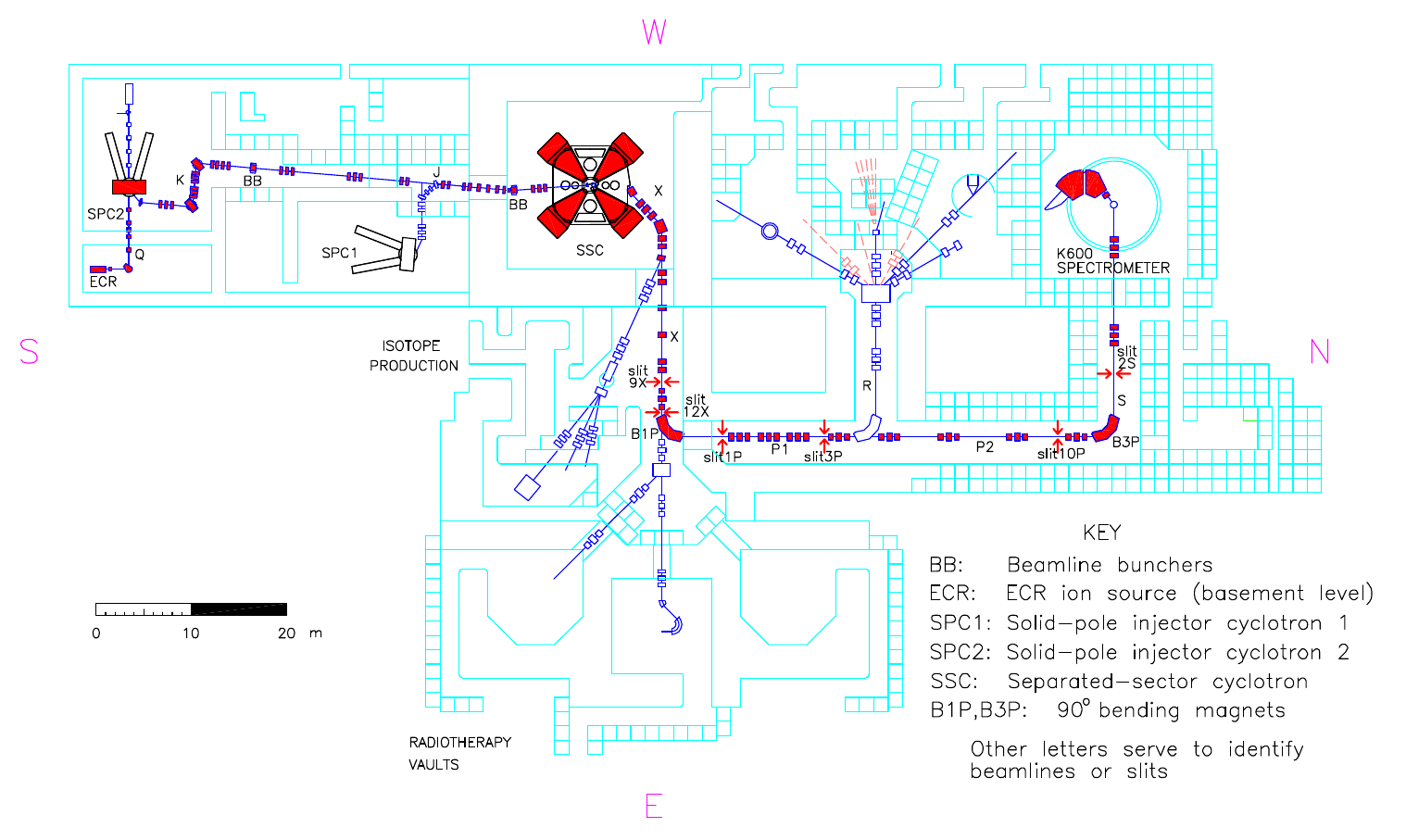}
\vspace{5cm}
}
\caption{
%Figure231. 
Overview of the cyclotron facility at iThemba LABS. 
For details see text.
Figure taken from ref.~\cite{nev11}.
} 
\label{fig231}  
\end{center} 
\end{figure*}
%%%%%%%%%%%%%

%%
\subsubsection{Data-acquisition system}
The data-acquisition system was developed at RCNP~\cite{tam96} and has been gradually updated.
The charge and the timing of the signals from each photomultiplier of the trigger plastic scintillators and HS-X are digitized by using the LeCroy FERA system.
The digitized data are transferred to LeCroy 1190 dual port memory modules in a VME crate.
The signals from the VDCs are pre-amplified and discriminated by REPIC RPA 220 cards.
Their timing information is digitized by CAEN V1190A multi-hit TDCs placed in another VME crate and recorded in their memory buffer.
The hit-pattern of the MWPCs is encoded by LeCroy PCOS-III system and recorded in the LeCroy 1190 memory modules placed in the former VME crate.

The data from the two VME crates as well as the data of the LAS spectrometer are transferred in parallel to the server computer via ethernet and stored on hard disks.
On-line event building and data analysis are performed on the server computer.
The data acquisition is operated in a multi-event buffering mode without dead time of the software.
Typical dead times of the hardware are 30~$\mu$sec for each event.

%--------------------------------------------------%
\subsection{Experimental setup at iThemba LABS}
\label{subsec23}
%--------------------------------------------------%

%%
\subsubsection{Accelerators and beam line}

An overview of the cyclotron facility at iThemba LABS is shown in Fig.~\ref{fig231}.
A proton beam is accelerated by the $K =8$ Solid Pole injector Cyclotron (SPC2)
and by the $K=200$ Separated-Sector Cyclotron (SSC) up to 200 MeV.
In this review we only discuss the case of a proton beam at 200 MeV.
Single-turn extraction from the two cyclotrons is fully realized. 
The transmission efficiency of the SSC is 50-60\%. 
This relatively low transmission efficiency may cause a larger beam emittance and a beam halo.
Thus, the beam is shaped by a collimator system after the acceleration by the SSC.
This is one of the main differences from the operation at RCNP.
The horizontal (vertical) slits 9X define the size of the beam at the object point to 1 (2) mm.
A 1 mm lip is added to the 63 mm thick slits for minimizing the background caused by slit-edge scattering.
The beam divergence is reduced to less than 1.7 mrad by the horizontal and vertical slits 12X.
The energy spread of the beam is limited to $\sim$80 keV by slit 1P.
Further slits, 3P, 10P and 2S, in the beam line are positioned close to the double-focus points and are used to clean up the halo part of the beam.
As a result, 8\% of the beam from the SSC reaches the target.

\subsubsection{Spectrometer}
%%
%%%% --- Figure ---%%%%
\begin{figure}
\begin{center}
\resizebox{0.48\textwidth}{!}{\includegraphics{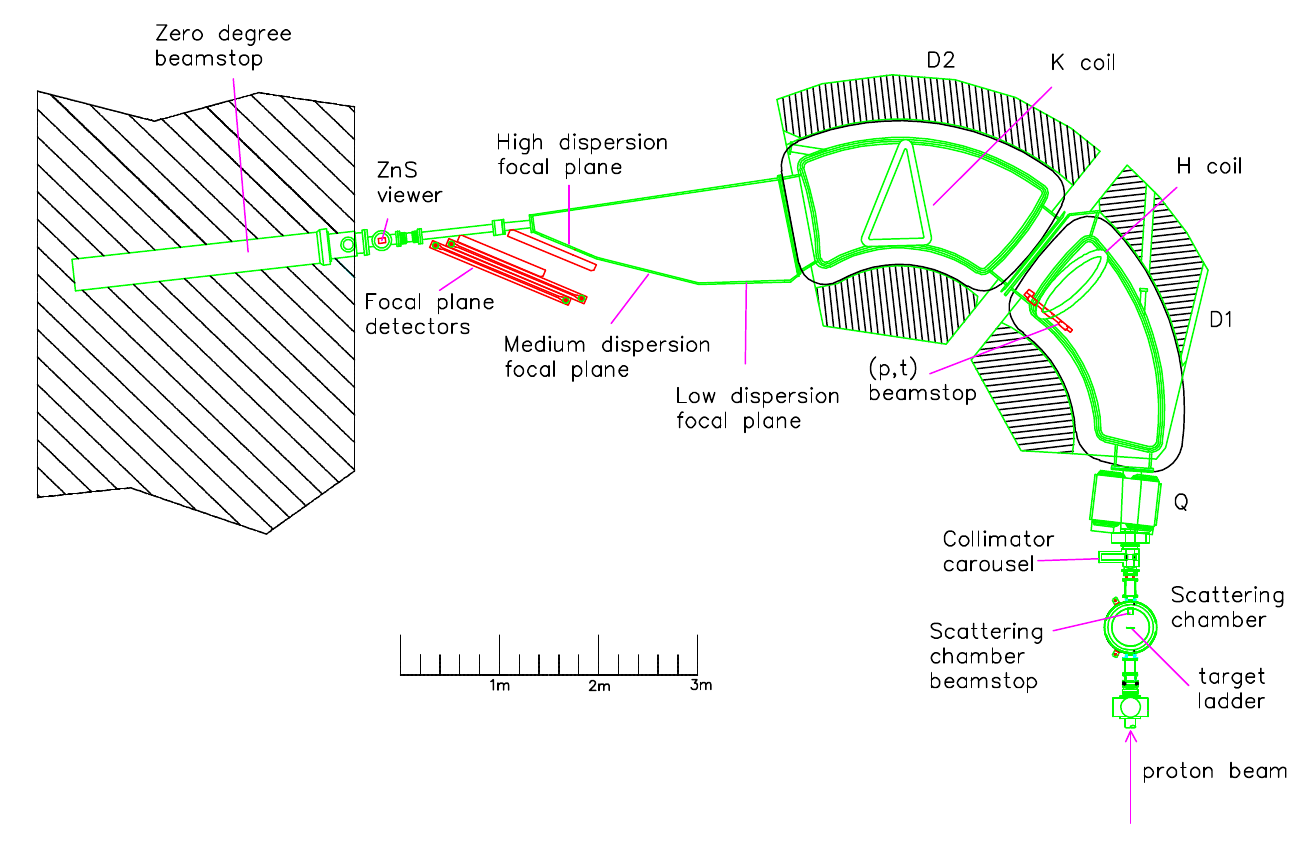}}
\caption{
%Figure232. 
Layout of the zero-degree facility at the K600 spectrometer at iThemba LABS. 
For details see text.
Figure taken from ref.~\cite{nev11}.} 
\label{fig232}  
\end{center} 
\end{figure}
%%%%%%%%%%%%%
The K600 spectrometer in the zero-degree setup is shown in Fig.~\ref{fig232}.
Out of the three available focal planes (high, medium, low dispersion), the high dispersive focal-plane is used for the zero-degree mode ta achieve a good separation of the beam and the scattered protons at the focal plane as well as for a good energy resolution.
A high dispersion mode with a dispersion of 10.9~m was developed for this purpose. 
The beam is stopped in the FC embedded in the concrete wall of the experimental area.
A luminescence viewer in front of the beam stop is used for optimizing the beam transportation through the beam line.

Two VDCs are placed at the focal plane for tracking the scattered protons.
The distance from the beam to the sensing region of the VDCs is 10~cm.
The covered excitation energy range for a 200 MeV proton beam is $8.75-24.5$ MeV.
The position resolution of the VDCs in the dispersive direction is $\sim$0.35 mm (FWHM).
Two plastic scintillators with thicknesses of 12.7 and 6.35 mm, respectively, are placed behind the VDCs for generating trigger signals and measuring the energy loss for particle identification.
The time-of-flight information is effective to remove the background particles originating from the beam-halo produced upstream of the target.

The shape of the collimator at the entrance of the spectrometer is optimized to achieve the lowest instrumental background condition.
In contrast to the situation at RCNP, the background increases when no collimator is installed.
The collimator has 49~mm diameter with an 11 mm thick brass lip, tapered to the angular acceptance.
The corresponding vertical and horizontal angular acceptance is $\pm 1.91^\circ$ compared to the full acceptance of $\pm 2.51^\circ$.

Typical target thicknesses are 1-2 mg/cm$^2$.
The low-momentum side of the target frame is removed for minimizing background from particle scattering from the frame.
The dispersion-matching condition between the beam line and the spectrometer is optimized by applying the faint beam method \cite{fuj02}.

\subsubsection{Data-acquisition system}
The signals from the photomultiplier tubes attached at both ends of the two plastic scintillators are digitized by a 12 bit current integrating QDC, CAEN model V792NC.
The signals from the sense wires of the VDCs are pre-amplified and discriminated by 16-channel electronic cards, Technoland model P-TM 005, and digitized by multi-hit TDCs, CAEN model V1190A.
The digitized data are recorded by the MIDAS data-acquisition system~\cite{MIDAS} developed originally at PSI and TRIUMF.

%--------------------------------------------------%
\subsection{Background subtraction}
\label{subsec24}
%--------------------------------------------------%
%%%% --- Figure ---%%%%
\begin{figure}[b]
\begin{center}
\resizebox{0.4\textwidth}{!}{\includegraphics{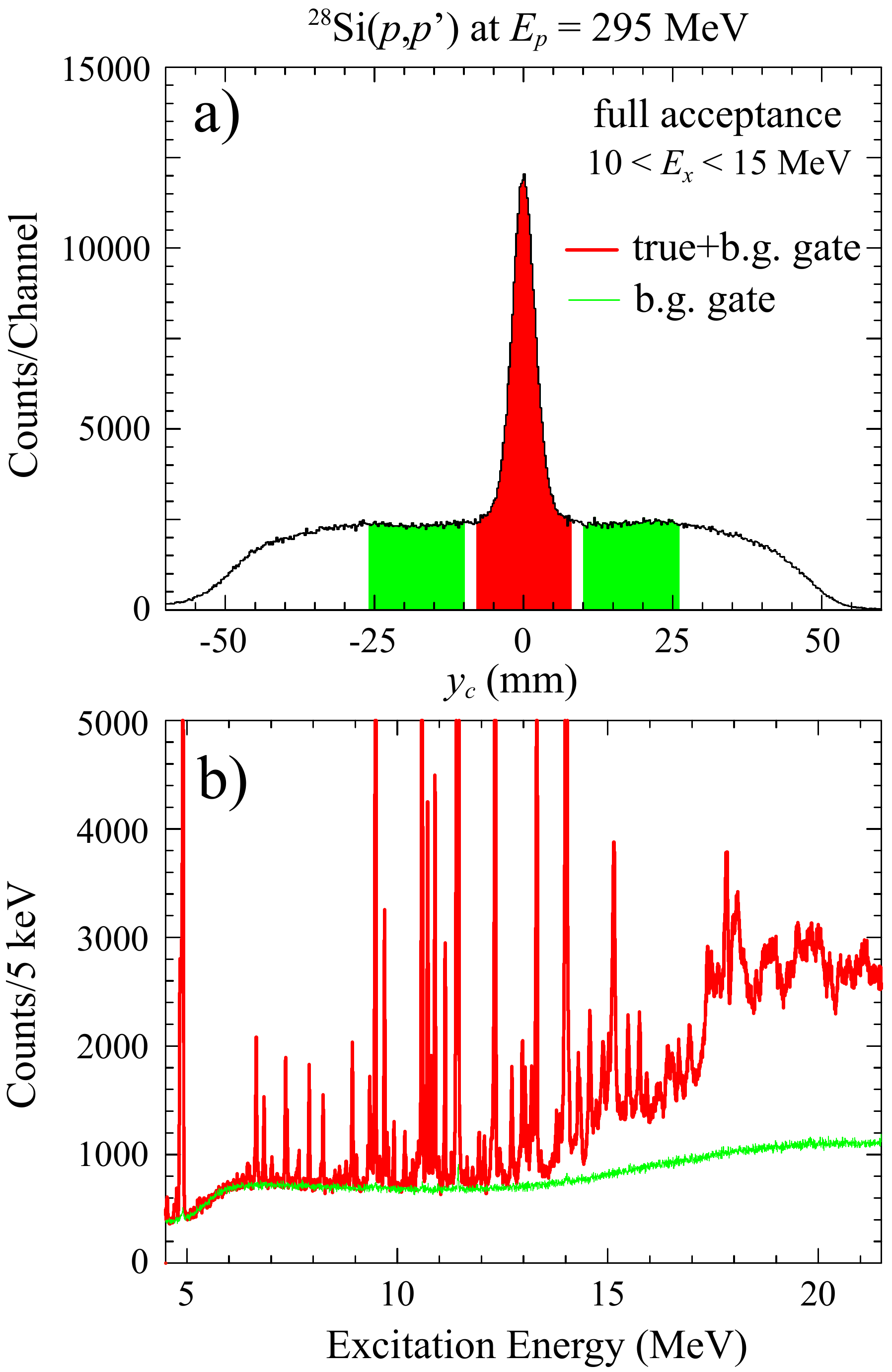}}
\caption{
%Figure241.
Background subtraction procedure for the Grand Raiden spectrometer at RCNP illustrated for the $^{28}$Si(p,p$^\prime$) reaction at 295 MeV.
Upper part:  Vertical-position spectrum at the focal plane.
Lower part: Excitation-energy spectrum corresponding to the central area of the vertical position spectrum (red) and to the average of the side parts (green).
The solid angle of the spectrometer gradually decreases below $E_{\rm x} =6.5$ MeV. 
Figure taken from ref.~\cite{tam09}.} 
\label{fig241}  
\end{center} 
\end{figure}
%%%%%%%%%%%%%
Instrumental background protons are unavoidable even with the best beam from the cyclotrons. 
The background protons originate from multiple Coulomb scattering in the target and from secondary scattering in the aperture of the spectrometer.
Thus, the amount of the background increases as the target atomic number and the target thickness increase.
One of the key points of the successful zero-degree measurement was the development of the background subtraction technique~\cite{tam99}.
A basic assumption in this method is that the physical events focus at the vertical focus point after the spectrometer while the background particles are spread nearly uniformly in the vertical direction.
%The validity of the assumption has been well confirmed in many experiments.

%%%% --- Figure ---%%%%
\begin{figure}
\begin{center}
\resizebox{0.48\textwidth}{!}{\includegraphics{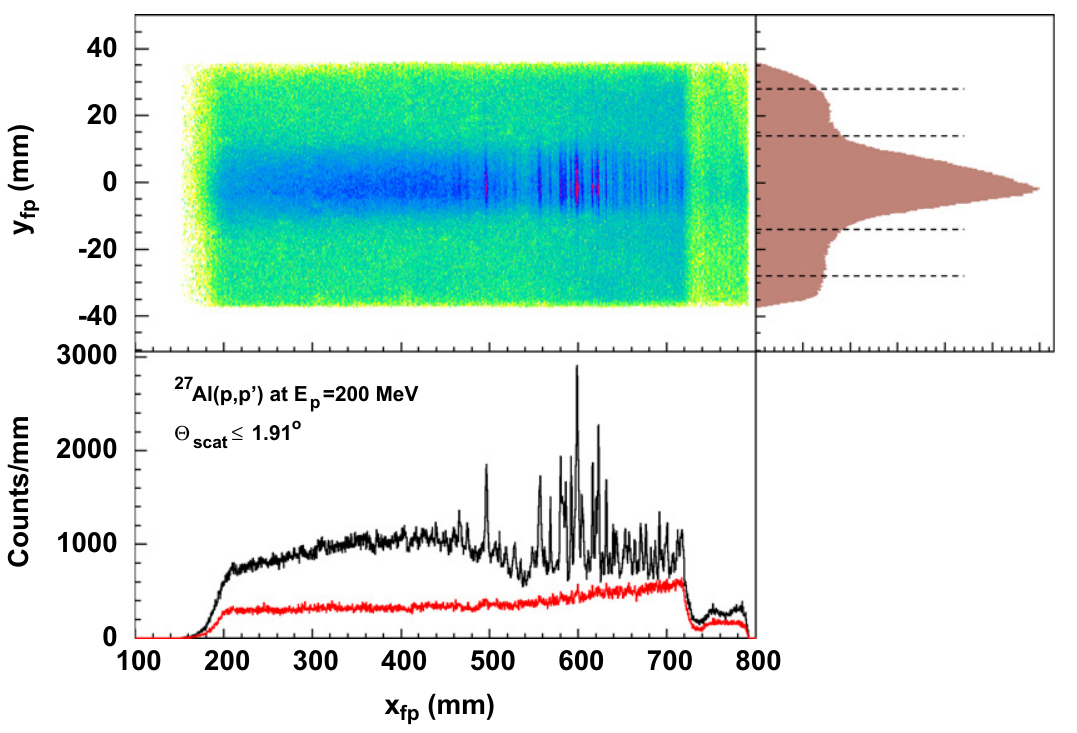}}
\caption{
%Figure242.
Background subtraction procedure for the K600 spectrometer at iThemba LABS illustrated for the $^{27}$Al(p,p$^\prime$) reaction at 200 MeV. 
Upper part: horizontal ($x_{\rm fp}$) vs.\ vertical ($y_{\rm fp}$) focal-plane position and projection on the $y$ axis (r.h.s.).
Lower part: Excitation-energy spectrum of the central gate in the $y_{\rm fp}$ distribution (black) and of the sum of the left and right gates (red).     
Figure taken from ref.~\cite{nev11}.} 
\label{fig242}  
\end{center} 
\end{figure}
%%%%%%%%%%%%%

The basic background subtraction procedure for the Grand Raiden spectrometer at RCNP is shown in Fig.~\ref{fig241} and for the K600 spectrometer at iThemba LABS in Fig.\ref{fig242}.
In both cases, the central part of the position at the vertical focal plane is selected for the physical events while the side parts are used for the determination of the contribution from the flat background contained in the central region.

\begin{figure}
\begin{center}
\resizebox{0.4\textwidth}{!}{\includegraphics{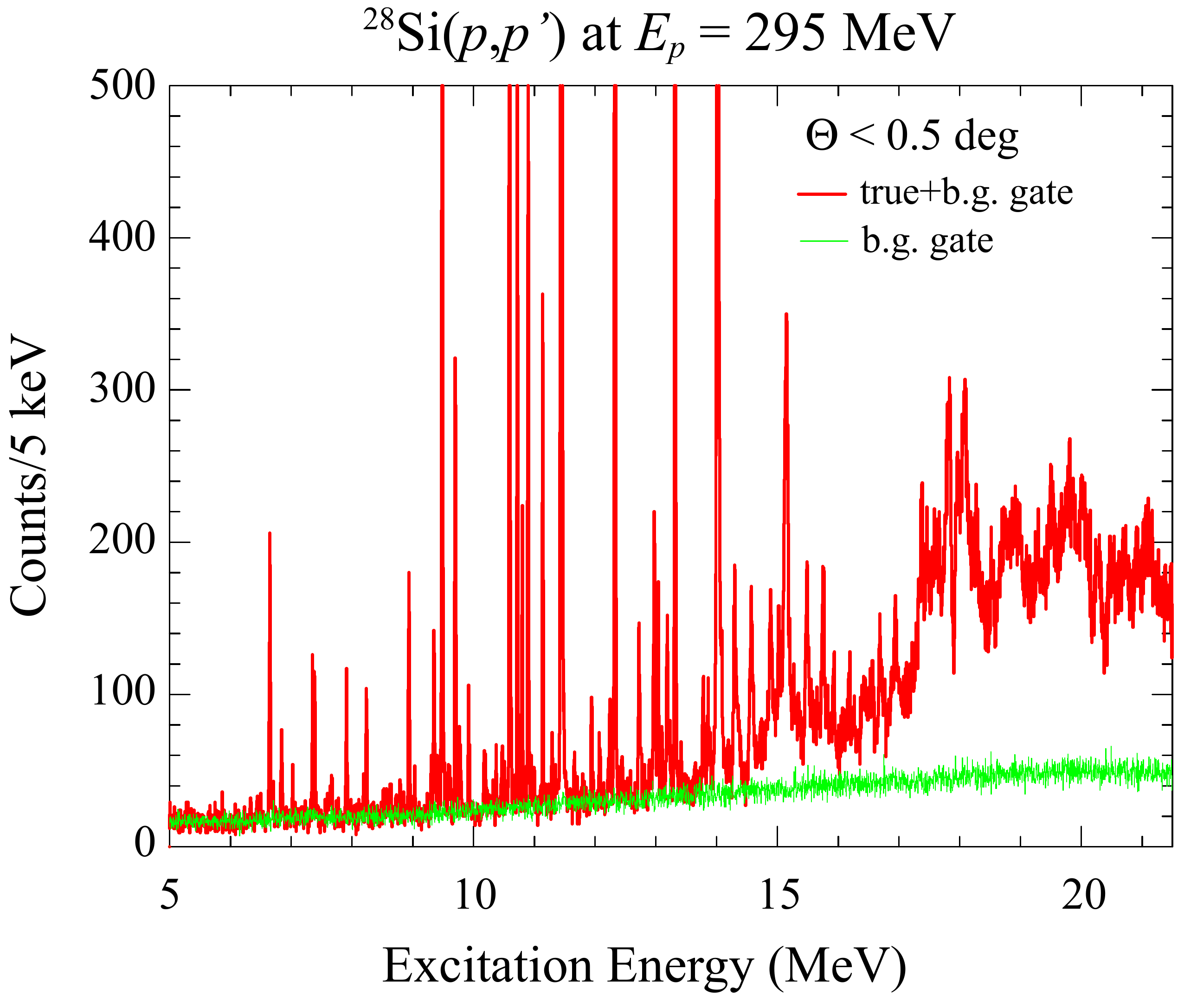}}
\caption{
%Figure243.
Application of the extended subtraction method as well as a cut to scattering angles $\leq 0.5^\circ$ to the data of fig.~\ref{fig241}.
Figure taken from ref.~\cite{tam09}.} 
\label{fig243}  
\end{center} 
\end{figure}

However, the method is only applicable for the analysis of the full solid-angle data since the vertical scattering angle is correlated with the vertical trajectory at the focal plane.
To enable a simultaneous background subtraction and vertical scattering-angle determination, a {\em mild under-focus mode} of the spectrometer Grand Raiden and an {\em extended subtraction method} were developed~\cite{tam09}.
The essence of the extended background subtraction is to shift the vertical position data of each event instead of changing the gate on the vertical position.
Figure~\ref{fig243} shows an application of the extended subtraction method as well as a cut to scattering angles $\leq 0.5^\circ$ to the data of fig.~\ref{fig241}.

%%%% --- Figure ---%%%%
%%%%%%%%%%%%%

%--------------------------------------------------%
%\subsubsection{Scattering angle calibration}
%--------------------------------------------------%

%--------------------------------------------------%
%\subsubsection{Differential cross section analysis}
%--------------------------------------------------%

%--------------------------------------------------%
\subsection{Polarization-transfer measurement at RCNP}
\label{subsec25}
%--------------------------------------------------%

%--------------------------------------------------%
\subsubsection{Polarization-transfer coefficients of zero-degree proton scattering}
\begin{figure}
\begin{center}
\resizebox{0.39\textwidth}{!}{\includegraphics{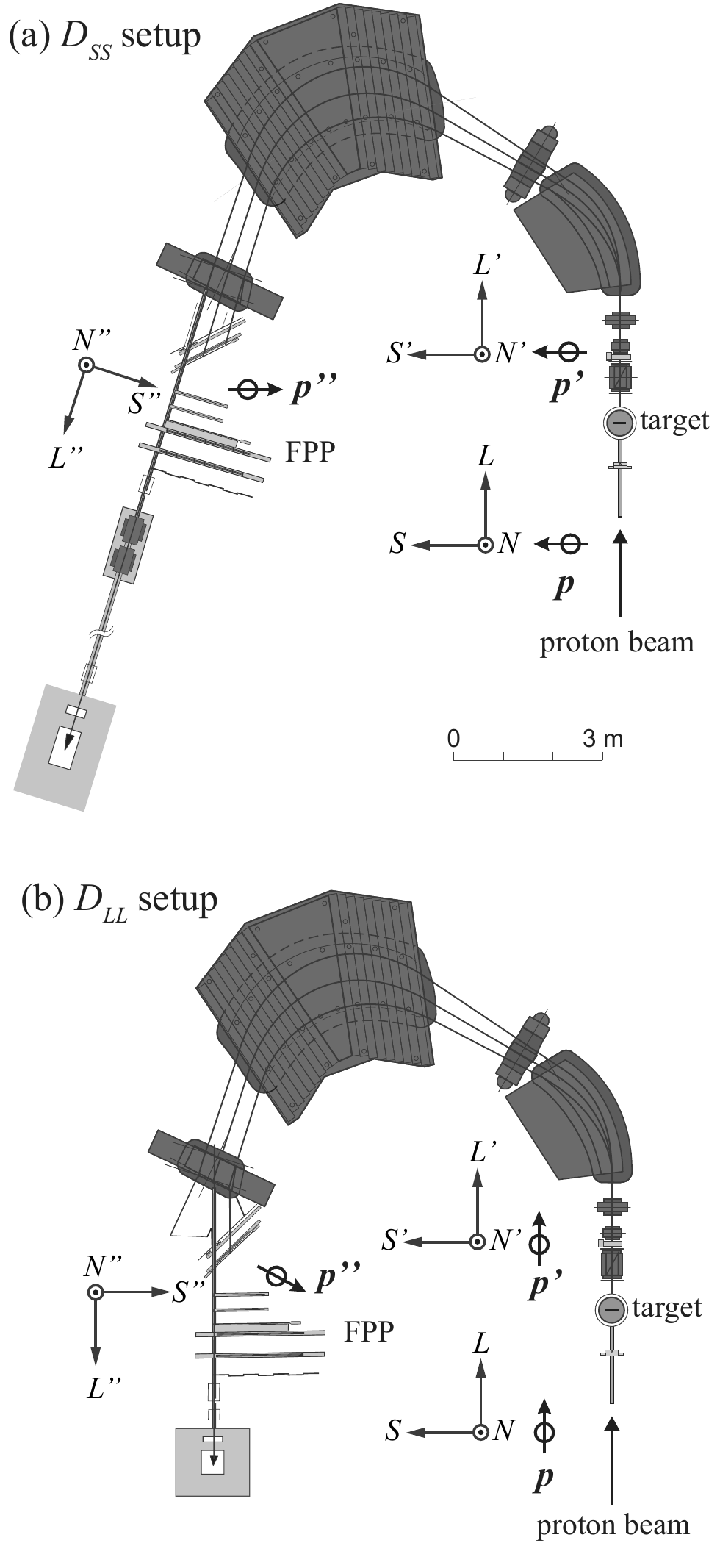}}
\caption{
%Figure251. 
Definition of the coordinate axes of the proton polarization vector before ($S$,$N$,$L$) and after ($S'$,$N'$,$L'$) the scattering process and at the detector position ($S''$,$N''$,$L''$) for (a) the $D_{SS}$ measurement setup with a bending angle of $\chi_s=162^\circ$ and (b) the $D_{LL}$ measurement setup with $\chi_s=180^\circ$. 
The beam polarization axis, {\boldmath$p$}, is adjusted to be close to the $S$($L$) direction for the $D_{SS}$($D_{LL}$) measurements. 
The $S'$($L'$) component of the polarization vector, {\boldmath$p'$}, is oriented in the direction shown by the arrow {\boldmath$p''$} at the focal plane polarimeter (FPP) after the precession in the spectrometer according to eq.~(\ref{eq2504}).
}
\label{fig251}  
\end{center} 
\end{figure}

%--------------------------------------------------%
Cartesian coordinate systems ($S$ $N$,$L$) and ($S'$,$N'$,$L'$) are defined for the incoming and outgoing protons, respectively, in the laboratory system as shown in Fig.~\ref{fig251}. 
Here, $S$($S'$), $N$($N'$), and $L$($L'$) correspond to the axes of the sideways, normal and longitudinal components of the proton polarization following the Madison convention~\cite{ohl72}.
Polarization-transfer coefficients, $D_{ji}(\theta)$, are described by the proton polarization vectors under the condition of the parity invariance as
\begin{eqnarray}
\left(\begin{array}{c}
   p'_{S'} \\
   p'_{N'} \\
   p'_{L'} \\
  \end{array}\right)  
  &=&
  \frac{1}{1+p_N A_N(\theta)}\nonumber\\
&\times&
\left\{
  \left(\begin{array}{ccc}
    D_{S'S}(\theta) & 0 &  D_{S'L}(\theta) \\
    0 & D_{N'N}(\theta) &  0 \\
    D_{L'S}(\theta) & 0 &  D_{L'L}(\theta) \\
  \end{array}\right)  
  \left(\begin{array}{c}
    p_{S} \\
    p_{N} \\
    p_{L} \\    
  \end{array}\right)  
  \right.\nonumber\\
&+&
\left.
  \left(\begin{array}{c}
    0 \\
    P_N(\theta) \\
    0 \\
  \end{array}\right)
\right\},
\label{eq2501}
\end{eqnarray}
where $p_{S,N,L}$ and $p'_{S',N',L'}$ are the components of polarization vectors of the incoming and outgoing protons, respectively, $A_N(\theta)$ is the vector analyzing power and $P_N(\theta)$ denotes the induced polarization of the scattered protons.
Since $A_N(\theta)$, $P_N(\theta)$, $D_{S'L}(\theta)$ and $D_{L'S}(\theta)$ are odd functions of the scattering angle  $\theta$, they vanish at a $0^\circ$ scattering angle. 
Thus, eq.~(\ref{eq2501}) reduces to 
\begin{equation}
\left(\begin{array}{c}
   p'_{S'} \\
   p'_{N'} \\
   p'_{L'} \\
  \end{array}\right)  
  =
  \left(\begin{array}{ccc}
    D_{SS}(0^\circ) & 0 &  0 \\
    0 & D_{NN}(0^\circ) &  0 \\
    0 & 0 &  D_{LL}(0^\circ) \\
  \end{array}\right)  
  \left(\begin{array}{c}
    p_{S} \\
    p_{N} \\
    p_{L} \\
  \end{array}\right).
\label{eq2502}
\end{equation}
Note that the coordinate systems of the incoming and outgoing protons coincide with each other at $0^\circ$.
Thus, one can write $D_{S'S}(0^\circ)$ as $D_{SS}(0^\circ)$ etc.
In addition, the sideways and normal polarization-transfer coefficients are equal to each other due to rotational symmetry with respect to the beam axis
\begin{equation}
    D_{SS}(0^\circ) = D_{NN}(0^\circ).
\label{eq2503}
\end{equation}
%%%% --- Figure ---%%%%
%%%%%%%%%%%%%

%--------------------------------------------------%
\subsubsection{Spin-precession in the spectrometer}
%--------------------------------------------------%
\label{sec:spin}
The polarization vector after scattering, $p'$, precesses in the magnetic field of the Grand Raiden spectrometer.
The coordinate system of the polarization vector, $p''$,  at the detector position of Grand Raiden is defined with respect to the direction of the central orbit of the spectrometer as shown in Fig.~\ref{fig251}.
The direction of the polarization vector relative to the momentum precesses by an angle $\chi_p$ in the spectrometer
\begin{equation}
\chi_p = \gamma(\frac{g_p}{2}-1)\chi_s,
\label{eq2504}
\end{equation}
where $\gamma$ is the Lorentz factor,  $g_p=5.586$ the $g$-factor of the proton magnetic moment and $\chi_s$ the bending angle of the proton in the spectrometer.
The bending angles of the central orbit of the Grand Raiden spectrometer are $\chi_s$=162$^\circ$ ($\chi_s$=180$^\circ$ ) for the $D_{SS}$($D_{LL}$) setups, see figs.~\ref{fig251}(a) and (b).

Each component of the polarization vector, $p''$, at the detector position is described by the components of the polarization vector, $p'$, at the entrance of the spectrometer by
%\begin{eqnarray}
%p''_{S''} &=& p'_{S'} \cos\chi_p + p'_{L'} \cos\chi_p \nonumber\\
%p''_{N''} &=& p'_{N'} \nonumber\\
%p''_{L''} &=& p'_{L'} \cos\chi_p - p'_{S'} \cos\chi_p 
%\end{eqnarray}
\begin{equation}
\left(\begin{array}{c}
   p''_{S''} \\
   p''_{N''} \\
   p''_{L''} \\
  \end{array}\right)  
  =
  \left(\begin{array}{ccc}
    \cos\chi_p & 0 &  \sin\chi_p \\
    0 & 1 &  0 \\
    -\sin\chi_p & 0 &  \cos\chi_p  \\
  \end{array}\right)  
  \left(\begin{array}{c}
    p'_{S'} \\
    p'_{N'} \\
    p'_{L'} \\
  \end{array}\right).
\label{eq2505}
\end{equation}

The FPP is sensitive to $p''_{S''}$ and $p''_{N''}$ but not  to $p''_{L''}$. 
Combining  eqs.~(\ref{eq2501}) and (\ref{eq2505}), one can express $p''_{S''}$ and $p''_{N''}$ as
\begin{eqnarray}
p''_{S''} &=& \{D_{S'S}(\theta)p_S+D_{S'L}(\theta)p_L\}\cos\chi_p\nonumber\\
&+&\{D_{L'S}(\theta)p_S+D_{L'L}(\theta)p_L\}\sin\chi_p
\label{eq2506}
\end{eqnarray}
and
\begin{equation}
p''_{N''} = D_{NN}(\theta)p_N.
\label{eq2507}
\end{equation}
The angular acceptance of Grand Raiden is only $|\theta|<0.05$ rad in the zero-degree setup.
The terms containing $D_{S'L}(\theta)$ and $D_{L'S}(\theta)$ vanish in the first order due to their property of being an odd function of $\theta$ and due to the symmetry of the Grand Raiden angular acceptance.
$D_{S'S}(\theta)$, $D_{N'N}(\theta)$ and $D_{L'L}(\theta)$ are equal to their values at $\theta=0^\circ$ in first order due to their property of being even functions of $\theta$.
Then eqs.~(\ref{eq2506}) and (\ref{eq2507}) can be simplified as
\begin{equation}
p''_{S''} = D_{SS}(0^\circ)p_S\cos\chi_p+D_{LL}(0^\circ)p_L\sin\chi_p
\label{eq2508}
\end{equation}
and
\begin{equation}
p''_{N''} = D_{NN}(0^\circ)p_N,
\label{eq2509}
\end{equation}
 respectively.

As an example, in the case of proton scattering measurements at $E_0=295$ MeV, the central orbit corresponds to an excitation energy $E_x\simeq$7 MeV.

For the corresponding proton energy of 288 MeV,
$\chi_p$=379.6$^\circ$, $\cos\chi_p=0.942$, $\sin\chi_p=0.336$ for the
$D_{SS}$ setup and $\chi_p$=421.7$^\circ$, $\cos\chi_p=0.473$,
$\sin\chi_p=0.881$ for the $D_{LL}$ setup.
The bending angle in the spectrometer is determined for each detected proton. 
The angle depends on the horizontal position at the detector with respect to the proton momentum and on the incidence angle at the detector with respect to the horizontal scattering angle at the target.

%--------------------------------------------------%
\subsubsection{Principle of the polarization analysis}
\begin{figure}
\begin{center}
\resizebox{0.35\textwidth}{!}{\includegraphics{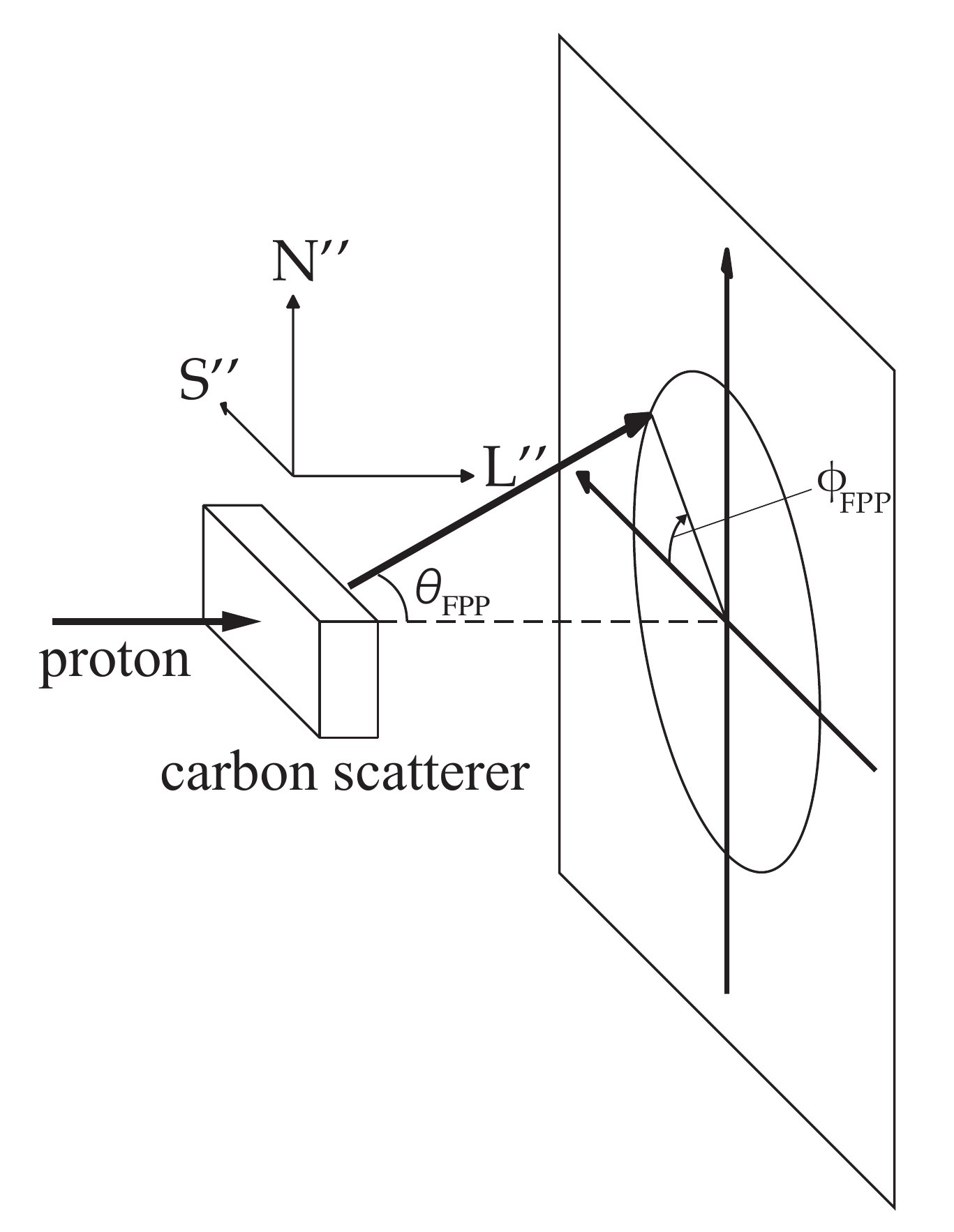}}
\caption{
%Figure252. 
Polar coordinate system ($\theta_{\rm FPP}$, $\phi_{\rm FPP}$) for protons scattered by the carbon scatterer at the FPP and coordinate axes $S''$,$N''$,$L''$ of the proton polarization vector.
}
\label{fig252}  
\end{center} 
\end{figure}

%--------------------------------------------------%
The polarization vector of scattered protons is measured by the FPP for the components in the plane perpendicular to the proton momentum, i.e.\ $p''_{S''}$ and $p''_{N''}$.
The scattering direction of each event is defined by the relative angle between the proton momenta before and after the carbon scatterer using a polar coordinate system ($\theta_{\rm FPP}$, $\phi_{\rm FPP}$) as shown in fig.~\ref{fig252}.
%%%% --- Figure ---%%%%
%%%%%%%%%%%%%

The intensity of protons scattered at angles ($\theta_{\rm FPP}$, $\phi_{\rm FPP}$) is described by
\begin{eqnarray}
&&I(\theta_{\rm FPP}, \phi_{\rm FPP})\nonumber\\
&&=I_0\sigma^{p\rm C}(\theta_{\rm FPP})\{1+p''_{\perp}A_y^{p\rm C}(\theta_{\rm FPP})\cos( \phi_{\rm FPP}-\phi''+90^\circ)\},\nonumber\\
&&
\label{eq2510}
\end{eqnarray}
where $\sigma^{p\rm C}(\theta_{\rm FPP})$ and $A_y^{p\rm C}(\theta_{\rm FPP})$ are the $p$+C inclusive scattering cross section and analyzing power, respectively.
They depend on the proton energy and the thickness of the carbon scatterer. 
$I_0$ is a normalization constant. 
$p''_{\perp}$ is the amplitude of the polarization vector in the plane perpendicular
to the proton momentum before scattering and $\phi''$ is its azimuthal angle from the $S''$ axis with respect to the $L''$  axis.
Thus, the sideways and normal components
of the proton polarization can be described as
\begin{equation}
p''_{S''} = p''_\perp\cos\phi''
\label{eq2511}
\end{equation}
and
\begin{equation}
p''_{N''} = p''_\perp\sin\phi''.
\label{eq2512}
\end{equation}

The angular range of $\theta_{\rm FPP}$ is chosen to maximize the figure of merit of measuring the polarization of the protons.
The scattered events with small angles have negligible analyzing power due to multiple Coulomb scattering in the carbon slab.
Thus, a typical minimum scattering angle is $\theta^{\rm min}_{\rm FPP}=6^\circ$. 
The maximum is determined by the angular acceptance of the FPP and is typically $\theta^{\rm max}_{\rm FPP}=18^\circ$.
After integration of the events from $\theta^{\rm min}_{\rm FPP}$ to $\theta^{\rm max}_{\rm FPP}$, the polarization of the protons can be expressed by the following equations
\begin{eqnarray}
I(\phi_{\rm FPP}) &\equiv& \int_{\theta^{\rm min}_{\rm FPP}}^{\theta^{\rm max}_{\rm FPP}}  I(\theta_{\rm FPP},\phi_{\rm FPP}) \sin\theta_{\rm FPP} d\theta_{\rm FPP}\nonumber\\
&=& I'_0\{1+p''_{\perp} \langle A_y^{p\rm C} \rangle \sin( \phi_{\rm FPP}-\phi'')\}
\label{eq2513}
\end{eqnarray}
\begin{equation}
I'_0 = I_0  \int_{\theta^{\rm min}_{\rm FPP}}^{\theta^{\rm max}_{\rm FPP}}  \sigma^{p\rm C}(\theta_{\rm FPP}) \sin\theta_{\rm FPP}d\theta_{\rm FPP}
\label{eq2514}
\end{equation}
\begin{equation}
 \langle A_y^{p\rm C} \rangle  = \frac
{\int_{\theta^{\rm min}_{\rm FPP}}^{\theta^{\rm max}_{\rm FPP}}  \sigma^{p\rm C}(\theta_{\rm FPP}) A_y^{p\rm C}(\theta_{\rm FPP}) \sin\theta_{\rm FPP}d\theta_{\rm FPP}}
{\int_{\theta^{\rm min}_{\rm FPP}}^{\theta^{\rm max}_{\rm FPP}}  \sigma^{p\rm C}(\theta_{\rm FPP}) \sin\theta_{\rm FPP}d\theta_{\rm FPP}}
\label{eq2515}
\end{equation}
Thus, from the $\phi_{\rm FPP}$ distribution of the scattered events given by eq.~(\ref{eq2513}), $p''_{S''}$ and $p''_{N''}$ in eqs.~(\ref{eq2511}) and (\ref{eq2512}), respectively,  can be determined using the knowledge of the p+C effective analyzing power $ \langle A_y^{\rm pC} \rangle$. 
The values of $\sigma^{p\rm C}(\theta_{\rm FPP})$ and $A_y^{\rm pC}(\theta_{\rm FPP})$ are tabulated as a function of proton energy $E_0$ at the carbon center, $\theta_{\rm FPP}$ and thickness of the carbon slab \cite{apr83,mcn85}. 
After integration, $\langle A_y^{\rm pC} \rangle$ ranges from 0.5 at $E_0=200$ MeV to 0.3 at 400 MeV with a typical uncertainty of 0.02.
Note that the absolute value of $\langle A_y^{p\rm C} \rangle$  cancels out in the analysis of polarization transfer coefficients at zero degrees as described in the next subsections, and thus does not affect the final results.

%--------------------------------------------------%
\subsubsection{Estimator method}
%--------------------------------------------------%
In the early stage of the measurements \cite{tam09,tam99,kaw02}, the {\em sector method} was applied for the determination of the polarization of the scattered protons. 
In this method, the number of scattered events in leftward (L), rightward (R), downward (D) and upward (U) regions of $\phi_{\rm FPP}$ are counted to extract $p''_{S''}$ from the asymmetry $(L-R)/(L+R)$ and $p''_{N''}$ from $(D-U)/(D+U)$~\cite{tam09}.
Later, starting with the analysis of the $^{208}{\rm Pb}$ data \cite{tam11}, the {\em estimator method} \cite{bes79} has been applied.
Furthermore, a way to remove the dependence on the effective analyzing power of the FPP has been developed as described below.

The unbiased efficiency estimator of the polarization
\begin{equation}
\hat{\epsilon} = 
\left(
\begin{array}{c}
\hat{\epsilon}_N\\
\hat{\epsilon}_S\\
\end{array}
\right)
\label{eq2516}
\end{equation}
is defined as~\cite{bes79}
\begin{equation}
\hat{\epsilon} \equiv F^{-1}B,
\label{eq2517}
\end{equation}
\begin{equation}
B = 
\left(
\begin{array}{c}
\sum \cos\phi_{\rm FPP}\\
\sum \sin\phi_{\rm FPP}\\
\end{array}
\right),
\label{eq2518}
\end{equation}
\begin{equation}
F = 
\left(
\begin{array}{cc}
\sum \cos^2\phi_{\rm FPP} & \sum \sin\phi_{\rm FPP}\cos\phi_{\rm FPP}\\
\sum \sin\phi_{\rm FPP}\cos\phi_{\rm FPP} & \sum \sin^2\phi_{\rm FPP}\\
\end{array}
\right),
\label{eq2519}
\end{equation}
where the sums are taken for events in the proper $\theta_{\rm FPP}$ angular range.
The $N''$ and $S''$ components of the polarization vector are given by \cite{bes79}
\begin{equation}
\left(
\begin{array}{cc}
p''_{N''}\\
p''_{S''}\\
\end{array}
\right)
\simeq
\frac{1}{\langle A_y^{p\rm C} \rangle}
\left(
\begin{array}{cc}
\hat{\epsilon}_{N''}\\
-\hat{\epsilon}_{S''}\\
\end{array}
\right)
\label{eq2520}
\end{equation}
The negative sign in the lower component originates from the definition of eq.~(2) in ref.~\cite{bes79}.
The statistical uncertainties are determined from the covariance matrix
\begin{equation}
V(\hat{\epsilon}) = F^{-1}.
\label{eq2521}
\end{equation}
For the data taken with the beam polarization direction flipped (Beam-Spin\#2), the angle $\phi_{\rm FPP}$ in the above equations is replaced by $\phi_{\rm FPP}+180^\circ$ and the sums are taken together with the non-flipped data (Beam-Spin\#1).

%--------------------------------------------------%
\subsubsection{Removal of the instrumental-background contribution}
%--------------------------------------------------%

The contribution from the instrumental background is subtracted for the determination of the proton polarization with the following method.
At first the counts of the true signal ($N_t$) due to scattering from the target and the counts of the instrumental background ($N_b$) are determined with the same method as described for the differential cross sections in sec.~\ref{subsec24}. 
Here, we consider only events that satisfy the condition $\theta_{\rm FPP}^{\rm min}\leq\theta_{\rm FPP}\leq\theta_{\rm FPP}^{\rm max}$.

The polarization of the events ($p''^{t+b}_k$) in the {\em true-signal gate} is a mixture of the polarization of the true signal ($p''^t_k$) and that of the background ($p''^b_k$) that are given by the ratio of $N_t$ and $N_b$.
Here, $k$ denotes either $S''$ or $N''$.
Thus one gets
\begin{equation}
p''^{t+b}_k = \frac{N_t p''^t_k + N_b p''^b_k}{N_t +N_b}.
\label{eq2522}
\end{equation}
The ratio of the polarization of the true signal to that of the background is given by
\begin{eqnarray}
\frac{p''^t_k}{p''^b_k} &=& \frac{N_t+N_b}{N_t}\frac{p''^{t+b}_k}{p''^b_k}-\frac{N_b}{N_t}\nonumber\\
&=& \frac{N_t+N_b}{N_t}\frac{\hat{\epsilon}^{t+b}_k}{\hat{\epsilon}^b_k}-\frac{N_b}{N_t}.
\label{eq2523}
\end{eqnarray}
The right-hand side of the equation consists of experimental quantities only. 
The effective analyzing power cancels out since both the signal and background polarizations are measured by the same polarimeter for the same proton energy bin.

The instrumental-background protons originate from slit scattering following multiple Coulomb scattering in the target and have no depolarization, i.e.\
\begin{eqnarray}
D_{SS}(0^\circ)=D_{NN}(0^\circ)=D_{LL}(0^\circ)=1.
\label{eq2524}
\end{eqnarray}
This fact represents a big advantage in the polarization analysis.
First, the background polarization can be assumed to be the same as the beam polarization.
The number of instrumental-background events is larger than the signal events and the beam-line polarimeter events and thus has a smaller statistical uncertainty.
Second, the background polarization is simultaneously measured with the signal polarization. 
Any fluctuation in the degree of beam polarization cancels out.
Third, the background polarization and the signal polarization are measured by the same detector system. 
Since only the ratio between the two polarizations is required as shown in eq.~(\ref{eq2523}), knowledge of the absolute value of the effective analyzing power $ \langle A_y^{p\rm C} \rangle$ is not required. 
Therefore, the measurements are independent of any variations of the detector efficiency and the proton beam energy.
Furthermore, no absolute calibration of the beam-line polarimeters is needed.
%Only the direction of the beam polarization needs to be measured.
%Fourth, the background protons and signal protons have the same precession effect of the polarization vector in the spectrometer. 

We can now rewrite the measured polarizations in eqs.~(\ref{eq2508}) and (\ref{eq2509}) for the true signal ($t$) and the background events ($b$) as
\begin{eqnarray}
p''^t_{S''} &=& D_{SS}(0^\circ)p_S\cos\chi_p+D_{LL}(0^\circ)p_L\sin\chi_p\nonumber\\
p''^t_{N''} &=& D_{NN}(0^\circ)p_N\nonumber\\
p''^b_{S''} &=& p_S\cos\chi_p+p_L\sin\chi_p\nonumber\\
p''^b_{N''} &=& p_N.
\label{eq2525}
\end{eqnarray}
The l.h.s.\ of eq.~(\ref{eq2523}) can be expressed in terms of the polarization-transfer coefficients 
\begin{eqnarray}
\frac{p''^t_{S''}}{p''^b_{S''}} &=& \frac{D_{SS}(0^\circ)+c_S D_{LL}(0^\circ)}{1+c_S}\nonumber\\
&=& \frac{D_{LL}(0^\circ)+c_L D_{SS}(0^\circ)}{1+c_L}\label{eq2526}\\
\frac{p''^t_{N''}}{p''^b_{N''}} &=& D_{NN}(0^\circ)\label{eq2527}\\
c_S &=& c_L^{-1}\equiv\frac{p_L}{p_S}\tan\chi_p
\label{eq2528}
\end{eqnarray}
Note that the ratio, ${p_L}/{p_S}=\tan\phi_b$ depends only on the angle of the beam polarization vector in the $S$-$L$ plane, and is independent of the magnitude of the beam polarization.
The angle $\phi_b$ can be determined with high accuracy by the BLPs.
The spectrometer bending angle ($\chi_p$) and the beam polarization direction ($\phi_b$) are optimized to have a small value of $c_S$($c_L$), typically $\le0.02$, for the $D_{SS}$($D_{LL}$) measurement.
The two independent coefficients, $D_{SS}(0^\circ) = D_{NN}(0^\circ)$ and $D_{LL}(0^\circ)$, can be determined with two measurements, either by the combination of the $D_{SS}$ and $D_{LL}$ measurements with changing the spectrometer setup \cite{tam11,has15}, or by the combination of the $D_{LL}$ and $D_{NN}$ measurements in the same setup but changing the beam polarization direction \cite{mar17,kru15}.
Some representative results of the polarization-transfer analysis are shown in sec.~\ref{subsec26}.

The above-described method of analysis is valid for the measurement centered at zero degrees that has a symmetric acceptance with respect to the rotation by 180$^\circ$. 
In a finite-angle setup of the spectrometer, the induced polarization, $P_N(\theta)$, and the analyzing power, $A_N(\theta)$, in eq.~(\ref{eq2501}) break the rotational symmetry. 
Also, the contribution from the off-diagonal components of the polarization-transfer coefficients, $D_{L'S}(\theta)$ and $D_{S'L}(\theta)$, must be taken into account.

%--------------------------------------------------%
%\subsubsection{Gamma-coincidence measurements}
%--------------------------------------------------%

%--------------------------------------------------%
%\subsubsection{CAGRA, gamma-coincidence measurement}
%--------------------------------------------------%

%%%%%%%%%%%%%%%%% Modified by AT - end %%%%%%%%%%%%%%%%%%%%%%%

%\section{Experimental techniques}
%\label{sec2}
 %
%\subsection{Experimental setup} 
%\label{subsec21}
%\begin{figure}
%\begin{center}
%\resizebox{0.4\textwidth}{!}{%
% \includegraphics{Figure211.pdf}
%% If not, use
%%\vspace{5cm}       % Give the correct figure height in cm
%}
%\caption{Figure211. 
%Scheme of the experimental setup for polarized proton scattering experiments at RCNP. 
%For details see text and ref.~\cite{tam09}.}
%\label{fig21}  
%\end{center} 
%\end{figure}
%
%\subsection{Data analysis}
%\label{subsec22}

\subsection{Spectra and polarization-transfer observables}
\label{subsec26}

Experimental observables derived from the $(\vec{\rm p},\vec{\rm p}^{\prime})$ experiments are double-differential cross sections as a function of excitation energy and scattering angle and the polarization-transfer coefficients $D_{SS}(0^\circ)$ or $D_{NN}(0^\circ)$ and $D_{LL}(0^\circ)$ discussed above.

\begin{figure}
\begin{center}
\resizebox{0.45\textwidth}{!}{%
  \includegraphics{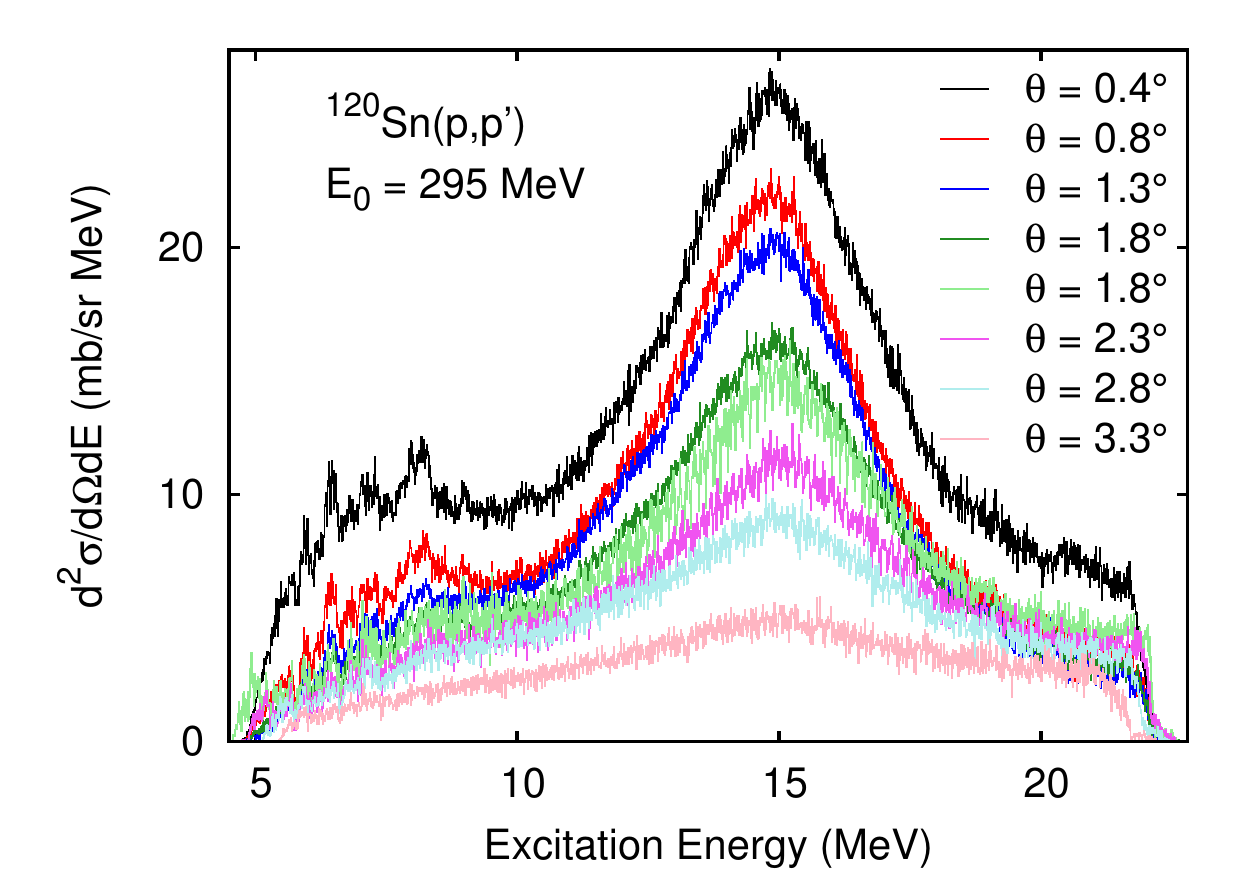}
%\vspace{5cm}       % Give the correct figure height in cm
}
\caption{
%Figure261.
Experimental cross sections of the $^{120}$Sn(p,p$^{\prime}$) reaction at $E_0 = 295$~MeV for different angle cuts. 
The top four spectra originate from a measurement with the Grand Raiden spectrometer angle set to $0^\circ$, whereas the lower four were taken at $2.5^\circ$.
Figure taken from ref.~\cite{kru15}.}
\label{fig261} 
\end{center} 
\end{figure}

A typical example of the cross-section spectra and their angular dependence observed for heavy nuclei is displayed in fig.~\ref{fig261}, which shows data of the $^{120}$Sn(p,p$^{\prime}$) reaction \cite{kru15}.
The top four spectra originate from a measurement with the Grand Raiden spectrometer angle set to $0^\circ$ applying software cuts on the reconstructed scattering angle, whereas the lower four result from a measurement with the spectrometer placed at $2.5^\circ$.
At most forward angles one observes a prominent structure around 15 MeV identified as excitation of the IVGDR.
Furthermore, a resonance-like structure is visible at lower energies ($\approx 6 - 8$ MeV) which contains E1 (PDR) and M1 (spin-flip resonance) parts. 
Both structures show a clear decrease of the cross sections with increasing angle consistent with   the properties of relativistic E1 Coulomb excitation or $\Delta L = 0$ angular distributions as expected for the spin-flip M1 transitions.   
Since the angular acceptance of the two spectrometer settings overlaps one can make a comparison of the cross sections at $\theta_{\rm lab} = 1.8^\circ$ (dark green and lime green spectra), which agree within systematic uncertainties.

\begin{figure}
\begin{center}
\resizebox{0.42\textwidth}{!}{%
  \includegraphics{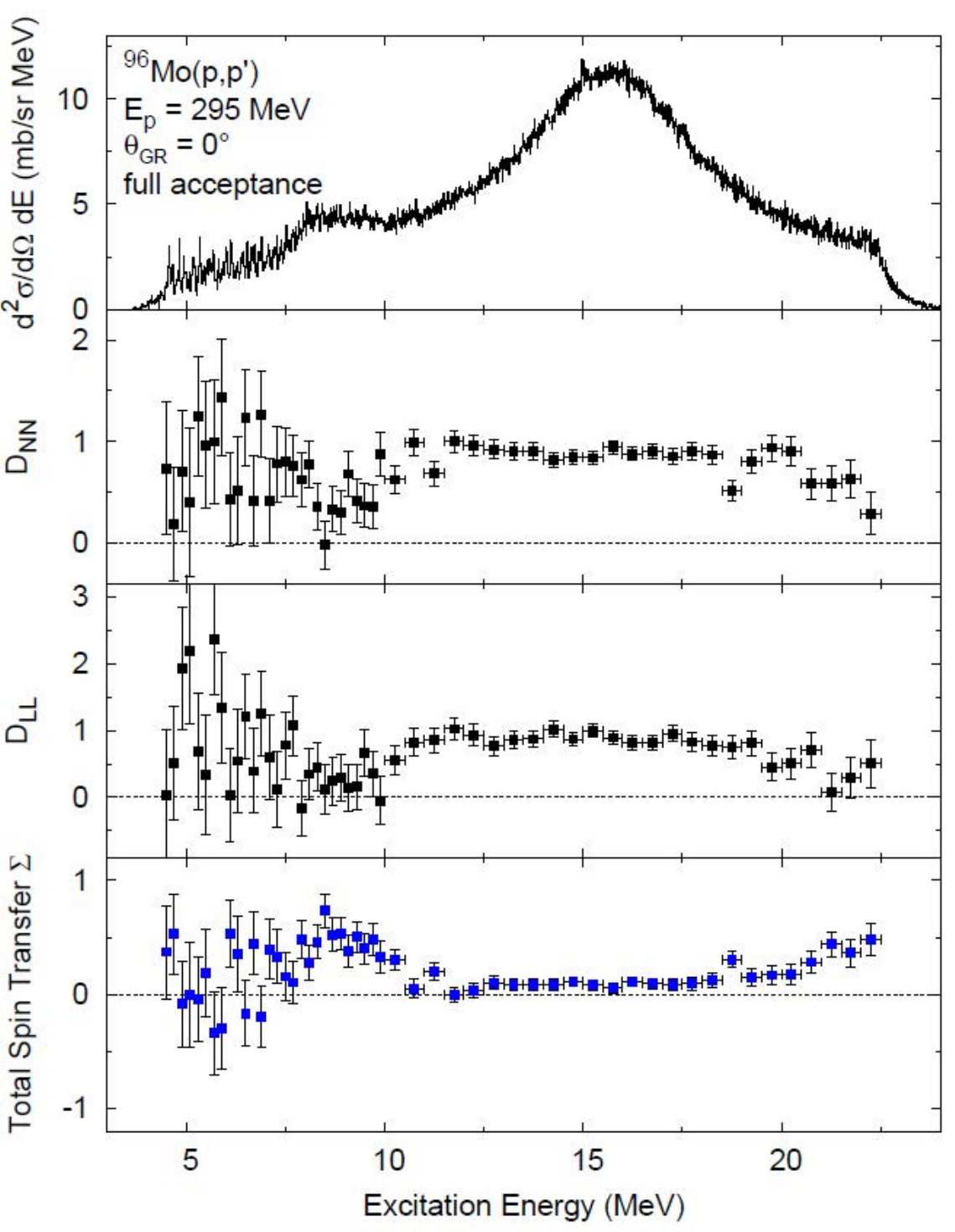}
%\vspace{5cm}       % Give the correct figure height in cm
}
\caption{
%Figure262.
Top frame: Spectrum of the $^{96}$Mo(p,p$^{\prime}$) reaction at $E_0 = 295$~MeV and $\theta_{\rm lab} = 0^\circ - 2.5^\circ$.
Second and third frames: Polarization-transfer observables $D_{\rm NN}$ and $D_{\rm LL}$. 
For the definition see ref.~\cite{ohl72}.
Bottom frame: Total spin transfer $\Sigma$, eq.~(\ref{eq321}).}
\label{fig262}  
\end{center} 
\end{figure}

As discussed in sec.~\ref{subsec32}, a combination of spin transfer observables provides information on the spin-flip character of a transition in the $(\vec{\rm p},\vec{\rm p}^{\prime})$ reaction.
%At $0^\circ$ it is sufficient to measure the combination of $D_{LL}$ and $D_{NN}$ or $D_{SS}$, where $L,N,S$ denote longitudinal, normal and sideward polarization with respect to the beam axis, respectively.  
Figure \ref{fig262} presents a measurement on $^{96}$Mo as an example \cite{mar17}.
The spectrum (top frame) shows again the characteristic double-hump structure due to excitation of the IVGDR and the PDR (+ spin-flip M1 resonance).
The second and third frames present the measured $D_{NN}$ and $D_{LL}$ values.
While they show strong variations at low excitation energies, about constant and equal values are found in the IVGDR region. 
As a result, their contributions cancel for the determination of the total spin transfer $\Sigma$, cf.\ Eq.~(\ref{eq321}), shown in the bottom frame.
For further interpretation, see sec.~\ref{subsec32}. 

\begin{figure}
\begin{center}
\resizebox{0.42\textwidth}{!}{%
  \includegraphics{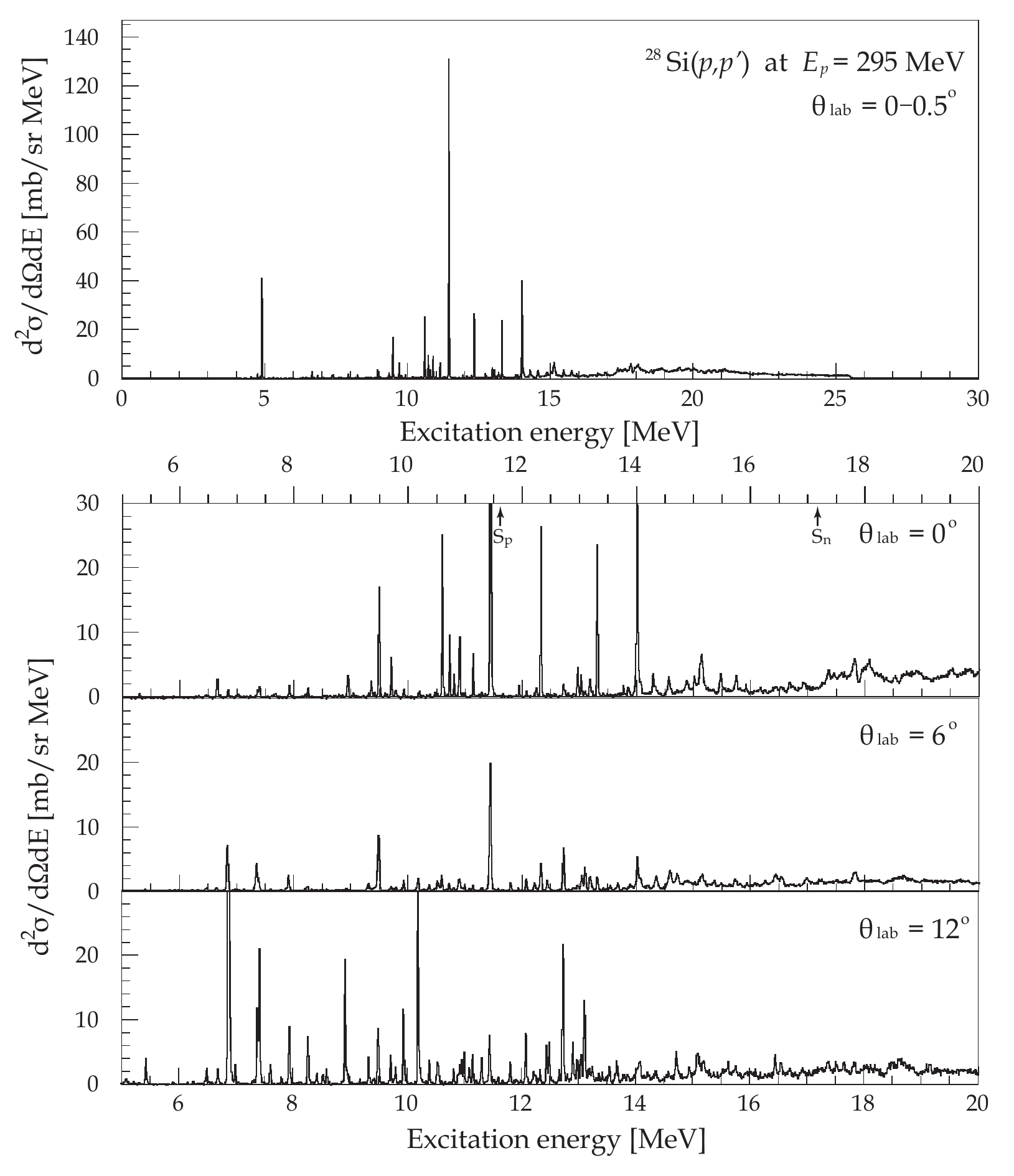}
%\vspace{5cm}       % Give the correct figure height in cm
}
\caption{
%Figure263.
Spectra of the $^{28}$Si(p,p$^{\prime}$) reaction at $E_0 = 295$~MeV and 
$\theta_{\rm lab} = 0^\circ$, $6^\circ$ and $12^\circ$ (lower part) and at very forward ($0^\circ - 0.5^\circ$) angles (upper part).}
\label{fig263}  
\end{center} 
\end{figure}

The spectra in light nuclei, where the level density is low, are dominated by individual resolved transitions.
Taking $^{28}$Si as an example, one can resolve individual transitions at very forward angles ($0^\circ - 0.5^\circ$, top part of fig.~\ref{fig263}) up to excitation energies well above 10 MeV \cite{mat15}.
The vast majority of these transitions shows M1 character.
The IVGDR lies at higher excitation energies and the cross sections are about an order of magnitude smaller than in heavy nuclei (see fig.~\ref{fig261} for an example) because of the target charge dependence of the Coulomb cross sections \cite{ber88}.
The lower part of fig.~\ref{fig263} compares spectra at $\theta_{\rm lab} = 0^\circ$, $6^\circ$ and $12^\circ$
The large differences in the variation of cross sections with angle for different peaks indicate that the multipole character of the observed transitions can be well distinguished in a multipole-decomposition analysis (see sec.~\ref{subsec31}).

\begin{figure}[b]
\begin{center}
\resizebox{0.45\textwidth}{!}{%
  \includegraphics{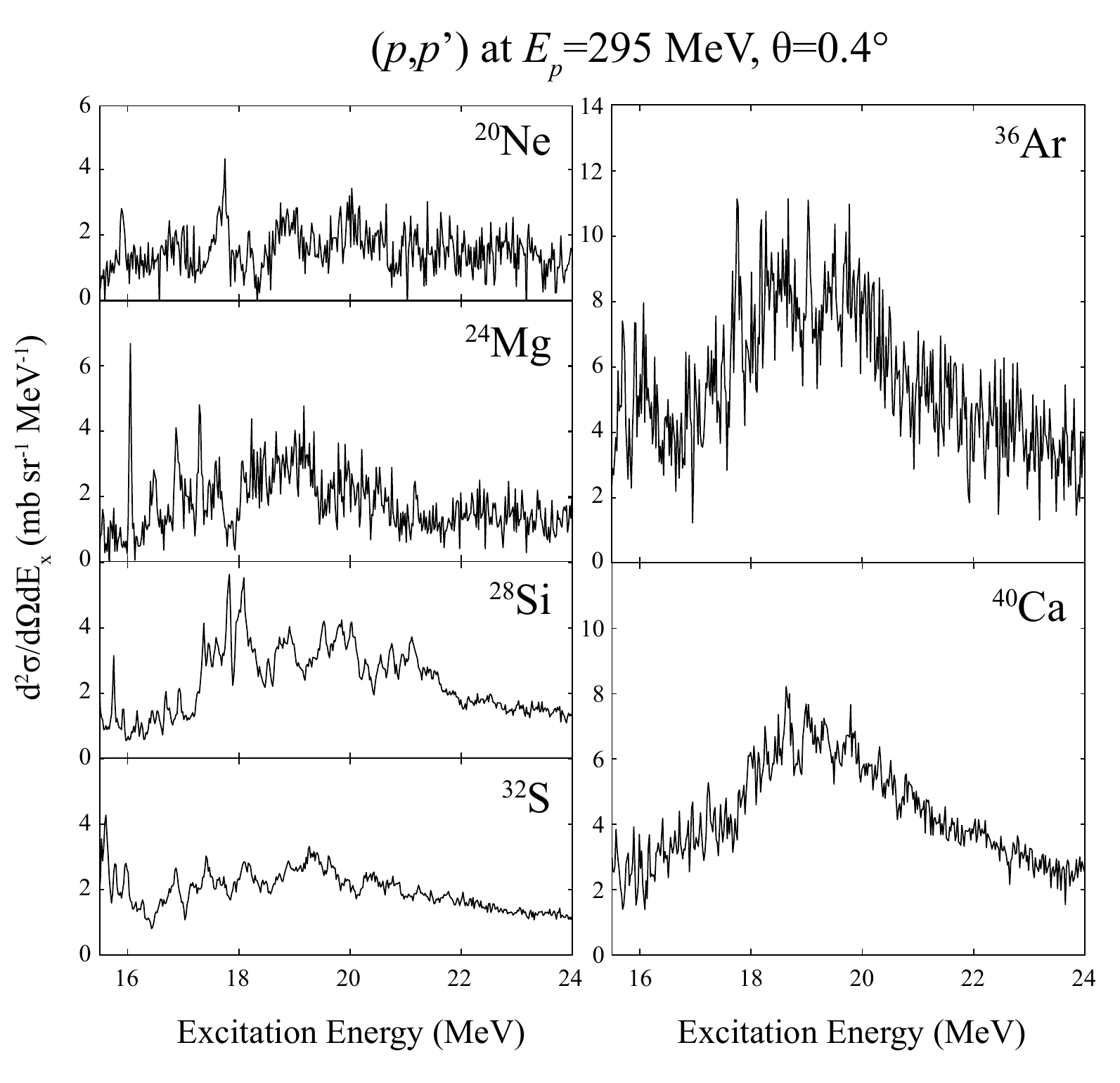}
%\vspace{5cm}       % Give the correct figure height in cm
}
\caption{
%Figure264.
 Spectra of the (p,p$^{\prime}$) reaction at $E_0 = 295$~MeV and $\theta_{\rm lab} = 0^\circ - 2.5^\circ$ on $N = Z$ nuclei from $^{20}$Ne to $^{40}$Ca in the excitation region of the IVGDR.}
\label{fig264}  
\end{center} 
\end{figure}
Figure \ref{fig264} shows a zoom into the higher-energy region of the $0^\circ$ spectra for various $N = Z$ nuclei from $^{20}$Ne to $^{40}$Ca.
The data for $^{20}$Ne, $^{24}$Mg, $^{28}$Si and $^{32}$S exhibit a strong fragmentation of the IVGDR, which can be related to deformation and $\alpha$ clustering \cite{fea18}, cf.\ sec.~\ref{sec6}. 
It is also known that a non-negligible part of the IVGDR strength resides at even higher excitation energies \cite{era86}.
In contrast, a compact resonance starts to form in $^{36}$Ar and is clearly observed in the doubly magic nucleus $^{40}$Ca.

\section{Extraction of E1 and M1 strength}
\label{sec3}

A separation of the E1 and M1 cross sections dominating the spectra at very forward angles from each other and from the excitation of other multipoles or nuclear processes was achieved with two independent methods.
These are a multipole-decomposition analysis of the cross section angular distributions with the aid of DWBA calculations and an analysis based on the measurement of the polarization-transfer coefficients.  
We also explain the methods used to convert the separated cross sections into E1 and M1 transitions strengths. 

\subsection{Multipole-decomposition analysis (MDA)}
\label{subsec31}

MDA, based on model predictions of the angular distribution shapes, is commonly used in the analysis of complex spectra from hadronic reactions.
It has been applied, e.g., for an extraction of B(GT) strengths in charge-exchange reactions \cite{wak97} or isoscalar giant resonance strength distributions from inelastic $\alpha$-particle scattering \cite{li09}, and also for inelastic electron scattering form factors of electric \cite{str00} and magnetic \cite{vnc99} giant resonances.
In the present approach, theoretical proton scattering cross sections are calculated using the code DWBA07 \cite{dwba07} with RPA amplitudes and single-particle wave functions from the quasiparticle-phonon model (QPM) (see, e.g., ref.~\cite{rye02}) as input. 
The interference of Coulomb and nuclear contributions to the cross sections is taken into account for E1 transitions.
The $t$-matrix parameterization of Franey and Love at 325~MeV \cite{fra85} was used to generate an effective projectile-target interaction.
For each discrete transition or excitation energy bin the experimental angular distributions are fitted by means of the least-square method to a sum of the
calculated angular distributions weighted with coefficients $a^{E/M \lambda}$ (with the condition $a^{E/M \lambda} \geq 0$), where $\lambda = 0,1,2,...$ is the multipolarity of the transition.
Additionally, we include a phenomenological angular distribution accounting for other nuclear processes (mainly quasi-free scattering).
In heavy nuclei, its shape was determined from the data at high excitation energies above the IVGDR (see ref.~\cite{pol11} for the case of $^{208}$Pb).

Some approximations are necessary to make the MDA tractable. 
Experimental data, although available up to angles $> 10^\circ$ in some cases, were restricted to scattering angles $\theta_{\rm lab}\leq 5^\circ$ because of the increasing complexity of contributions from different multipoles at higher momentum transfers \cite{hof07}.
Isovector spin-$M1$ excitations are represented by a single characteristic curve for each nucleus, justified by the identical angular dependence of the cross section for all transitions of this type in the angular range considered. 
Furthermore, in the calculations the largest part of the E1 strength resides in a few transitions only (typically less than 10) and only these are taken into account.
\begin{figure}[b]
\begin{center}
\resizebox{0.4\textwidth}{!}{%
  \includegraphics{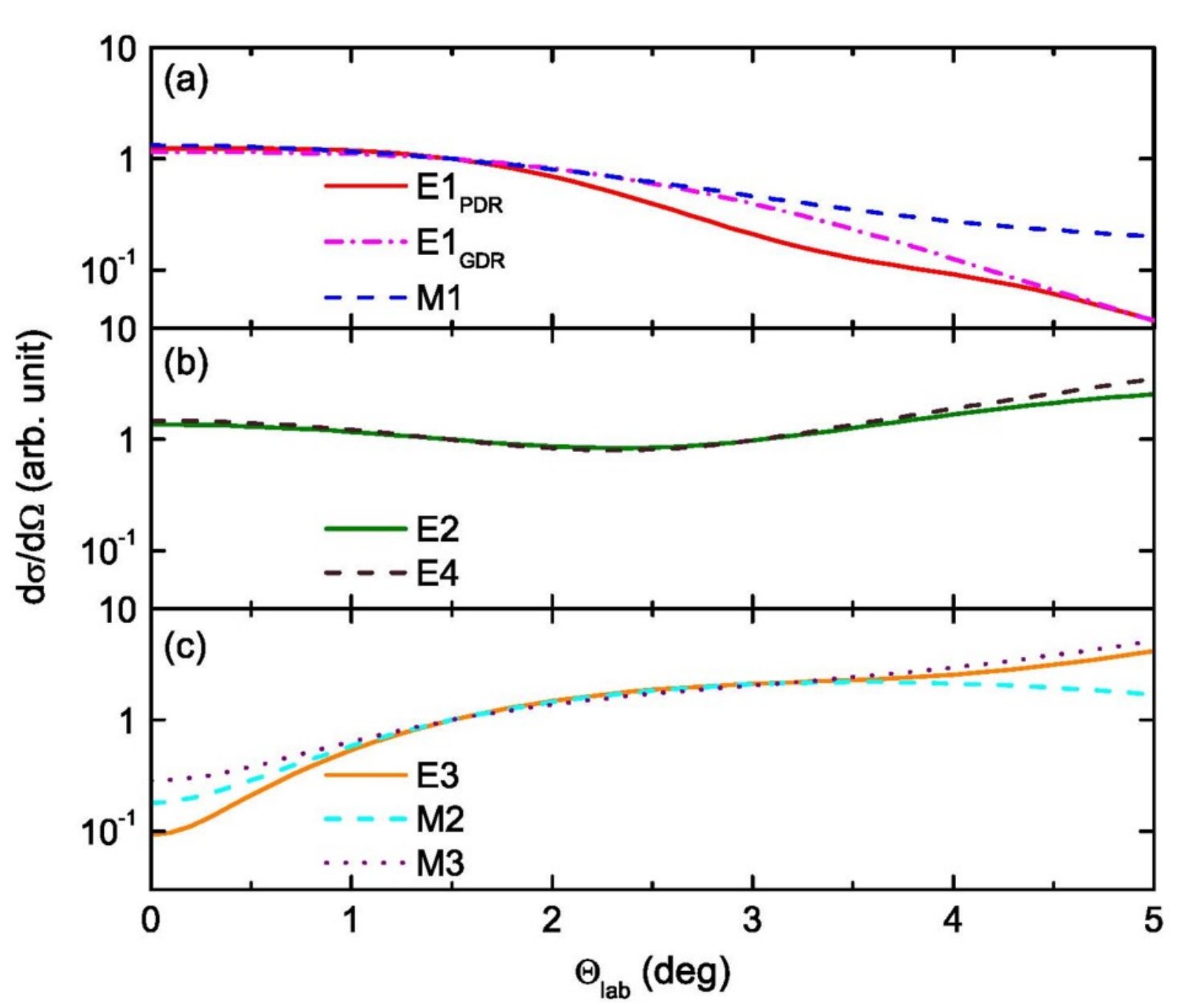}
%\vspace{5cm}       % Give the correct figure height in cm
}
\caption{
%Figure311.
Comparison of theoretical angular distribution shapes used in the MDA analysis of the $^{208}$Pb(p,p$^\prime$) reaction: (a) $M1$ and representative examples of $E1$ transitions in the PDR and IVGDR excitation-energy region, (b) $E2$ and $E4$, (c) $E3$, $M2$ and $M3$. 
All curves are normalized at $\theta_{\rm lab} = 1.5^\circ$.
Figure taken from ref.~\cite{pol12}.}
\label{fig311}  
\end{center} 
\end{figure}
Figure \ref{fig311}(a) compares the shape of the isovector spin-M1 transition with representative examples of E1 transitions to states of the PDR and IVGDR, respectively, for the case of $^{208}$Pb \cite{pol12}.
Indeed, the latter can be distinguished not only from the M1 case but also from each other.
All other contributions to the cross sections are substituted by angular distributions of either $E2$ or $E3$ transitions, whose shapes were taken to be that of the most collective transition of each type. 
Other multipolarities of potential relevance like $M2$, $M3$ or $E4$ exhibit very similar angular distribution shapes to either $E2$ or $E3$ as demonstrated in figs.~\ref{fig311}(b) and (c). 
The QPM results predict that isoscalar monopole transitions are only weakly excited in proton scattering at these incident energies with a contribution of less than 3\%  of the cross sections at $0^\circ$ at the maximum of the giant monopole resonance for $^{208}$Pb.
Therefore, possible contributions from $E0$ transitions were neglected in the MDA.

\begin{figure}
\begin{center}
\resizebox{0.48\textwidth}{!}{%
  \includegraphics{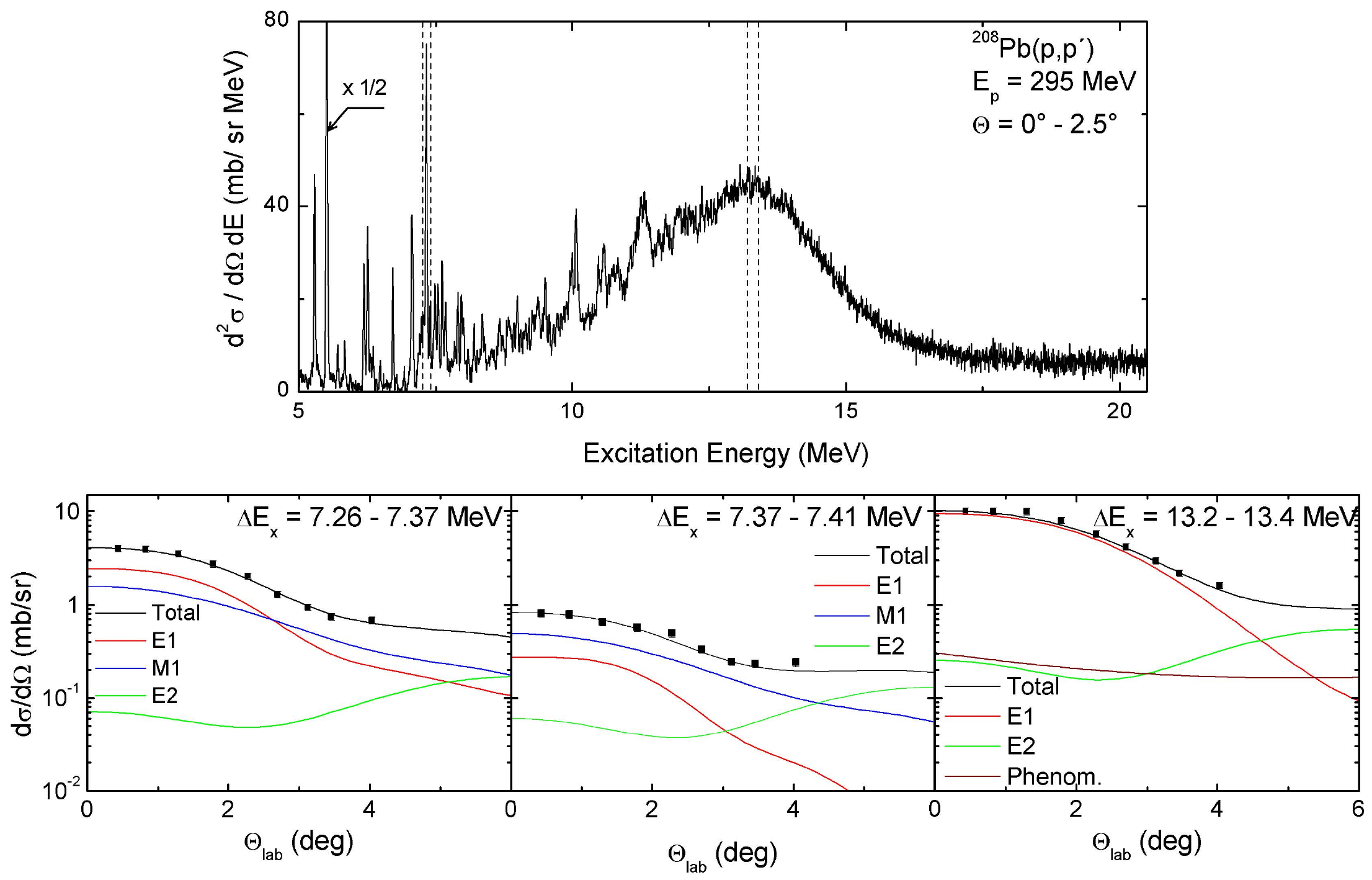}
%\vspace{5cm}       % Give the correct figure height in cm
}
\caption{
%Figure312.
Example of the MDA analysis of the $^{208}$Pb(p,p$^{\prime}$) data.
Upper part: full acceptance spectrum with the spectrometer placed at $0^\circ$. 
Lower part: MDA for two adjacent energy bins in the energy region of overlapping levels near neutron threshold and for an energy bin at the maximum of the IVGDR indicated by vertical dashed lines in the spectrum, respectively.}
\label{fig312}  
\end{center} 
\end{figure}

The final coefficients are obtained by computing the MDA for all possible combinations of $E1$, $M1$ and $E2$ (or $E3$) transitions plus the phenomenological background taking the $\chi^2$-weighted average of all individual $a^{E/M \lambda}$ values.
Examples of fits are displayed for the $^{208}$Pb(p,p$^{\prime}$) data in the bottom part of fig.~\ref{fig312}. 
The corresponding energy bins are indicated by the vertical dashed lines in the spectrum taken with the spectrometer placed at $0^\circ$ shown in the top part of fig.~\ref{fig312}. 
The two adjacent energy bins in the excitation-energy region of the PDR demonstrate the sensitivity of the MDA to distinguish E1 and M1 contributions to the cross sections with a dominance of E1 in the former and M1 in the latter bin, respectively.
Figure \ref{fig312}(c) presents a fit for an energy bin near the maximum of the IVGDR, illustrating the expected dominance of E1 cross sections.

\begin{figure}
\begin{center}
\resizebox{0.4\textwidth}{!}{%
  \includegraphics{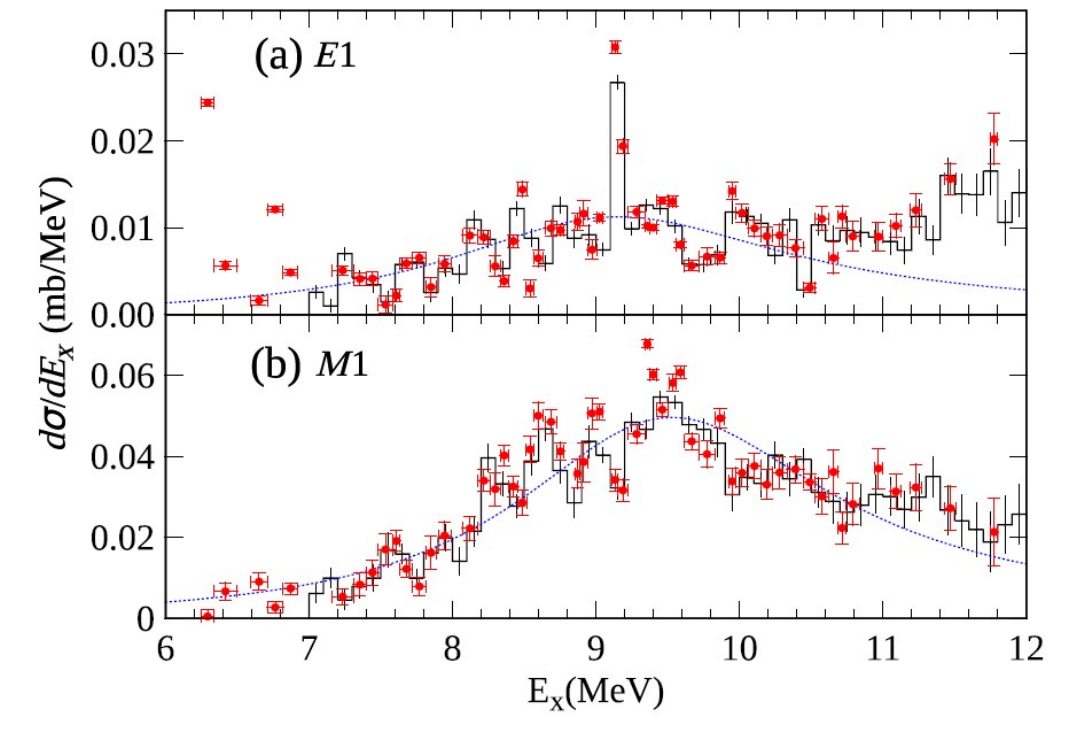}
%\vspace{5cm}       % Give the correct figure height in cm
}
\caption{
%Figure313.
(a) E1 and (b) M1 cross sections in $^{90}$Zr between 6 and 12 MeV from a MDA. 
The histogram shows the cross section distribution in steps of 100 keV and the solid circles the magnitude of 55 selected peaks.
The dotted lines represent best-fit Lorentzian functions for the PDR and spin-M1 resonance.
Figure taken from ref.~\cite{iwa12}.}
\label{fig313}  
\end{center} 
\end{figure}
The decomposition of E1 and M1 cross sections for the example of $^{90}$Zr \cite{iwa12} in the excitation energy region from 6 to 12 MeV is shown in fig.~\ref{fig313} for 100 keV bins. 
One observes strong fluctuations of the E1/M1 ratio from bin to bin.
The high resolution of the experiment is sufficient to resolve many individual transitions and the result of their MDA is shown as full red circles. 
The dotted lines show the fits for the PDR and spin-M1 resonance cross sections in $^{90}$Zr assuming Lorentzian shapes.
We note, however, that $^{208}$Pb and $^{90}$Zr are the only cases investigated so far where individual transitions could be resolved in heavy nuclei.
In general the level density is too large even in the energy region below the IVGDR (see, e.g., fig.~\ref{fig261}).

\begin{figure}[b]
\begin{center}
\resizebox{0.45\textwidth}{!}{%
  \includegraphics{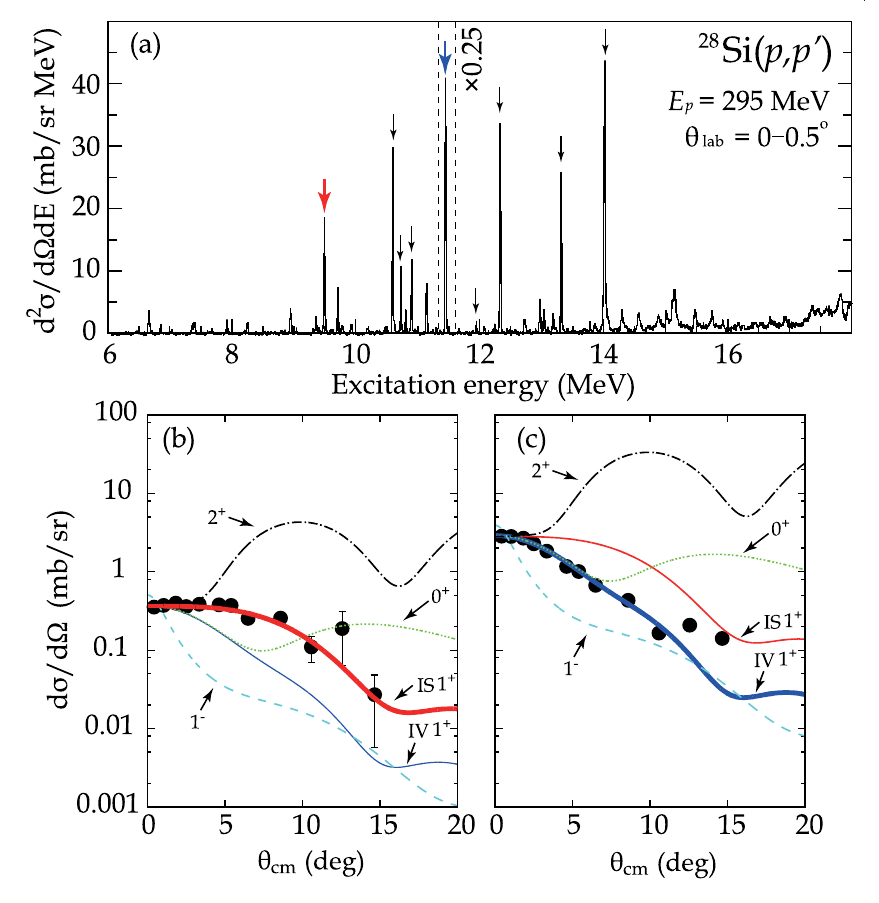}
%\vspace{5cm}       % Give the correct figure height in cm
}
\caption{
%Figure314.
(a) Excitation-energy spectrum of the $^{28}$Si(p,p$^{\prime}$) reaction at $E_0 = 295$~MeV and $\theta_{\rm lab} = 0^\circ - 0.5^\circ $.
The arrows indicate states with a definite $J^\pi = 1^+$ assignment.  
(b) and (c) Examples of angular distributions of an IS and an IV spin-M1 transition marked, respectively, by red and blue arrows in (a) in comparison with theoretical angular distributions for several multipolarities.
Figure taken from ref.~\cite{mat15}.}
\label{fig314}  
\end{center} 
\end{figure}
The situation is different in light nuclei.
The upper part of fig.~\ref{fig314} shows a spectrum of the $^{28}$Si(p,p$^\prime$) reaction at extreme forward angles $< 0.5^\circ$ \cite{mat15}.
Transitions up to $E_{\rm x}\approx 15$ MeV can be clearly resolved. 
All prominent excited states have been identified as $J^\pi = 1^+$ states (marked with arrows).

The experimental angular distributions for each state were compared to distorted-wave impulse approximation (DWIA) calculations for identifying the $0^+ \rightarrow 1^+$ transitions. 
The DWIA calculations were performed with the code DWBA07 \cite{dwba07} using one-body transition densities obtained from shell-model calculations with the code NuShellX@MSU \cite{bro14} incorporating the USD interactions \cite{bro87,bro06,ric08}.
Predictions for different multipoles are shown in figs.~\ref{fig314}(b) and (c) in comparison with experimental data for the transitions marked with red and blue arrow in fig.~\ref{fig314}(a).
All theoretical curves are normalized to the experimental data at the most forward angle.
IS and IV M1 transitions can be clearly distinguished from other multipoles but also from each other, since the former angular distribution is flatter than the latter due to the contribution of an exchange tensor component in the effective NN interaction \cite{lov81}.

\subsection{Polarization-transfer analysis (PTA)}
\label{subsec32}

\begin{figure}
\begin{center}
\resizebox{0.45\textwidth}{!}{%
  \includegraphics{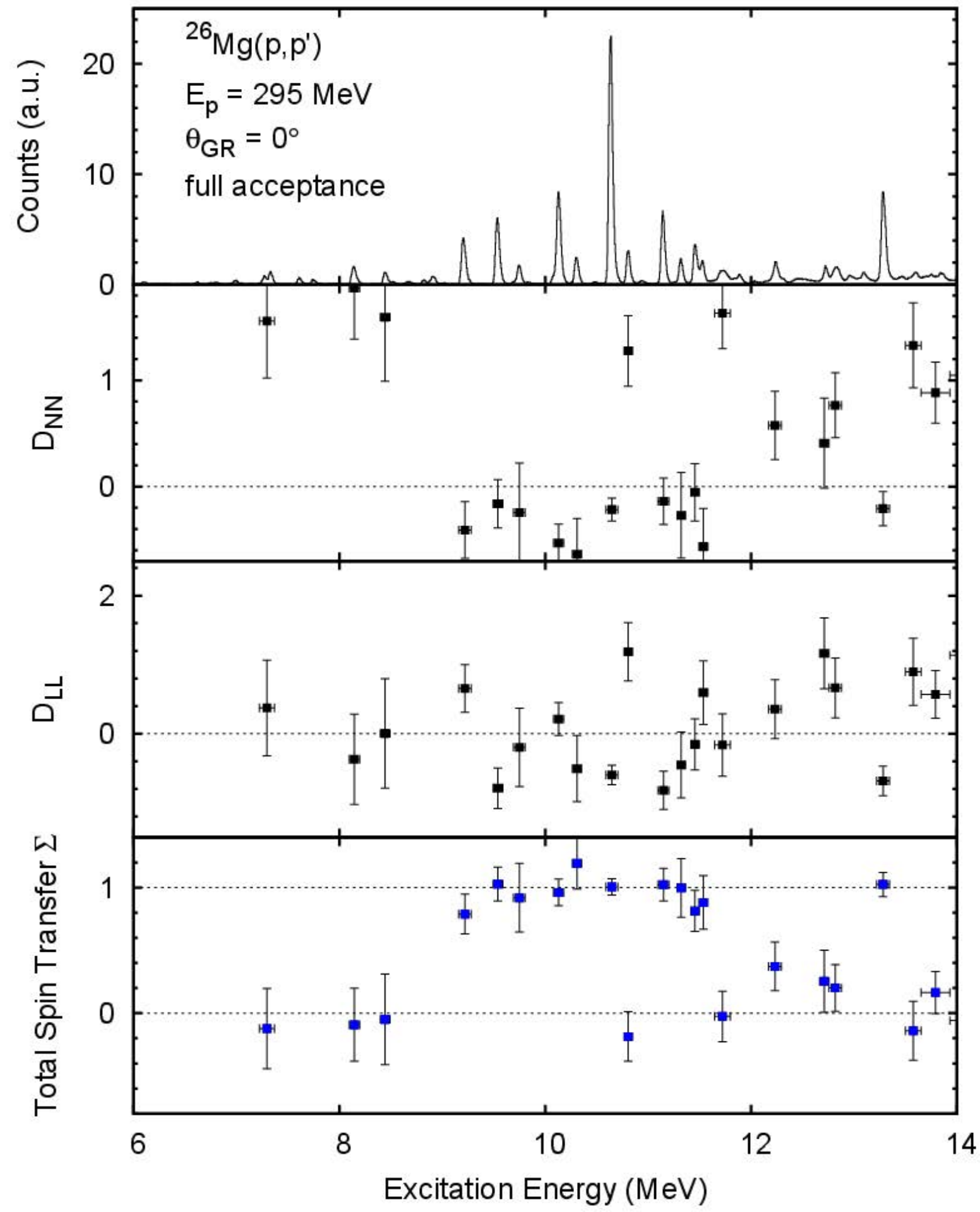}
}
%\vspace{5cm}       % Give the correct figure height in cm
\caption{
%Figure321.
PTA of the  $^{26}$Mg$(\vec{\rm p},\vec{\rm p}^{\prime})$ reaction at $E_0 = 295$~MeV and $\theta_{\rm lab} = 0^\circ - 2.5^\circ$.
From top to bottom: spectrum, $D_{NN}$, $D_{LL}$, and total spin transfer $\Sigma$; see eq.~(\ref{eq321}).
}
\label{fig321}  
\end{center} 
\end{figure}

An independent separation of E1 and M1 parts of the cross sections, particularly useful in the overlap region of the PDR and the spin-M1 resonance, can be achieved by a decomposition of spin-flip and non-spin-flip cross sections.
This information can be extracted from the polarization-transfer observables $D_{LL}$, $D_{SS}$ and $D_{NN}$ introduced in sec.~\ref{subsec25}.
As pointed there, $D_{SS}$ and $D_{NN}$ are indistinguishable at $0^\circ$, thus only one of them needs to be measured.
It is convenient to introduce the total spin transfer
\begin{equation}
\label{eq321}
\Sigma = \frac{3-2D_{SS}(\, {\rm or} \, D_{NN})-D_{LL}}{4},
\end{equation}
which takes values of zero for non-spin-flip and one for spin-flip transitions \cite{suz00}.
The application of eq.~(\ref{eq321}) is illustrated in fig.~\ref{fig321} with data of the $^{26}$Mg $(\vec{\rm p},\vec{\rm p}^{\prime})$ reaction.
The top part of fig.~\ref{fig321} displays the spectrum measured at a spectrometer angle of $0^\circ$.
All transitions observed are known from the literature \cite{ensdf} and the more prominent excitations all have spin-M1 character \cite{cra89}.
The second and third frames present the measured $D_{NN}$ and $D_{LL}$ values, respectively, which both show a wide scattering for the individual transitions.
However, their $\Sigma$ values shown in the bottom frame of fig.~\ref{fig321} sort into two distinct groups with values close to either 0 or 1.
Excitations with $\Sigma \approx 0$ all correspond to transitions to natural-parity states, while those with $\Sigma \approx 1$ belong to the spin-M1 resonance exciting unnatural-parity ($1^+$) states.
The few cases with intermediate $\Sigma$ values observed between 12 and 13 MeV correspond to weak bumps in the spectrum, which most likely result from the unresolved superposition of non-spin-flip and spin-flip excitations.

\begin{figure}[b]
\begin{center}
\resizebox{0.45\textwidth}{!}{%
  \includegraphics{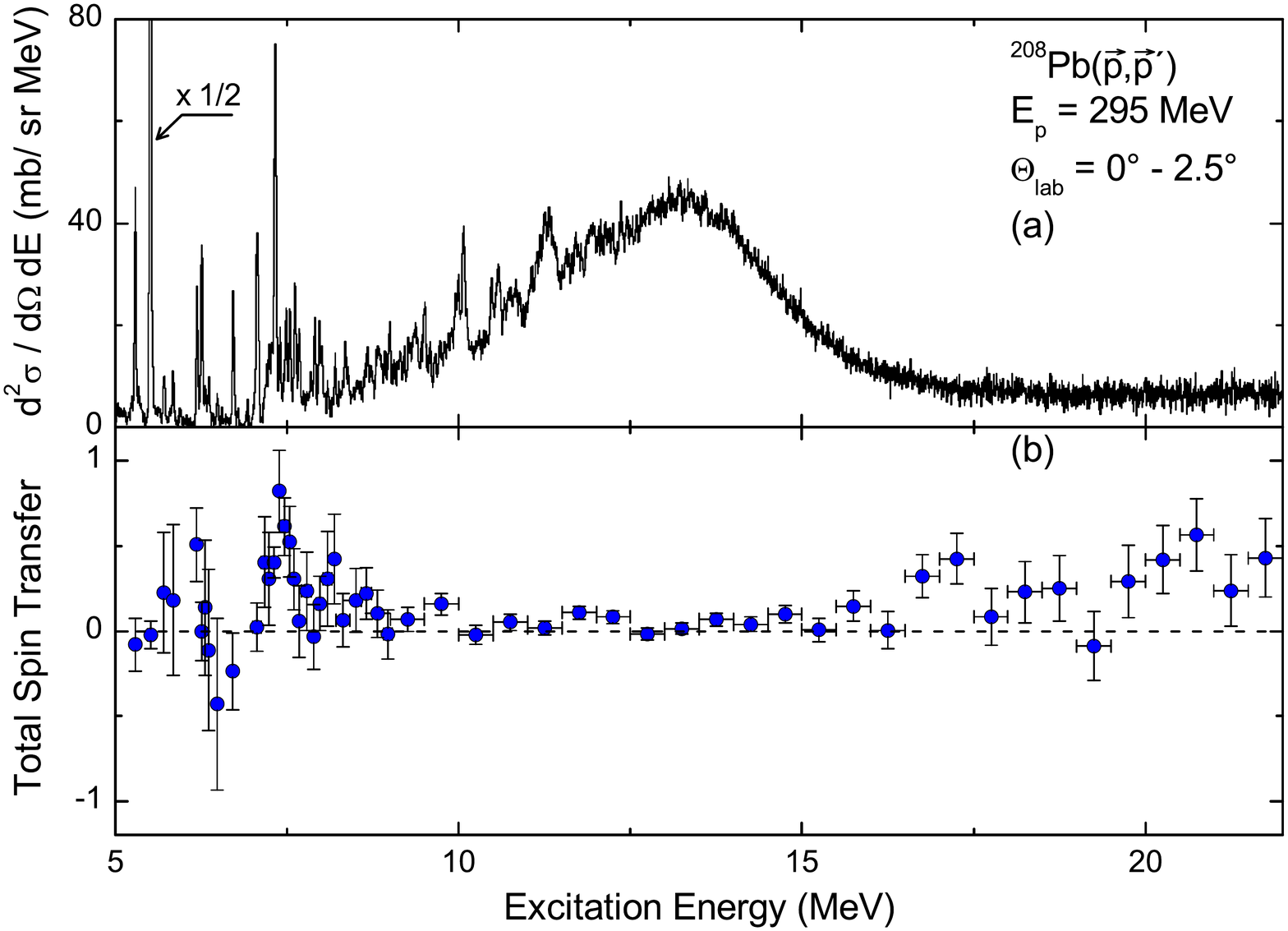}
%\vspace{5cm}       % Give the correct figure height in cm
}
\caption{
%Figure322.
PTA of the  $^{208}$Pb$(\vec{\rm p},\vec{\rm p}^{\prime})$ reaction at $E_0 = 295$~MeV and $\theta_{\rm lab} = 0^\circ - 2.5^\circ$.
(a) spectrum, (b) total spin transfer $\Sigma$, eq.~(\ref{eq321}).
Figure taken from ref.~\cite{tam11}.
}
\label{fig322}  
\end{center} 
\end{figure}
In heavier nuclei, Coulomb excitation becomes an increasingly important contribution to the cross sections besides excitation of the spin-M1 resonance due to the strong spin-isospin-flip part of the proton-nucleus interaction at small momentum transfers.
Because of the different reaction mechanisms, non-spin-flip and spin-flip excitations in the PTA can be identified with E1 and M1 transitions, respectively.
As an example for a heavy nucleus, the PTA of the $^{208}$Pb$(\vec{\rm p},\vec{\rm p}^{\prime})$ reaction in the energy region $E_x = 5 - 22$ MeV is presented in fig.~\ref{fig322}, where (a) shows the spectrum measured at $0^\circ$ and (b) the total spin transfer $\Sigma$. 
Values of $\Sigma$ between 0 and 1 result from a summation over partially unresolved transitions with different spin-flip character. 
The data reveal a competition of spin-flip strength in the energy region $7 - 9$ MeV, where the spin-$M1$ resonance in $^{208}$Pb is located \cite{hey10}, with non-spin-flip excitation of the PDR.
The bump between 10 and 16 MeV has $\Delta S = 0$ character consistent with an excitation of the IVGDR.
The $\Delta S = 1$ strength at high excitation energies may result from the spin-flip part of quasi-free scattering \cite{bak97} dominating the cross sections above the IVGDR energy region.

\begin{figure}
\begin{center}
\resizebox{0.45\textwidth}{!}{%
  \includegraphics{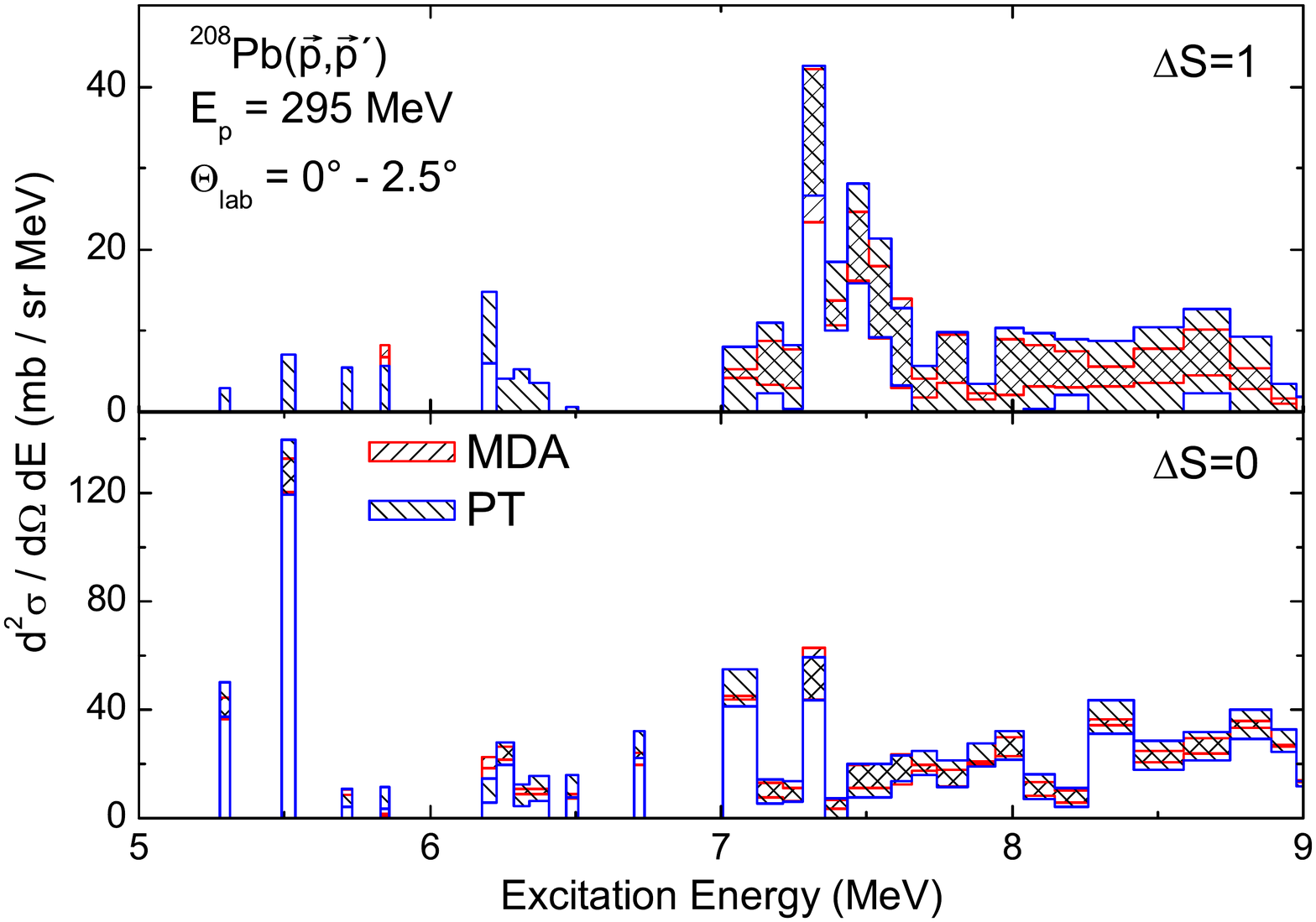}
%\vspace{5cm}       % Give the correct figure height in cm
}
\caption{
%Figure323.
Decomposition of non-spin-flip (E1, $\Delta S = 0$) and spin-flip (M1, $\Delta S = 1$) cross-section parts of the  $^{208}$Pb$(\vec{\rm p},\vec{\rm p}^{\prime})$ reaction based on the MDA and PTA, respectively, in the excitation energy region $5 - 9$ MeV. 
The hatched areas indicate the experimental uncertainties.
Figure taken from ref.~\cite{tam11}.}
\label{fig323}  
\end{center} 
\end{figure}

E1 and M1 parts from the MDA and $\Delta S = 0$ and 1 parts from the PTA applied to the $^{208}$Pb data are compared in fig.~\ref{fig323} for $E_{\rm x} < 9$~MeV \cite{tam11}. 
Within the experimental uncertainties the correspondence between the two completely independent decomposition methods is excellent.
Similar agreement between the two methods was achieved in $^{120}$Sn \cite{has15} and $^{96}$Mo \cite{mar17}. 
This put confidence in the MDA results for heavy nuclei, where individual transitions typically cannot be resolved. 
The MDA of cross sections provides much better resolution because of the superior statistics compared to a double scattering measurement of polarization-transfer observables. 
As a consequence, later experiments have been restricted to the measurement of cross sections only.
In the IVGDR region no direct comparison between the two methods is possible because of the unknown $\Delta S$ content of the phenomenological background. 
However, both methods agree that $\Delta S =1 $ contributions are very small.

\subsection{Conversion of Coulomb excitation to photoabsorption cross sections}
\label{subsec33}

\begin{figure}
\begin{center}
\resizebox{0.45\textwidth}{!}{%
  \includegraphics{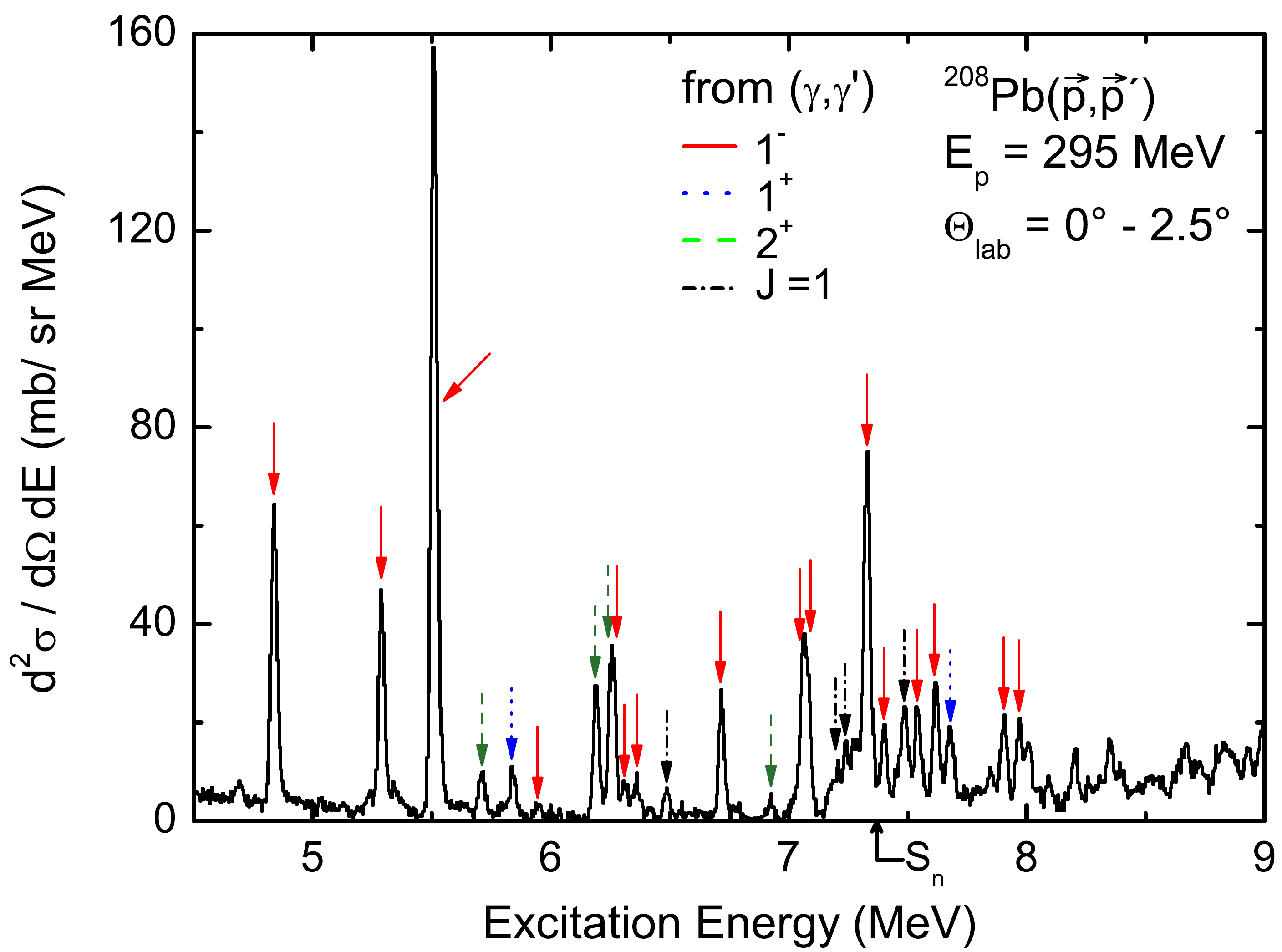}
%\vspace{5cm}       % Give the correct figure height in cm
}
\caption{
%Figure331.
Low-energy part of the spectrum of the $^{208}$Pb(p,p$'$) reaction at E$_{\rm p}$ = 295~MeV and $\theta_{\rm lab} = 0^\circ - 2.5^\circ$. 
The arrows indicate transitions with dipole character also observed in $^{208}$Pb($\gamma$,$\gamma^\prime$) experiments~\cite{rye02,end03,shi08,sch10}.
Figure taken from ref.~\cite{pol12}.}
\label{fig331}  
\end{center} 
\end{figure}

The low-energy part of the $^{208}$Pb spectrum (fig.~\ref{fig322}) up to 9 MeV is displayed in fig.~\ref{fig331}.
Arrows indicate dipole and quadrupole transitions known from previous work, mainly from ($\gamma,\gamma^\prime$) experiments \cite{rye02,end03,shi08,sch10}.
One observes a one-to-one correspondence, and the largest transitions in the proton scattering experiment all have E1 character.
This points to a dominance of Coulomb cross sections in the (p,p$^\prime$) scattering at very forward angles, since nuclear excitation cross sections of E1 transitions are weak.    

\begin{figure}[b]
\begin{center}
\resizebox{0.45\textwidth}{!}{%
  \includegraphics{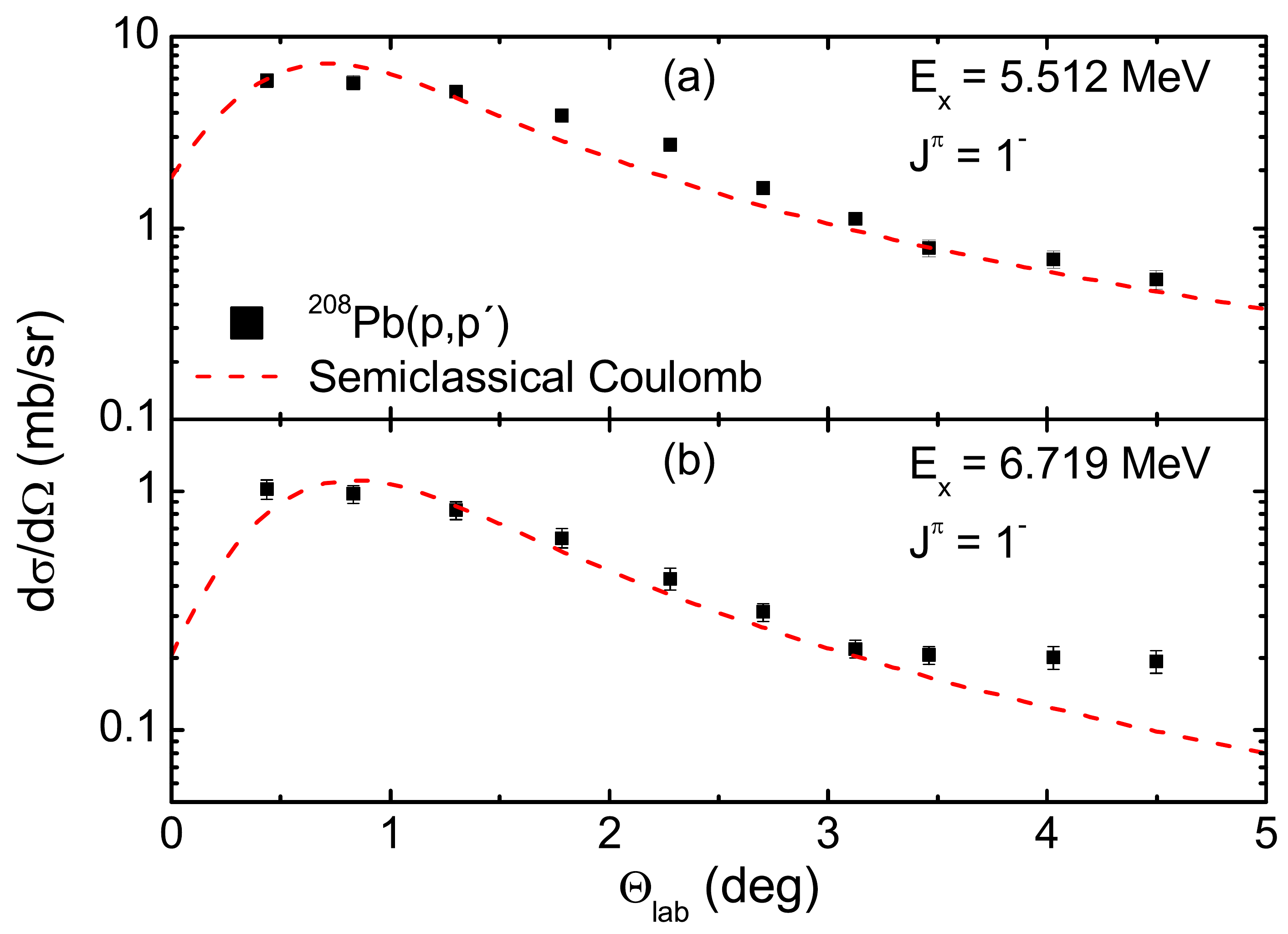}
%\vspace{5cm}       % Give the correct figure height in cm
}
\caption{
%Figure332.
Angular distributions of the 1$^-$ states at $E_{\rm x}=5.512$ MeV and 6.719 MeV prominently excited in the $^{208}$Pb(p,p$'$) reaction at $E_0 = 295$ MeV. 
The dashed lines are predictions of Coulomb-excitation cross sections based on the semi-classical approach \cite{ber88}.
Figure taken from ref.~\cite{pol12}.}
\label{fig332}  
\end{center} 
\end{figure}
\begin{figure*}
\begin{center}
%\resizebox{0.48\textwidth}{!}{%
%  \includegraphics{Figure333a.pdf}
\resizebox{0.85\textwidth}{!}{%
  \includegraphics{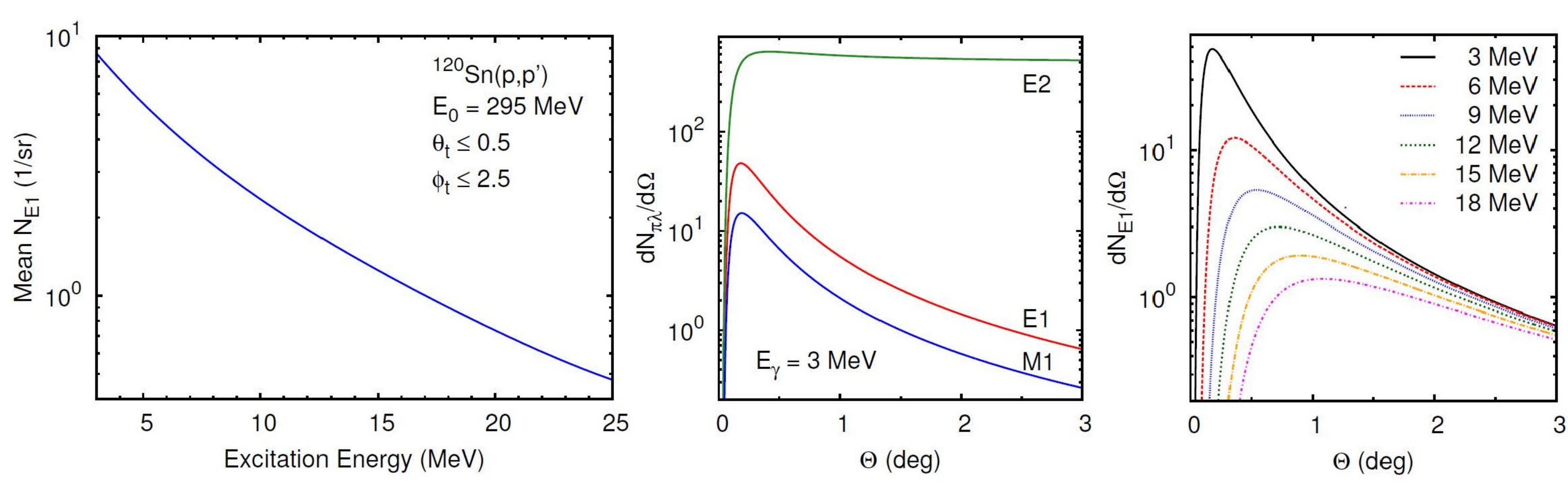}
%\resizebox{0.3\textwidth}{!}{%
 % \includegraphics{Figure333a.pdf}}
  %\resizebox{0.3\textwidth}{!}{%
  %\includegraphics{Figure333b.pdf}}
  %\resizebox{0.3\textwidth}{!}{%
  %\includegraphics{Figure333c.pdf}
%\vspace{5cm}       % Give the correct figure height in cm
}
\caption{
%Figure333.
Left: Virtual-photon number for E1 transitions averaged over the solid angle for the $^{120}$Sn(p,p$^{\prime}$) reaction at $E_0 = 295$~MeV as a function of excitation energy. 
Middle: Virtual-photon numbers per unit solid angle for E1, E2 and M1 transitions at $E_{\gamma} = 3$ MeV as a function of the scattering angle $\theta_{\rm lab}$. 
Right: Same for E1 transitions at different $\gamma$ energies.}
\label{fig333}  
\end{center} 
\end{figure*}
To verify the assumption, Coulomb-excitation predictions for the angular distributions of the prominent transitions to 1$^-$ states at $E_{\rm x} =5.512$~MeV and 6.719~MeV were calculated based on a semi-classical model \cite{ber88} shown as dashed lines in fig.~\ref{fig332}.
Because of the finite angular resolution of the Grand Raiden spectrometer they were convoluted with Gaussian functions with widths corresponding to the vertical and horizontal angular acceptance  of the detector system. 
The shape of the experimental angular distributions (solid squares) is well described and their absolute magnitudes can be reproduced when the calculations are normalized to the B(E1) strengths deduced from the ($\gamma$,$\gamma^\prime$) experiments \cite{rye02,end03,shi08,sch10}. 
The remaining deviations at angles larger than $2^\circ$ are attributed to effects of Coulomb-nuclear interference and in case of the transition to the state at 6.719 MeV to possible contributions from an unresolved transition with higher multipolarity.

Since the E1 cross sections for angles $< 2^\circ$ are shown to be of pure Coulomb nature, one can convert them to equivalent photoabsorption cross sections or B(E1) strengths with the virtual-photon method \cite{ber88}
\begin{equation}
\frac{d^2{\sigma}}{d{\Omega}dE_{\gamma}}={\frac{1}{E_{\gamma}}} {\frac{dN_{E1}}{d{\Omega}}} \sigma_{\gamma} (E_{\gamma}).
\label{eq331}
\end{equation}
relating the experimental to the photoabsorption cross section $\sigma_\gamma$.

As an example, the virtual-photon numbers for E1 transitions averaged over the solid angle for the  $^{120}$Sn(p,p$^{\prime}$) reaction at $E_0 = 295$~MeV as a function of  energy is presented in l.h.s.\ of fig.~\ref{fig333}. 
It exhibits a strong energy dependence amounting to about an order of magnitude for the excitation-energy region of interest.
The middle part of fig.~\ref{fig333} presents virtual-photon numbers per unit solid angle for E1, E2 and M1 transitions at $E_{\gamma} = 3$ MeV as a function of the scattering angle $\theta_{\rm lab}$. 
M1 transitions show an angular distribution similar to E1, but smaller virtual-photon numbers.
Taking into account the different excitation probability in real-photon experiments, the M1 Coulomb-excitation cross sections are negligible compared with nuclear excitation of the spin-M1 resonance discussed in sec.~\ref{subsec34}.
For E2 transitions large virtual-photon numbers are observed.
However, the corresponding contribution to $\sigma_\gamma$ is nevertheless negligible.
All virtual-photon spectra show a steep minimum at $0^\circ$ due to the assumption of a sharp cutoff of the impact parameter (semi-classical approximation).
The dominance of the E1 part of the photo-response is well known and independent of excitation energy (see, e.g., ref.~\cite{bas16}).
Finally, the $\gamma$ energy dependence of the angular distribution of ${\rm d}N_{E1}/{\rm d}\Omega$ is shown in the r.h.s.\ of fig.~\ref{fig333}. 
The maximum moves towards larger angles with increasing energy.
For a meaningful comparison with experiment it is important to integrate the cross sections over an angular region which includes the first maximum of the virtual-photon spectrum to avoid a dependence on the artificial minimum of the semi-classical approximation.

\begin{figure}[b]
\begin{center}
\resizebox{0.45\textwidth}{!}{%
  \includegraphics{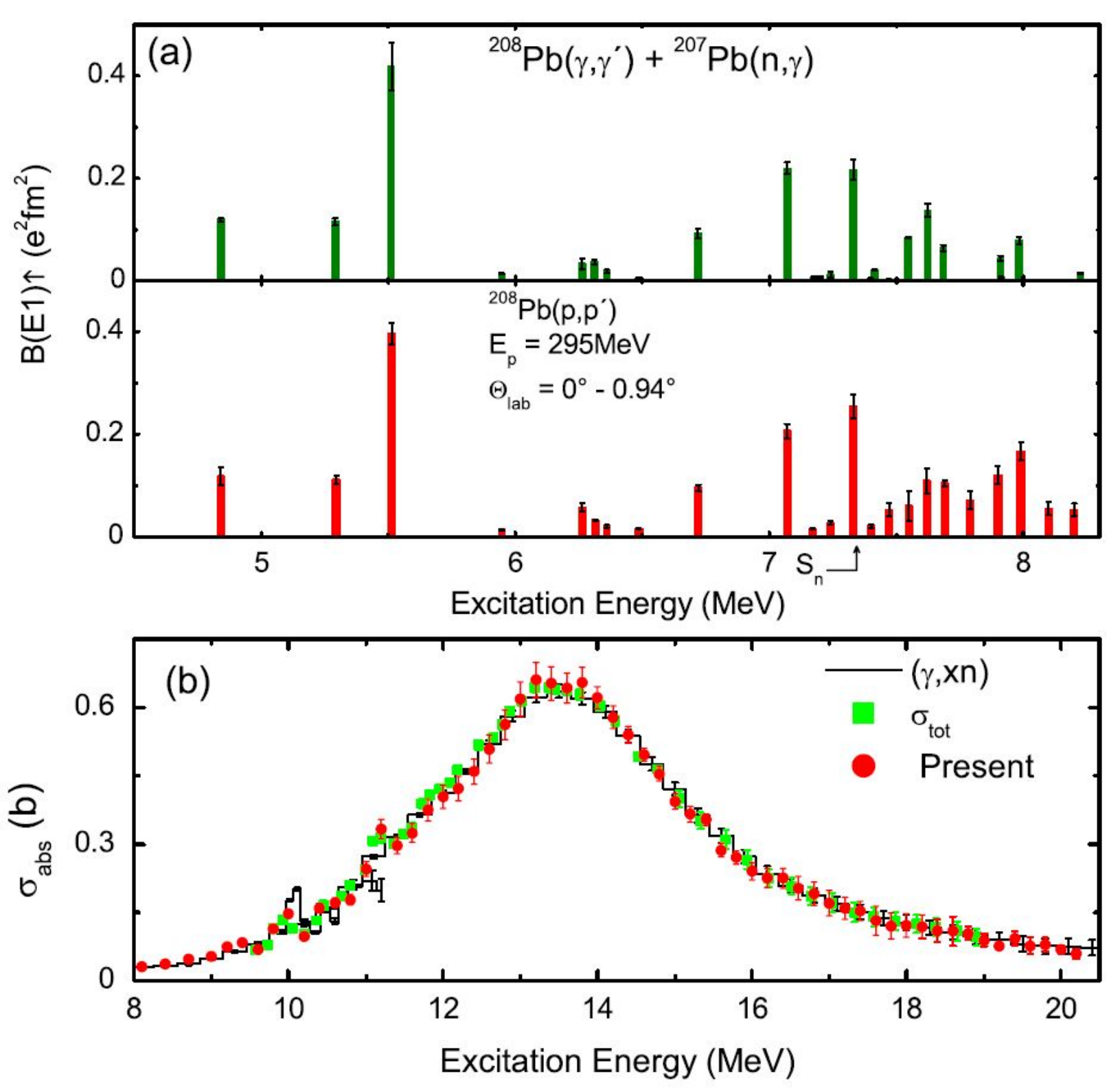}
%\vspace{5cm}       % Give the correct figure height in cm
}
\caption{
%Figure334.
(a) B(E1) strength in $^{208}$Pb in the excitation-energy region $E_x \simeq 4.8 - 8.2$ MeV deduced from the (p,p$^\prime$) experiment in comparison with $(\gamma,\gamma')$  \cite{rye02,end03,shi08,sch10} and (n,$\gamma$) \cite{koh87} experiments.
(b) Photoabsorption cross sections in the IVGDR region compared to ($\gamma$,xn) \cite{vey70} and elastic photon scattering \cite{sch88} measurements.
Figure taken from ref.~\cite{tam11}.
}
\label{fig334}  
\end{center} 
\end{figure}
\begin{figure}
\begin{center}
\resizebox{0.4\textwidth}{!}{%
  \includegraphics{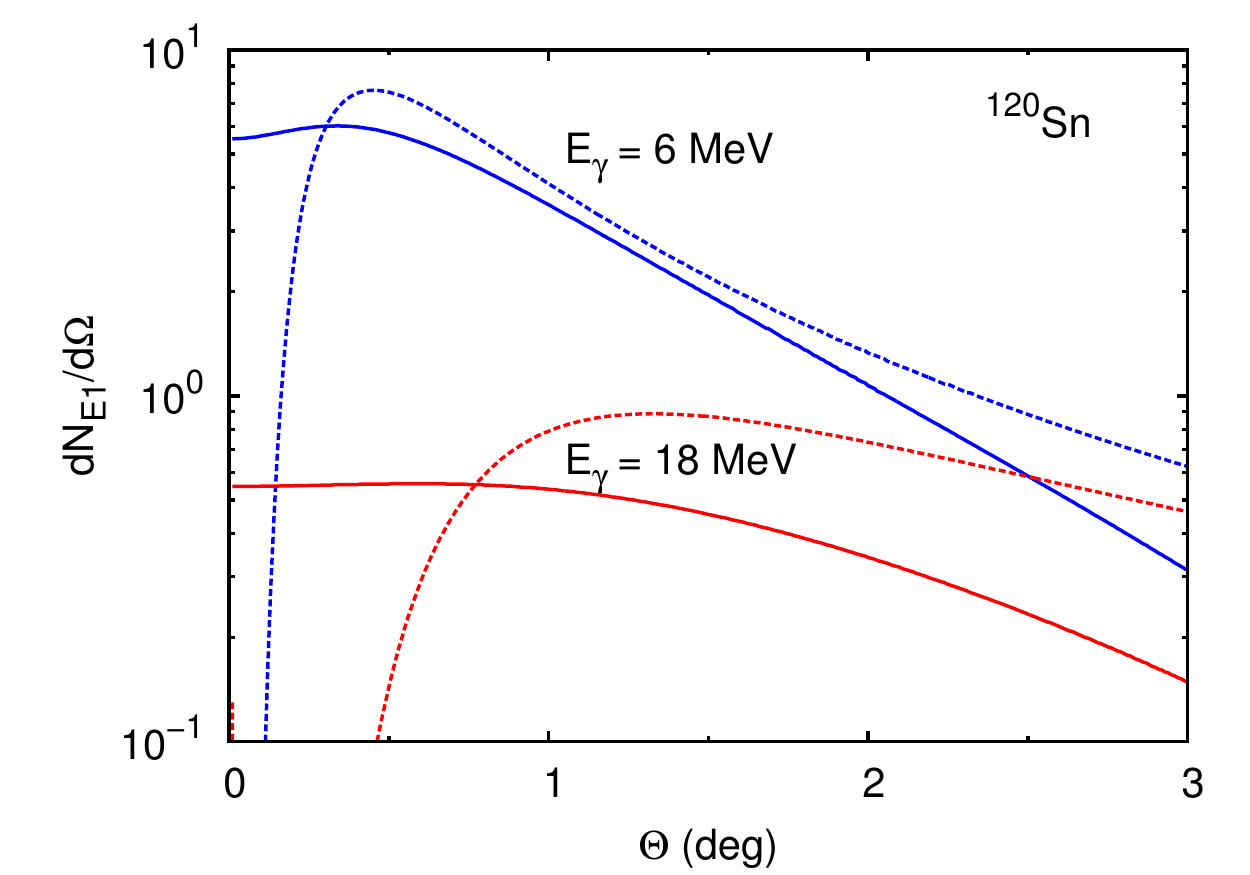}
}
\caption{
%Figure335.
Comparison of angular distributions of virtual-photon numbers for E1 transitions calculated with the semi-classical \cite{ber88} (dotted lines) and the eikonal \cite{ber93} (full lines) approaches for the $^{120}$Sn(p,p$^{\prime}$) reaction at $E_0 = 295$~MeV and $\gamma$ energies of 6 (blue) and 18 (red) MeV, respectively. 
}
\label{fig335}  
\end{center} 
\end{figure}

The application of the virtual-photon method to the $^{208}$Pb is presented in fig.~\ref{fig334}.
The upper part compares the extracted B(E1) strength distribution in the region $E_x \simeq 4.8 - 8.2$ MeV with $(\gamma,\gamma')$ \cite{rye02,end03,shi08,sch10} and (n,$\gamma$) \cite{koh87} experiments. 
Excellent agreement is obtained up to the neutron threshold $S_{\rm n}$. 
The excess strength in the (p,p$'$) data above the neutron threshold can be attributed to previously unknown neutron decay widths of the excited $1^-$ states, which modify the branching ratios in the $\gamma$-decay experiments and thus the extracted B$(E1)$ values. 
The lower part shows the photoabsorption cross sections in the IVGDR region together with results from a ($\gamma$,xn) \cite{vey70} and an elastic photon scattering \cite{sch88} experiment. 
Again, very satisfactory agreement of all three measurements is observed.
While the older analyses of $^{208}$Pb \cite{tam11,pol12} and $^{120}$Sn  \cite{has15,kru15} were based on the semi-classical model, in more recent work \cite{mar17,bir17} an eikonal approach \cite{ber93} was used for the calculation of the virtual-photon spectra.
It allows for a proper treatment of relativistic and retardation effects and provides more realistic angular distributions due to taking into account absorption on a diffuse nuclear surface. 
The differences between the two approaches are illustrated in fig.~\ref{fig335} again for the example of $^{120}$Sn.
The dotted and solid lines compare the angular dependence of the virtual-photon spectra obtained with the semi-classical and eikonal approaches, respectively, for energies of 6 (blue) and 18 (red) MeV. 
One clearly sees that the unphysical minimum at zero degrees obtained in the semi-classical model vanishes when applying the eikonal approach and the resulting angular distributions are much closer to the experimental observations (cf.\ fig.~\ref{fig332}).
However, when integrating over the angular region where Coulomb/nuclear interference effects can be neglected (up to about $2^\circ$ in the example shown in fig.~\ref{fig335}), differences of the resulting virtual-photon flux are small between both models, typically less than 10\% in heavy nuclei.
In contrast, differences become sizable in lighter nuclei and photoabsorption cross sections extracted using the virtual-photon flux calculated with the semi-classical model are found to be way too small compared to results obtained with real-photon probes \cite{bas14}.

\subsection{Unit cross-section method}
\label{subsec34}

The extraction of spin-M1 matrix elements from the M1 (p,p$^{\prime}$) cross sections poses a problem.
While the forward-peaked $\Delta L = 0$ angular distribution of isovector spin-flip M1 (ISVM1) transitions can be well described independent of details of the DWBA calculation, absolute predictions of cross sections show a large uncertainty depending on the choice of the effective proton-nucleus interaction \cite{hof07}.  
Therefore, the concept of unit cross section developed for the extraction of Gamow-Teller (GT) strength from charge-exchange (CE) reactions \cite{tad87,zeg07} is employed and a similar relation is derived for the (p,p$'$) reaction.
For CE reactions the cross section at scattering angle $\theta =0^\circ$ can be written as
\begin{equation}
\frac{\mathrm{d}\sigma}{\mathrm{d}\Omega}({\rm CE}, 0^\circ) = \hat{\sigma}_\mathrm{GT} F(q,\omega) B(\mathrm{GT}), 
\label{eq341}
\end{equation}
where $\hat{\sigma}_\mathrm{GT}$ is a nuclear-mass dependent factor (the unit cross section), $F(q,\omega)$ a kinematical factor normalized to $F(0,0)=1$ correcting for non-zero momentum and energy transfer, and $B$(GT) the reduced GT transition strength. 
The total energy transfer $\omega = E_{\rm x} - Q$, where $Q$ denotes the reaction $Q$ value.
One can define a corresponding relation for the inelastic scattering cross sections 
\begin{equation}
\frac{\mathrm{d}\sigma}{\mathrm{d}\Omega}({\rm p},{\rm p}', 0^\circ) = \hat{\sigma}_\mathrm{M1} F(q,E_x) B(\mathrm{M1}_{\sigma \tau}), 
\label{eq342} 
\end{equation}
where $B(\mathrm{M1}_{\sigma\tau})$ denotes the reduced IVSM1 transition strength.
Obviously, the kinematical factors in eqs.~(\ref{eq341}) and (\ref{eq342}) differ for isobaric analog states.
The kinematical correction factor depends on the energy transfer and is determined by DWBA calculations comparing the predicted cross sections for a given transition with and without energy transfer. 
Extrapolation to the cross section at $0^\circ$ from experimental data at finite angles is achieved with the aid of theoretical angular distributions. 

The reduced GT and IVSM1 transition strengths from a $J^\pi = 0^+$ ground state to a $J^\pi = 1^+$ excited state can be expressed as
\begin{eqnarray}
B(\mathrm{GT})  & = & \frac{C_{\mathrm{GT}}^2}{2(2T_f+1)}  |\langle f||| \sum_k^A \sigma_k \tau_k ||| i \rangle |^2 \label{eq343} \\
B(\mathrm{M1}_{\sigma \tau}) & = & \frac{C_{\mathrm{M1}}^2}{4(2T_f+1)}  |\langle f||| \sum_k^A \sigma_k \tau_k ||| i \rangle |^2.\label{eq344}
\end{eqnarray}
Here, $\sigma_k$ and $\tau_k$ are the spin and isospin operators acting on the $k^{\rm th}$  nucleon, $\langle|||\sigma \tau||| \rangle$ denotes a matrix element reduced in spin and isospin, and $i,f$ are initial and final states with isospin $T_i,T_f$.
The Clebsch-Gordan (CG) coefficients $C_{\mathrm{GT/M1}}$ depend on the reaction and on the $T_i$,$T_f$ values \cite{fuj11}. 
The $(p,n)$ reaction can excite GT transitions to states with isospin $T_f = T_i-1,T_i,T_i+1$ and the corresponding strength is commonly termed $B(\mathrm{GT}_-)$, $B(\mathrm{GT}_0)$, $B(\mathrm{GT}_+)$. 
The $\beta$ decay transitions used to determine the parameters of Eq.~(\ref{eq341}) possess $T_f = T_i -1$ while the IVSM1 resonance observed in the $(p,p')$ reaction has $T_f = T_i$.
$T_i + 1$ states can also be excited in principle but are well separated in excitation energy and strongly suppressed for large values of $T_i$ \cite{fuj11}).

At the small momentum transfers of inelastic scattering close to $0^\circ$, isospin symmetry requires $\hat{\sigma}_{\mathrm{M1}} \simeq \hat{\sigma}_\mathrm{GT}$.
The systematics of $\hat{\sigma}_\mathrm{GT}$  has been studied in ref.~\cite{sas09} for the (p,n) reaction at a beam energy comparable to the present experiments (297 MeV). 
They find a parameterization of the mass dependence
\begin{equation}
\hat{\sigma}_\mathrm{GT} = 3.4(3)\mathrm{exp} \left[ -0.40(5) \left( A^{1/3} - 90^{1/3} \right) \right].
\label{eq345}
\end{equation} 

The assumption of equal unit cross sections leads to
\begin{equation} 
B({\mathrm M1_{\sigma \tau}}) = \frac{1}{2} \frac{T_i}{T_i + 1}  B({\mathrm GT}_-) 
\label{eq346}
\end{equation}
and for the case of an analog GT transition with $T_f = T_i$ 
\begin{equation} 
B({\mathrm M1_{\sigma \tau}}) = \frac{1}{2} T_i  B({\mathrm GT}_0). 
\label{eq347}
\end{equation}
Equations (\ref{eq346},\ref{eq347}) imply that the IVSM1 matrix elements can also be derived from studies of the GT strength with the $(p,n)$ reaction in the same kinematics.
However, in particular for the study of discrete transitions the poor energy resolution of the (p,n) reaction of several hundred keV (FWHM) represents a severe limitation.
Furthermore, for nuclei with $T_i \gg 0$, the corresponding CG coefficients suppress the cross sections \cite{fuj11}.  

%The factor $1/2$ arises from the different coupling of the isospin operator in Eq.~(\ref{eqme}) for $(p,p')$ and $(p,n)$ reactions.
%It has been experimentally verified \cite{doz09} by a comparison of both reactions on $^{12}$C for the [$1p_{3/2}^{-1}1p_{1/2}$] transition populating the $1^+$ state at 15.11 MeV and the isobaric analog state of $^{12}$N (the ground state). 
%The isospin-dependent factor in Eq.~(\ref{eqratio}) stems from the ratio of the CG coefficients in Eq.~(\ref{eqme}).

A number of approximations was made in the derivation of eqs.~(\ref{eq346},\ref{eq347}), whose validity is discussed below.
Several effects can break the equality of eqs.~(\ref{eq346},\ref{eq347}).
In contrast to the purely IV CE reactions, the (p,p$'$) cross sections contain IS contributions.
However, because of the dominance of the $\sigma\tau$ over the $\sigma$ part of the effective interaction \cite{lov81} these are typically $\leq 5$\% and energetically separated in heavy nuclei \cite{hey10}.
Differences of exchange terms contributing to the (p,p$'$) and (p,n) cross sections and Coulomb effects can affect the extrapolation of cross sections to $q = 0$.
However, estimates based on the Love-Franey effective interaction \cite{lov81,fra85} indicate that these are negligible.

A general problem of (p,p$'$) as well as CE reactions are incoherent and coherent $\Delta L = 2$ contributions to the angular distributions of M1 excitations, the latter due to the tensor part of the proton-nucleus interaction.
Because of the difference of angular distribution shapes, the incoherent $\Delta L=2$ cross sections are effectively taken into account in a MDA of the data, while the coherent part requires explicit knowledge of the excited-state wave function.
In ref.~\cite{zeg06}, a shell-model study has been performed for the case of the $^{26}$Mg($^3$He,t) CE reaction at 140 MeV/u indicating $10 - 20$\% changes of individual transition strengths with decreasing importance for stronger transitions and random sign. 
Thus, for the total strength the uncertainties should be smaller than 10\%.

With a few additional assumptions one can also derive the corresponding electromagnetic B(M1) transition strength
\begin{equation}
B(\mathrm{M1}) = \frac{3}{4\pi} |\langle f|| g_l^{\mathrm{IS}}\vec{l} + \frac{g_s^{\mathrm{IS}}}{2}\vec{\sigma} - (g_l^{\mathrm{IV}} \vec{l} + \frac{g_s^{\mathrm{IV}}}{2}\vec{\sigma})\tau_0||i \rangle |^2\, \mu_\mathrm{N}^2,
\label{eq348}
\end{equation}
which contains spin and orbital contributions for the isoscalar (IS) and isovector (IV) parts.
For small orbital and IS contributions $B(\mathrm{M1})$ and $B({\mathrm M1_{\sigma \tau}})$ can be related by  
\begin{equation}
B(\mathrm{M1}) \cong  \frac{3}{4\pi}\left(g_s^{\mathrm{IV}} \right)^2 B(\mathrm{M1}_{\sigma \tau}) \, \mu_\mathrm{N}^2.
\label{eq349}
\end{equation}

The approximations made when going from eq.~(\ref{eq348}) to eq.~(\ref{eq349}) are justified by the following arguments.
Because of the anomalous proton and neutron $g$ factors the IS spin part is small [$(g_s^{\rm IS})^2 \approx 0.035 (g_s^{\rm IV})^2$] and can usually be neglected (see, however, the special case of $^{48}$Ca discussed below).
Furthermore, orbital M1 strength is related to deformation \cite{end05} and thus can be neglected in nuclei near closed shells.
However, one can even argue that eq.~(\ref{eq349}) should approximately hold in general, at least for the total strength.
For light deformed nuclei the spin-orbital interference can be sizable for individual transitions (see e.g., refs.~\cite{fuj97,hof02}) but the overall strength is weakly modified ($\leq 10$\%) because of the random mixing sign \cite{ric90,fay97}.
In heavy deformed nuclei, spin and orbital M1 strengths are energetically well separated and mixing is predicted to be weak \cite{hey10}.

Finally, meson-exchange current contributions can differ for electromagnetic and hadronic reactions.
These differences are relevant in light nuclei and have, e.g., been observed in the comparison of M1 and GT strengths in $sd$-shell nuclei \cite{ric90,lue96,vnc97}.
However, for $A \geq 40$ the available data indicate that the quenching factors in microscopic calculations are the same \cite{ric85}, consistent with theoretical expectations \cite{tow87,tok80}.    Overall, the approximations discussed above limit the accuracy of $B$(M1) strengths derived from the (p,p$'$) data to about 10\% near closed-shell nuclei and about 15-20\% in deformed nuclei. 

\begin{figure}
\begin{center}
\resizebox{0.5\textwidth}{!}{%
  \includegraphics{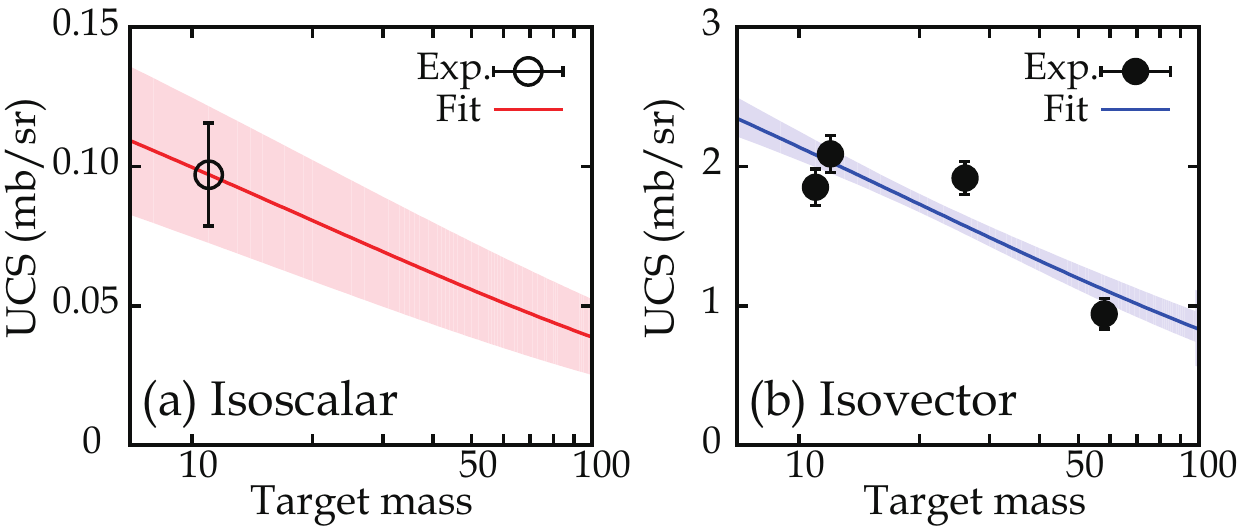}
%\vspace{5cm}       % Give the correct figure height in cm
}
\caption{
%Figure341.
Mass dependences of the unit cross section for (a) IS and (b) IV spin-M1 transitions in light nuclei.
Figure taken from ref.~\cite{mat15}.}
\label{fig341}  
\end{center} 
\end{figure}

Since the mass dependence of $\hat{\sigma}_\mathrm{GT}$ in ref.~\cite{sas09} was derived from $\beta$-decay data of medium-mass and heavy nuclei only, an independent analysis was performed for the study of the (p,p$^\prime$) reaction in light nuclei assuming the same functional form as in eq.~(\ref{eq345}) 
The result is shown in the r.h.s\ of fig.\ref{fig341}.  
The resulting parameters are in very good agreement with those for heavier nuclei \cite{mat15}.

\begin{figure}
\begin{center}
\resizebox{0.45\textwidth}{!}{%
  \includegraphics{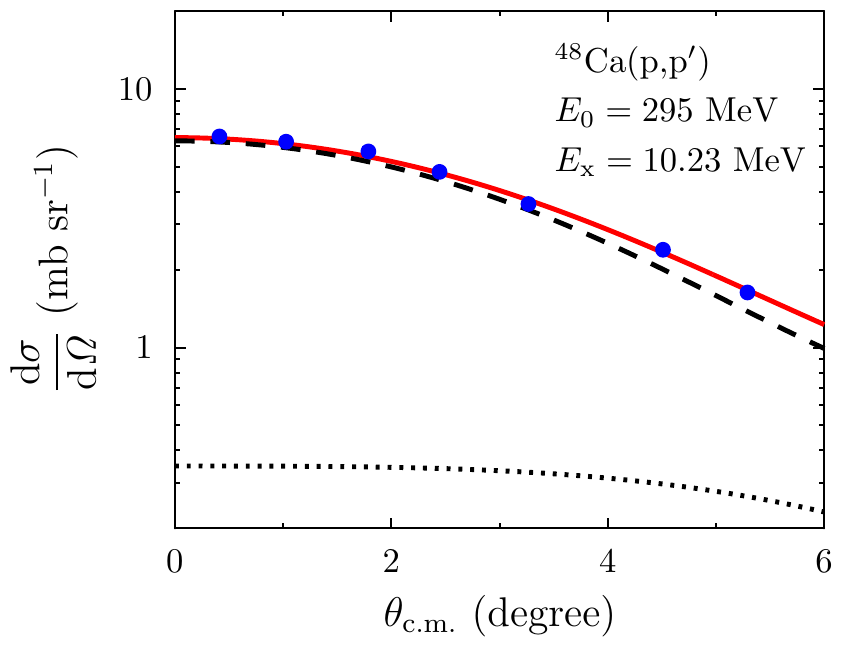}
%\vspace{5cm}       % Give the correct figure height in cm
}
\caption{
%Figure342.
Angular distribution of the peak at $E_{\rm x} =10.23$~MeV (full circles) excited in the $^{48}$Ca(p,p$^\prime$) reaction in comparison to model calculations with the code DWBA07 for a neutron (solid line), IS (dotted line), and IV (dashed line) spin-flip $1f_{7/2}$$\rightarrow$$1f_{5/2}$ transition using the Love-Franey effective proton-nucleus interaction \cite{lov81,fra85}. 
Figure taken from ref.~\cite{bir16}.}
\label{fig342}  
\end{center} 
\end{figure}

As mentioned above, the (p,p$^\prime$) cross sections contain an IS part.
It can usually be neglected for nuclei with a large g.s.\ isospin, but even in cases where the IS contribution is enhanced, it remains on the level of a few \%.
This is illustrated in fig.~\ref{fig342} for the case of the pure neutron $1f_{7/2} \rightarrow 1f_{5/2}$ transition in $^{48}$Ca \cite{bir16} discussed below.
The figure compares the measured angular distribution with DWBA calculations assuming a neutron (full red line), pure IS (dotted line) and pure IV (dashed line) transition. 
The IS contribution is relevant at larger angles but close to $0^\circ$ it amounts to 2\% of the total cross section only.

However, in nuclei with $T = 0$ g.s.\ isospin pure IS M1 transitions can be excited and experimentally distinguished from IV transitions (cf.\ fig.~\ref{fig314}).
Utilizing a relation analogous to eq.~(\ref{eq342}), one can extract the reduced IS spin-M1 strength (called B$_\sigma$).
The corresponding unit cross section has been determined using $^{11}$B($p,p'$) data \cite{kaw04} and an IS $M1$ strength deduced from the $\gamma$-decay lifetimes of the first-excited mirror states in $^{11}$B and $^{11}$C \cite{ensdf}. 
Note that the linear combination of the $\gamma$-decay strengths is proportional to the B$_\sigma$ value because of isospin symmetry. 
Further details on the determination of the UCSs in light nuclei can be found in ref.~\cite{mat10}.
The l.h.s.\ of fig.~\ref{fig341} illustrates that the IS unit cross section at comparable mass number is about a factor of 20 smaller.        

\section{Electric dipole response}
\label{sec4}

\subsection{Polarizability, neutron skin and parameters of the symmetry energy}
\label{subsec41}

Besides their relevance for the test of nuclear structure models, experiments providing the E1 response over an energy range covering the PDR and the IVGDR are of particular interest for a determination of the static dipole polarizability $\alpha_\mathrm{D}$  \cite{boh81}
\begin{equation}
\label{eq411}
  \alpha_\mathrm{D}
  =
  \frac{\hbar c}{2\pi^{2}}  
  \int_0^\infty \frac{\sigma_\mathrm{abs}}{E_{\rm x}^{2}}{\rm d}E_{\rm x} 
  = 
  \frac{8 \pi}{9} \int_0^\infty \frac{\rm B(E1)}{E_{\rm x}}{\rm d}E_{\rm x}.
\end{equation}
While the low-energy E1 strength typically amounts to at most a few \% of the energy-weighted sum rule (EWSR) only, it is important for $\alpha_{\rm D}$ because of the inverse energy weighting.
As discussed in sec.~\ref{subsec42} the low-energy E1 strengths derived from different experimental methods show large variations, mainly because they depend on unknown properties of the $\gamma$ decay.
The importance of the (p,p$^\prime$) experiments lies in the fact that they measure the absorption probability and are thus independent of knowledge or modeling of the $\gamma$-decay branching ratios.  

Systematic studies within energy density functional (EDF) theory have shown that there are two experimental observables which can serve as a measure of the parameters of the symmetry energy, viz.\ the neutron-skin thickness and $\alpha_{\rm D}$ \cite{rei10,pie12}. 
The neutron-skin thickness $r_\mathrm{skin}=\langle{r}\rangle_{\rm n}-\langle{r}\rangle_{\rm p}$ is defined as
the difference of the neutron and proton root-mean-square radii $\langle{r}\rangle_{\rm n,p}$.
For stable nuclei, the proton radii are experimentally well known \cite{ang13} and the neutron radii can be inferred from measurements of the 
matter radius by coherent photoproduction of $\pi^{0}$ mesons \cite{tar14}, antiproton annihilation \cite{klo07,bro07} or proton elastic scattering \cite{sak17}.
A nearly model-independent determination of the neutron skin is possible by measuring the weak form factor of nuclei with parity-violating elastic electron scattering \cite{hor01}.  
Such an experiment has been performed for $^{208}$Pb  \cite{abr12} but the statistical uncertainties are still too large for serious constraints of the neutron skin.

\begin{figure}
\begin{center}
\resizebox{0.35\textwidth}{!}{%
  \includegraphics{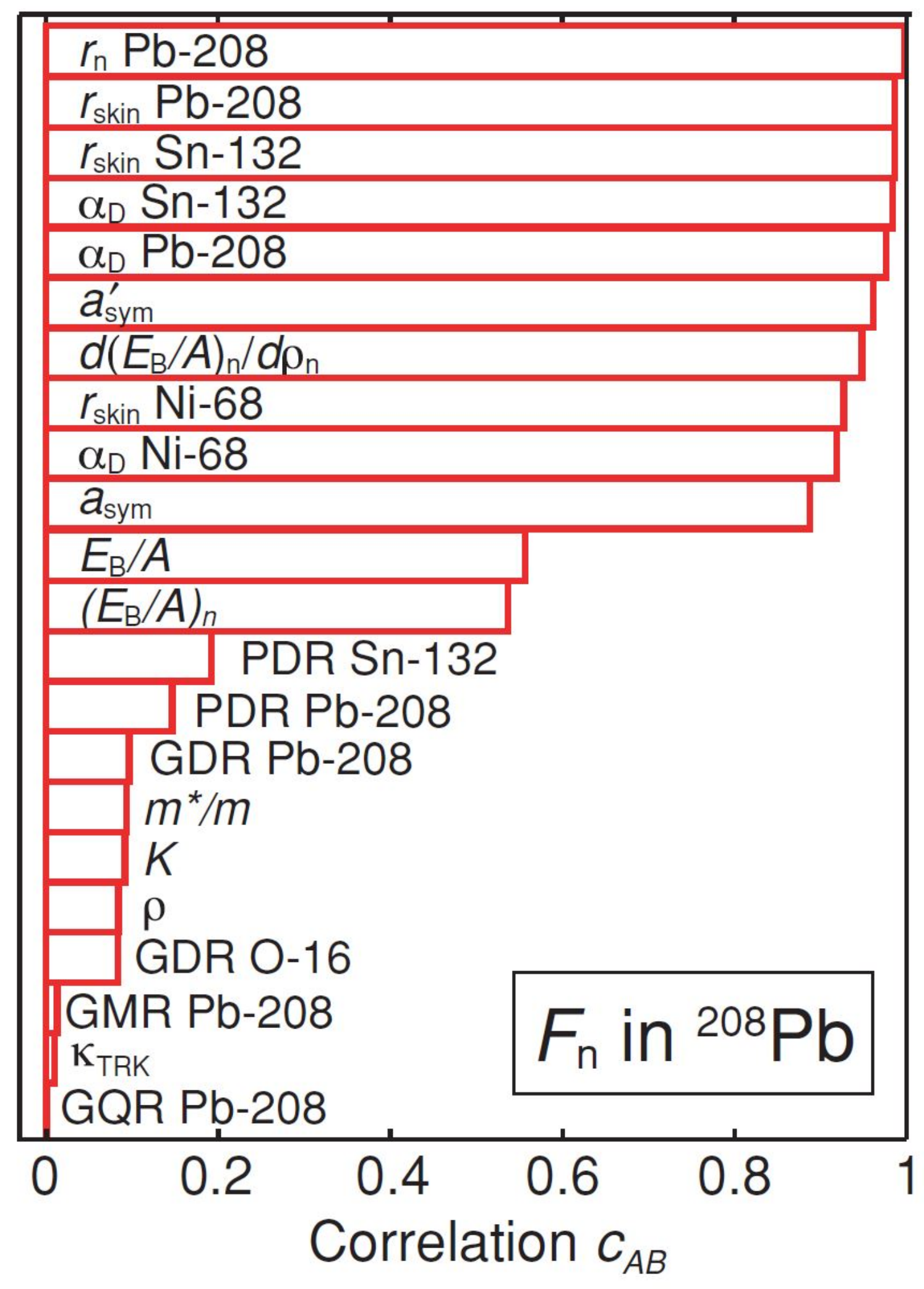}
%\vspace{5cm}       % Give the correct figure height in cm
}
\caption{
%Figure411. 
Correlation of various observables with the neutron form factor $F_{\rm n}$ ($q = 0.45$ fm$^{-1})$ in $^{208}$Pb.
For details see text.
Figure taken from ref.~\cite{rei10}.
}
\label{fig411}  
\end{center} 
\end{figure}

Figure~\ref{fig411} shows the correlation (defined in ref.~\cite{rei10}) of the neutron form factor at the momentum transfer of the experiment of ref.~\cite{abr12} with other bulk properties of nuclei for a systematic variation of EDF model parameters.
The correlation is almost perfect for the neutron-skin thickness and the polarizability of neutron-rich nuclei as well as the parameters of the symmetry energy (called $a_{\rm sym}$ and $D(E_{\rm B/A})/d\rho_{\rm n}$ here).
All those quantities can thus be viewed as good indicators of isovector properties of nuclei.
Bulk properties like isoscalar and isovector effective mass, incompressibility and
saturation density as well as the energies of ISGMR, IVGDR and ISGQR show a poor correlation.
The results of ref.~\cite{rei10} also suggest that the PDR is not a good observable to constrain the symmetry energy properties in contrast to other findings \cite{car10}.
However, this may depend on the choice of the EDF (SV-min \cite{klu09} for the results of fig.~\ref{fig411}). 
In particular, some relativistic mean field (RMF) functionals tend to predict a larger collectivity of the PDR and a higher correlation with the symmetry energy parameters \cite{kli07,roc18}. 

While calculations with different EDFs all confirm the strong correlation of $\alpha_{\rm D}$ with the neutron skin and the symmetry-energy parameters, the absolute values  can differ considerably \cite{pie12}.
It is therefore important to compare experimental values systematically for many nuclei before conclusions on $J$ and $L$ can be drawn.
The situation resembles the extraction of the nuclear-matter incompressibility from EDF calculations of the isoscalar giant monopole and dipole resonance energies \cite{gar18}.  
Values for $\alpha_{\rm D}$ from $(\vec{\rm p},\vec{\rm p}^{\prime})$ experiments have been extracted for $^{120}$Sn \cite{has15} and $^{208}$Pb \cite{tam11}.
In principle, the integral in eq.~(\ref{eq411}) runs to infinity, but the inverse energy weighting limits the contribution from energies above the giant resonance region to less than 5\%.
Furthermore, the strength at high energies is dominated by non-resonant processes (the so-called quasi-deuteron mechanism \cite{lev51}) which is not included in the nuclear-structure calculations.
It was thus suggested to neglect it for the comparison between experiment and theory \cite{roc15}. 

\begin{figure}
\begin{center}
\resizebox{0.5\textwidth}{!}{%
  \includegraphics{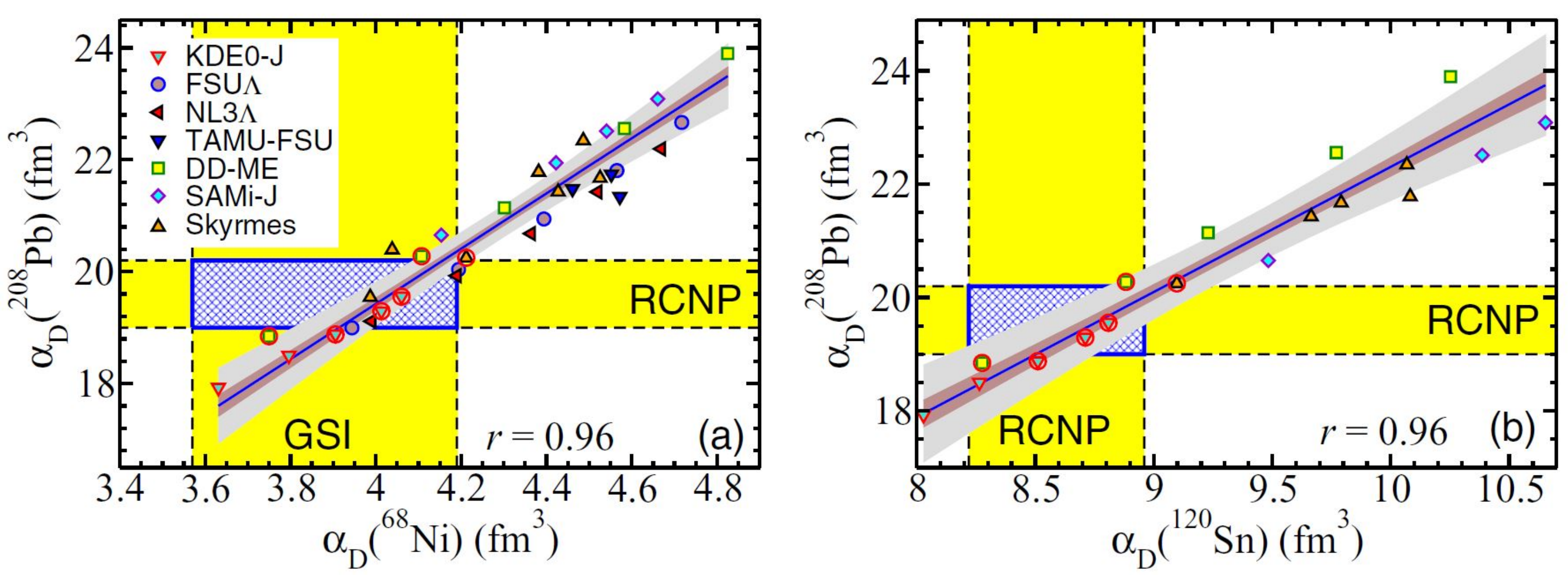}
%\resizebox{0.7\textwidth}{!}{%
 % \includegraphics{Figure412.pdf}
%\vspace{5cm}       % Give the correct figure height in cm
}
\caption{
%Figure412.
Comparison of the theoretical results from ref.~\cite{roc15} for the dipole polarizability of $^{68}$Ni, $^{120}$Sn and $^{208}$Pb with the experimental data. 
(a) $^{68}$Ni vs.\ $^{208}$Pb. 
(b) $^{120}$Sn vs.\ $^{208}$Pb.  
The symbols encircled in red correspond to models compatible with all three experimental results.
Figure taken from ref.~\cite{roc15}.}
\label{fig412}  
\end{center} 
\end{figure}

Roca-Maza {\it et al}.~\cite{roc15} used these corrected data together with a measurement of $\alpha_{\rm D}$ in the exotic nucleus $^{68}$Ni \cite{ros13} to constrain the wealth of EDFs (both of Skyrme and RMF type) by requesting reproduction of all three experimental values within error bars. 
Figure~\ref{fig412} illustrates that only a handful of functionals can simultaneously describe all data.
New constraints on $J$, $L$ and $r_\mathrm{skin}$ could be derived from averaging over the predictions of these selected calculations.  

\begin{figure}[b]
\begin{center}
\resizebox{0.4\textwidth}{!}{%
  \includegraphics{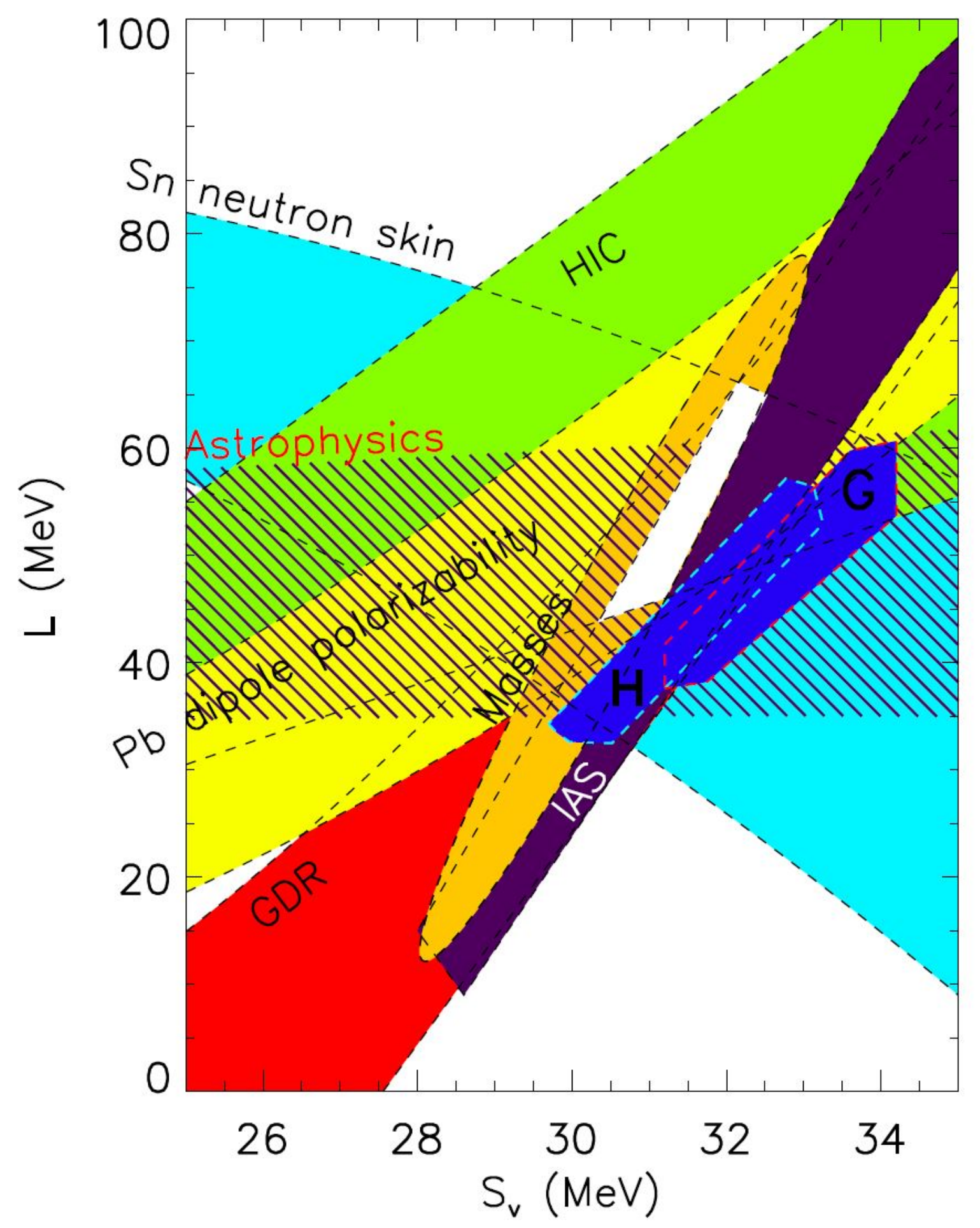}
%\vspace{5cm}       % Give the correct figure height in cm
}
\caption{%
Constraints of symmetry-energy parameters from different experimental methods.
Note that the leading parameter of the density expansion of the symmetry energy is called here $S_{\rm V}$ rather than $J$.
The result from the dipole polarizability is based on the data for $^{208}$Pb converted with the relation derived in ref.~\cite{roc13}.
G and H refer to the neutron-matter studies of refs.~\cite{heb10} and \cite{gan12}, respectively.
See ref.~\cite{lat16} for further discussion and references to the experimental data and interpretation. 
Figure taken from ref.~\cite{lat16}.}
\label{fig413}  
\end{center} 
\end{figure}
A variety of experimental methods permits to extract information on the symmetry energy with different dependence on the density \cite{roc18}.
Besides astrophysical observations of the mass-radius relation of neutron stars \cite{lat07}, constraints on the symmetry-energy parameters can be extracted from heavy-ion collisions, masses, isobaric analogue states, measurements of the difference of proton and neutron radii and from properties of the IVGDR and PDR (see, e.g., ref.~\cite{tsa12}).
Current results together with those of neutron-matter studies of refs.~\cite{heb10} and \cite{gan12} are summarized in fig.~\ref{fig413} \cite{lat16}.
Concerning the measurements of $\alpha_{\rm D}$, the droplet model predicts a relation of the form $\alpha_{\rm D} \cdot J \propto L$, which holds independent of the choice of a particular EDF functional \cite{roc13}.
Using the results of refs.~\cite{roc15,roc13}, one can derive the yellow band of allowed $J,L$ values from the polarizability result for $^{208}$Pb in fair agreement with the constraints from most other experimental methods.  
Since the extraction of the symmetry energy parameters is model-dependent, it is important to estimate the theoretical uncertainties for future improvement \cite{naz14}.    

\begin{figure}
\begin{center}
\resizebox{0.4\textwidth}{!}{%
  \includegraphics{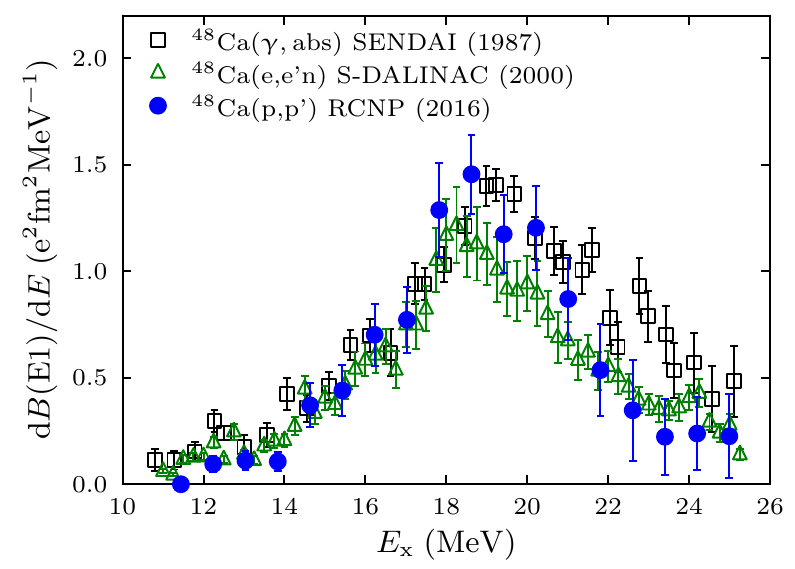}
%\vspace{5cm}       % Give the correct figure height in cm
}
\caption{
%Figure414.
Comparison of B(E1) strength distributions in $^{48}$Ca deduced from ref.~\cite{oke87} (black squares), ref.~\cite{str00} (green triangles), and from the (p,p$^\prime$) reaction (blue circles).
Figure taken from ref.~\cite{bir17}.}
\label{fig414}  
\end{center} 
\end{figure}

Recently, it has become possible to study such correlations in medium-mass nuclei with ab-initio calculations based on chiral effective field theory ($\chi$EFT) interactions \cite{hag16a,mio16}. 
By using a set of chiral two- plus three-nucleon interactions \cite{heb11,eks15} and exploiting correlations between $\alpha_D$ and the proton and neutron radii, Hagen {\it et al}.~\cite{hag16a} predicted the electric dipole polarizability and a neutron-skin thickness for $^{48}$Ca.
Since the analysis of ref.~\cite{roc15} also provides a prediction of $\alpha_{\rm D}$ from EDF theory, a study of the dipole polarizability in $^{48}$Ca is of special interest.
The B(E1) strength in $^{48}$Ca in the IVGDR energy region was extracted from (e,e$^\prime$n) \cite{str00} and photoabsorption \cite{oke87} experiments with conflicting results at higher excitation energies as shown in fig.~\ref{fig414} by the green triangles and black squares, respectively. 
The result of a corresponding (p,p$^\prime$) study \cite{bir17} shown as blue circles agrees within error bars with the data of ref.~\cite{str00}.
A possible explanation for the differences with ref.~\cite{oke87} is the unfolding procedure with a not well-known bremsstrahlung spectrum as input necessary for the extraction of $\sigma_{\rm abs}$, leading to sizable systematic uncertainties not reflected in the quoted error bars.

\begin{figure}
\begin{center}
\resizebox{0.45\textwidth}{!}{%
  \includegraphics{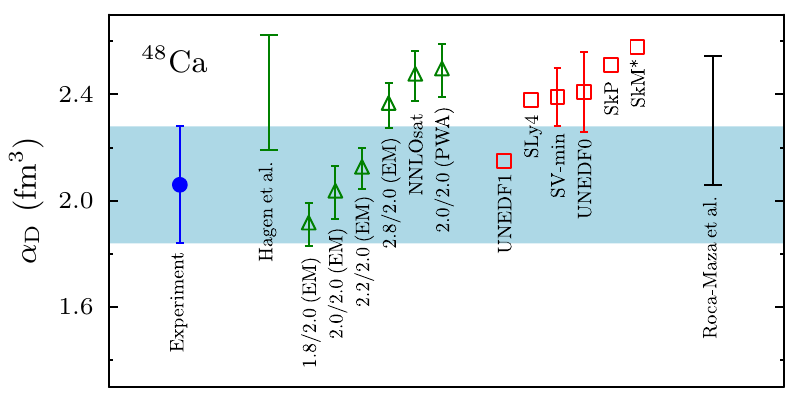}
%\vspace{5cm}       % Give the correct figure height in cm
}
\caption{
Experimental electric dipole polarizability in $^{48}$Ca (blue band) from ref.~\cite{bir17} and predictions from $\chi$EFT (green triangles) and EDFs (red squares, for details on the functionals see~\cite{hag16a}, error bars from ref.~\cite{nazpc}).  
The green and black bars indicate the $\chi$EFT predictions selected to reproduce the $^{48}$Ca charge radius~\cite{hag16a} and the range of $\alpha_D$ predictions~\cite{roc15} from EDFs 
providing a consistent description of polarizabilities in $^{68}$Ni~\cite{ros13}, $^{120}$Sn~\cite{has15}, and $^{208}$Pb~\cite{tam11}, respectively.
Taken from ref.~\cite{bir17}.}
\label{fig415}  
\end{center} 
\end{figure}

The dipole polarizability of $^{48}$Ca was extracted  from the combined (p,p$^\prime$) and  (e,e$^\prime$n) data plus  photoabsorption results of ref.~\cite{ahr75} on a natural Ca target at energies above the IVGDR region.
E1 strength below the neutron threshold was measured with the $(\gamma,\gamma^\prime)$
reaction~\cite{har02}.  
Unlike in heavy nuclei, where the low-energy strength is a sizable correction, the contribution is negligibly small in $^{48}$Ca (about 0.05\%). 
The experimental result is compared in fig.~\ref{fig415} with predictions from state-of-the-art $\chi$EFT (green triangles) and EDFs (red squares) functionals (for details see ref.~\cite{bir17} and references therein).
It provides a clear constraint on the parameters of the $\chi$EFT interaction and excludes most of the EDFs discussed in ref.~\cite{hag16a}, while the prediction based on EDFs describing the previously available $\alpha_{\rm D}$ data \cite{roc15} agrees within uncertainties.
Taking only the interactions and functionals in fig.~\ref{fig415} consistent with the experimental range implies a small neutron skin in $^{48}$Ca, where the lower values stem from the ab-initio calculations.
In the ab-initio results the small neutron skin is related to the strong $N=28$ shell closure leading to practically identical charge radii for $^{40}$Ca and $^{48}$Ca \cite{hag16a}.
The ab-initio results also provide symmetry-energy parameter ranges $J = 28.5 - 33.3$ MeV and $L =  43.8 - 48.6$ MeV.
These constraints are highly competitive, in particular the value of $L$ if compared with the constraints from different methods shown in fig.~\ref{fig413}.  
The EDF results show larger scattering, in particular for the density dependence, but this may be reduced if the analysis of \cite{roc15} is extended to include the $^{48}$Ca result.

\subsection{Pygmy dipole resonance}
\label{subsec42}

Extraction of the properties of the PDR in nuclei with neutron excess is a complex problem both experimentally and theoretically.
In models relating the PDR to the formation of a neutron skin, transitions are expected to have dominant neutron character.
Thus, excitation of the states forming the PDR is expected for both IS and IV probes.
Because of the dominance of Coulomb excitation at forward angles, the (p,p$^\prime$) experiments discussed here investigate the electromagnetic strength, i.e.\ the IV channel. 
$\alpha$ \cite{sav13}, heavy-ion \cite{bra15} and very recently also proton scattering  \cite{sav18} at beam energies, where nuclear excitation dominates over Coulomb excitation, have been used to measure the IS response.
Since these reactions are less selective, the identification of PDR transitions requires a coincidence measurement with $\gamma$ decay to the g.s., which serves as a unique identification of the E1 character.

Theoretically, PDR transitions can be identified by the specific form of their transition densities \cite{paa07} with proton and neutron contributions approximately in-phase (i.e.\ isoscalar) in the interior and a pronounced peak of the neutron density at the surface.
In contrast, transitions populating the IVGDR exhibit out-of-phase proton and neutron contributions (i.e.\ isovector).
However, in the energy region between the peaks of the PDR and IVGDR one also finds transitions with mixed character and there is no clear procedure for a decomposition \cite{tso08}.
The structure of the E1 transitions can only be inferred indirectly from the comparison of predictions of experimental observables like the IS and IV E1 strengths.  
 
\begin{figure} 
\begin{center}
\resizebox{0.5\textwidth}{!}{%
  \includegraphics{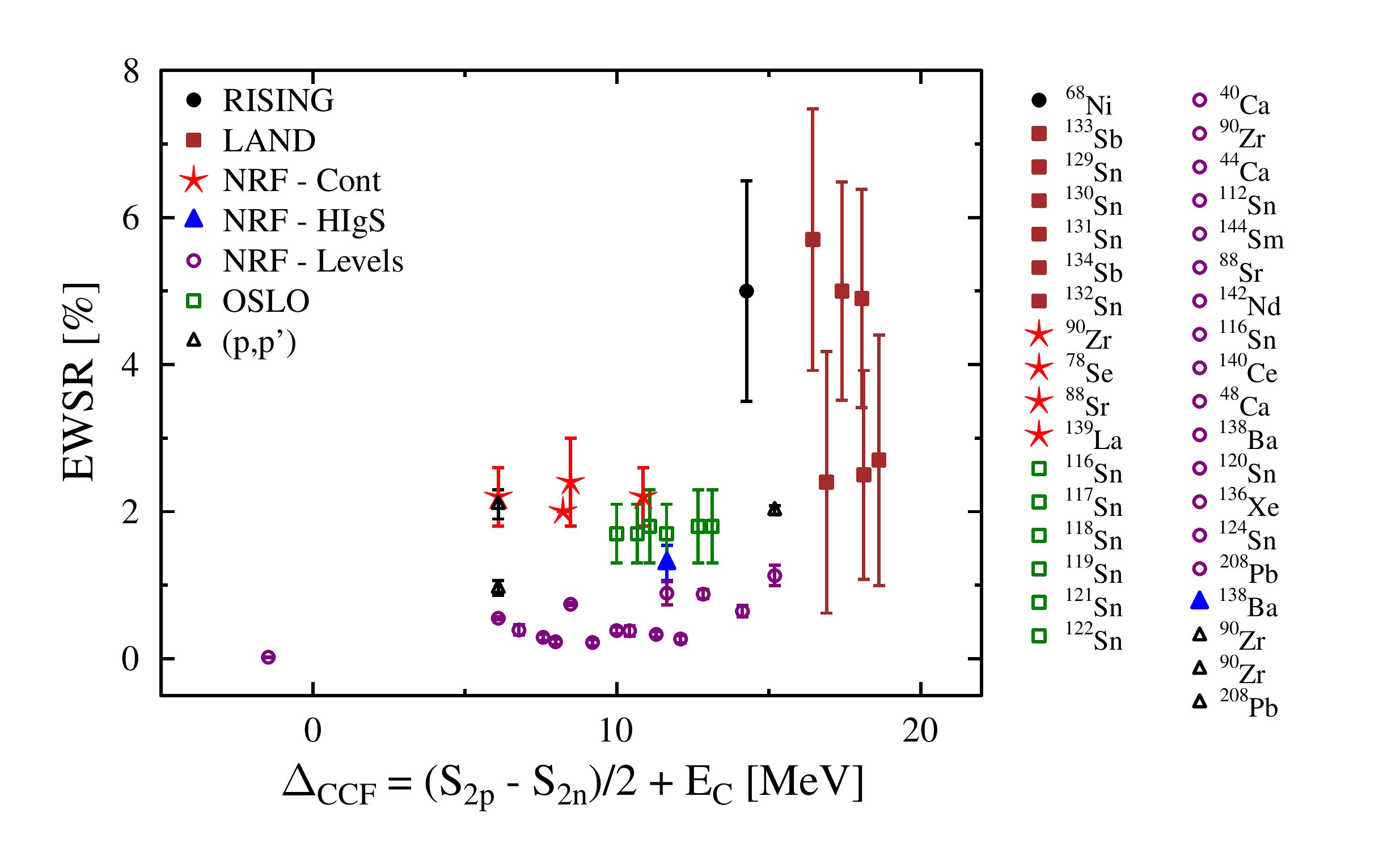}
%\vspace{5cm}       % Give the correct figure height in cm
}
\caption{
%Figure421.
Fraction of the E1 EWSR exhausted by the PDR as reported by various experiments as a
function of the parameter $\Delta_{\rm CCF}$ of the corresponding nuclei. 
The nuclei are listed separately for the individual methods. 
For details and a list of the original references see ref.~\cite{sav13}.
Figure taken from ref.~\cite{sav13}.}
\label{fig421}  
\end{center} 
\end{figure}

Most experimental techniques for the investigation of the IS or IV E1 strength are based on the measurement of $\gamma$ or (above threshold) particle decay.
An extraction of the corresponding reduced transition strengths requires knowledge of the branching ratio to the g.s., which -- apart from exceptional cases like $^{208}$Pb \cite{pol12} -- cannot be measured for the $\gamma$ decay.
Thus, a variety of correction methods has been developed based on branching ratios calculated within the statistical model \cite{rus09} or assuming different classes of states either decaying statistically or directly to the g.s.\ \cite{ton10}. 
Theoretically, a correlation of the total PDR strength with the neutron-skin thickness is expected in many models \cite{roc18}.
Figure \ref{fig421} presents a summary of the available data on the EWSR fraction of the E1 strength identified to belong to the PDR as a function of the parameter $\Delta_{\rm CCF}$
\begin{equation}
 \Delta_{\rm CCF} = \frac{S_{\rm 2p} - S_{\rm 2n}}{2} + E_{\rm C},
 \label{eq421}
\end{equation}
where $S_{\rm 2p}$ and $S_{\rm 2n}$ are the two-proton and two-neutron separation energies, respectively, and $E_{\rm C}$ is the height of the Coulomb
barrier at the nuclear radius (CFF = Coulomb-Corrected Fermi energy).
This quantity is well correlated with the neutron-skin thickness in EDF calculations \cite{sav13}. 

The experimental results do not clearly favor a correlation between the two quantities as theoretically predicted.
Also, results obtained at comparable values of $\Delta_{\rm CCF}$ in different experiments do show a large scattering.
(We note that the compilation shown in fig.~\ref{fig421} is several years old but inclusion of newer data would probably not change the picture significantly.)

\begin{figure}
\begin{center}
\resizebox{0.4\textwidth}{!}{%
  \includegraphics{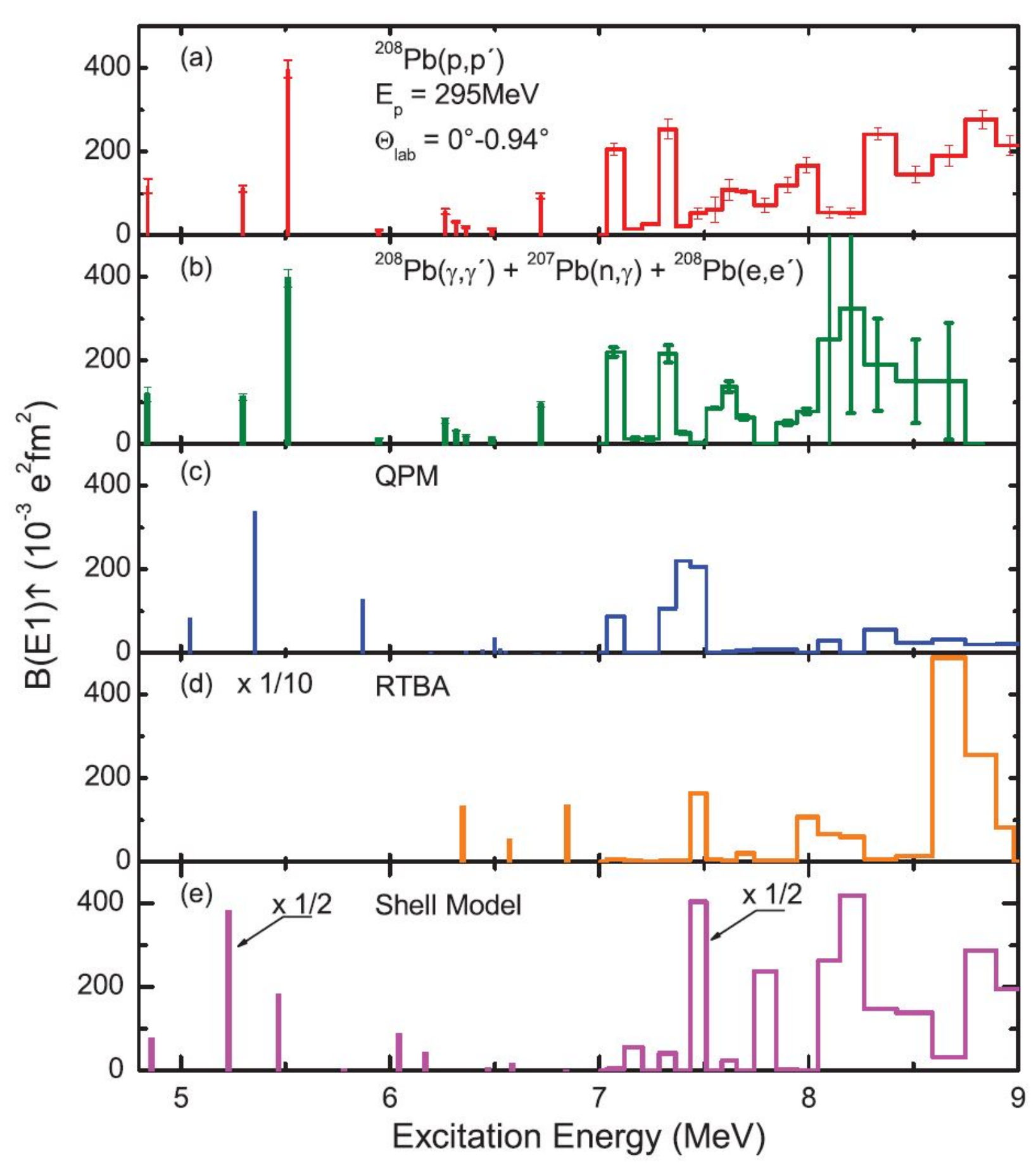}
%\vspace{5cm}       % Give the correct figure height in cm
}
\caption{
%Figure422.
B(E1) strength distributions in $^{208}$Pb between 4.8~MeV and 9~MeV from (a) the (p,p$^\prime$) \cite{tam11,pol12} and (b) $(\gamma,\gamma')$  \cite{rye02,end03,shi08,sch10} and (n,$\gamma$) \cite{koh87} experiments in comparison to theoretical calculations with (c) QPM \cite{rye02},  (d) RTBA  \cite{lit07} and (e) shell model \cite{sch10}. 
Note the scale reduction by a factor of 10 for the RTBA results.
Figure taken from ref.~\cite{pol12}.}
\label{fig422}  
\end{center} 
\end{figure}

Results from the (p,p$^\prime$) experiments play a potentially important role because they are independent of the knowledge of branching ratios.
The measured cross sections are directly related to the reduced B(E1) strengths, cf.\ eq.~(\ref{eq331}).
However, so far only a limited number of results are available.
For $^{208}$Pb, where decay to excited states can be neglected because of their large energy gap to the g.s., good agreement with $(\gamma,\gamma^\prime)$ studies is obtained \cite{pol12,sch10} as can be seen in figs.~\ref{fig422}(a) and (b).
It also represents a case of special interest for theory, because the double shell closure allows the application of refined methods including complex configurations beyond 1p-1h excitations considered in RPA.
These have been shown to be important for a realistic description of the low-energy E1 strength (see, e.g.,\ ref.~\cite{oze14}).

Figure \ref{fig422} compares the experimental B(E1) strengths between 4.8 and 9 MeV with theoretical results from (c) QPM \cite{rye02}, (d) RTBA \cite{lit07} and (e) shell-model \cite{sch10} calculations.
The QPM calculation, including a model space up to 3-phonon states, provides a good description up to 7.5 MeV. 
At higher excitation energies the strength is too small. 
The RTBA calculation includes only extra configurations of the type $1p1h\otimes{\rm phonon}$, and thus the fragmentation is insufficient to describe the data. 
(Note the reduction by a factor of 10 in fig.~\ref{fig422}(d) to bring it on the scale of the other results). 
The integrated strength up to 9 MeV is more than a factor two too large. 
However, the model has been further developed to include full multiphonon coupling \cite{lit15} and one may expect an improved description within such an approach.
The SM provides a good description with a fair agreement for the total strength but a slight indication that correlations beyond the present 2p-2h truncation of the model space are relevant.  

\begin{figure}
\begin{center}
\resizebox{0.4\textwidth}{!}{%
  \includegraphics{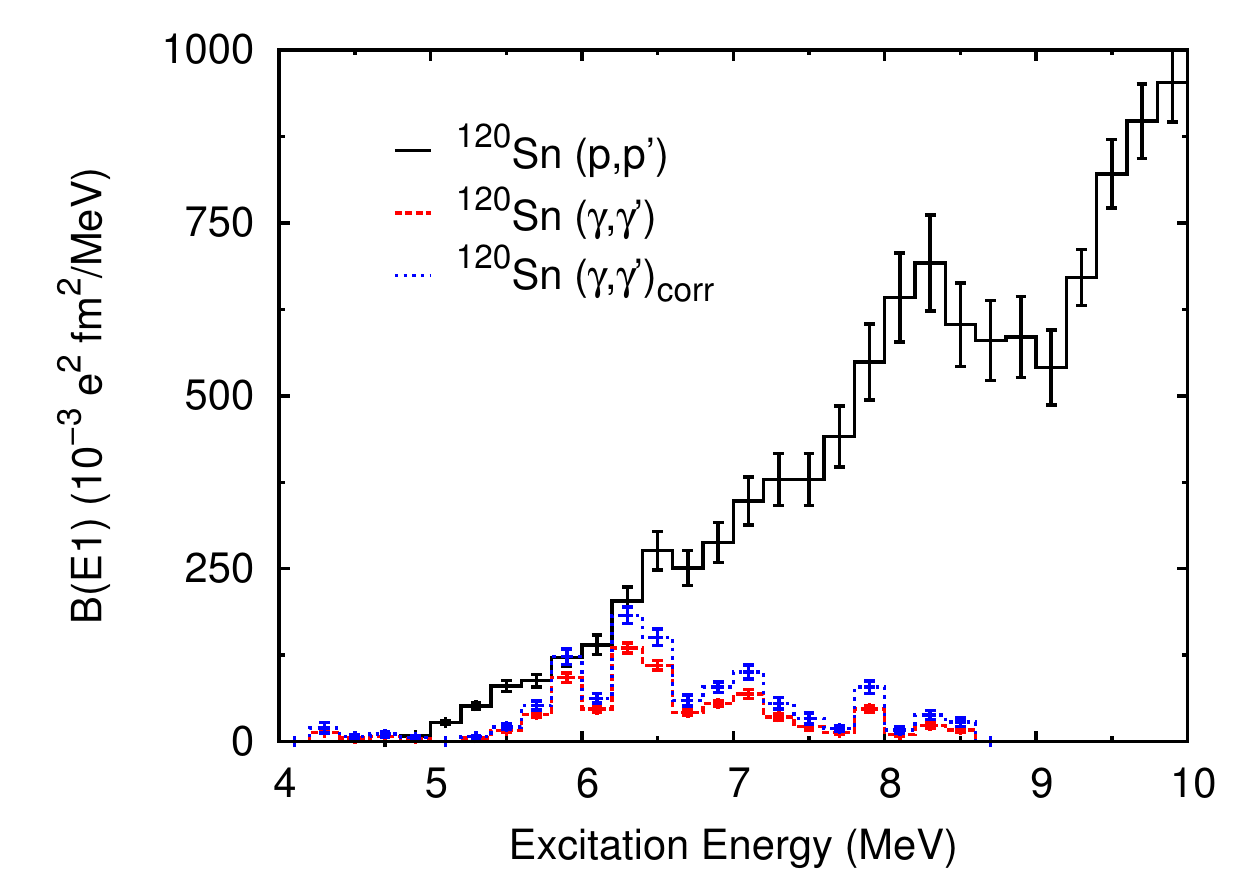}
%\vspace{5cm}       % Give the correct figure height in cm
}
\caption{
%Figure423.
B(E1) strength distribution of $^{120}$Sn in 200 keV bins from the (p,p$^\prime$) \cite{kru15} and $(\gamma,\gamma')$  \cite{oze14} reactions. 
$^{120}$Sn$(\gamma,\gamma')_\textnormal{corr}$ denotes the strength after correction for g.s.\  branching ratios from a statistical-model calculation~\cite{rus08} using the level-density parameterization of ref.~\cite{rau97}.
Figure taken from ref.~\cite{kru15}.}
\label{fig423}  
\end{center} 
\end{figure}

For the case of $^{120}$Sn, fig.~\ref{fig423} presents the B(E1) strength distributions in 200~keV bins deduced from  the (p,p$^\prime$) \cite{kru15} and $(\gamma,\gamma^\prime)$ \cite{oze14} experiments shown as black histogram and red circles, respectively.
Note that the $(\gamma,\gamma^\prime)$ measurement is limited to energies below the neutron threshold ($S_{\rm n}=9.1~$MeV). 
Both results show reasonable agreement at excitation energies around 6 MeV although the absolute B(E1) strength from the (p,p$^\prime$) data is on the average 30\% larger.
Above 6.3~MeV the results diverge. 
In the (p,p$'$) result a resonance-like structure at about 8.3 MeV is visible not seen in the $(\gamma,\gamma')$ data. 

The blue squares in fig.~\ref{fig423} depict the $(\gamma,\gamma')$ strength distribution after correction for the unknown ground-state branching ratios by a statistical-model calculation. 
The method is described in ref.~\cite{rus08} and details of the application to $^{120}$Sn in ref.~\cite{oze14}. 
The result shown uses the level-density parameterization of ref.~\cite{rau97}, but the dependence on the choice of the level-density model is weak \cite{oze14}.
Inclusion of the statistical-model correction brings both results in fair agreement in the energy region around 6 MeV.
Remaining differences at energies up to 6.5 MeV may be related to non-negligible unresolved strength in the $(\gamma,\gamma')$ data \cite{oze14} and for the region around 5.5 MeV to problems of the background subtraction in the (p,p$^\prime$) data. 
However, these cannot explain the sizable differences at higher excitation energies with more than an order-of-magnitude difference for the peak at 8.3 MeV.
While the increase of the E1 strength due to the statistical-model corrections can be large in more deformed nuclei \cite{rus08,mas14}, in the semi-magic nucleus $^{120}$Sn it is limited to about 40\% independent of the choice of parameters.

The large discrepancy suggests the existence of two classes of $1^-$ states with distinct wave functions, viz.\ a number of selected states, in the present case around 6.5 MeV, with large g.s.\ decay probability \cite{loe16}, while the larger part of the B(E1) strength seems to come from states with non-negligible ground-state decay width $\Gamma_0$ but ground-state branching ratios decreasing with excitation energy \cite{sch13}.
The bump around 6.5 MeV may be considered as the 'true' PDR.
This is supported by investigations of the isospin structure using isoscalar probes \cite{end10,pel14}. 
Although these were performed for $^{124}$Sn, the similarity of the IV strength distributions in the stable Sn isotope chain \cite{bas18} suggest a comparable picture for $^{120}$Sn.
The nature of the pronounced peak around 8.3 MeV is presently unclear.
A possible interpretation as local concentration on the low-energy tail of the GDR is discussed in ref.~\cite{oro98}.

A similar comparison of dipole strengths from (p,p$^\prime$) and $(\gamma,\gamma^\prime)$ experiments for the deformed nucleus $^{96}$Mo is presented in sec.~\ref{subsec72}.
In this case, the E1 strength deduced from the $(\gamma,\gamma^\prime)$  data is larger but dominated by the statistical-model correction \cite{rus09}.

\subsection{IVGDR in deformed nuclei}
\label{subsec43}

As pointed out above, the (p,p$^\prime$) experiments do not only provide information on the PDR but also on the IVGDR.
While good agreement with photoabsorption cross sections deduced from  ($\gamma$,xn) experiments was reported for the closed-shell nuclei $^{120}$Sn \cite{has15} and $^{208}$Pb \cite{tam11}, unexpected differences are observed for deformed nuclei.

The chains of stable even-even neodymium and samarium isotopes are known to comprise a transition from spherical to deformed ground states for heavier isotopes and thus represent an excellent test case to study the influence of deformation on the properties of the IVGDR. 
The simultaneous measurement of the partial photonuclear cross sections $\sigma(\gamma,\mathrm{n})+\sigma(\gamma,\mathrm{pn})$ and $\sigma(\gamma,2\mathrm{n})$ at Saclay using a monochromatic photon beam provides total photoabsorption cross sections in the IVGDR energy region for the stable even-even neodymium \cite{car71} and samarium \cite{car74}  isotopic chains.
The results show a width increasing with deformation evolving into a pronounced double-hump structure in the most deformed nuclides, considered to be a textbook example of $K$ splitting  \cite{boh75} owing to oscillations along the different axes of the quadrupole-deformed ground state. 

\begin{figure}
\begin{center}
\resizebox{0.38\textwidth}{!}{%
  \includegraphics{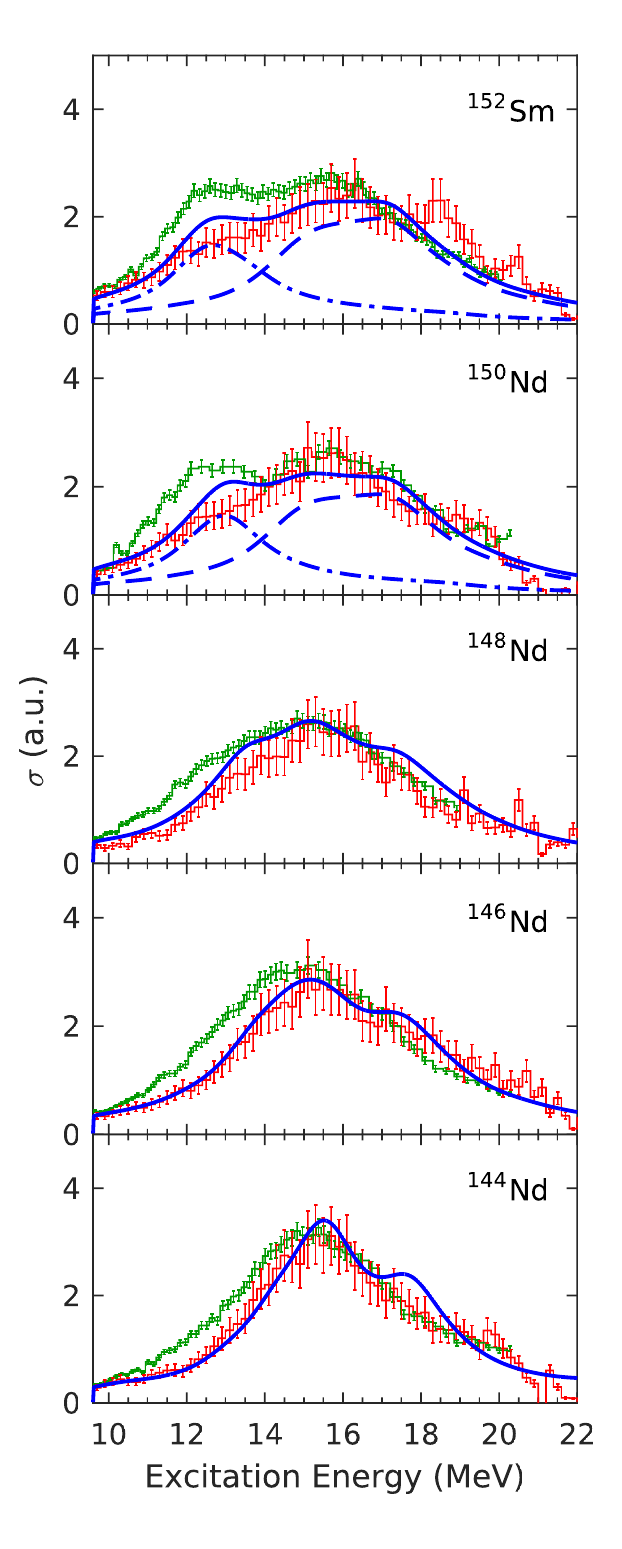}
%\vspace{5cm}       % Give the correct figure height in cm
}
\caption{
%Figure431. 
Photoabsorption cross sections deduced from the (p,p$^\prime$) data (red histograms) \cite{don18} normalized to the maximum of the preexisting ($\gamma$,xn) results (green histograms) \cite{car71,car74}. 
The blue lines show the results of RPA calculations with the SLy6 force described in ref.~\cite{don18}  (dash-dotted: $K=0$ part, dashed: $K= 1$ part, solid: sum).
Figure taken from ref.~\cite{don18}.}
\label{fig431}  
\end{center} 
\end{figure}

New photoabsorption cross sections for $^{144,146,148,150}$Nd and $^{152}$Sm extracted from 200 MeV proton scattering experiments at iThemba LABS have been reported \cite{don18} with results differing significantly from those of refs.~\cite{car71,car74}. 
No MDA could be performed with these data, thus the main background contributions in the energy region of the IVGDR from excitation of the ISGMR and ISGQR were subtracted using experimental strength functions \cite{ito03} (for details see ref.~\cite{don18}).
Since the present configuration of the $0^\circ$ facility at iThemba LABS \cite{nev11} allows only to determine the energy dependence but not the magnitude of the photoabsoprtion cross sections, the results are plotted in fig.~\ref{fig431} normalized to the maximum of the Saclay data to facilitate a comparison of the evolution of the shape of the IVGDR with increasing deformation.

Carlos et al.~\cite{car71,car74} (green histograms) observed a spreading of the IVGDR as the nuclei become softer followed by a splitting of the IVGDR into two distinct dipole modes for $^{150}$Nd and $^{152}$Sm, which were interpreted as $K=0$ and $K=1$ components.
The equivalent photoabsorption cross sections from the new work (red histograms) display a similar trend, i.e., a general broadening of the IVGDR with increasing deformation.
However, for the most deformed $^{150}$Nd and $^{152}$Sm, the resonance becomes skewed with increased strength on the low-energy side, but no splitting into two distinct components is observed. 

Model calculations of the IVGDR evolution in the Nd and Sm isotope chains were performed 
in the framework of the Skyrme Separable Random-Phase Approximation (SSRPA) approach \cite{nes06} using the SLy6 \cite{cha98} interaction which was shown to provide a good description of the IVGDR in medium-heavy, deformed nuclei \cite{kle08}.
The resulting photoabsorption cross sections are shown in fig.~\ref{fig431} as blue lines.
They are normalized to the data at the high-energy flank of the IVGDR, where both sets of data agree reasonably well.
The separation into $K = 0$ and $K = 1$ components is additionally shown for the most deformed nuclei, $^{150}$Nd and $^{152}$Sm as dash-dotted and dashed lines, respectively.
For the spherical and transitional nuclei, $^{144,146,148}$Nd, the calculations are in better agreement with the present results, i.e. favoring smaller cross sections on the low-energy flank.  
For $^{150}$Nd and $^{152}$Sm, the SSRPA results display a double-hump structure, but again with a lower $K = 0$ component than observed in the Saclay results and total cross sections somewhat closer to the present data.

\begin{figure}[b]
\begin{center}
\resizebox{0.45\textwidth}{!}{%
  \includegraphics{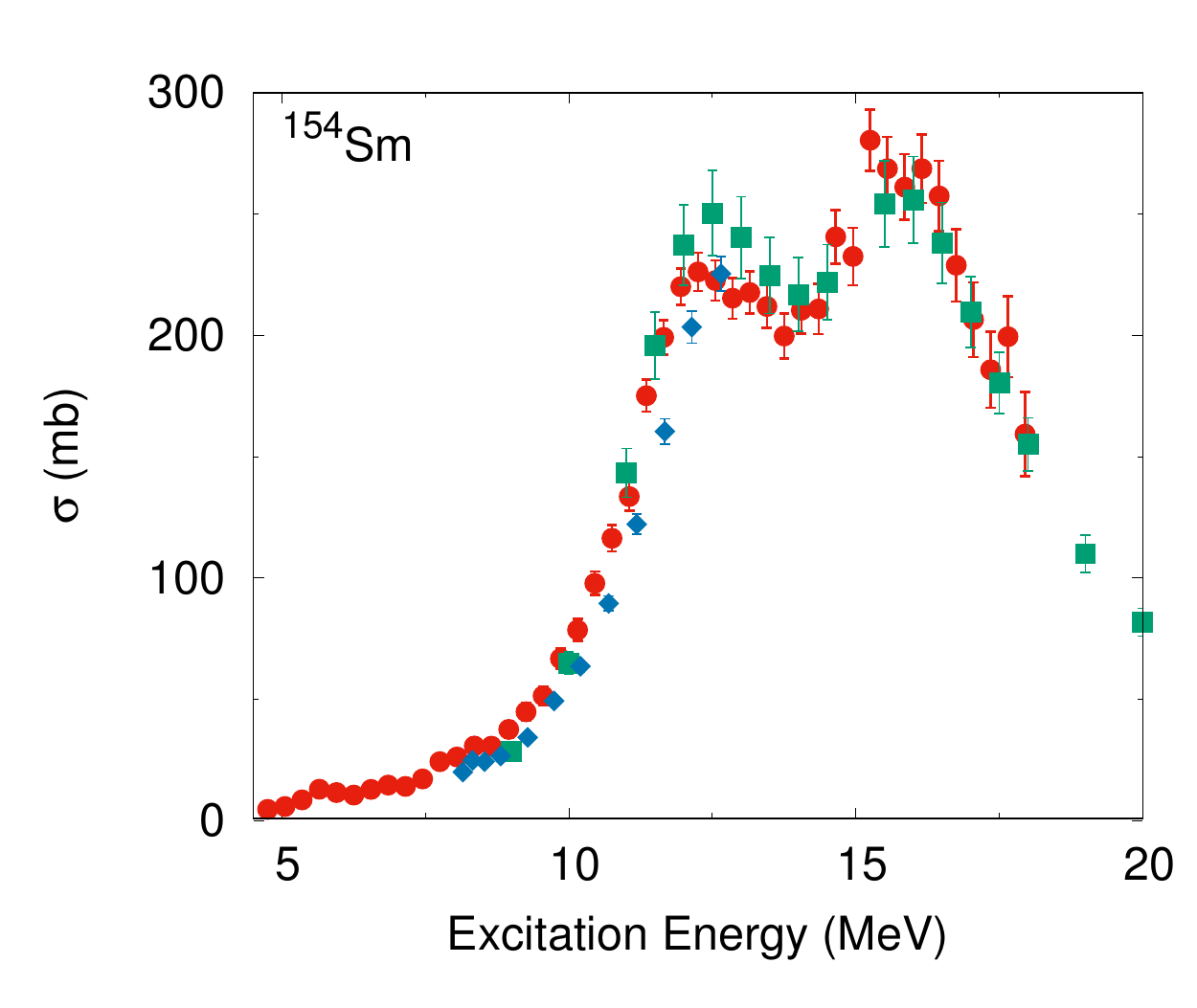}
%\vspace{5cm}       % Give the correct figure height in cm
}
\caption{
%Figure432.
Comparison of photoabsorption cross sections deduced from the $^{154}$Sm(p,p$^\prime$) (red) \cite{kru14}, ($\gamma$,xn) (green) \cite{car74} and ($\gamma$,n) (blue) \cite{fil14} reactions.
%Taken from ref.~\cite{kru18}.
}
\label{fig432}  
\end{center} 
\end{figure}
One possible interpretation of the unexpected result is that it may be related to the special structure of these two nuclei which are predicted to lie near the critical point \cite{cas01} of a shape phase transition from spherical to quadrupole-deformed ground states characterized by a soft potential in the $\beta$ degree of freedom \cite{iac01}.
The corresponding shape fluctuations may thus enhance the width of the resonance peaks which, in turn, may hinder a clear discrimination of the $K=0$ and 1 branches.
This cannot be modeled presently in the SSRPA calculations, which start from a well-defined $\beta$ deformation.
It is therefore of interest to extend the experiments to $^{154}$Sm with a well-defined prolate-deformed g.s.\ predicted by all models.  
Figure \ref{fig432} compares the result of a $^{154}$Sm(p,p$^\prime$) experiment performed at RCNP (red) \cite{kru14}  with photoabsorption cross sections from ($\gamma$,xn) (green) \cite{car74} and ($\gamma$,n) (blue) \cite{fil14} experiments.
The (p,p$^\prime$) result does show a double-hump structure and agrees within error bars, but with a slightly reduced maximum of the $K = 0$ peak and slightly increased $K = 1$ peak compared to the Saclay data.
One should note that the analysis of the (p,p$^\prime$) data is based on PTA data and still contains the contributions from the ISGMR and ISGQR.
A corrected analysis using the subtraction procedure described in ref.~\cite{don18} is in preparation. 
The finding of a reduced $K = 0/K =1$ ratio is qualitatively consistent with a global reanalysis of data taken with the Saclay method, which indicates that the ($\gamma$,n) cross sections are systematically too large and the ($\gamma$,2n) cross sections too small \cite{var14}.
We also note that new measurements of ($\gamma$,xn) cross sections for the deformed nuclei $^{181}{\rm Ta}$ and $^{159}{\rm Tb}$ with laser-Compton backward-scattering gamma rays at New SUBARU \cite{utspc} show different relative strengths between the two $K$ components compared with previous results \cite{ber75}.

\section{Spin-magnetic dipole response}
\label{sec5}

\subsection{Spin-M1 resonance in heavy nuclei}
\label{subsec51}

\begin{figure}
\begin{center}
\resizebox{0.4\textwidth}{!}{%
  \includegraphics{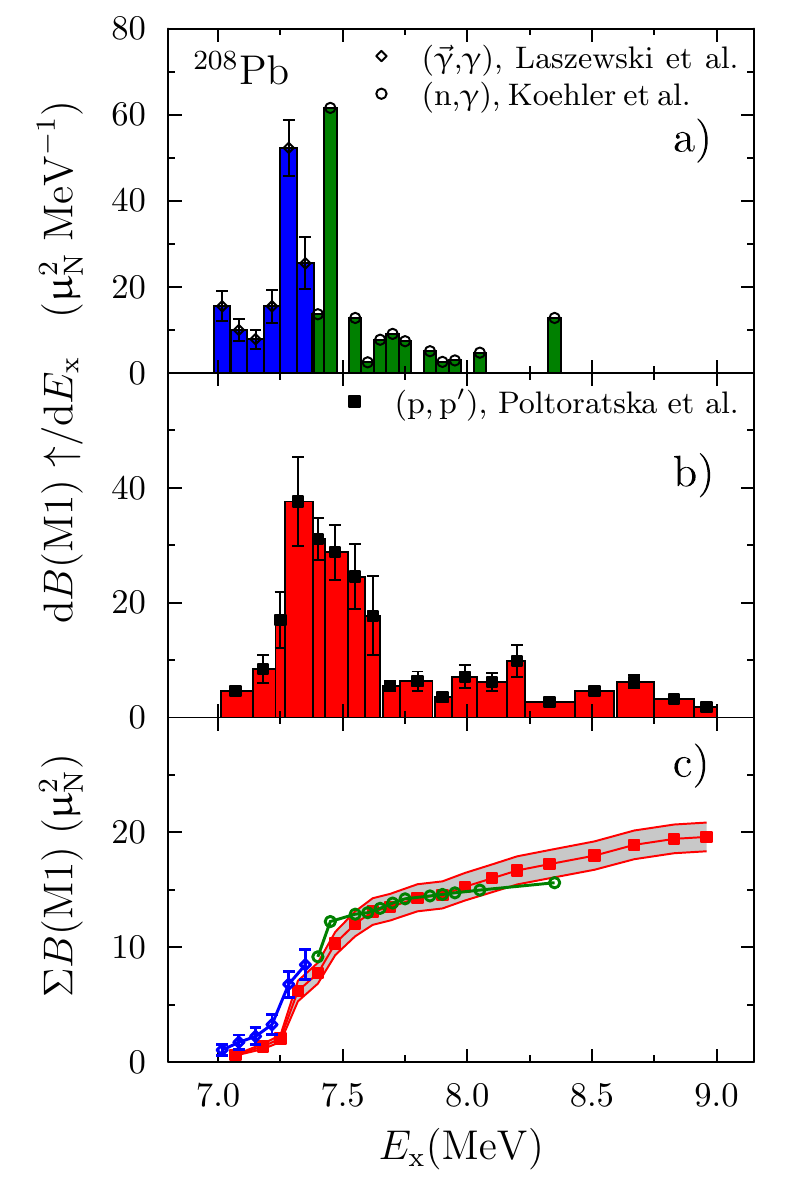}
%\vspace{5cm}       % Give the correct figure height in cm
}
\caption{
%Figure511.
B(M1) strength distribution in $^{208}$Pb between 6.5 and 9 MeV from a) refs.~\cite{koh87,las88} and b) the M1 proton scattering cross sections of ref.~\cite{pol12} applying the method described in ref.~\cite{bir16}. 
c) Comparison of the running sums of a) and b).
Figure taken from ref.~\cite{bir16}.
}
\label{fig511}  
\end{center} 
\end{figure}
A test case for the approach presented in sec.~\ref{subsec34} is $^{208}$Pb, where information on the electromagnetic M1 strength is claimed to be complete \cite{las88}.
Figure~\ref{fig511}(a) presents the combined data of $(\vec{\gamma},\gamma')$ \cite{las88} and  (n,$\gamma$) \cite{koh87} experiments providing information below and above threshold, respectively.
Note that ref.~\cite{las88} claims some M1 strength below 7 MeV, but the error bars are close to 100\%.
Furthermore, it is excluded by subsequent NRF experiments \cite{rye02,shi08,sch10} with much improved sensitivity. 
The B(M1) strength distribution extracted from the M1 part resulting from the MDA of the $^{208}$Pb(p,p$^\prime$) cross sections \cite{pol12} is given in Fig.~\ref{fig511}(b).
The agreement of the energy distribution and total strength is excellent. 
The seeming discrepancies around 7.5 MeV result from the different binning of the two data sets, as one can see from the running sum [Fig.~\ref{fig511}(c)].
The summed strength up to 8 MeV in ref.~\cite{las88} of $14.8^{+1.9}_{-1.5}~\mu_\mathrm{N}^2$  is to be compared with 16.0(1.2)~$\mu_\mathrm{N}^2$ from the $(p,p')$ data. 
In the energy region between 8 and 9 MeV, previous experiments had limited sensitivity (cf.\ fig.~6 in ref.~\cite{koh87}), which most likely explains why the strength seen in the $(p,p')$ experiment was missed.
A total strength of $\sum B(\mathrm{M1})= 20.5(1.3)$~$\mu_\mathrm{N}^2$ is deduced for the spin-M1 resonance in $^{208}$Pb \cite{bir16}.

\begin{figure}[b]
\begin{center}
\resizebox{0.4\textwidth}{!}{%
  \includegraphics{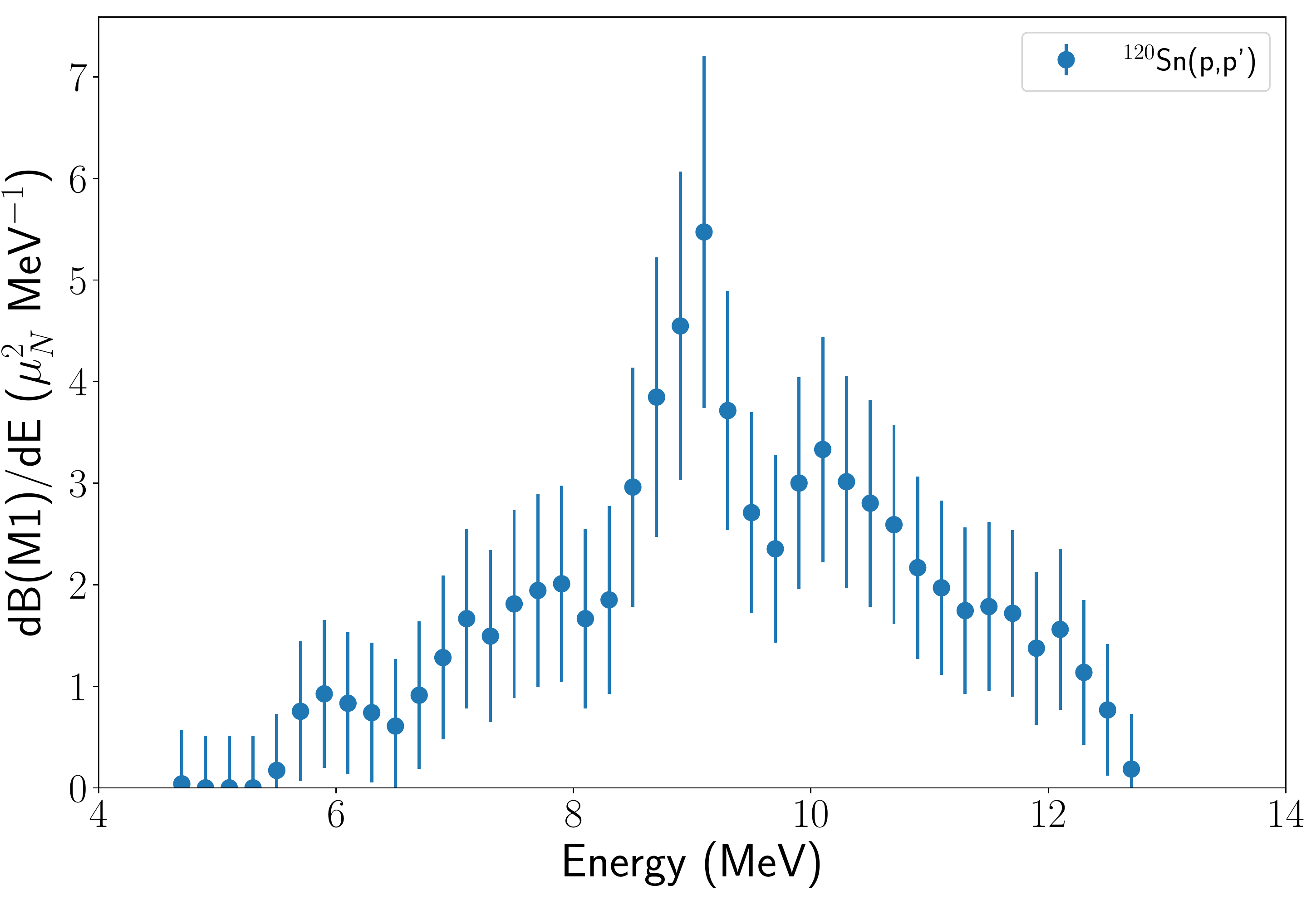}
%\vspace{5cm}       % Give the correct figure height in cm
}
\caption{
%Figure512.
B(M1) strength distribution in $^{120}$Sn from the M1 (p,p$^\prime$) cross sections deduced with the MDA described in ref.~\cite{kru15} and applying the UCS method explained in sec.~\ref{subsec34}.
}
\label{fig512}  
\end{center} 
\end{figure}

The result of the $^{120}$Sn(p,p$^{\prime}$) experiment is presented in fig.~\ref{fig512}.
The distribution peaks at about 9 MeV with a width of about 1.5 MeV (FWHM), which is comparable to $^{208}$Pb but with broader tails.  
The total strength amounts to $\sum {\rm B(M1)} = 14.7 (1.2)$ $\mu_N^2$.

\begin{figure}
\begin{center}
\resizebox{0.4\textwidth}{!}{%
  \includegraphics{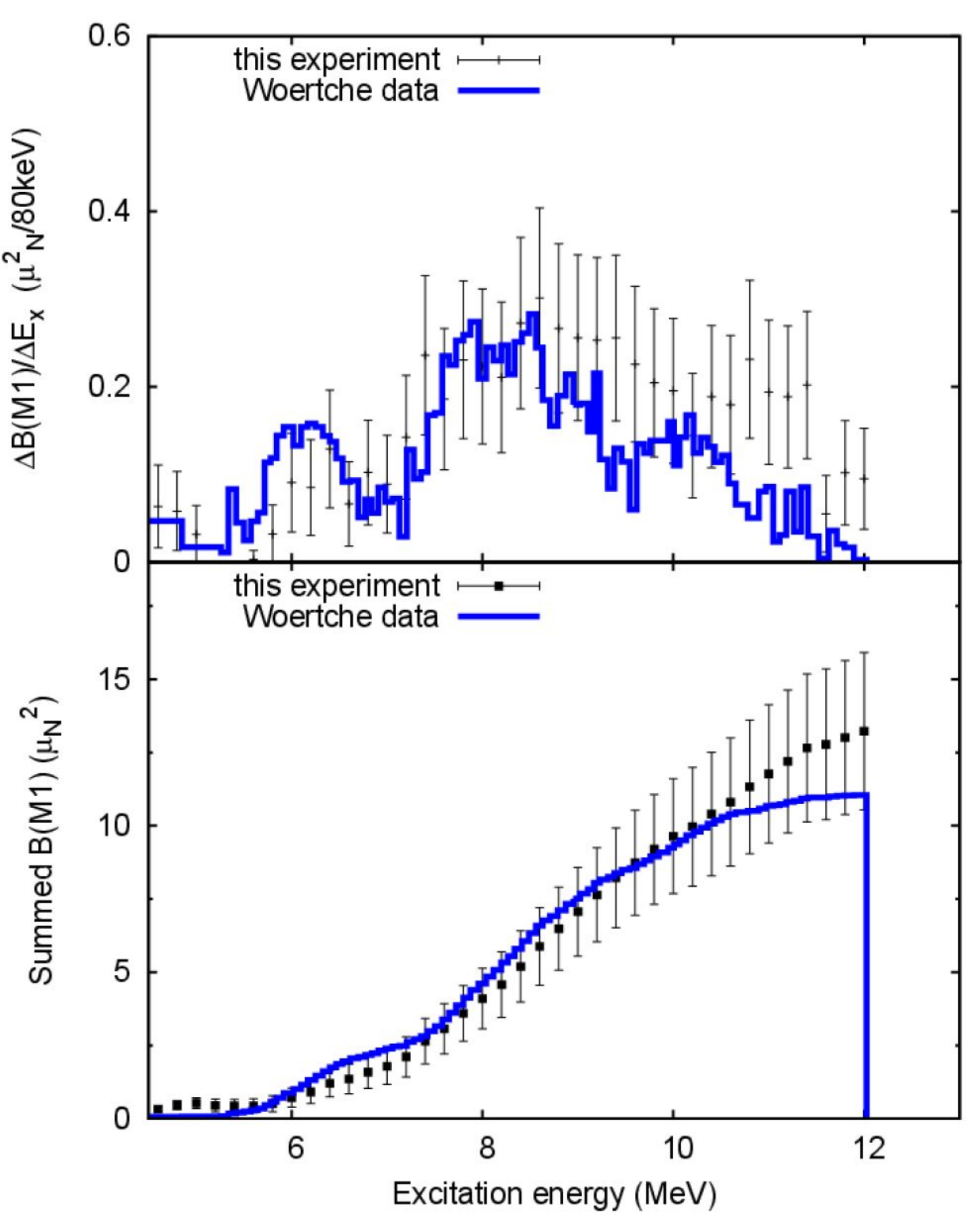}
%\vspace{5cm}       % Give the correct figure height in cm
}
\caption{
%Figure513.
Top: B(M1) strength distribution from a  $^{154}$Sm(p,p$^{\prime}$) experiment at RCNP  deduced with the PTA described in ref.~\cite{kru14} and applying the UCS method explained in sec.~\ref{subsec34} compared with the results of a forward-angle (p,p$^{\prime}$) experiment at TRIUMF \cite{hey10,woe94}.
Bottom: Running sums.}
\label{fig513}   
\end{center} 
\end{figure}

The spin-M1 resonance in heavy deformed nuclei has been studied in forward-angle (p,p$^{\prime}$) experiments at TRIUMF and shows a broad structure in the excitation region $E_{\rm x} \approx 5 - 10$ MeV \cite{hey10,fre90,woe94}. 
A comparison of their result for $^{154}$Sm shown in fig.~\ref{fig513} is in good agreement with those from the experiment at RCNP \cite{kru14}.
For the latter data only a PTA analysis in 200 keV bins was performed.
The large error bars are dominated by the limited statistics.
The bottom part of fig.~\ref{fig513} compares the running sums. 
No error bars are shown for the TRIUMF result.
In this case, the experimental errors are certainly below 10\%, but there is a hard-to-quantify additional systematic uncertainty from the model chosen to describe the quasi-free background in the MDA \cite{woe94}.

\subsection{Quenching of the isovector strength}
\label{subsec52}

It is well known that the axial coupling constant in weak decay -- and thus the GT strength -- is modified in the nuclear medium with respect to its free value \cite{tow87}.
Studies of the GT strength in CE reactions have shown that the Ikeda sum rule for nuclei with mass $A \geq 50$  is systematically quenched by a factor of two with respect to shell-model or QRPA predictions \cite{ich06}.
A similar behavior is expected for the isospin-analog, the isovector spin-M1 (IVSM1) resonance.     
GT strengths derived from the IVSM1 strength with the aid of eq.~(\ref{eq349}) do show significant differences from the results of CE reactions in lighter nuclei.
These have been interpreted to arise from the different role of meson-exchange currents in vector and axial coupling \cite{ric90,lue96,vnc97}.
However, in $fp$-shell nuclei a quenching with a factor of two comparable to GT $\beta$ decay strength \cite{mar96} has been found for the IVSM1 resonance \cite{vnc98} when comparing to shell-model calculations.

In heavy nuclei with shell closures, recent polarized photon scattering experiments at HI$\gamma$S have been able to extract the detailed spin-M1 strength distribution below neutron threshold \cite{ton10,rus13,ton17}.
A comparable quenching is observed in this energy region with respect to QPM calculations.
Nevertheless, some uncertainty remains because the IVSM1 resonance normally extends across the neutron threshold.
The (p,p$^\prime$) data on $^{208}$Pb and $^{120}$Sn discussed in sec.~\ref{subsec51}  demonstrate that indeed a considerable fraction of the full IVSM1 strength lies above threshold.
The same is true in heavy deformed nuclei \cite{hey10}.
 
The IVSM1 resonance in $^{48}$Ca has long been considered as a reference case for the quenching of the spin-isospin response because its strength is largely concentrated in the excitation of a single state at 10.23 MeV.
It was first observed in inelastic electron scattering \cite{ste83} with a reduced transition strength B(M1) = 3.9(3)~$\mu_N^2$.
Its $[\nu 1f_{7/2}^{-1} 1f_{5/2}]$ particle-hole structure is particularly simple allowing to investigate the role of second-order effects (see, e.g.,\ ref.~\cite{tak88} and references therein).
Furthermore, shell-model calculations including the $fpg$ valence space are possible providing all correlations due to the one-body interaction. 

\begin{figure}
\begin{center}
\resizebox{0.45\textwidth}{!}{%
  \includegraphics{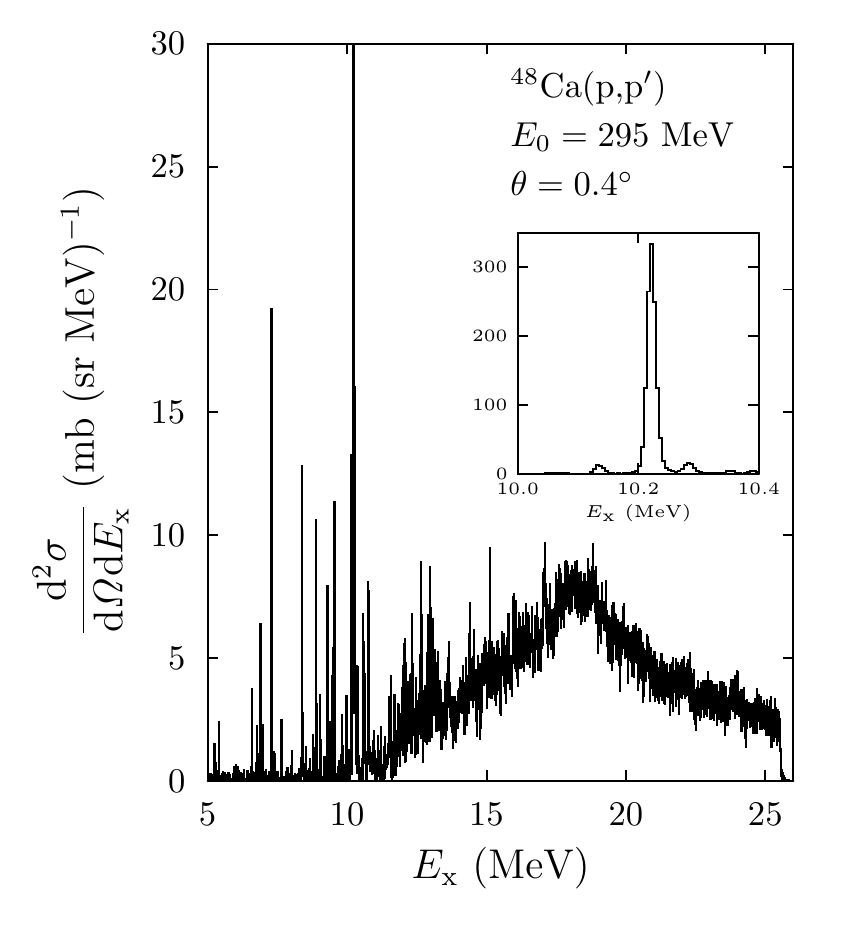}
%\vspace{5cm}       % Give the correct figure height in cm
}
\caption{
%Figure521.
Spectrum of the $^{48}$Ca(p,p$'$) reaction at $E_0 = 295$~MeV and $\theta_{\rm lab} = 0.4^\circ$.
The inset shows the spectral region in the vicinity of the dominating transition at $E_{\rm x} =10.23$~MeV. 
Note the factor-of-ten difference in the Y axis.
Figure taken from ref.~\cite{bir16}.
}
\label{fig521}  
\end{center} 
\end{figure}

Recently, a new result from a $^{48}$Ca($\gamma$,n) measurement at the HI$\gamma$S facility has been reported \cite{tom11}. 
The deduced B(M1) strength B(M1) of 6.8(5)~$\mu_N^2$ was almost two times larger.
If correct, this value would question our present understanding of quenching in microscopic models, since the consistent shell-model quenching factors of the IVSM1  in $fp$-shell nuclei (including $^{48}$Ca) \cite{vnc98} and GT $\beta$ decay strength \cite{mar96}, successfully applied to the modeling of weak interaction processes in stars \cite{lan03}, would be challenged. 
The $^{48}$Ca(p,p$^\prime$) experiment discussed in sec.~\ref{subsec41} provides an independent result to resolve this discrepancy \cite{bir16}. 
The excitation of the $1^+$ state at 10.23 MeV is by far the strongest line in all spectra as shown by way of example in fig.~\ref{fig521} for $\theta_{\rm lab} = 0.4^\circ$.  
The broad structure peaking at about 18.5 MeV is identified as the IVGDR \cite{bir17}.
The inset of fig.~\ref{fig521} expands the energy region around the 10.23 MeV peak (note the factor-of-ten difference in the Y axis).

The pure neutron character of the strong M1 transition in $^{48}$Ca predicted by the independent particle model was experimentally demonstrated by comparing with its excitation using an IS probe ($\pi$ scattering) \cite{deh84}.
In this particular case, the $\vec{\sigma}$ term of the electromagnetic operator of M1 transitions [eq.~(\ref{eq348})] cannot be neglected for the extraction of the $B(\mathrm{M1})$ value because of the interference term.   
The IS contribution to the (p,p$^\prime)$ cross sections was estimated by fitting the theoretical angular distributions for IS and IV $1f_{7/2} \rightarrow 1f_{5/2}$ transitions shown in fig.~\ref{fig342} to the experimental angular distribution. 

\begin{figure}
\begin{center}
\resizebox{0.45\textwidth}{!}{%
  \includegraphics{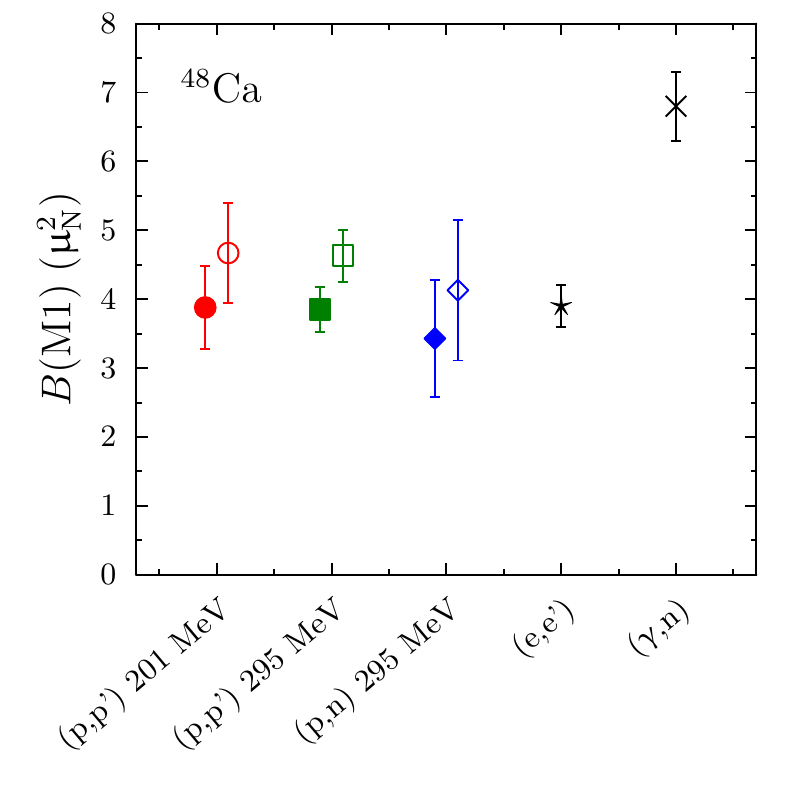}
%\vspace{5cm}       % Give the correct figure height in cm
}
\caption{
%Figure522.
B(M1) strengths for the transition to $J^\pi = 1^+$ state at 10.23 MeV in $^{48}$Ca deduced from different experiments \cite{bir16,ste83,tom11,cra83,yak09}.
The dependence on the unknown quenching of the IS part in the reactions using hadronic probes is illustrated assuming no quenching (full symbols) or taking the value as for IV quenching (open symbols). 
Figure taken from ref.~\cite{bir16}. 
}
\label{fig522}  
\end{center} 
\end{figure}

Extraction of the analog electromagnetic strength requires the inclusion of quenching conveniently implemented in microscopic calculations by effective  $g$ factors $g_{s,\mathrm{eff}}^{\rm IS/IV} = q^{\rm IS/IV} \times g_{s}^{\rm IS/IV}$ in eq.~(\ref{eq348}), where $q$ denotes the magnitude of quenching. 
As discussed above, a quenching factor $q^{\rm IV} = 0.75(2)$ should be applied for the IV strength in $fp$-shell nuclei. The quenching factor for the IS part is less well known and one may assume that $q^{\rm IS} =q^{\rm IV}$. 
However, it is generally expected that isoscalar spin-M1 strength (ISSM1) is less quenched \cite{lip84}.
A recent study in a series of $sd$-shell nuclei discussed in sec.~\ref{subsec53}  indicates that shell-model calculations can describe the ISSM1 strength without the need for a quenching factor \cite{mat15}, i.e.\ $g^{\mathrm{IS}}_{s,\mathrm{eff}} = g^{\mathrm{IS}}_s$.
Taking these two extremes one gets a range of possible transition strengths $B(\mathrm{M1}) = 3.85(32) - 4.63(38)$~$\mu_\mathrm{N}^2$.
The same analysis was applied to older data for the $^{48}$Ca(p,p$^\prime)$ reaction at $E_0 = 200$ MeV \cite{cra83} with very similar results summarized in fig.~\ref{fig522}.

Because of the large spin matrix element, the isobaric analog state of the level at 10.23 MeV in $^{48}$Ca is prominently excited in the $^{48}$Ca(p,n) \cite{yak09} and ($^3$He,t) \cite{gre07} reactions at high excitation energies in the forward angle spectra.
The assumption of equal unit cross sections for (p,p$^\prime$) and CE reactions based on isospin symmetry provides a relation between the spin-M1 transition strength (B(M1)$_{\sigma \tau}$) and the GT strength 
\begin{equation} 
\mathrm{B(M1)}_{\sigma \tau} = \frac{1}{2} T_i  \mathrm{B(GT_0)}, 
\label{eq521}
\end{equation}
where $T_i$ denotes the isospin of the initial state.
Application of  eq.~(\ref{eq521}) leads to B(M1) strengths from the $^{48}$Ca(p,n) reaction ranging between 3.45(85) and 4.1(1.0) $\mu_\mathrm{N}^2$, depending on the assumption about the IS quenching in the determination of B(M1)$_{\sigma \tau}$.
The summary of all results in fig.~\ref{fig522} shows that the B(M1) strengths deduced from all three hadronic reactions agree well with each other and with the result from the $(e,e^\prime)$ experiment, in particular if no or little IS quenching is assumed.
Even considering the uncertainty due to the unknown magnitude of IS quenching, the large value from the $(\gamma,n)$ experiment is inconsistent with these results.

\begin{figure}[b]
\begin{center}
\resizebox{0.45\textwidth}{!}{%
  \includegraphics{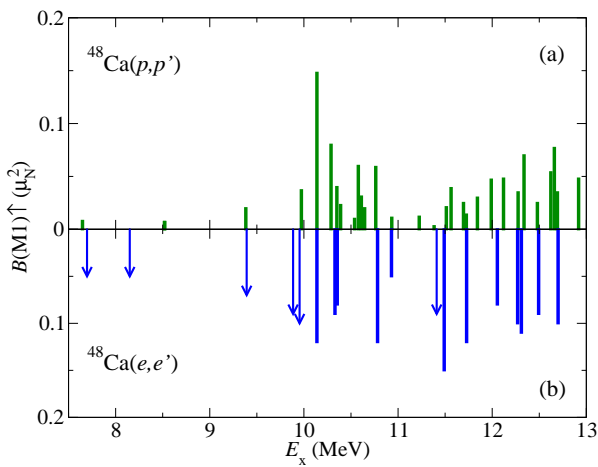}
%\vspace{5cm}       % Give the correct figure height in cm
}
\caption{
%Figure523.
B(M1) strength distributions in $^{48}$Ca deduced from (a) the (p,p$'$) \cite{mat17} and (b) the (e,e$'$) \cite{ste83} reactions.
Arrows indicate upper limits.  
The prominent transition to the state at $E_\mathrm{x}=10.23$~MeV is excluded.
Figure taken from ref.~\cite{mat17}.}
\label{fig523}  
\end{center} 
\end{figure}
For a quantitative interpretation of quenching in microscopic models the full IVSM1 strength must be known experimentally. 
Eighteen additional M1 transitions in $^{48}$Ca were identified in (e,e$^\prime$) scattering \cite{ste83}, whose strength distribution is shown in fig.~\ref{fig523}(b).
Although individually weak ($\leq 0.15$~$\mu_{\rm N}^2$), they sum up to about 1.2~$\mu_{\rm N}^2$ corresponding  to roughly 25\% of the total B(M1) strength.
Most of these transitions were close to the detection limit of the (e,e$^\prime$) experiment, and there is considerable uncertainty about possible unobserved strength below the detection limit set by the radiative background and the high level density in the spectra at excitation energies above 10~MeV. 
The $^{48}$Ca(p,p$^\prime$) data are free of instrumental background.
Besides the excitation of the E1 and spin-M1 modes, at momentum transfers close to $q = 0$  there is only a small contribution from quasi-free scattering, which sets in above the neutron threshold $(S_{\rm n} = 9.9$ MeV).
A MDA of the $^{48}$Ca(p,p$^\prime$) data was performed to extract the spin-flip M1 cross sections \cite{mat17}.
The resulting B(M1) strength distribution is displayed in fig.~\ref{fig523}(a).
Arrows indicate experimental upper limits. 

The strengths from electron scattering tend to be larger but are still consistent within error bars in many cases.
Possible differences between the strengths may be related to the assumptions underlying the analysis of the (p,p$^\prime$) data discussed in sec.~\ref{subsec34}. 
For example, orbital contributions -- although shown to be weak \cite{ric85} -- could lead to a systematic enhancement of the $B(M1)$ strength by constructive interference with the spin part, since the dominant shell-model configurations are the same in all $1^+$ states.
For the same reason one could also speculate about a systematic reduction of $B(M1)_{\sigma \tau}$ due to the interference of $\Delta L =2$ contributions to the angular distributions.     
While the shell-model study of $^{26}$Mg showed a random sign of the interference in an open-shell nucleus \cite{zeg06}, this may be different here because the main components of the wave functions of all excited $1^+$ states are similar.   

The present analysis finds 30 M1 transitions compared to 18 seen in ref.~\cite{ste83}.
This may be related to the different sensitivity thresholds in both experiments.
For the (e,e$'$) data, a statistical limit due to the radiative tail in the spectra and difficulties to distinguish  $M1$ and $M2$ form factors for weak transitions dominate the uncertainties.
The limits for the (p,p$^\prime$) data come from the model assumptions of the MDA. 

\begin{figure}
\begin{center}
\resizebox{0.45\textwidth}{!}{%
  \includegraphics{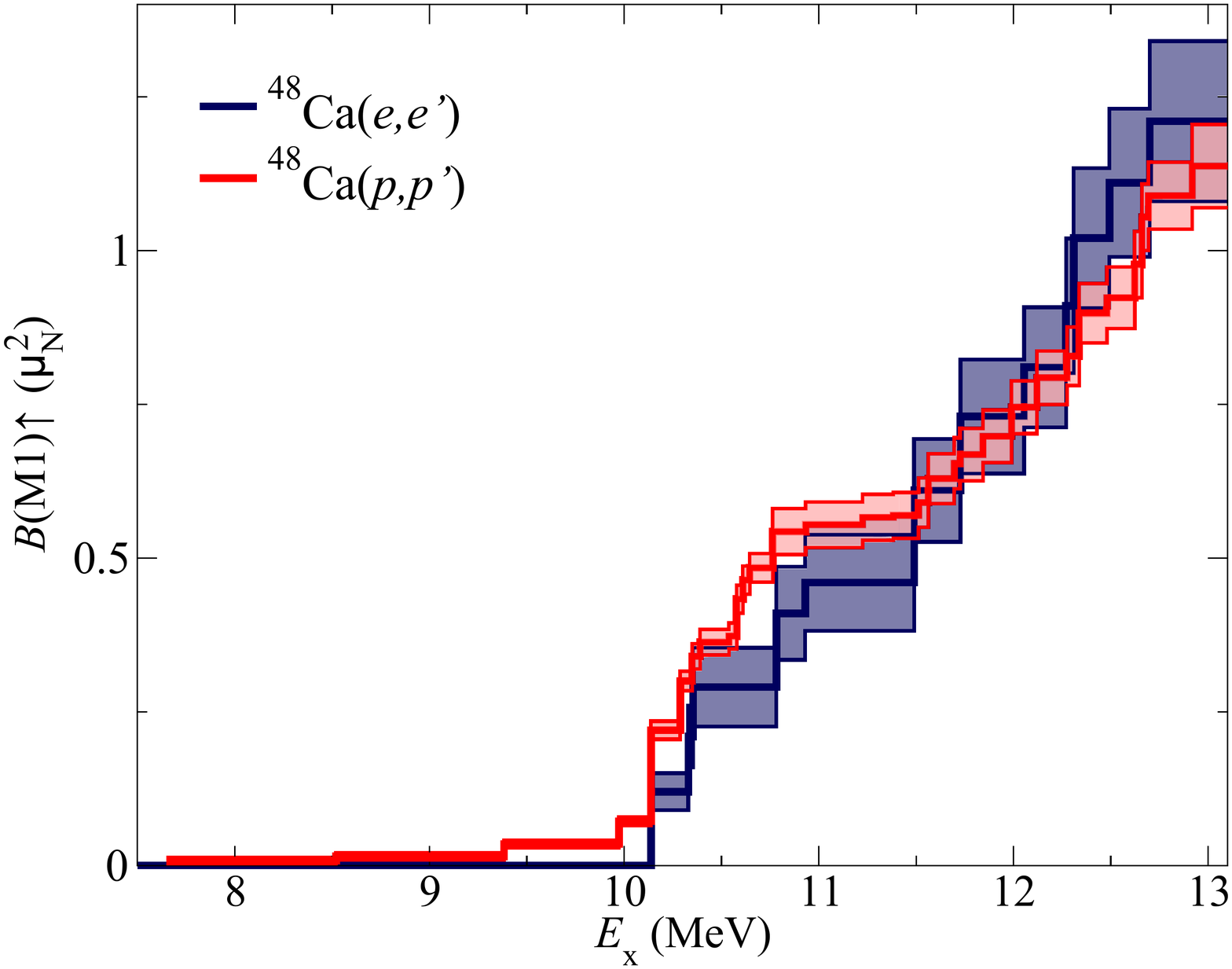}
%\vspace{5cm}       % Give the correct figure height in cm
}
\caption{
%Figure524.
Running sums of the B(M1) strength in $^{48}$Ca between 7 and 13~MeV (excluding the prominent transition to the state at $E_\mathrm{x}=10.23$~MeV) from the (p,p$'$) \cite{mat17}  and the (e,e$'$) \cite{ste83} reactions. 
The bands indicate the experimental uncertainties.
Figure taken from ref.~\cite{mat17}.}
\label{fig524}  
\end{center} 
\end{figure}
Figure~\ref{fig524} compares the running sums of the B(M1) strengths from both experiments.
They exhibit a similar slope and agree within error bars except for the region between 10.5 and 11.5~MeV, where the present analysis finds a number of weaker transitions not observed in ref.~\cite{ste83}.
However, considering that most of the transitions are near the limits of experimental sensitivity in both experiments and taking into account the effects which may modify their relative ratio discussed above, the agreement is good.   
The good correspondence suggests that there is little additional fragmented strength hidden in the data. 
Accordingly, the quenching factor $g^{\rm IV}_{s,{\rm eff}} \simeq 0.75$ for $M1$ strength deduced from the comparison of  large-scale shell-model calculations with (e,e$^\prime$) data on the IVSM1 resonance in $N =28$ nuclei \cite{lan04,vnc98} is confirmed.

\subsection{Quenching of the isoscalar strength}
\label{subsec53}

While the quenching of IV spin-flip transitions has been studied extensively as discussed in the previous section, much less is known for IS transitions.
In most cases IS and IV contributions to spin-M1 transitions compete in (p,p$^\prime$) scattering and because of the dominance of the $\sigma \tau$ over the $\sigma$ part of the effective proton-nucleus interaction \cite{lov81}, the IS cross sections are small (see fig.~\ref{fig342} for an example) and thus hard to determine.
A clear distinction is possible though for targets with g.s.\ isospin $T = 0$, where isospin symmetry requires $T = 0$ and $ T = 1$ final state isospins for IS and IV transitions, respectively.
Pure IS and IV spin-M1 transitions can be distinguished by their different angular distributions (cf.\ fig.~\ref{fig314}).  

Forward-angle inelastic proton scattering by the nuclear interaction is the preferred probe for experimental investigations of the ISSM1 response because of the additional orbital contributions of M1 transitions excited in electromagnetic reactions.
$sd$-shell nuclei are particularly suited for experiments since effective shell-model interactions provide an excellent description of the magnetic dipole moments and transition strengths.
Various theoretical studies \cite{bro87,tow87,ari87} suggest that the quenching of IS spin-$M1$ transitions in $sd$-shell nuclei should be similar to that of the IV ones. 
Several experimental studies have been performed with conflicting results \cite{ana84,cra89}.
It was not clear to what extent they were caused by limitations with respect to the sensitivity of the $J^{\pi}$ assignments of excited states or by the model for the conversion from the differential cross sections to transition strengths. 

\begin{figure}
\begin{center}
\resizebox{0.45\textwidth}{!}{%
  \includegraphics{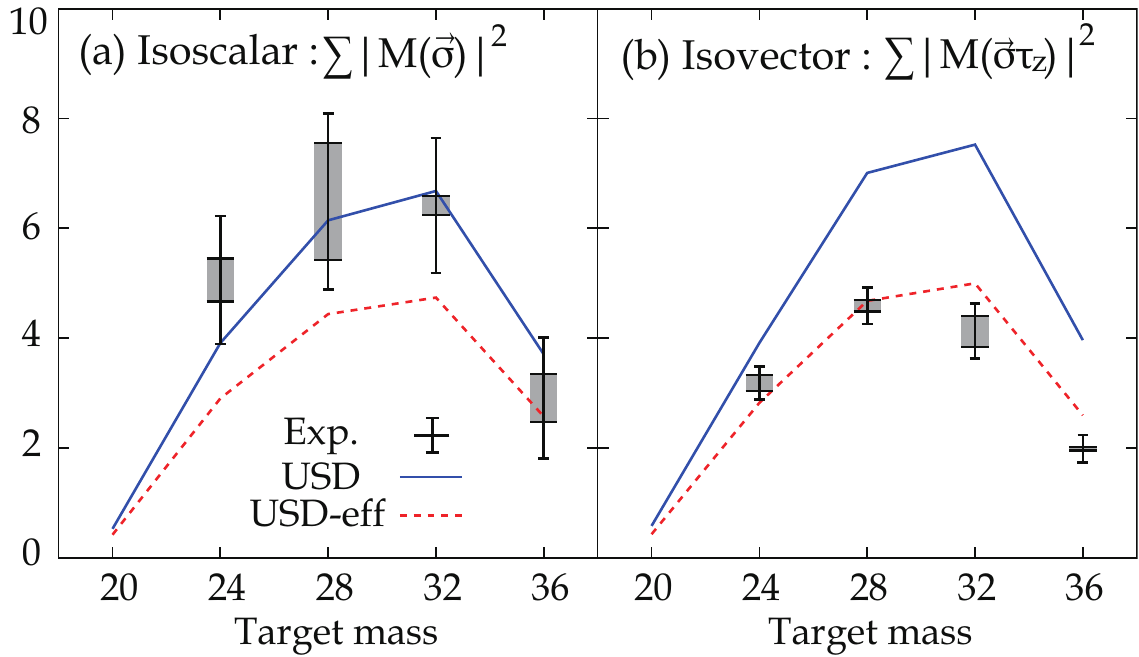}
%\vspace{5cm}       % Give the correct figure height in cm
}
\caption{%
%Figure531.
Accumulated sums of the spin-$M1$ nuclear matrix elements  for (a) IS and (b) IV transitions in $^{24}$Mg, $^{28}$Si, $^{32}$S and $^{36}$Ar nuclei up to $E_{\rm x}=16$~MeV. 
The error bars and gray bands indicate the total experimental uncertainties and the partial uncertainties from the spin assignments, respectively. 
The solid and dotted lines are the predictions of shell-model calculations using the USD interaction \cite{bro87} with bare and effective $g$-factors, respectively.
Figure taken from ref.~\cite{mat15}.
}
\label{fig531}  
\end{center} 
\end{figure}

The (p,p$^\prime$) setup at RCNP has been used to systematically study the spin-M1 strength in $N = Z$ nuclei from $^{12}$C to $^{40}$Ca \cite{mat10}.
While results for the IVSM1 strength in $sd$-shell nuclei agrees fairly well with previous studies \cite{ric90,cra89}, significant differences are found for the ISSM1 strength \cite{mat15}.
 Figure \ref{fig531} presents the sum of the squared spin-M1 matrix elements, defined for IS (operator $\sigma$) and IV (operator $\sigma\tau_z$) transitions as
 \begin{eqnarray}
\sum \left|M(\sigma)\right|^2&=&\frac{1}{2J_i+1}  |\langle f|| \sum_k^A \sigma_k || i \rangle |^2 \label{eq531}\\
\sum \left|M(\sigma\tau_z)\right|^2&=&\frac{1}{2J_i+1}  |\langle f|| \sum_k^A \sigma_k \tau_{z,k} || i \rangle |^2 \label{eq532}
  \end{eqnarray}
 for $^{24}$Mg, $^{28}$Si, $^{32}$S and $^{36}$Ar up to $E_{\rm x}=16$~MeV. 
The initial state spin ($J_i$) is zero for all target nuclei.
Shell-model calculations with the USD interaction \cite{bro87} are shown by the solid lines.
The dashed lines are scaled by the ratio of the effective spin $g$ factors to their free values.
(We note that the difference between predictions of the summed $M1$ strengths using the USD and the newer USDA/B interactions \cite{bro06,ric08} is less than 10\% and thus not significant in the following discussion). 
Averaging over the nuclei measured, one obtains quenching factors $q_{\rm IS} = 1.01(5)$ and $q_{\rm IV} =  0.78(4)$ for the IS and IV spin-$M1$ strengths, respectively. 
The quenching factor of the IS spin-$M1$ transitions is consistent with unity within error bars, while that of the IV ones is significantly smaller and in agreement with the studies of the analogous GT transitions \cite{and87,and91}. 
The present result shows that the widely-used effective $g$-factors lead to an over-quenching for  IS spin-$M1$ transitions in the $sd$-shell. 
Effective IS $g$-factors were introduced to reproduce the diagonal spin matrix element $ \langle S \rangle$ of the ground state, see eq.~(20) in ref.~\cite{bro83}. 
Experimental $ \langle S \rangle $ values were obtained from the IS magnetic moments of mirror nuclei and subtracting the contribution of the total angular momentum $J$. 
Although the quenching of $ \langle S \rangle $  in nuclei with closed-$LS$-shell plus/minus one nucleon is obvious \cite{bro87,tow87,ari87}, the quenching in the middle of the $sd$-shell seems to be insignificant \cite{bro83}. 
The finding is consistent with the present observation of no quenching of ISSM1 transitions in open $sd$-shell nuclei.

In order to shed further light on these observations, one can consider the difference $\Delta_{\rm spin}$  between the summed ISSM1 and IVSM1 transition strengths
\begin{equation}
  \Delta_{\rm spin} = \frac{1}{16} \left [ \sum {|M(\sigma)|^2} - \sum {|M(\sigma\tau_z)|^2} \right ],
\label{eq533}
\end{equation}
where it is assumed that summing the experimental results up to $E_{\rm x}= 16$ MeV exhausts the strengths.
%Being a measure of the spin correlation at the ground state,
%this $\Delta_{\rm{spin}} (E_{\rm x})$ may be suggestive to mechanism of the quenching.
With the total spin operators for protons (neutrons)
\begin{equation}
\vec{S}_{p(n)} = \frac{1}{2}\sum_{i=1}^{Z(N)}\vec{\sigma}_{p(n),i},
\label{eq534}
\end{equation}
the spin-M1 transition strengths can be rewritten as \cite{mat15}
\begin{equation}
  \langle(\vec{S}_p  + \vec{S}_n)^2\rangle  =   \frac{1}{4}  |M(\sigma)|^2,  \,\,  \langle(\vec{S}_p  - \vec{S}_n)^2\rangle  =   \frac{1}{4} |M(\sigma\tau_z)|^2,
  \label{eq535}
\end{equation}
%\begin{eqnarray}
  %\langle(\vec{S}_p  + \vec{S}_n)^2\rangle &=&  \frac{1}{4}  {\rm B(M1}_\sigma) \nonumber\\
  %\langle(\vec{S}_p  - \vec{S}_n)^2\rangle &=&  \frac{1}{4}  {\rm B(M1}_{\sigma \tau},
 % \label{eq533}
%\end{eqnarray}
where the expectation values are taken for the 0$^+$ ground state. 
Insertion into eq.~(\ref{eq533}) leads to 
\begin{equation}
  \Delta_{\rm spin}  = \langle\vec{S}_p \cdot \vec{S}_n\rangle,
 \label{eq536}
\end{equation}
i.e., the difference of total ISSM1 and IVSM1 strengths is related to the spin correlation between protons and neutrons in the ground state.

\begin{figure}
\begin{center}
\resizebox{0.5\textwidth}{!}{%
  \includegraphics{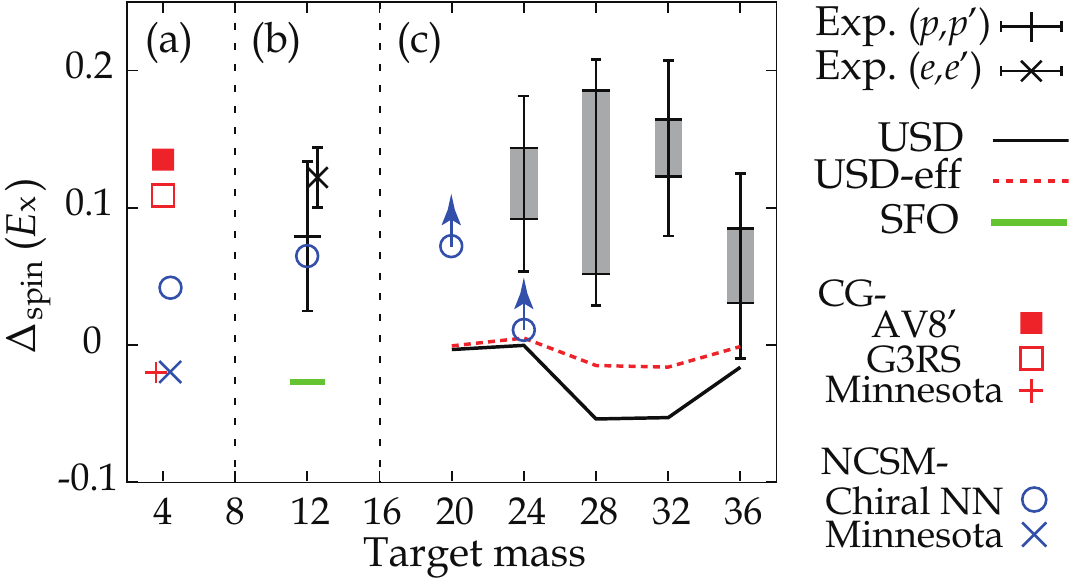}
%\vspace{5cm}       % Give the correct figure height in cm
}
\caption{
%Figure532.
Experimental $\Delta_{\rm spin}$ values [eq.~(\ref{eq533})] and theoretical predictions of $\langle\vec{S}_p\cdot\vec{S}_n\rangle$ [eq.~(\ref{eq536})] for (a) $^4$He, (b) $^{12}$C, and (c) $sd$-shell nuclei. 
The arrows of the NCSM results in (c) indicate lower limits. 
Figure taken from ref.~\cite{mat15}.}
\label{fig532}  
\end{center} 
\end{figure}

Figure~\ref{fig532} presents experimental and theoretical results for eq.~(\ref{eq536}) in several shell regions. 
$\langle\vec{S}_p\cdot\vec{S}_n\rangle$ values from state-of-the-art nuclear structure calculations for $^4$He using the  Correlated Gaussian (CG) method \cite{suz08} and No-Core Shell Model (NCSM) \cite{bar13} are displayed in fig.~\ref{fig532}(a).
Positive values are obtained for realistic AV8' \cite{pud97}, G3RS \cite{tam68}) and chiral \cite{ent03} NN forces due to the inclusion of tensor correlations, in contrast to the Minnesota \cite{tho77} interaction, which does not contain the tensor force.
Figure~\ref{fig532}(b) compares experimental results for $\Delta_{\rm spin}$ in $^{12}$C derived from (p,p$'$) \cite{mat10} and (e,e$'$) \cite{vnc00} experiments with $\langle\vec{S}_p\cdot\vec{S}_n\rangle$ values obtained from shell-model calculations. 
Both the experiments and the NCSM with realistic forces show positive values while a calculation using the effective interaction of ref.~\cite{suz03} gives a slightly negative value.
Finally, fig.~\ref{fig532}(c) shows  $\Delta_{\rm spin}$ values derived from the (p,p$^\prime$) data for $sd$-shell nuclei in comparison to the shell-model calculations using the USD interaction discussed above. 
The data exhibit positive values as in $^{12}$C and are comparable to the values predicted with realistic forces for lower-mass nuclei. 
In contrast, the shell-model calculations are unable to reproduce the experimental results independent of the correction by the ratio of the effective spin $g$-factors to their free values.

However, predictions by the NCSM (open blue circles) indicate positive $\langle\vec{S}_p\cdot\vec{S}_n\rangle$ values for $^{20}{\rm Ne}$ ($N_{\rm MAX}=4$) and $^{24}{\rm Mg}$ ($N_{\rm MAX}=2$). 
Here, $N_{\rm MAX}$ defines the maximal allowed harmonic-oscillator excitation energy above the unperturbed ground state \cite{bar13} and hence represents a measure of the model space.
The results ($-0.007$, 0.028, and 0.072 for $N_{\rm MAX}$ = 0, 2, and 4 for $^{20}{\rm Ne}$ and $-0.018$ and 0.011 for $N_{\rm MAX}$=0 and 2 for $^{24}$Mg, respectively) show a clear correlation with the size of $N_{\rm MAX}$ but represent a lower boundary only because they are not converged yet for the present $N_{\rm MAX}$ values. 
The increase of  $\langle\vec{S}_p\cdot\vec{S}_n\rangle$ with increasing $N_{\rm MAX}$ implies that mixing with higher-lying orbits due to tensor correlations is important for reproducing $\Delta_{\rm spin} > 0$ values.  
This should be verified in future NCSM calculations, which might be possible up to $N_{\rm MAX}=8$ for the nuclei under investigation \cite{rothpc}. 

We note that both the observed isoscalar and isovector strengths, and thus $\Delta_{\rm spin}$, are reproduced in a shell-model calculation by enhancing the isoscalar spin-triplet pairing interaction and by introducing the $\Delta$-hole coupling effect~\cite{sag16,sag18}.

\section{Fine structure of the IVGDR}
\label{sec6}

\begin{figure}
\begin{center}
\resizebox{0.45\textwidth}{!}{%
  \includegraphics{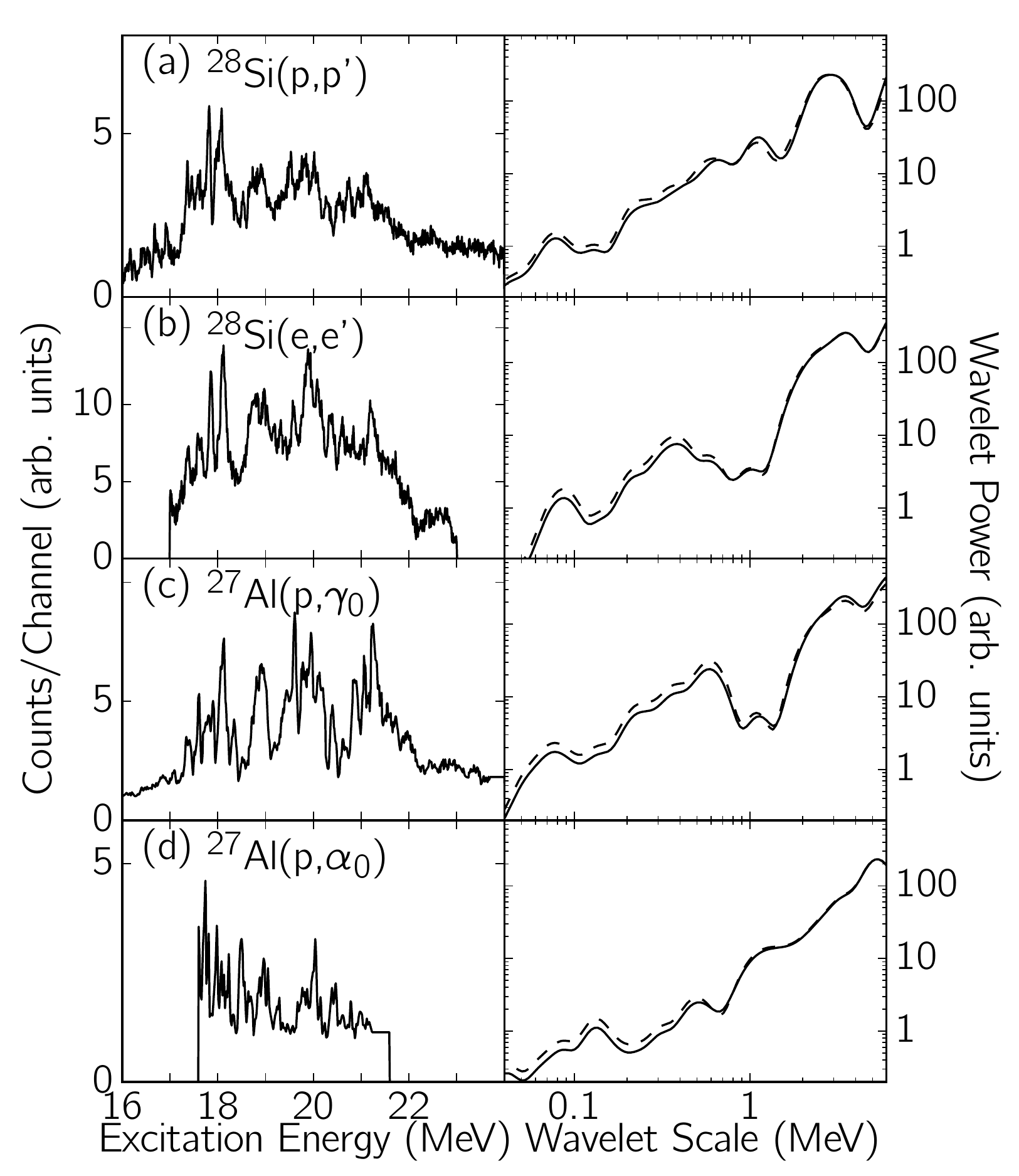}
%\vspace{5cm}       % Give the correct figure height in cm
}
\caption{
Left: Spectra of the high-resolution $^{28}$Si(p,p$^\prime$) \cite{fea18}, $^{28}$Si(e,e$^\prime$) \cite{ric85}, $^{27}$Al$(p,\gamma)$ \cite{sin65} and the $^{27}$Al$(p,\alpha_0)$ \cite{law65,put68} reactions populating the energy region of the IVGDR in $^{28}$Si.
Right: Corresponding power spectra from a wavelet analysis summed over excitation-energy regions $16-24$ MeV (solid line) and $17-23$ MeV (dashed line).}
\label{fig601}  
\end{center} 
\end{figure}

Fine structure of the giant resonances in heavy nuclei has been observed for many different modes like the IVGDR \cite{iwa12,pol14}, the ISGQR  \cite{she04}, the GT resonance \cite{kal06}, or the magnetic quadrupole resonance \cite{vnc99}.
For the IVGDR \cite{fea18,bir17,jin18} and the ISGQR \cite{she09,usm11,usm16} it has been demonstrated to appear across the nuclear chart.
The phenomenon of fine structure is also independent of the exciting probe provided the reaction is selective for the mode under investigation.
This is illustrated in the l.h.s.\ fig.~\ref{fig601} for the case of the IVGDR  (an example for the ISGQR is shown in ref.~\cite{kam97}), which compares spectra of $^{28}$Si in the IVGDR energy region from different probes \cite{fea18}.  
These sets of high-resolution data are  from (a) the $^{28}$Si(p,p$^\prime$) reaction, (b) the $^{28}$Si$(e,e')$ reaction \cite{ric85}, (c) the  $^{27}$Al$(p,\gamma)$ reaction \cite{sin65}, and (d) the $^{27}$Al$(p,\alpha_0)$ reaction \cite{law65,put68}.
It is expected that reactions (a)-(c) predominantly excite the IVGDR and indeed all three reactions show very similar structures.
Reaction (d) favors isospin $T = 0$ states in $^{28}$Si and is therefore not selective towards $1^-$ levels.  
The spectrum clearly shows different patterns compared with spectra (a)-(c).  

\subsection{Quantitative characterization with a wavelet analysis}
\label{subsec61}

Wavelet transforms are an established tool to analyze different types of signals hidden in fluctuating quantities, e.g., as a function of time or energy. 
They are used in diverse areas, such as image processing and data compression \cite{dau92,byr92}, meteorology \cite{tor98}, astrophysics \cite{ire99} or accelerator physics \cite{fed98}.
Wavelet analysis can be regarded as an extension of the Fourier analysis, which allows to conserve the correlation between the observable and its transform.

In the present case, energy spectra of nuclear giant resonances are analyzed.
The coefficients of the wavelet transform are then defined as
\begin{equation}
   C\left( {\delta E,E_{\rm x}} \right) = \int\limits_{ - \infty}^\infty  \sigma \left( E \right) \Psi \left( \delta E,E_{\rm x} \right) dE.
   \label{eq611}
\end{equation}
They depend on two parameters, the scale $\delta E$ stretching and compressing the wavelet function $\Psi$, and the position E$_{\rm x}$ shifting the wavelet function in the spectrum $\sigma(E)$. 
The variation of the variables can be carried out using continuous or discrete steps called continuous (CWT) and discrete (DWT) wavelet transform, respectively. 
The analysis of the fine structure of giant resonances is performed using CWT, where the fit procedure can be adjusted to the required precision. 
Applications of the CWT to high-resolution nuclear spectra of giant resonances are described in refs.~\cite{she04,she09,fea18,pol14,usm16,pet10,kur18}.
Further details and a comparison with other techniques for the analysis of fine structure in nuclear giant resonances can be found in ref.~\cite{she08}. 

The choice of the wavelet function plays an important role in the analysis. 
In order to achieve an optimum representation of the signal using wavelet transformation one has to select a function $\Psi$ which resembles the properties of the studied signal $\sigma$. 
In fact, the better the correspondence between the shape of $\Psi$ and the signal $\sigma$ is, the larger is the wavelet coefficient.
A maximum of the wavelet coefficients at certain value $\delta$E indicates a correlation in the signal at the given scale, called characteristic scale.
The best resolution for nuclear spectra is obtained with the so-called Morlet wavelet (cf.\ fig.~9 in ref.~\cite{she08}) because the detector response is typically close to the Gaussian line shape and the Morlet wavelet is a product of Gaussian and cosine functions
\begin{equation}
\psi_{Morlet} (x) = \pi^{-1/4} e^{ikx} e^{-x^2/2}.
\label{eq612}
\end{equation}

As an example, a CWT analysis of a $^{208}$Pb(p,p$^\prime$) excitation spectrum in the energy region of the IVGDR and for scattering angles $\theta =0^\circ - 0.94^\circ$ is discussed.
At these extreme forward angles $E1$ Coulomb excitation dominates the cross sections and nuclear transitions are suppressed with the exception of the isovector spin-flip $M1$ resonance. 
The excitation energy region below 9~MeV, where the spin-$M1$ mode is located and contributes significantly to the cross sections (cf.\ sec.~\ref{subsec51}), is thus excluded.
In order to search for characteristic scales it is helpful to construct the power spectrum of the signal, i.e.\ the projection of the absolute values of the wavelet coefficients on the scale axis. 

\begin{figure}[b]
\begin{center}
\resizebox{0.5\textwidth}{!}{%
  \includegraphics{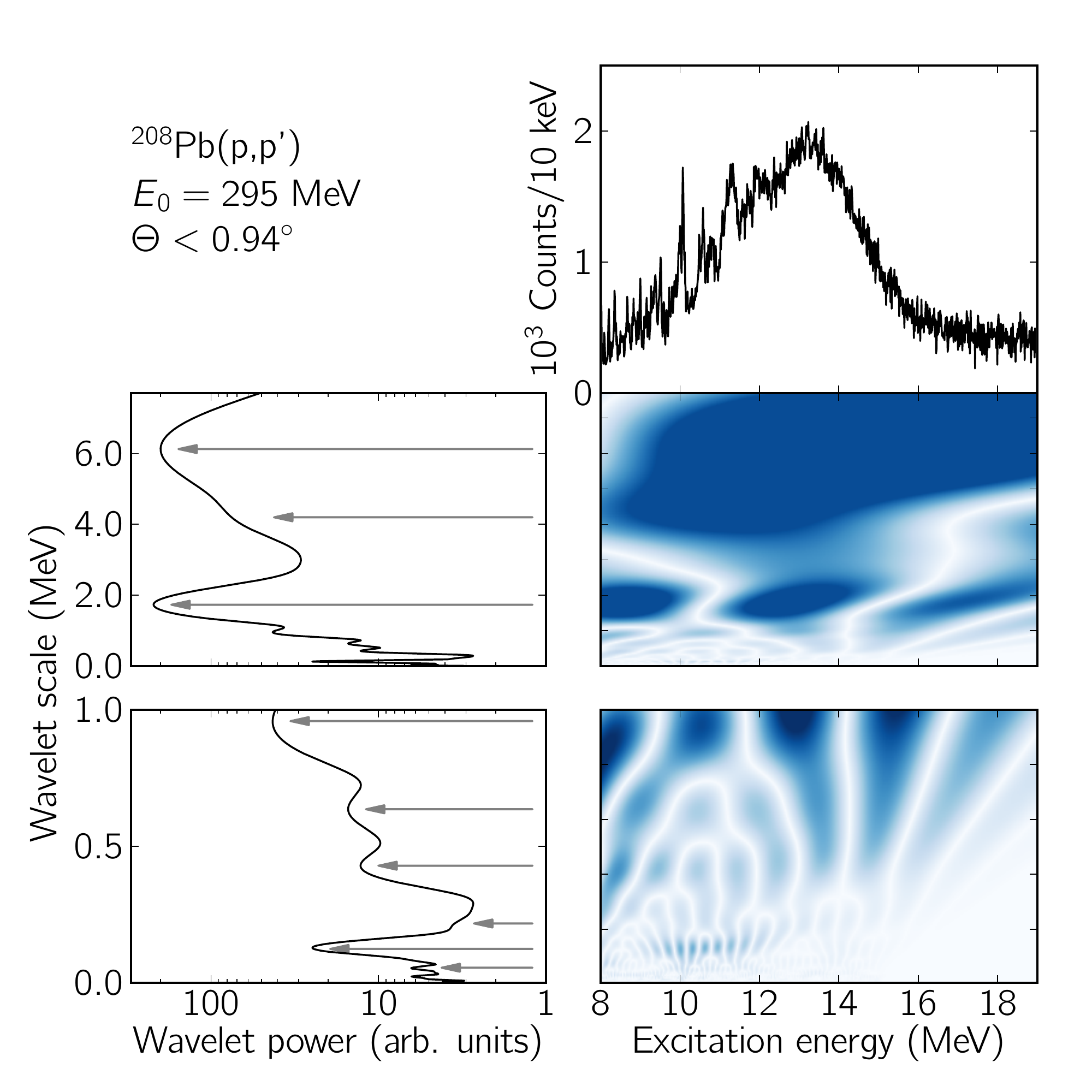}
%\vspace{5cm}       % Give the correct figure height in cm
}
\caption{
%Figure611.
CWT analysis of the excitation-energy spectrum of the $^{208}$Pb(p,p$^\prime$) reaction at E$_0 =295$~MeV and $\theta_{\rm lab} < 0.94^\circ$. 
Top-right: Spectrum of the reaction in the IVGDR energy region.
Middle: Absolute values of the wavelet coefficients (right) and power spectrum (left). Bottom: Enlarged picture for the region of wavelet scales below 1~MeV.
White color corresponds to smallest wavelet coefficients, while dark (blue) regions indicate the largest values. 
Arrows indicate the positions of local maxima in the power spectrum identified as characteristic scales.
Figure taken from ref.~\cite{pol14}.}
\label{fig611}  
\end{center} 
\end{figure}
In Fig.~\ref{fig611} the excitation energy spectrum (upper right) and corresponding absolute values of the wavelet coefficients (middle and lower right) are plotted. 
White regions indicate the smallest values of the wavelet coefficients, while dark (blue) ones denote maxima, i.e.\ characteristic scales.
One identifies scale values where the absolute values of the wavelet coefficients show a local maximum, albeit with a characteristic minimum/maximum variation as a function of excitation energy induced by the oscillating wavelet function.
For a better recognition of such characteristic scales, power spectra are plotted (middle- and lower-left). 
The power values are divided by the corresponding scale in order to remove a trivial increase with increasing scale \cite{liu07}. 
The middle panel shows the scale region up to 7 MeV, while the lower panel gives an enlarged view of the region below 1~MeV. 
Local maxima indicated by the horizontal arrows in the power spectra are identified as characteristic scales.
Note that these can appear as a peak but also as a shoulder on the flank of another, more prominent scale, like the one at about 4 MeV.  
The relation of characteristic scales to specific giant resonance decay mechanisms is discussed in sec.~\ref{subsec62}.

Another application of the CWT is presented on the r.h.s.\ of fig.~\ref{fig601} discussed above.
The different reactions selective towards population of the IVGDR in $^{28}$Si exhibit very similar power spectra and characteristic scales, while the power spectrum  of reaction (d) is distinctively different. 
The impact of choosing different excitation-energy windows for the CWT analysis is illustrated by the dashed ($17-23$ MeV) and full ($16 - 24$ MeV) lines and found to be small.

\begin{figure}[b]
\begin{center}
\resizebox{0.48\textwidth}{!}{%
  \includegraphics{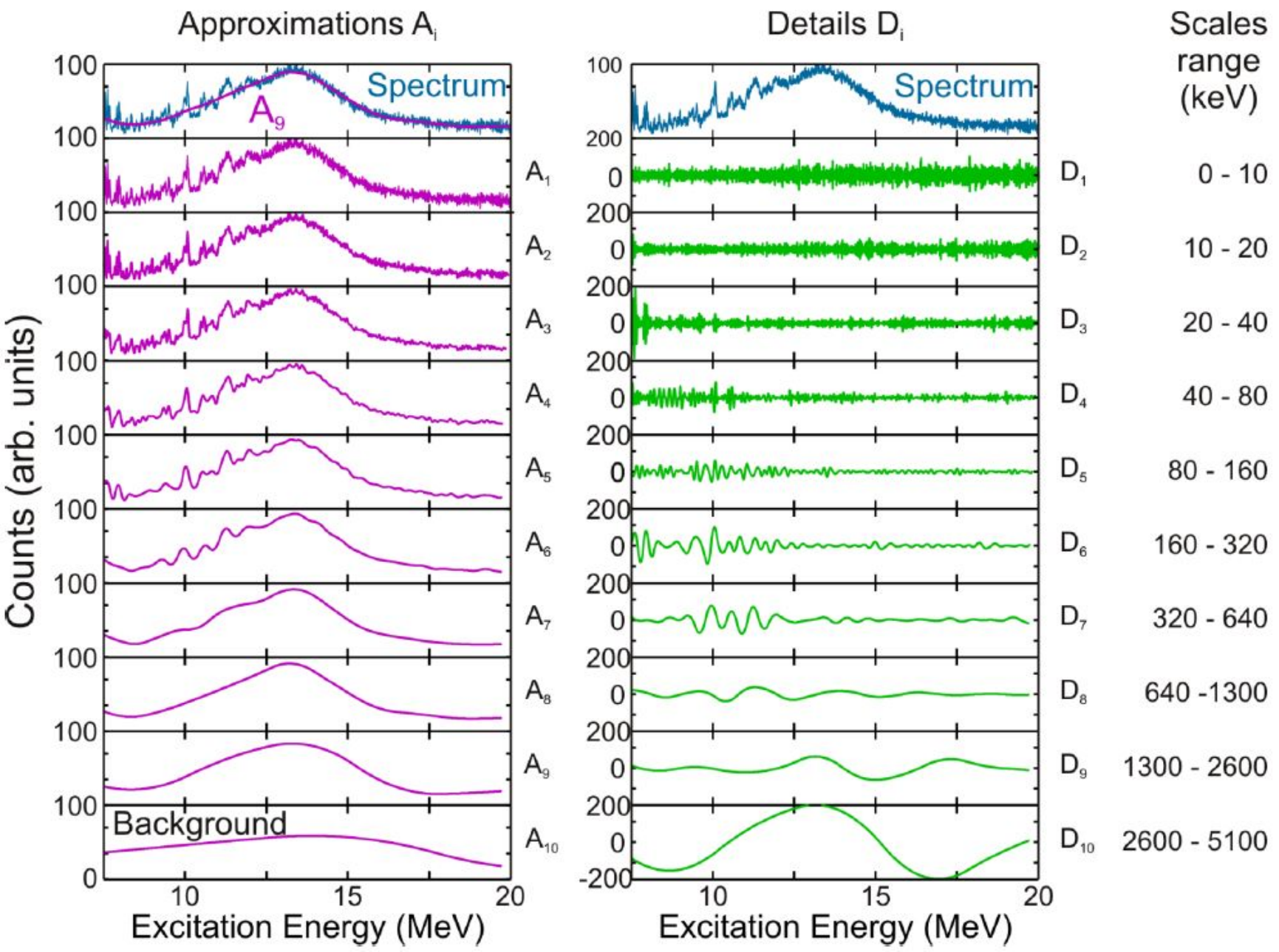}
%\vspace{5cm}       % Give the correct figure height in cm
}
\caption{
%Figure612. 
Decomposition of the $^{208}$Pb(p,p$^\prime$) spectrum with a DWT analysis into approximations $A_i$ and details $D_i$, eq.~(\ref{eq613}). 
Approximation $A_9$ describes the total width of the GDR, thus $A_{10}$ can be adopted as background shape.
Figure taken from ref.~\cite{pol14}.}
\label{fig612}  
\end{center} 
\end{figure}
An exact spectrum decomposition can be achieved with the DWT, where scales and positions in the wavelet analysis are varied by powers of two.
It works by filtering and provides two signals in each step $i$, approximation $A_i$ and detail $D_i$. 
Application of the method to the spectrum of the $^{208}$Pb(p,p$^\prime$) reaction is shown in fig.~\ref{fig612}. 
The approximation is the large-scale or low-frequency component of the signal, and the detail corresponds to the small-scale or high-frequency part for a given scale region analogue to the effect of high- and low-pass filters in an electric circuit. 
In each step $i$ of the decomposition, the initial signal $\sigma(E)$ can be reconstructed as
\begin{equation}
\sigma(E)=A_i + \sum D_i.
\label{eq613}
\end{equation}
This operation can be repeated until the individual detail consists of a single bin. 

A DWT can only be performed with a certain class of wavelets possessing a so-called scaling function \cite{she08}.
This is not the case for the Morlet wavelet, thus the Bior3.9 wavelet function \cite{mal98} is used in the example, which has a similar form (cf.\ fig.~9 in ref.~\cite{she08}).
Each wavelet function can be characterized by its number of vanishing moments,
\begin{equation}
   \label{eq:vanishingm}
   \int\limits_{ - \infty }^\infty  {E^n \Psi \left( E \right)dE =
   0,\;\; n = 0,1...m}.
\end{equation}
For Bior3.9 the number is equal to three, i.e.\ any background in the spectrum that can be approximated by a quadratic polynomial function does not contribute to the wavelet coefficients. 

The largest characteristic scale in the spectrum is given by the total width of the IVGDR and should thus show up in the approximations.
Indeed, approximation $A_9$ provides a very good description of the broad structure in the experimental spectrum displayed in the top row of  fig.~\ref{fig612}.
Thus, the next approximation A$_{10}$ can be considered to describe the shape of non-resonant contributions to the spectrum.

\subsection{Characteristic scales and giant resonance decay mechanisms}
\label{subsec62}

Giant resonances are elementary excitations of the nucleus and their understanding forms a cornerstone of microscopic nuclear theory.
They are classified according to their quantum numbers (angular momentum, parity, isospin). 
Gross properties like energy centroid and strength in terms of exhaustion of sum rules are fairly well described by microscopic models \cite{har01}.
However, a systematic understanding of the decay widths is still lacking. 

\begin{figure}
\begin{center}
\resizebox{0.48\textwidth}{!}{%
  \includegraphics{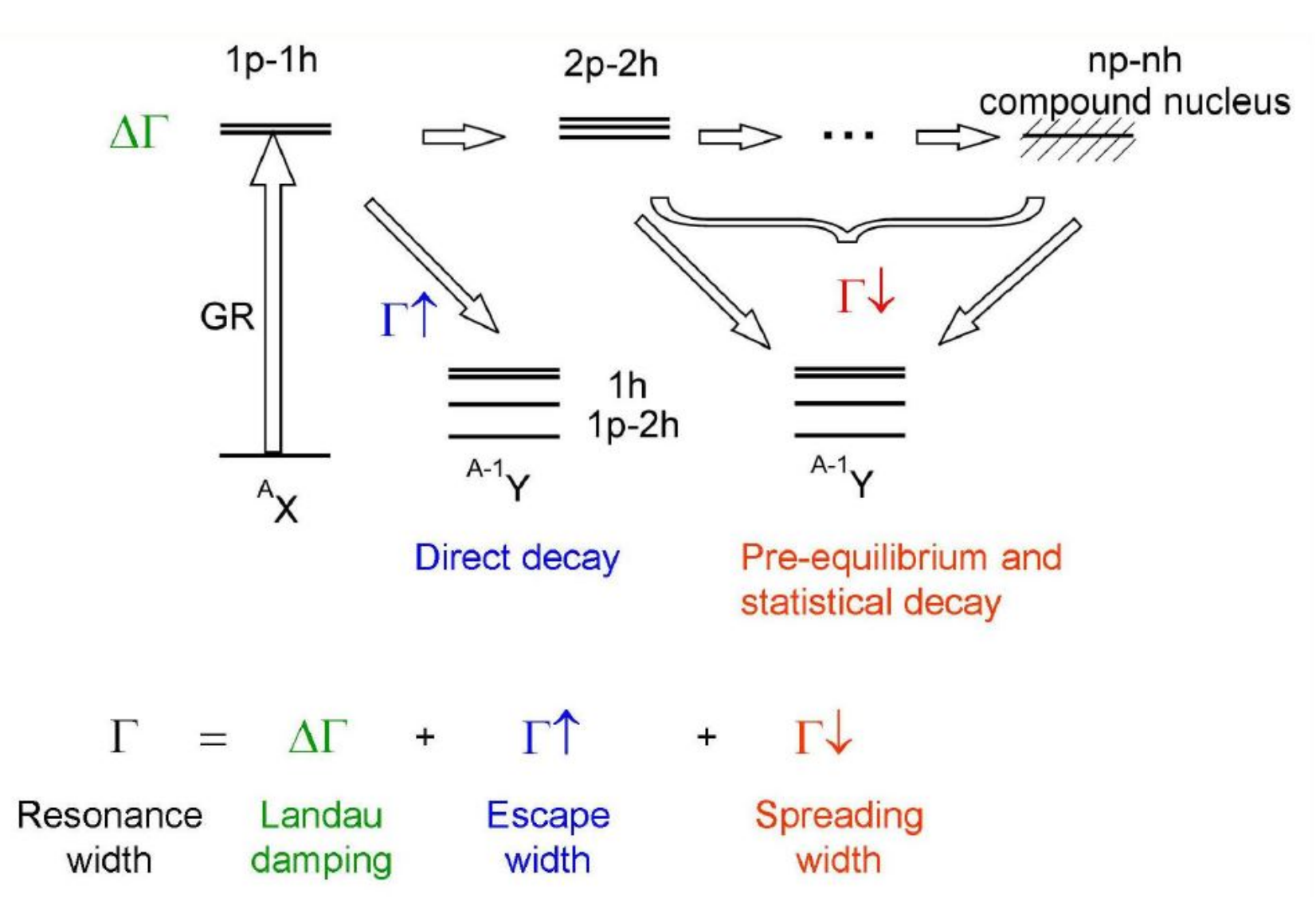}
%\vspace{5cm}       % Give the correct figure height in cm
}
\caption{
%Figure621.
Contributions to the decay width of a giant resonance.
For details see text.}
\label{fig621}  
\end{center} 
\end{figure}

The giant resonance width $\Gamma$ is determined by the interplay of different mechanisms illustrated in fig.~\ref{fig621}: fragmentation of the elementary 1p-1h excitations (Landau damping $\Delta \Gamma$), direct particle decay out of the continuum (escape width $\Gamma\!\uparrow$), and statistical particle decay due to coupling to 2p-2h and many particle-many hole (np-nh) states (spreading width  $\Gamma\!\downarrow$) 
\begin{equation}
\label{eq:width}
\Gamma = \Delta \Gamma + \Gamma\!\uparrow + \Gamma\!\downarrow.
\end{equation}
This scheme implies a hierarchy of widths and timescales resulting in a fragmentation of the giant resonance strength in a hierarchical manner \cite{bbb98}. 
An important theoretical problem is to explain the nature of couplings between the levels in this
hierarchy, and to predict the scales of the fragmentation of the strength which arise from it.

A powerful approach to investigate the role of the different components are coincidence experiments, where direct decay can be identified by the population of hole states in the daughter nucleus and the spreading width contribution can be estimated by comparison with statistical model calculations (see, e.g., refs.~\cite{str00,bol88,die94,hun03}).
Recently, alternative methods have been developed based on a quantitative analysis of the fine structure of giant resonances observed in high-resolution inelastic scattering and charge-exchange reactions.
These new methods for the extraction of such scales include the local scaling dimension approach \cite{aib99}, the entropy index method \cite{lac00}, and the use of wavelet techniques \cite{she08,hei10}.
Of these, wavelet analysis has been established as a particularly successful approach \cite{she08}. 

A direct interpretation of the wavelet scales as underlying widths, respectively a characteristic time scale, is not possible because the wavelet analysis is also sensitive to spacing, and both effects are intertwined in the power spectrum.
Therefore, one needs microscopic calculations of the giant-resonance strength including one or several of the possible giant-resonance decay mechanisms.
The CWT can be applied to the model results and the resulting scales compared to experiment. 
At present there are no calculations available including all three mechanisms. 
RPA results account for Landau damping and extensions to include the spreading width via coupling to the continuum or 2p-2h configurations at various levels of approximation have been formulated for nonrelativistic and relativistic interactions. 
 
Taking again the IVGDR in $^{208}$Pb as an example, the escape width is expected to make a minor contribution in heavy nuclei and the focus is on the possible role of the spreading width.   
The 2p-2h states, or more specifically the coupling to low-lying surface vibrations \cite{ber83}, was shown to be the mechanism responsible for the fine structure observed in the ISGQR \cite{she04,she09}. 
Results of the CWT analysis for microscopic calculations of the electric dipole response in $^{208}$Pb with the quasiparticle phonon model (QPM)  and relativistic RPA (RRPA) are discussed here.
Both models allow for the inclusion of complex configurations. 
Therefore, besides calculations on the 1p-1h level (called QPM 1-phonon and RRPA, respectively), also extensions including 2p-2h  states (called QPM and relativistic time-blocking approximation (RTBA), respectively) are considered.
A general description of the QPM can be found in ref.~\cite{sol92} and of the RTBA in ref.~\cite{lit07}.
Details of the present QPM calculations are discussed in refs.~\cite{rye02,tam11,pol12}. 

\begin{figure}
\begin{center}
\resizebox{0.5\textwidth}{!}{%
  \includegraphics{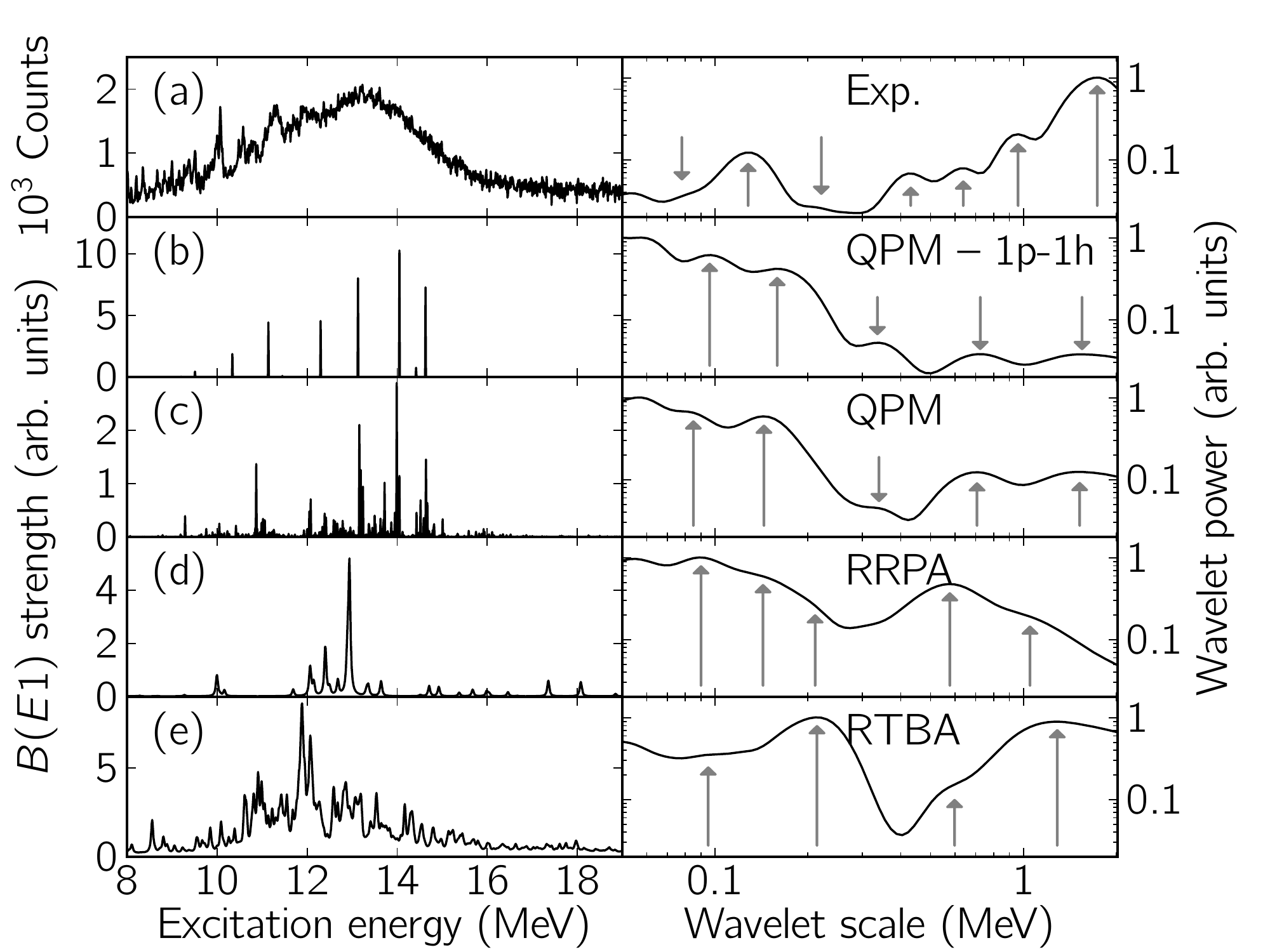}
%\vspace{5cm}       % Give the correct figure height in cm
}
\caption{
%Figure622.
(a) Experimental spectrum of the $^{208}$Pb(p,p$^\prime$) reaction (fig.~\ref{fig312}) in comparison with theoretical predictions of  the B(E1) strength distribution in $^{208}$Pb (l.h.s.) and the resulting power spectra from a CWT analysis (r.h.s.). 
Theoretical results are shown for the QPM with 1-phonon (b) and (1+2)-phonon (c) model spaces, (d) relativistic RPA and (e) RTBA. 
Characteristic scales in the power spectra are marked by arrows.
Figure taken from ref.~\cite{pol14}.}
\label{fig622}  
\end{center} 
\end{figure}

The experimental cross sections at $0^\circ$ (l.h.s.) and the power spectrum (r.h.s.) resulting from the CWT analysis are plotted in fig.~\ref{fig622}(a) together with the B(E1) strength distributions obtained with the different models in fig.~\ref{fig622}(b-e).
It should be noted that the experimental spectrum does not represent the B(E1) strength but the Coulomb-excitation cross section, which is modified by the virtual-photon spectrum discussed in sec.~\ref{subsec33}. 
Extraction of the B(E1) distribution is possible (cf.~Refs.~\cite{tam11,pol12}). 
However, the need to disentangle the E1 cross section from other contributions can only be achieved for larger energy bins, where the information on the fine structure is partially lost.
A conversion of the experimental Coulomb cross sections to B(E1) strength would lead to a slight shift ($< 5$\%) of the characteristic scales and a relative increase of the average power towards higher excitation energies.

As shown in fig.~\ref{fig622}(b), a QPM calculation on the RPA level results in a B(E1) strength distribution dominated by 5 transitions distributed between 11 and 15 MeV with a centroid  energy of 13.25 MeV (defined as $m_1/m_0$, where $m_i$ denotes the $i$th moment of the distribution).
The experimental centroid energy of 13.43 MeV is fairly well reproduced.
Inclusion of 2-phonon configurations, fig.~\ref{fig622}(c), leads to fragmentation but the dominant 1p-1h transitions remain and the centroid energy is unaffected.
A similar comparison of RRPA, fig.~\ref{fig622}(d), and RTBA, fig.~\ref{fig622}(e), results shows somewhat larger differences of the distributions although the centroid energy is hardly changed (13.01 MeV for RRPA and 13.06 MeV for RTBA, respectively). 

Since there is no absolute scale, the corresponding CWT power spectra shown on the r.h.s.\ of fig.~\ref{fig622} are normalized relative to each other. 
They provide a qualitative measure for the ability of different models to describe fine structure and characteristic scales.
Overall, both models broadly reproduce the variation of power with scale value but the scale values of power maxima and minima are better reproduced by the QPM.
However, the relative ratio of maxima at smaller and larger scales is predicted to decrease in the QPM while experiment shows an increase
Here, the RTBA result is closer to the data.  
The region of scales in the figure is restricted to 2 MeV because the theoretical calculations show limited power at even larger scale values, in contrast to the experiment.
This finding may be related to neglecting the coupling to complex states beyond the 2p-2h level in the models.

The comparison of figs.~\ref{fig622}(b,c) and (d,e) allows to extract information on the damping mechanism responsible for the fine structure.
Clearly, the QPM results show structure already at the 1-phonon level.
While the appearance of scales $\geq 1$ MeV can be easily understood by the spacing of the five dominant transitions, the wavelet analysis of the 1-phonon results (b) also finds characteristic scales with smaller values $<1$ MeV.  
The similarity between the power spectra and scales deduced from the QPM calculation for a 1-phonon model space with those including two-phonon states suggests that the fragmentation of 1p-1h transitions (i.e., Landau damping) is the most important mechanism leading to fine structure of the IVGDR in $^{208}$Pb. 
While the relative weight changes, major scales are also found at about the same energies in the CWT analysis of the RRPA (d) and RTBA (e) results.
The observation of characteristic scales in the RRPA calculation again supports an interpretation of Landau damping as a main cause of the fine structure of the IVGDR in $^{208}$Pb.
Similar conclusions are drawn in lighter nuclei \cite{jin18}.
An in-depth discussion of the relation between the wavelet analysis of spectra of the IVGDR and ISGQR and different damping mechanisms can be found in ref.~\cite{vnc19b}.  

A surprising result of the high-resolution (p,p$^\prime$) experiments is that the fine structure prevails in heavy deformed nuclei despite many-orders-of-magnitude larger LDs than in closed-shell or vibrational nuclei.
Indeed, wavelet analysis provided the first clear experimental signature for $K$ splitting of the ISGQR in heavy deformed nuclei \cite{kur18}.
Unlike the IVGDR, where $K$ splitting is already visible in the gross structure \cite{ber75}, the average separation of $K$ components  of the ISGQR is smaller than the average width and the width exhibits a broadening only with respect to spherical nuclei. 

\begin{figure}
\begin{center}
\resizebox{0.5\textwidth}{!}{%which 
  \includegraphics{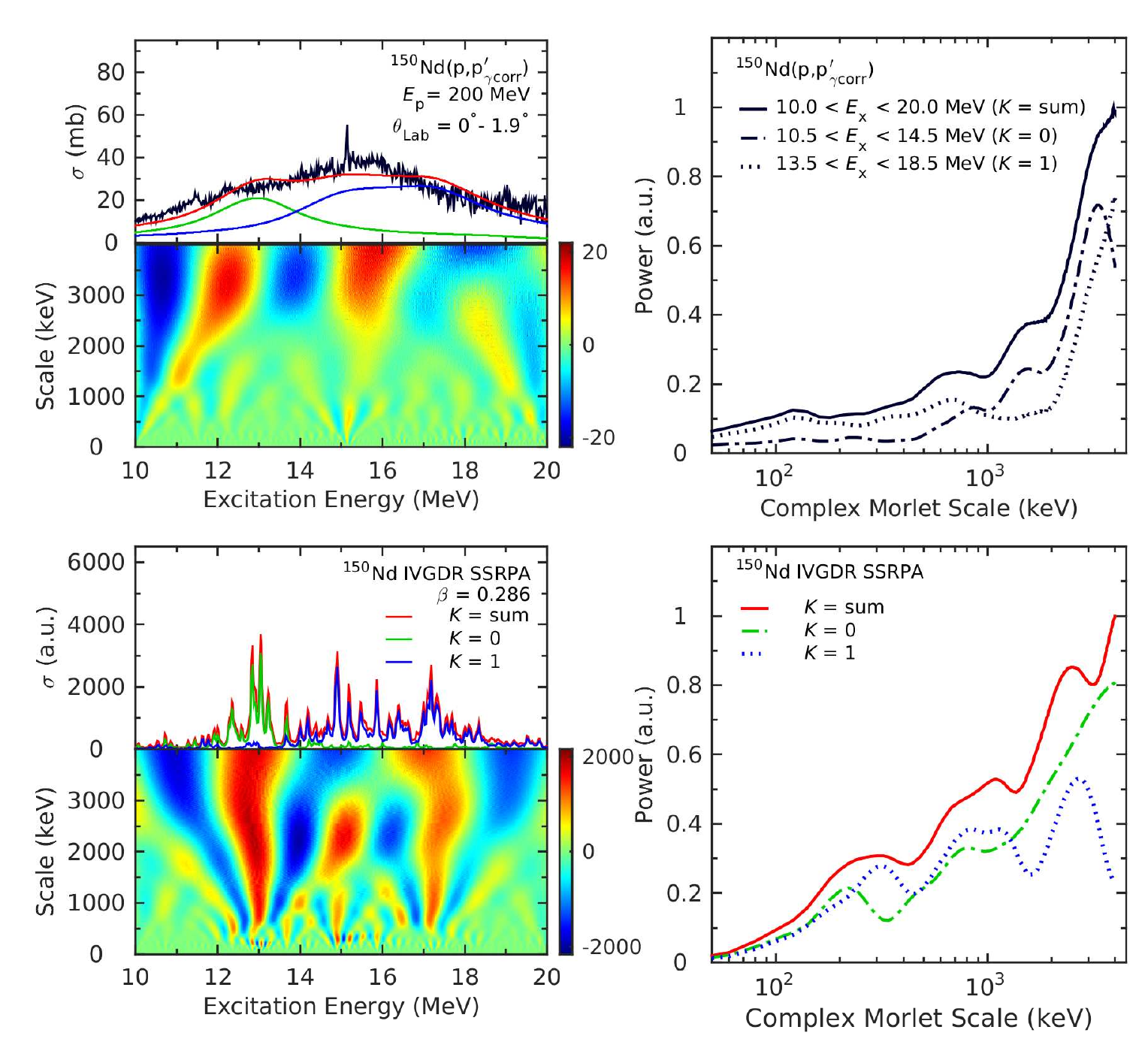}
%\vspace{5cm}       % Give the correct figure height in cm
}
\caption{
%Figure623.
Top row: CWT of a $^{150}$Nd(p,p$^\prime$) spectrum measured at $E_0 = 200$ MeV and $\theta_{\rm lab} = 0^\circ - 1.9^\circ$) (left) and corresponding power spectra (right) for the full (solid line) energy region of the IVGDR and  the energy region of $K = 0$ (dash-dotted line) and $K = 1$ (dotted line) dominance.
Bottom row:  Same for SSRPA calculations \cite{kle08}, but using the exact separation of $K = 0$ and $K = 1$ parts.
Taken from ref.~\cite{don16}.
}
\label{fig623}  
\end{center} 
\end{figure}

Here, we discuss whether characteristic scales of the fine structure in a heavy deformed nucleus ($^{150}$Nd) can be related to the $K$ components of the IVGDR.
The upper part of Fig.~\ref{fig623} shows on the l.h.s.\ a $0^\circ$ spectrum of the $^{150}$Nd(p,p$^\prime$)  reaction measured at iThemba LABS \cite{don18} and the corresponding CWT and on the r.h.s.\ the power spectra of the full data (solid line) and a separation into $K = 0$ (dash-dotted) and $K = 1$ (dotted) components. 
The latter were produced summing over parts of the spectrum ($10.5-14.5$ MeV for $K = 0$, $13.5-18.5$ MeV for $K = 1$), where one of the $K$ components dominates.
The power spectra do show distinctive differences in the two excitation regions with the strongest scales at about 1.5 MeV and 860 keV appearing in the low-energy($K = 0$) and at 680 keV in the high-energy ($K = 1$) part only.
The peak in the total power spectrum at about 740 keV seems to be the unresolved superposition of a $K = 0$ scale at about 860 keV and a $K = 1$ scale at about 680 keV.

The lower part of fig.~\ref{fig623} displays the results of an analogous analysis of a $^{150}$Nd  photoabsorption spectrum calculated with a Skyrme separable RPA (SSRPA)  approach \cite{kle08}.
It provides a good description of the experimental data (red line).
The corresponding power spectra show a picture very similar to the experimental results with different scales for the $K = 0$ and $K = 1$ components and bumps in the total power spectrum resulting from the superposition of $K = 0$ and 1 scales.  
Further details of such an analysis can be found in ref.~\cite{don16}.

\subsection{Level densities}
\label{subsec63}

In the following, we describe how level densities can be extracted from the (p,p$^\prime$)  spectra with a fluctuation analysis. 
The method was originally proposed to analyze $\beta$-delayed particle emission spectra \cite{jon76}, but later it was successfully adopted for the study of electron scattering data~\cite{mue83,kil87} and can be used in general for high-resolution spectra of nuclear reactions (see, e.g., refs.~\cite{kal07,kal06,end97}).  
The analysis takes advantage of the autocorrelation function in order to obtain a measure of the cross-section fluctuations with respect to a stationary mean value.

The method can be applied in an energy region where the mean level spacing $\langle D \rangle$ is smaller than the experimental energy resolution $\Delta E$. 
One has to distinguish between two possible cases: 
(i) $\left\langle \Gamma  \right\rangle \le \left\langle D \right\rangle$, i.e., the mean level width $\left\langle \Gamma \right\rangle$ is smaller than the average distance between levels and the fluctuations result from the density of  states and their incoherent overlap, and
(ii) $\left\langle \Gamma  \right\rangle > \left\langle D \right\rangle$, so-called Ericson fluctuations~\cite{eri60}, which result from the coherent overlap of the states.
In principle, it is possible to utilize the method in the Ericson regime, but the statistics has to be very high because of the large number of open decay channels. 
Thus, in practice one is usually limited to the regime $\left\langle \Gamma \right\rangle \le \left\langle D \right\rangle$.

The application of the fluctuation analysis is based on the assumption that the average spacing and intensity distributions can be described by random matrix theory (RMT)~\cite{wei09,mit10}, i.e., the average spacing can be described by a Wigner distribution and the transition strengths by a Porter-Thomas distribution.
This is motivated by the good description of nuclear excitations in the vicinity of the neutron separation energy by RMT \cite{haq82}.

% For one column wide figures use
\begin{figure}[b]
\begin{center}
\resizebox{0.45\textwidth}{!}{%
  \includegraphics{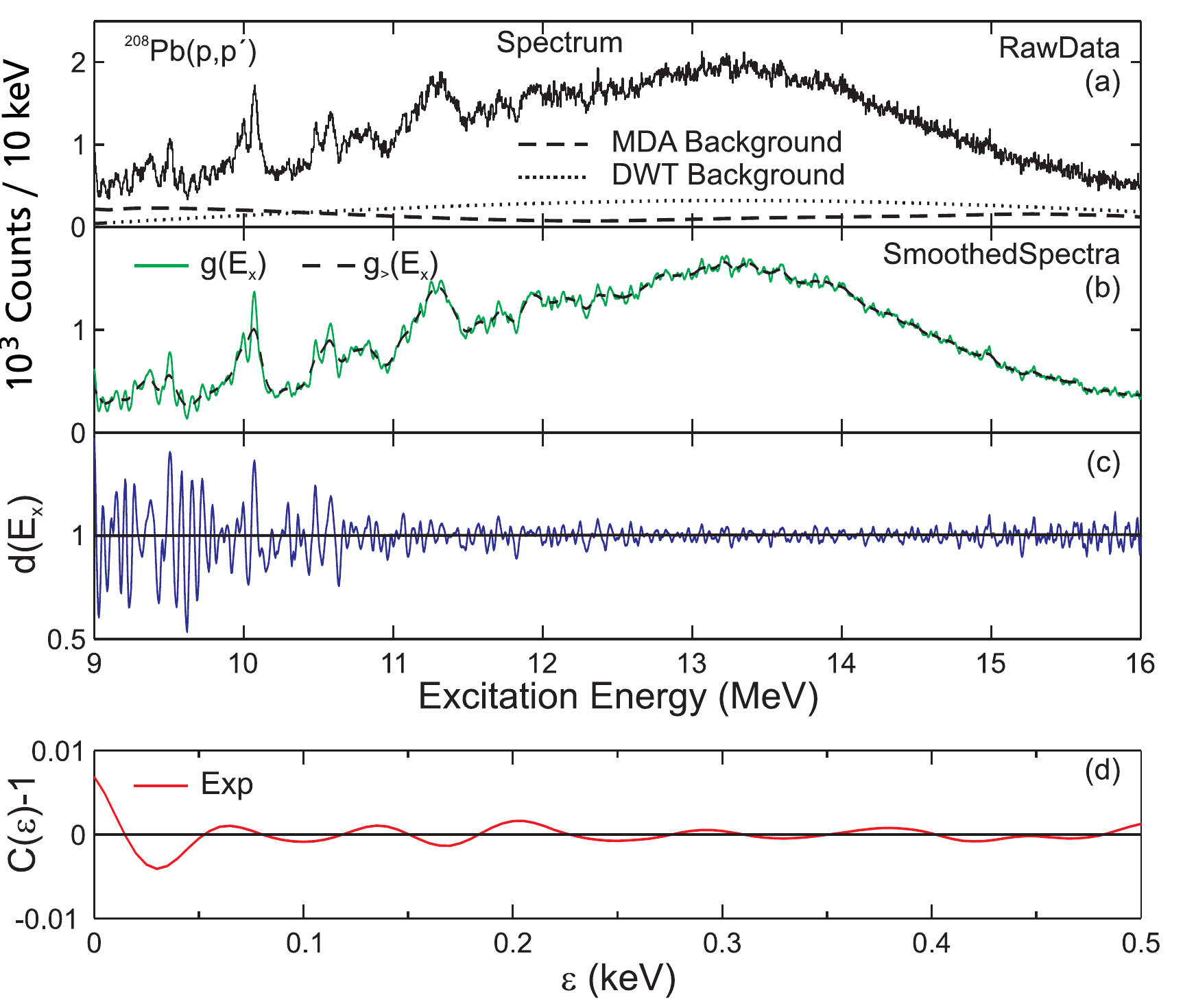}
%\vspace{5cm}       % Give the correct figure height in cm
}
\caption{
%Figure631.
Outline of the fluctuation analysis for the  example of the $^{208}$Pb(p,p$^\prime$) spectrum shown in fig.~\ref{fig611}.
(a) Spectrum and the background obtained from MDA (dashed line) \cite{pol12} and DWT (dotted line) \cite{pol14}. 
(b) Background-subtracted and smoothed spectra $g(E_{\rm x})$ and $g_>(E_{\rm x})$. 
(c) Stationary spectrum $d(E_{\rm x})$, eq.~(\ref{eq631}).  
(d) Experimental autocorrelation function, eq.~(\ref{eq632}).
Figure taken from ref.~\cite{pol14}.}
\label{fig631}  
\end{center} 
\end{figure}
The main steps of the fluctuation analysis are demonstrated for the $^{208}$Pb(p,p$^\prime$) data in the excitation-energy region of the IVGDR.
In order to obtain a spectrum containing only the information needed, one has to subtract any background not arising from excitations of the nuclear mode under investigation (a). 
In the present case, there are two independent results from the MDA (sec.~\ref{subsec31}) and DWT (sec.~\ref{subsec61}), fig.~\ref{fig631}(a).
After background subtraction, the spectrum contains the information on the fluctuations in the spectrum of the IVGDR. 
In order to eliminate the fluctuation contributions arising from finite statistics, the spectrum is folded with a Gaussian function with a width $\sigma$ chosen to be smaller than the experimental energy resolution. 
The resulting spectrum $g(E_{\rm x})$ is shown as green solid line in fig.~\ref{fig631}(b). 
Similarly, a second spectrum $g_>(E_{\rm x})$ is created by the convolution with a Gaussian function, whose width $\sigma_>$ is at least two times larger than the energy resolution in the experiment in order to remove gross structures from the spectrum (dashed line). 

The dimensionless stationary spectrum $d(E_{\rm x})$ defined by
\begin{equation}
    \label{eq631}
        d\left( {E_{\rm x}} \right) =
        \frac{{g_> \left( {E_{\rm x}} \right)}}{{g \left( {E_{\rm x}} \right)}}
\end{equation}
is shown in fig.~\ref{fig631}(c). 
As a result of the normalization on the local mean value, the energy dependence of the cross sections vanishes. 
The value of $d(E_{\rm x})$ is sensitive to the fine structure of the spectrum and distributed around an average intensity $\langle d(E_{\rm x})\rangle = 1$. 
With increasing excitation energy the mean level spacing is decreasing, and in turn the oscillations of $d(E_{\rm x})$ are damped.
A quantitative description is given by the autocorrelation function
\begin{equation}
    \label{eq632}
        C\left( \epsilon  \right) =
        \frac{{\left\langle {d\left( {E_{\rm x} } \right) \cdot d\left( {E_{\rm x}  +
        \epsilon } \right)} \right\rangle }}{{\left\langle {d\left( {E_{\rm x} }
        \right)} \right\rangle  \cdot \left\langle {d\left( {E_{\rm x}  +
        \epsilon } \right)} \right\rangle
        }}.
\end{equation}
The value $C(\epsilon = 0) - 1$ is nothing but the variance of $d(E_{\rm x})$
\begin{equation}
    \label{eq633}
        C\left( {\epsilon  = 0} \right) - 1 = \frac{{\left\langle {d^2
        \left( {E_{\rm x} } \right)} \right\rangle  - \left\langle {d\left( {E_{\rm x}
        } \right)} \right\rangle ^2 }}{{\left\langle {d\left( {E_{\rm x} }
        \right)} \right\rangle ^2 }}.
\end{equation}
According to ref.~\cite{jon76}, this experimental autocorrelation function (d) can be approximated by an expression
\begin{equation}
\label{eq634}
C(\epsilon) - 1 =  \frac{\alpha \cdot \langle \mbox{D} \rangle}{2 \sigma \sqrt{\pi}} \times f(\sigma,\sigma_>),
\end{equation}
%
%\begin{widetext}
%
%\begin{eqnarray}
%    \label{eq:autocorrtheo}
%        C(\epsilon) \; - \; 1 
%        \;\;\; = \;\;\; \frac{\alpha \cdot \langle \mbox{D} \rangle}{2
%       \sigma \sqrt{\pi}} \; \times & & \!\!\!\!\!\!\!\!\! \left\{ \exp
%        \left( -\frac{\epsilon^2}{4\sigma^2} \right) \; + \; \frac{1}{y}
%        \cdot \exp \left( -\frac{\epsilon^2}{4 \sigma^2 y^2} \right) \; -
%        \right. \nonumber \\ & & \nonumber \\ & & \!\!\!\!\!\!\!\!\!
%        \left. \sqrt{\frac{8}{1+y^{2}}} \cdot \exp \left( -
%        \frac{\epsilon^{2}}{4\sigma^2 (1+y^{2})} \right) \right\},
%\end{eqnarray}
%
%\end{widetext}
%
where the function $f$ depends on the experimental parameters (energy resolution, smoothing widths) only.
The value $\alpha$ is the sum of the normalized variances of the assumed spacing ($\alpha _{\rm D}$)  and transition width ($\alpha _{\rm I}$) distributions
\begin{equation}
    \label{eq635}
        \alpha  = \alpha _{\rm D}  + \alpha _{\rm I}.
\end{equation}
If only transitions with the same quantum numbers $J^\pi$ contribute to the spectrum, then $\alpha$ can be directly determined as the sum of the variances of the Wigner and Porter-Thomas distributions, $\alpha = \alpha_{\rm W} + \alpha_{\rm PT} = 0.273 + 2.0$. 
The mean level spacing $\langle {\rm D}\rangle$ is proportional to the variance of $d(E_{\rm x})$ and can be extracted from the value of $C(\epsilon = 0) - 1$. 
The nuclear level density can then be determined from the mean level spacing as $\rho(E)=1/\langle D\rangle$.

\begin{figure}
\begin{center}
\resizebox{0.4\textwidth}{!}{%
  \includegraphics{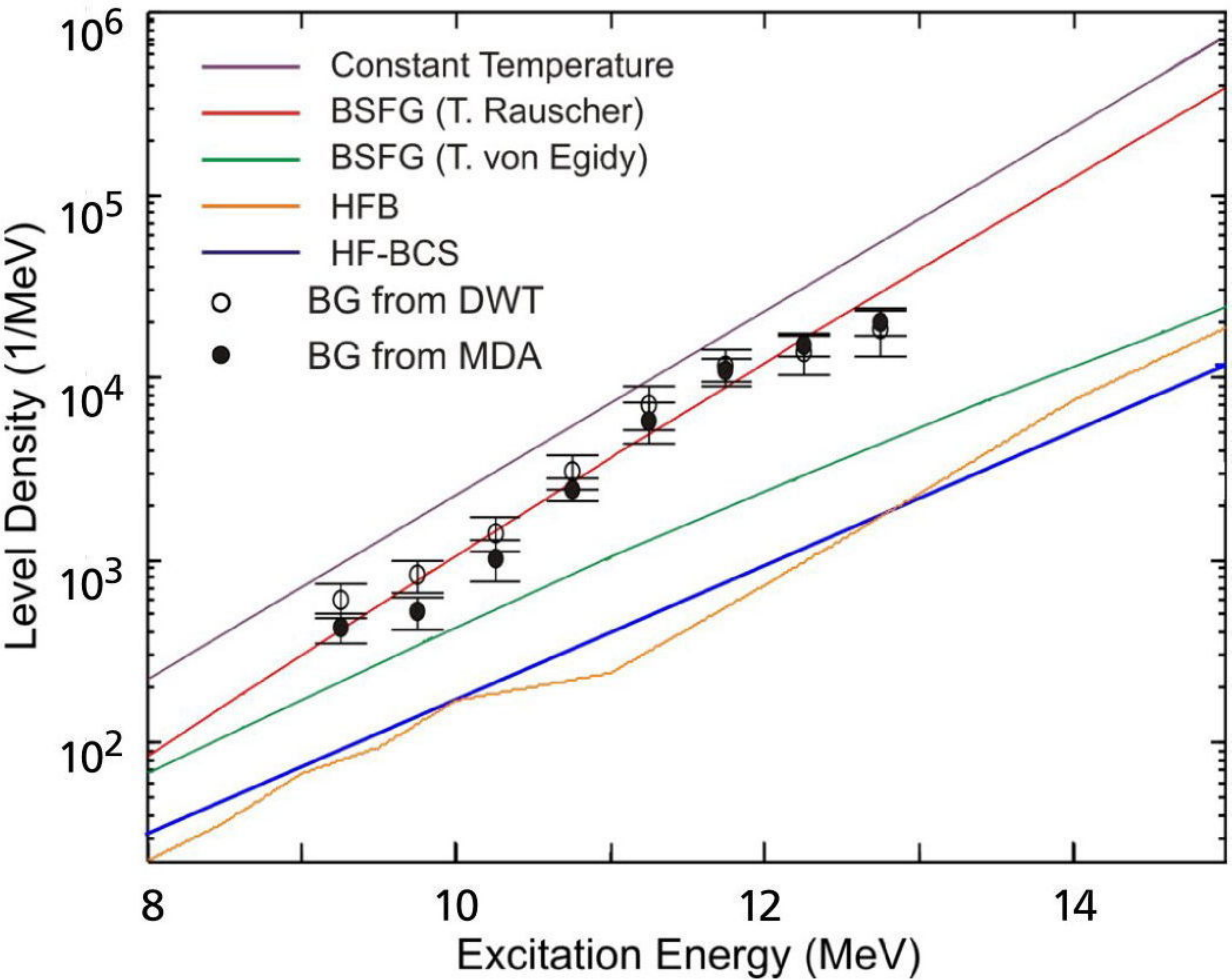}
%\vspace{5cm}       % Give the correct figure height in cm
}
\caption{
%Figure632.
Comparison of the experimentally obtained level densities of 1$^-$ states in $^{208}$Pb in the excitation-energy range from 9 to 12.5~MeV with predictions from the BSFG model using the parameterizations of ref.~\cite{rau97} (red line)  and \cite{egi05} (green line), the CT model \cite{egi09} (purple line), HFB-BCS \cite{dem01} (blue line) and HFB \cite{gor08} (orange line).
Figure taken from ref.~\cite{pol14}.}
\label{fig632}  
\end{center} 
\end{figure}

The experimental level densities of 1$^-$ states in $^{208}$Pb determined for the two different approaches of background subtraction are shown in fig.~\ref{fig632} as open and full circles, respectively.
For the fluctuation analysis the considered excitation-energy interval between 8.5~MeV and 16~MeV was split into subintervals of 0.5~MeV length and the mean level spacing was determined in each bin.
However, insufficient statistics of the experimental spectrum or the onset of the Ericson fluctuations lead to a drop of deduced level densities in the excitation-energy region above $~$12~MeV. 
The phenomenon has also been observed in a similar analysis of $M2$ resonances in $180^\circ$ electron scattering data~\cite{vnc99}. 
Repetition of the analysis using spectra for different angle bins suggests an excitation energy region $E_{\rm x} = 9 -12.5$ MeV in which the fluctuation analysis can be applied in the presented example.

The experimental results are compared in fig.~\ref{fig632} with different parameterizations of the back-shifted Fermi gas (BSFG)  \cite{rau97,egi05} and constant temperature (CT) models \cite{egi09} and with microscopic calculations performed in the framework of a Hartree-Fock-Bogoliubov (HFB) \cite{gor08} or a Hartree-Fock-BCS approach \cite{dem01}.
Good agreement with the BSFG parameterization of ref.~\cite{rau97} is found.
The constant temperature model of ref.~\cite{egi09} reproduces correctly the energy dependence but gives two times higher level densities.
All other models including the BSFG parameterization of ref.~\cite{egi05} and the microscopic HFB \cite{gor08} and HF-BCS \cite{dem01} approaches fail to reproduce the magnitude and the energy dependence of the experimental data.

\begin{figure}
\begin{center}
\resizebox{0.45\textwidth}{!}{%
  \includegraphics{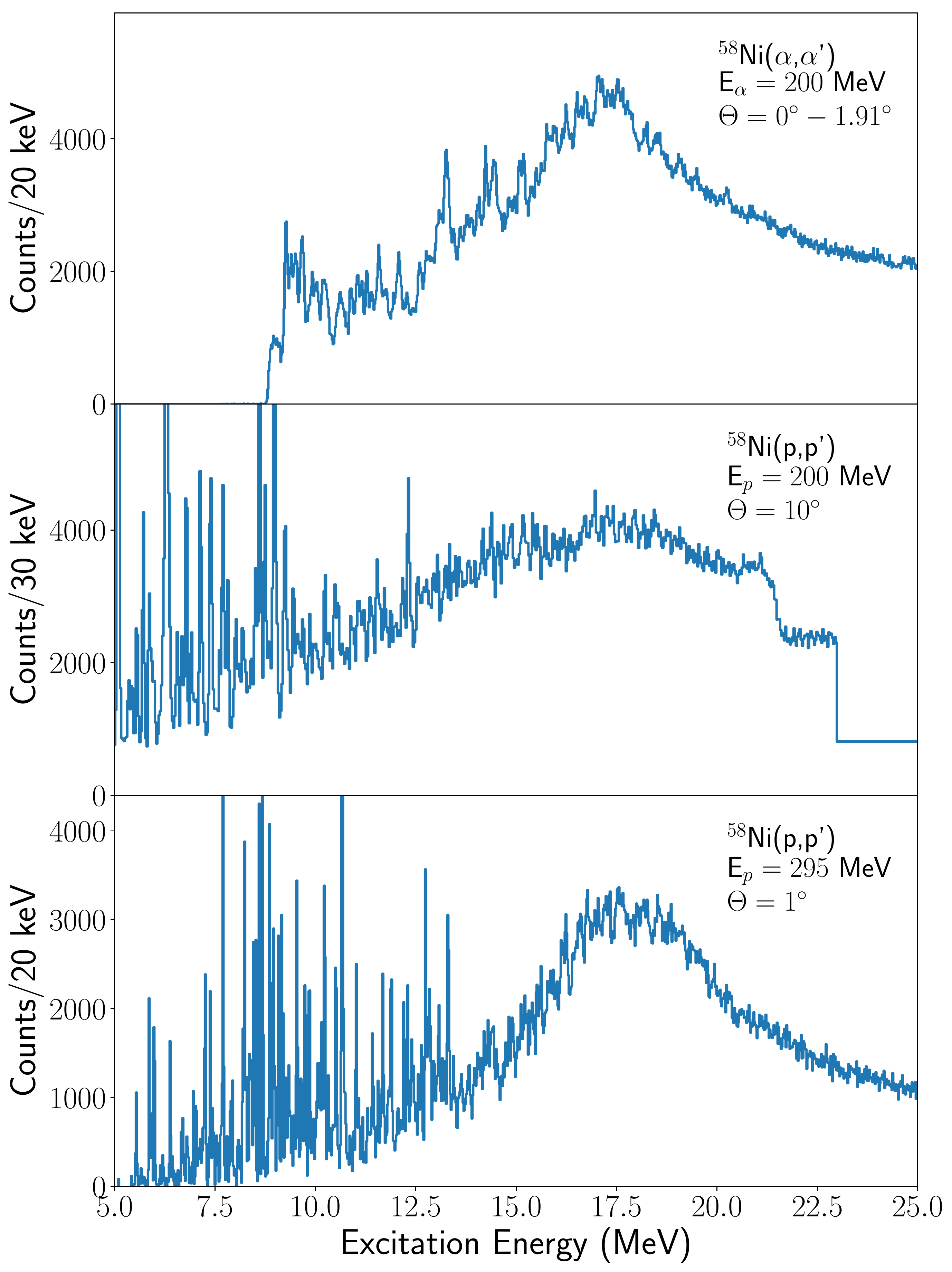}
%\vspace{5cm}       % Give the correct figure height in cm
}
\caption{
%Figure633.
High-resolution spectra of (p,p$^\prime$) and ($\alpha,\alpha^\prime$) scattering populating different giant resonances in $^{58}$Ni.
Top: ($\alpha,\alpha^\prime$ reaction at $E_0 = 200$ MeV and $\theta_{\rm lab} = 0^\circ-1.91^\circ$ measured at iThemba LABS (ISGMR).
Middle: (p,p$^\prime$) reaction at $E_0 = 200$ MeV and $\theta_{\rm lab} = 10^\circ$ measured at iThemba LABS \cite{she09} (ISGQR).
Bottom: (p,p$^\prime$) reaction at $E_0 = 295$ MeV and $\theta_{\rm lab} = 0.4^\circ$ measured at RCNP (IVGDR).
}
\label{fig633}  
\end{center} 
\end{figure}

As pointed out above, fine structure of giant resonances is a global phenomenon. 
This has been utilized to test predictions of a possible parity dependence of the LD \cite{nak97,hil06,hou09} with impact on astrophysical reaction network calculations by a fluctuation analysis of spectra of the ISGQR and M2 resonances populating $J^\pi = 2^+$ and $2^-$ states in the same nuclei, respectively \cite{kal07}.
In a similar way, data for different giant resonances in the same nucleus may contribute to improve our limited knowledge of the spin dependence \cite{vnc19a}.
As an example, a set of high-resolution spectra of the ISGQR \cite{she09}, the IVGDR measured at RCNP, and the ISGMR measured recently with $\alpha$ scattering at the $0^\circ$ setup iThemba LABS  are shown in fig.~\ref{fig633} for the nucleus $^{58}$Ni, providing information on the $J = 0,1,2$ LDs. 
Such combined data now exist for various nuclei and a systematic analysis is underway. 

\section{Gamma strength function}
\label{sec7}

The GSF describes the average $\gamma$-decay behavior of a nucleus.
%It depends on the level density at initial and final energy.
In general, all multipoles allowed for electromagnetic processes contribute but in practice E1 dominates. 
Thus, the isovector giant dipole resonance (IVGDR) dominates the GSF at higher excitation energies, but M1 (and E2) can also contribute to the total GSF.  
For the special case of $\gamma$ decay to the g.s.\ the GSF can be related to the photoabsorption cross section by the principle of detailed balance
\begin{equation}
	\label{eq71}
	f^{X\lambda}(E_\gamma, J) = \frac{1}{(\pi\hbar c)^2} \frac{2J_0+1}{2J+1}
	\frac{\left\langle\sigma_{abs}^{X\lambda}\right\rangle}{E_\gamma^{2\lambda-1}},
\end{equation}
where $J, J_0$ are the spins of excited and ground state (g.s.), respectively, $X$ denotes the electric or magnetic character, and $\lambda$ the multipolarity.
The brackets $\langle \rangle$ indicate averaging over an energy interval.
Thus, g.s.\ photoabsorption experiments are a possible source of information on the GSF.

\subsection{Test of global parameterizations}
\label{subsec71}

Many applications require information on the GSF for a large variety of nuclides and often no experimental data are available. 
Thus, a large effort is made to construct systematic parameterizations summarized in the RIPL-3 data library \cite{cap09}.       
In this section, we briefly demonstrate how the (p,p$^\prime$) data can contribute to critically test these parameterizations.
Again, we use the $^{208}$Pb results as a representative example.  
 
\begin{figure}
\begin{center}
\resizebox{0.45\textwidth}{!}{%
  \includegraphics{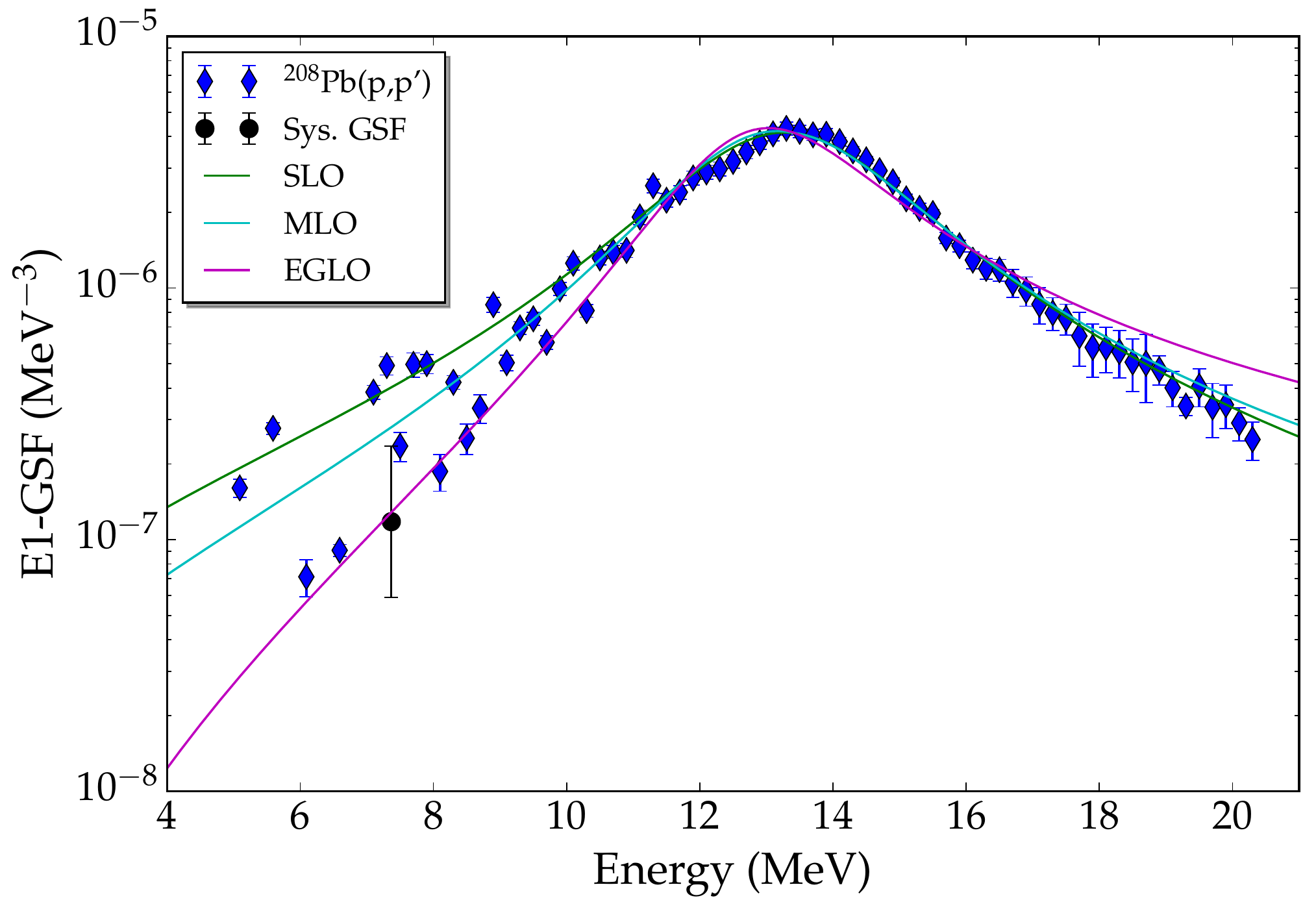}
%\vspace{5cm}       % Give the correct figure height in cm
}
\caption{
%Figure711.
E1 GSF of $^{208}$Pb deduced from the (p,p$^\prime)$ data \cite{tam11,pol12} (blue diamonds) in comparison with the SLO (green line), MLO (cyan line), and EGLO (magenta line) models explained in the text.
The black circle shows the prediction from experimental systematics at the neutron separation threshold \cite{cap09}.
Figure taken from ref.~\cite{bas16}.}
\label{fig711}  
\end{center} 
\end{figure}

The photoabsorption cross sections in $^{208}$Pb determined from the analysis described in sec.~\ref{subsec33} can be converted to the E1 GSF with the aid of eq.~(\ref{eq71}). 
Figure~\ref{fig711} displays the result in comparison with three widely used models and with a GSF value at the neutron separation threshold deduced from experimental systematics over a wide mass range~\cite{cap09}.
Two of the models for the description of the E1 GSF are a standard Lorentzian (SLO) and an enhanced generalized Lorentzian (EGLO).
The EGLO consists of two terms~\cite{kop93}, a Lorentzian with an energy- and temperature-dependent width $\Gamma_K(E,T)$ and a term describing the shape of the low-energy part of the GSF allowing for a finite GSF at zero temperature. 
The temperature dependence is estimated within Fermi liquid theory.
The modified Lorentzian model (MLO) is based on general relations between the GSF and the imaginary part of the nuclear response function predicting an enhancement of the GSF with increasing temperature. 
The energy- and temperature-dependent width is calculated with micro-canonically distributed initial states~\cite{plu99}.
For details of the models see ref.~\cite{cap09}.

The predictions for $^{208}$Pb are shown in Fig.~\ref{fig711} as green (SLO), magenta (EGLO), and cyan (MLO) curves.
In the region around the maximum of the IVGDR all models provide a good description of the data.
The high-energy tail of the IVGDR is well described by SLO and MLO while EGLO overestimates the photoabsorption cross sections.
In $^{208}$Pb, the low-energy tail of the IVGDR exhibits strong fluctuations which complicate the comparison with smooth strength functions. 
For excitation energies down to about 8 MeV, MLO describes the average behavior fairly well while SLO(EGLO) are roughly consistent with the upper(lower) limits of the fluctuations but over(under)estimate the average cross sections.
Between 6 and 8 MeV a resonance-like structure dominates the GSF identified as the PDR in $^{208}$Pb \cite{pol12}.
This low-energy resonance is not included in the models.
Finally, the GSF value expected at the neutron threshold ($S_n = 7.37$ MeV in the present case) from experimental systematics of neutron-capture cross sections (black circle) is almost an order of magnitude smaller than the experimental strengths in the PDR.
However, this may be an artefact of the unusually low level density in the doubly magic nucleus $^{208}$Pb with corresponding strong fluctuations of individual strengths at energies close to the neutron threshold (note that the experimental GSF values correspond to energy bins rather than to individual transitions for excitation energies above 7 MeV (cf.\ table I in ref.~\cite{pol12}). 

\begin{figure}
\begin{center}
\resizebox{0.45\textwidth}{!}{%
  \includegraphics{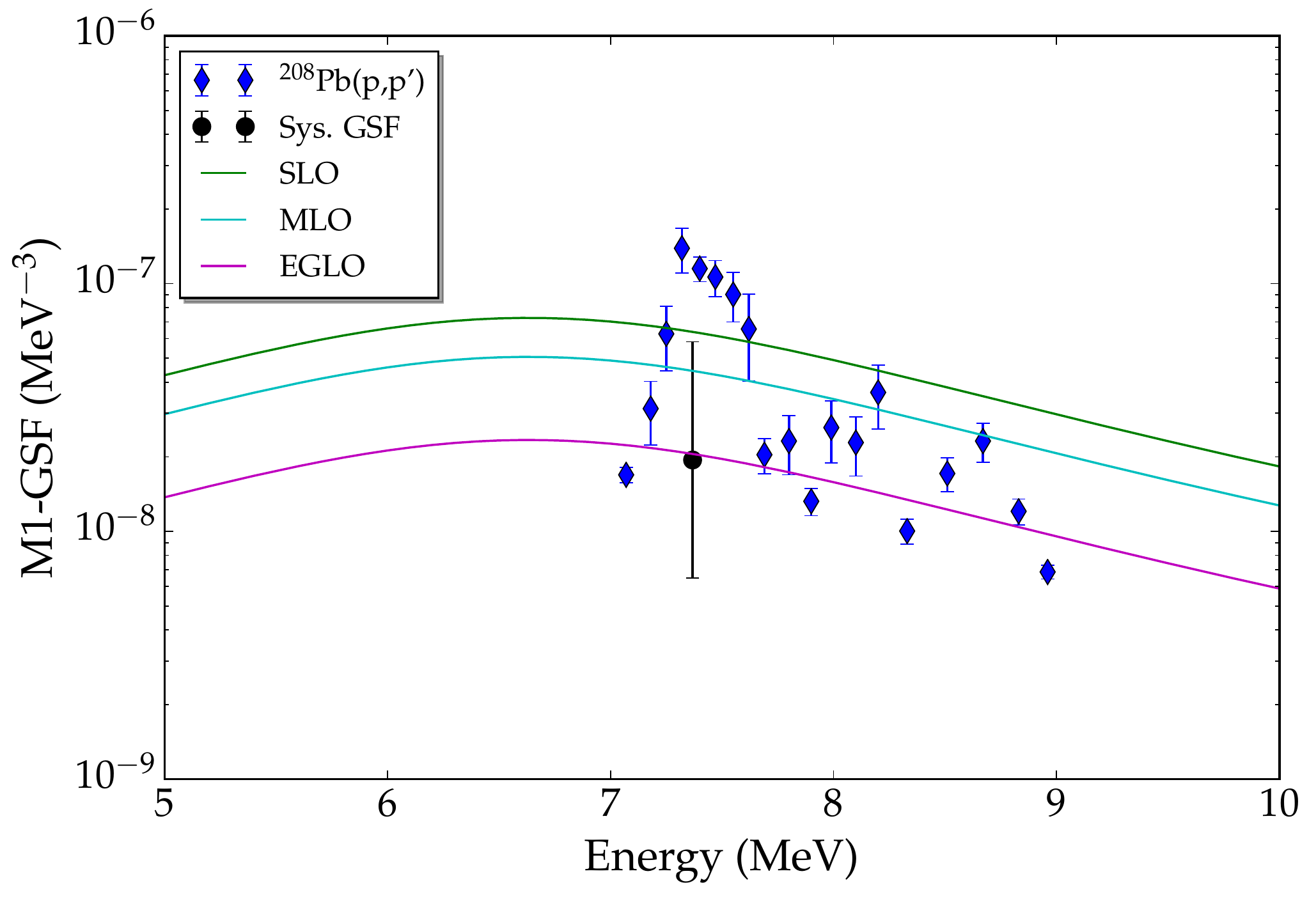}
%\vspace{5cm}       % Give the correct figure height in cm
}
\caption{
%Figure712.
Same as fig.~\ref{fig711} but for the M1 GSF.
%M1 GSF of $^{208}$Pb deduced from the (p,p$^\prime)$ data \cite{tam11,pol12} with the method described in sec.~\ref{subsec51}  (blue diamonds) in comparison with the SLO (green line), MLO (cyan line), and EGLO (magenta line) models explained in the text.
%The black circle shows the prediction from experimental systematics at the neutron separation threshold \cite{cap09}.
Figure taken from ref.~\cite{bas16}.}
\label{fig712}  
\end{center} 
\end{figure}
Figure~\ref{fig712} presents the M1 part of the GSF deduced with the method described in sec.~\ref{subsec34} in comparison with SLO, EGLO, and MLO model predictions for $^{208}$Pb.
The M1 GSF model results are derived from the E1 models by applying an empirical mass-dependent correction and assuming a SLO parameterization of the spin-flip M1 resonance with an energy centroid $E_r = 41 \cdot A^{-1/3}$ and $\Gamma_r = 4$ MeV \cite{kop87}.
Obviously, the assumed resonance properties represent a poor approximation of the data only.
The theoretical maxima are about 500 keV too low and the experimental width is grossly overestimated.
As a result, the predicted total strengths of the spin-flip M1 resonance exceed the experimental value $\sum B({\rm M1}) = 20.5(13) \, \mu_N^2$ \cite{bir16} by factors ranging from two (EGLO) to five (SLO).

\begin{figure}
\begin{center}
\resizebox{0.45\textwidth}{!}{%
  \includegraphics{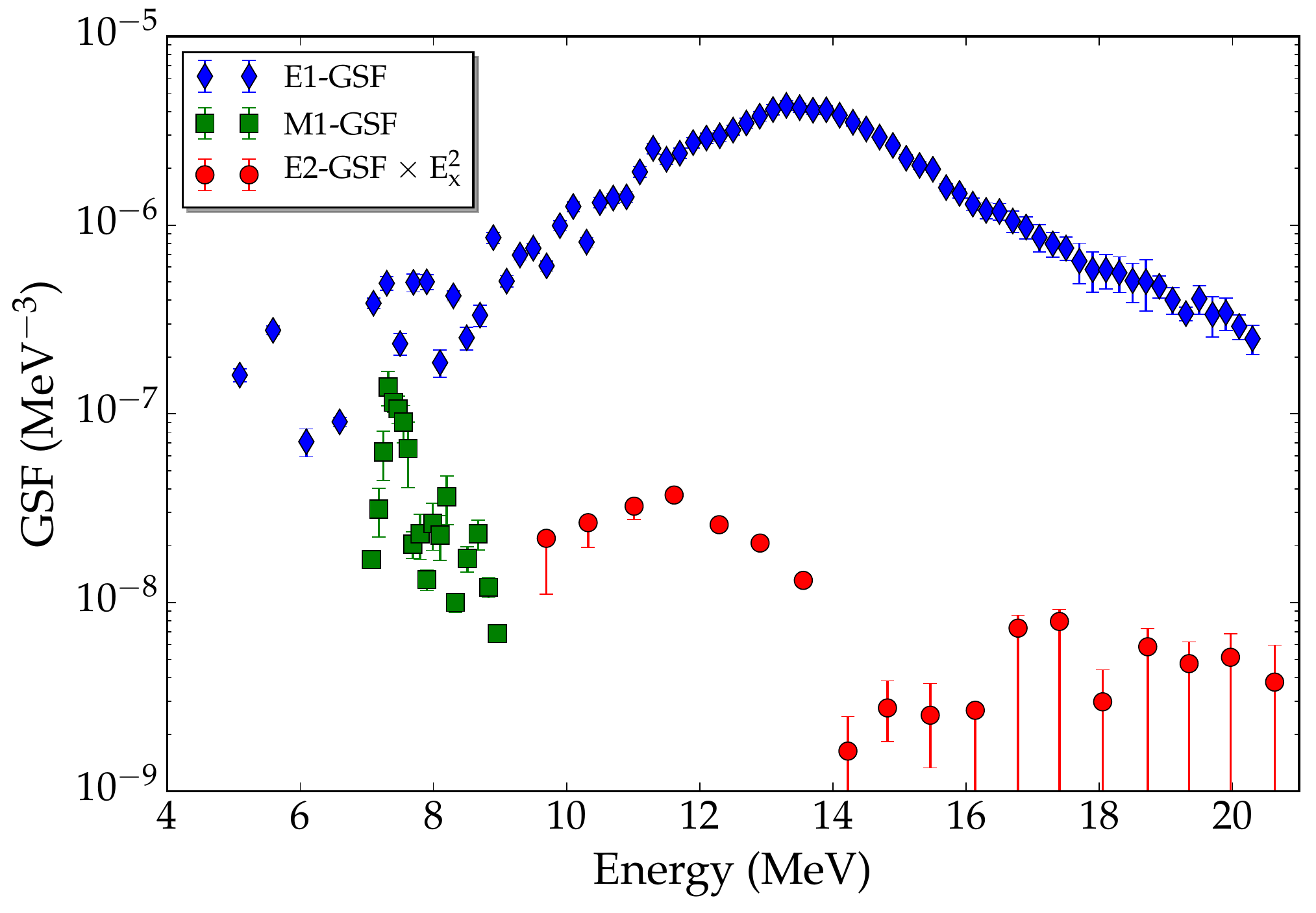}
%\vspace{5cm}       % Give the correct figure height in cm
}
\caption{
%Figure713.
 Comparison of E1, M1 and E2 contributions to the GSF of $^{208}$Pb.
Figure taken from ref.~\cite{bas16}.
}
\label{fig713}  
\end{center} 
\end{figure}

The E2 contribution to the GSF can be estimated using the MDA results of an $(\alpha$,$\alpha^\prime)$ scattering experiment \cite{you04}. 
The resulting ISGQR strength distribution exhausts $100\pm15\%$ of the energy weighted sum rule (EWSR). 
Figure~\ref{fig713} summarizes the E1, M1 and E2 contributions to the total GSF.
As can be seen, E1 strength dominates at all $\gamma$ energies. 
The M1 contribution is of the order of a few percent for excitation energies above 8 MeV and reaches about 10\% (up to 30\% in a single bin) near the peak of the M1 resonance. 
The E2 contribution is of comparable magnitude to M1 but located at higher excitation energies. 
Because of the simultaneous strong rise of the E1 part in the IVGDR energy region the E2 contribution to the GSF at the maximum of the E2 resonance is about 1\% only.  

\subsection{Validity of the Brink-Axel hypothesis in the PDR energy region}
\label{subsec72}

\begin{figure}
\begin{center}
\resizebox{0.4\textwidth}{!}{%
  \includegraphics{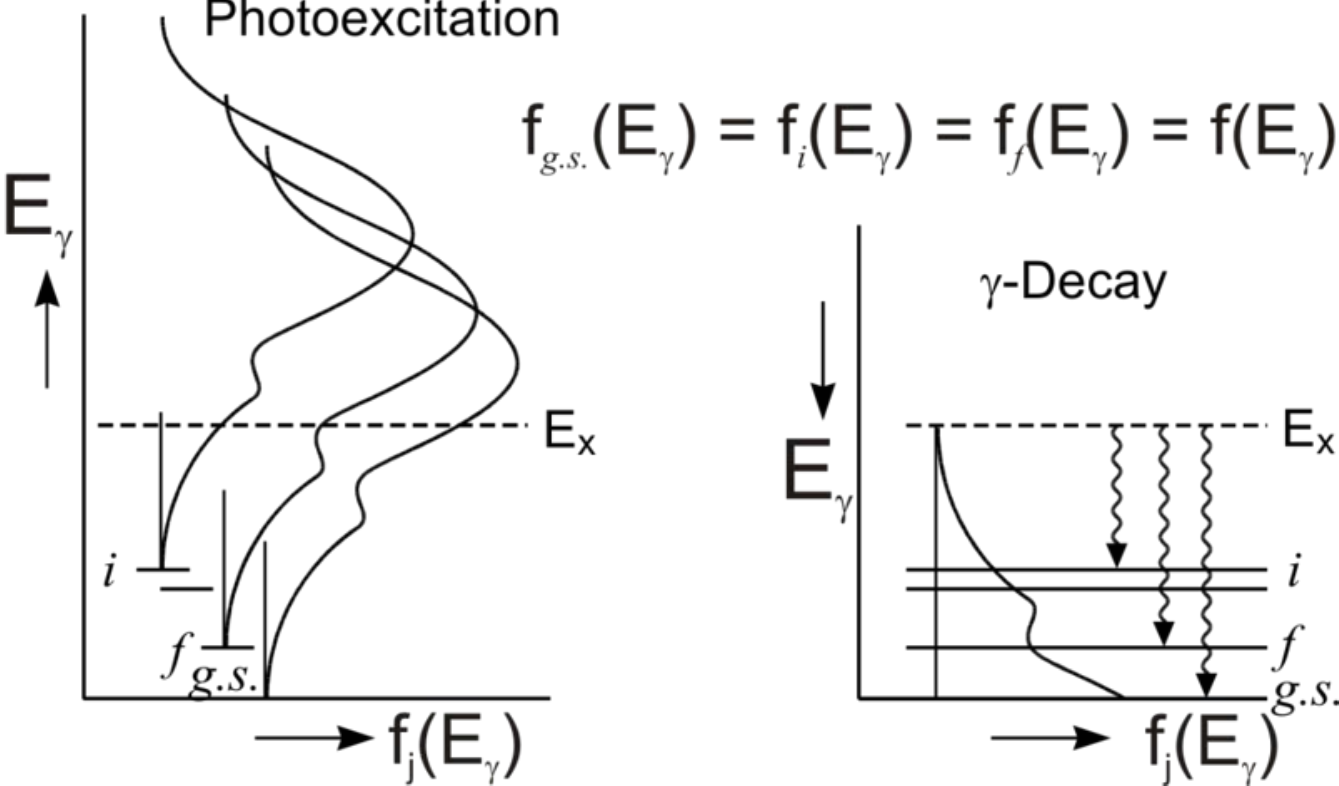}
%\vspace{5cm}       % Give the correct figure height in cm
}
\caption{
%Figure721.
Illustration of the generalized Brink-Axel hypothesis.
For explanations see text.}
\label{fig721}  
\end{center} 
\end{figure}

Knowledge of the GSF is required for calculations of statistical nuclear reaction in astrophysics \cite{arn07}, reactor design \cite{cha11}, and waste transmutation \cite{sal11}.
Most applications imply an environment of finite temperature, notably in stellar scenarios \cite{wie12}, and thus reactions on excited states (e.g., in a (n,$\gamma$) reaction) become relevant.
Their contributions to the reaction rates are usually estimated applying the Brink-Axel (BA) hypothesis \cite{bri55,axe62} illustrated in fig.~\ref{fig721}.
Historically, it was formulated for the IVGDR stating that the strength function built on an excited state is the same as on the g.s. (l.h.s.\ of fig.~\ref{fig721}).
It was shown to hold approximately for not too high temperatures \cite{bbb98}.
Later, it was generalized to hold for absorption and decay applying the principle of detailed balance (r.h.s.\ of fig.~\ref{fig721}).
Then, the GSF becomes independent of initial and final state energy and depends on the $\gamma$ energy only.
This is nowadays a commonly used assumption to calculate the low-energy E1 and M1 strength functions of astrophysical interest. 
However, recent theoretical studies \cite{joh15,hun17} put that into question demonstrating that the strength functions of collective modes built on excited states do show an energy dependence.
However, numerical results for E1 strength functions in light nuclei showed an approximate constancy consistent with the BA hypothesis \cite{joh15}. 

The so-called Oslo method, where primary spectra of $\gamma$ decay following compound nuclear reactions are extracted, is a major source of data on the GSF below the particle thresholds.
Since the $\gamma$ transmission probability is proportional to the product of the GSF and the final-state LD, assumption of the generalized BA hypothesis is a prerequisite of the analysis \cite{sch00}.     
Recent Oslo-type experiments have indeed demonstrated independence of the GSF from excitation energies and spins of initial and final states in a given nucleus in accordance with the BA hypothesis \cite{lar17,gut16,cre18}.
However, there are a number of results which clearly indicate violations in the low-energy region when comparing $\gamma$ emission and absorption experiments.
For example, the GSF in heavy deformed nuclei at excitation energies of $2 -3$ MeV is dominated by the orbital M1 scissors mode \cite{boh84} and potentially large differences in B(M1) strengths are observed between upward \cite{end05} and downward \cite{gut12,ang16} GSFs. 
Furthermore, at very low energies ($< 2$ MeV) an increase of GSFs is observed in Oslo-type experiments \cite{lar17,voi04}, which cannot have a counterpart in ground state absorption experiments on even-even nuclei because of the pairing gap.           

For the low-energy E1 strength in the region of the PDR, the validity of the BA hypothesis is far from clear when comparing results from the Oslo method with photoabsorption data.
Below particle thresholds most information on the GSF stems from nuclear resonance fluorescence (NRF) experiments, which suffer from the problem of unobserved branching ratios to excited states. 
These can be corrected, in principle, by Hauser-Feshbach calculations assuming statistical decay \cite{rus09}.
The resulting correction factors are sizable and show a strong dependence on the neutron threshold energy and the g.s.\ deformation.
On the other hand, there are clear indications of non-statistical decay behavior of the PDR from recent measurements \cite{rom15,loe16,isa19}.
Violation of the BA hypothesis was also claimed in a simultaneous study of the $(\gamma,\gamma^\prime)$ reaction and average ground-state branching ratios \cite{ang12} in $^{142}$Nd (see, however, the erratum \cite{ang15} to ref.~\cite{ang12}).   
Clearly, information on the GSF in the PDR energy region from independent experiments is called for.
The (p,p$^\prime$) experiments described in this review are almost ideal for this purpose, since they simultaneously provide the E1 and (spin-)M1 part of the GSF and -- with a completely independent method -- also the level density.

\begin{figure}
\begin{center}
\resizebox{0.45\textwidth}{!}{%
  \includegraphics{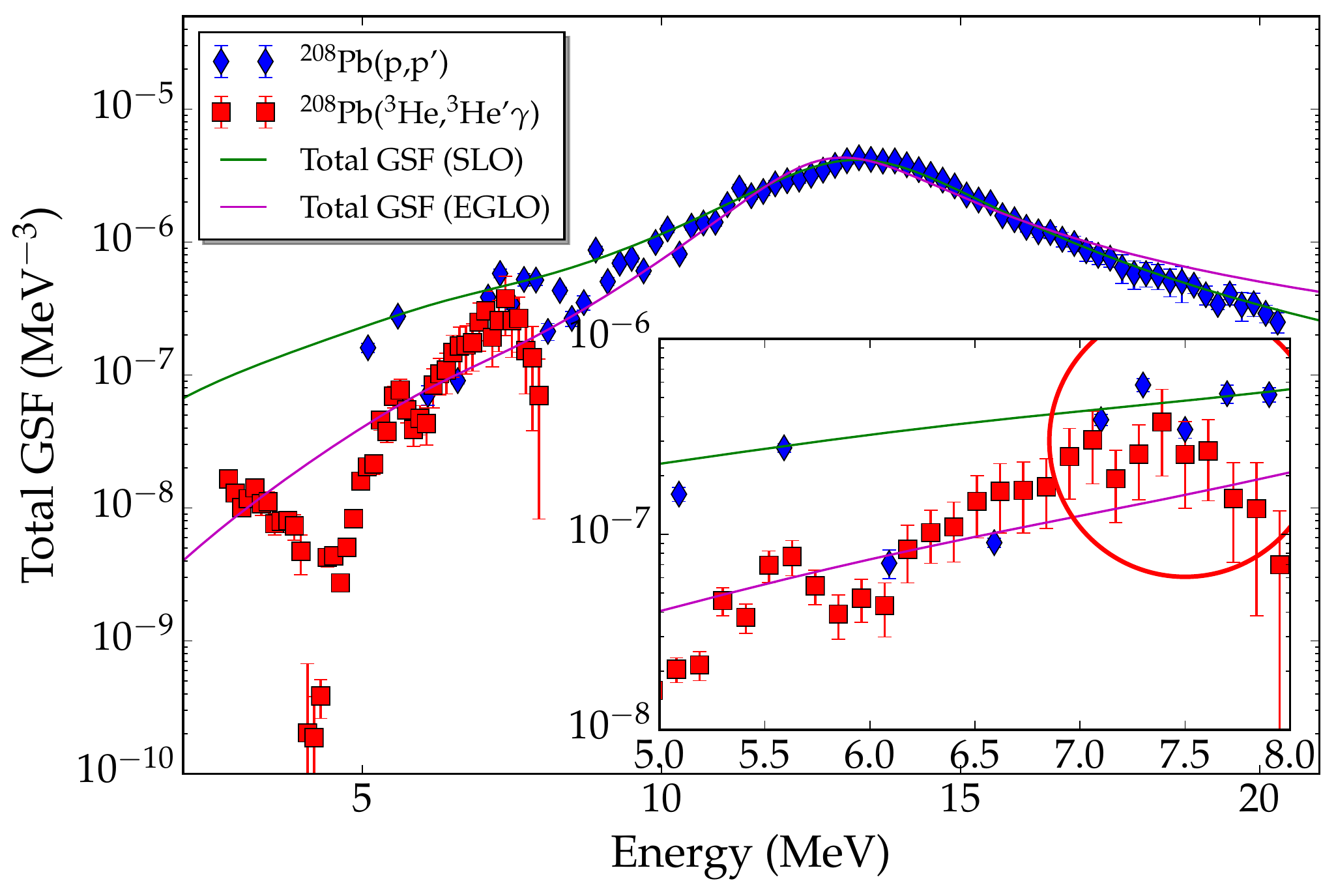}
%\vspace{5cm}       % Give the correct figure height in cm
}
\caption{
%Figure722.
Total GSF of $^{208}$Pb deduced from the (p,p$'$) data \cite{tam11,pol12} in comparison with the reanalyzed (for details see ref.~\cite{bas16}) results from the Oslo experiment \cite{sye09}.
The inset shows an expanded view of the low-energy region $5-8$ MeV.
Figure taken from ref.~\cite{bas16}.}
\label{fig722}  
\end{center} 
\end{figure}
The total GSF of $^{208}$Pb is displayed in fig.~\ref{fig722} (blue diamonds)  and compared with reanalyzed \cite{bas16} data derived with the Oslo method from a $^{208}$Pb($^3$He,$^3$He$^\prime \gamma$) experiment (red squares) \cite{sye09}.
There are overlapping results from both experiments in the energy region between 5 and 8 MeV (see inset of fig.~\ref{fig722}).
The GSF derived from the (p,p$^\prime)$ data is systematically higher in the PDR region although they seem still compatible within error bars in the peak region around the neutron threshold.
Between 6 and 7 MeV consistent results are found while below 6 MeV the strong transitions observed in ref.~\cite{pol12} exceed the average $\gamma$ strength in the Oslo data by factors 4 to 5.
However, one should be aware that single transitions are analyzed for excitation energies $E_{\rm x} < 7$ MeV \cite{pol12} and the level density of $1^-$ states excited from the ground state is probably too low to discuss an average behavior in the PDR region.
Rather, the upward GSF is dominated by Porter-Thomas intensity fluctuations. 

\begin{figure}
\begin{center}
\resizebox{0.45\textwidth}{!}{%
  \includegraphics{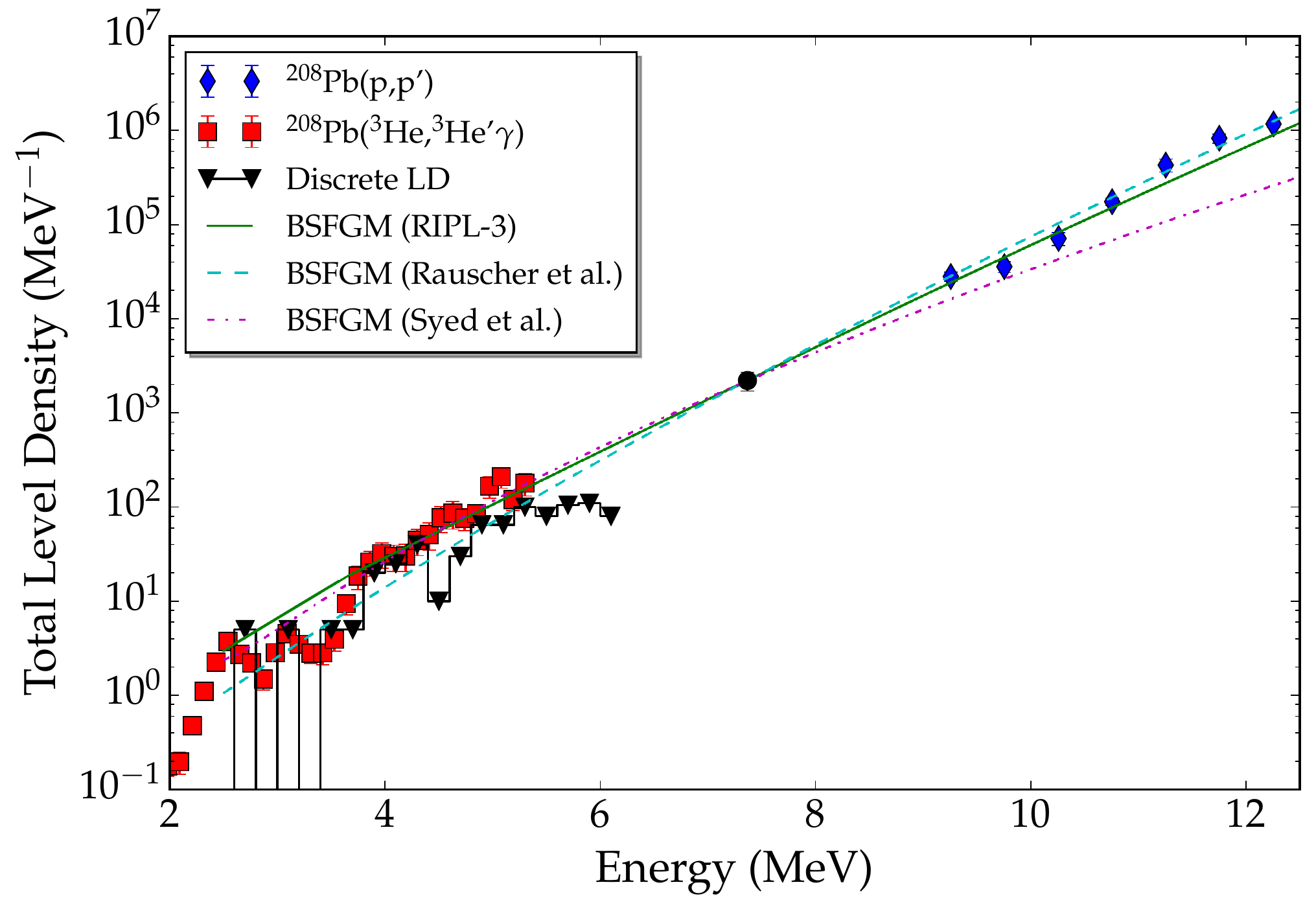}
%\vspace{5cm}       % Give the correct figure height in cm
}
\caption{
%Figure723.
Total LD in $^{208}$Pb from the (p,p$'$) data \cite{bas16} in comparison with the reanalyzed results from the Oslo experiment \cite{sye09}.
The black downward triangles are results from counting the levels identified in ref.~\cite{heu16} in 200 keV bins.
The magenta dash-dotted, green solid and cyan dashed lines are BSFG model predictions with the parameters of ref.~\cite{sye09}, ref.~\cite{cap09} and ref.~\cite{rau97}, respectively. 
Figure taken from ref.~\cite{bas16}.}
\label{fig723}  
\end{center} 
\end{figure}

The LD of $1^-$ states in $^{208}$Pb in the excitation energy region from 9 to 12.5~MeV was determined from the (p,p$^\prime$) data \cite{pol14} as discussed in sec.~\ref{subsec63}.
In order to compare with the results from the Oslo experiment, it must be converted to a total LD.
The corresponding spin distribution was calculated with the aid of systematic BSFG parameterizations and their variation was taken as a measure of the systematic uncertainty of the procedure (for details see ref.~\cite{bas16}).
Figure \ref{fig723} displays the resulting total LD (blue diamonds) together with results from the Oslo experiment (red squares), from level counting \cite{heu16} (black triangles) at lower energies and the data point at the neutron threshold from neutron capture \cite{cap09} (black square).  
Several BSFG model results are shown as solid \cite{cap09}, dashed \cite{rau97} and dotted \cite{sye09} lines, respectively. 
The RIPL-3 parameterization \cite{cap09} provides a very satisfactory description of all experimental data indicating that the decomposition into GSF and LD in the Oslo method is essentially correct.  

\begin{figure}
\begin{center}
\resizebox{0.45\textwidth}{!}{%
  \includegraphics{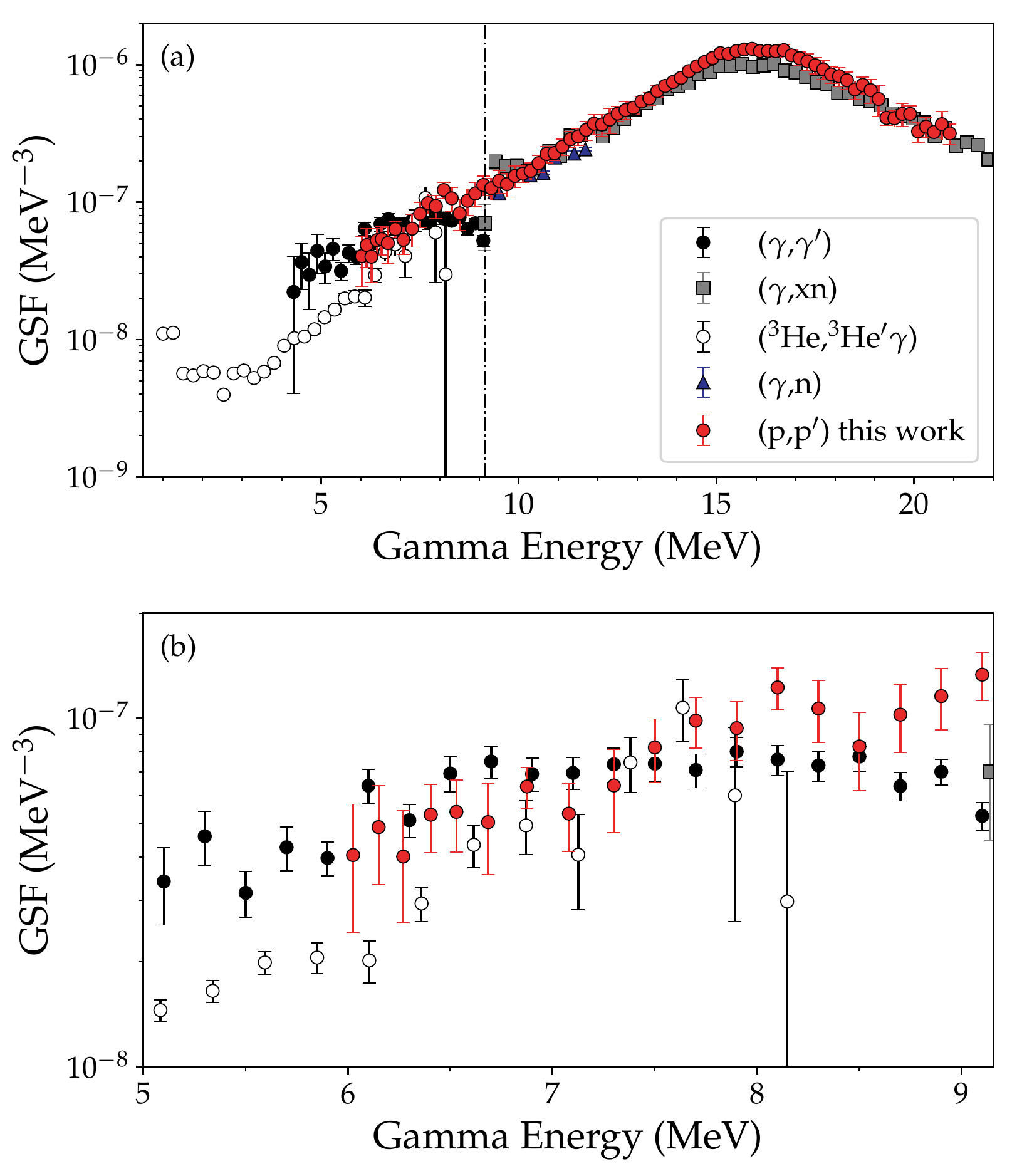}
%\vspace{5cm}       % Give the correct figure height in cm
}
\caption{
%Figure724.
(a) GSF of $^{96}$Mo from the (p,p$^\prime$) reaction \cite{mar17} (red circles) compared with ($^3{\rm He},^3{\rm He}^\prime \gamma$)  \cite{gut05,lar10} (open circles),  ($\gamma$,xn) \cite{bei74} (grey squares), ($\gamma,n$) \cite{uts13}, (blue upward triangles) and ($\gamma,\gamma^\prime$) data including a statistical-model correction for unobserved branching ratios \cite{rus09} (black circles).
(b) Expanded range from 5 MeV to neutron threshold.   
Figure taken from ref.~\cite{mar17}.}
\label{fig724}  
\end{center} 
\end{figure}

Another study of this type was performed for $^{96}$Mo \cite{mar17}, a considerably deformed nucleus with LDs high enough to permit a comparison with the GSF from a decay experiment averaging over appropriate energy intervals. 
The choice of $^{96}$Mo was motivated by the large discrepancies of GSFs derived from Oslo \cite{gut05,lar10} and NRF \cite{rus09} experiments.
The top part of Fig.~\ref{fig724} summarizes the available GSF data.
The energy region below neutron threshold is expanded in the bottom part showing the results from the Oslo (open circles), the NRF (black circles) and the (p,p$^\prime$) experiments (red circles).
For $\gamma$ energies between 6 and 8 MeV covered by all experiments, the GSF deduced from Coulomb excitation lies between the two other results but overall agrees better with the Oslo result.
The LD deduced from the (p,p$'$) data can again be described consistently with the Oslo results at lower energies using BSFG model parameterizations.
For details see ref.~\cite{mar17}. 

\section{Conclusions and outlook}
\label{sec8}

This review presents a new approach to study the electric and spin-magnetic dipole strength distributions in nuclei based on very forward-angle inelastic proton scattering including $0^\circ$ at energies of a few 100 MeV, where these modes dominate the cross sections.
Their decomposition can be achieved with two independent methods based on the momentum-transfer (or angle) dependence of the cross sections (MDA) or on polarization-transfer observables (PTA).
Since the cases studied so far do show good correspondence of the two methods, future experiments can restrict themselves to the MDA avoiding the long measurement times needed for statistical significance of polarization-transfer observables determined in a secondary scattering process.
Systematic uncertainties of the MDA at higher excitation energies are presently mainly determined by the assumptions about the continuum background, in particular for lighter nuclei.
It is dominated by quasi-free scattering and systematic parameterizations are available in the energy range of interest \cite{kal90}.
However, these do not include data at very small momentum transfers.       
Thus, future studies -- in particular in lighter nuclei, where the Coulomb-excitation cross sections are reduced -- would profit from experimental data at higher excitation energies permitting an improved description and reducing the MDA uncertainties. 

The few results from the present work on the polarizability already made a significant impact on our understanding of the density dependence of the symmetry energy, a key quantity to understand the formation of neutron skins in nuclei and the EOS of neutron-rich matter governing the properties of neutron stars.
Since the relation is model-dependent, systematic studies are needed to single out appropriate theoretical approaches. 
In heavy nuclei, these are based on mean-field models with either nonrelativistic (Skyrme- of Gogny-type) or relativistic effective interactions \cite{roc18}. 
The situation is reminiscent of the determination of the compressibility of nuclear matter from systematic investigations of the properties of compressional collective modes \cite{gar18}.  
Future experimental work needs to establish the impact of deformation and the role of neutron excess.
The latter is best studied in isotopic chains and an investigation of the stable Sn isotopes is especially promising \cite{bas18}.

Recently, ab-initio calculations based on chiral effective field theory
($\chi$EFT) interactions have been developed to study the correlations between $\alpha_{\rm D}$, neutron skins and symmetry-energy properties in medium-mass nuclei \cite{hag16b,mio16}. 
The present work on $^{48}$Ca \cite{bir17} demonstrates that the polarizability also provides important constraints on the chiral two- plus three-nucleon interactions \cite{heb11,eks15} used to predict neutron-matter properties. 
The coupled-cluster approach used to calculate the dipole strength distribution \cite{hag14} has been extended to heavier nuclei \cite{hag16b} and calculations for heavier magic and semi-magic nuclei are within reach \cite{hag16b,gys19}.
Recent work also indicates the importance of np-nh correlations for a calculation of the dipole polarizability \cite{mio18}.
This can be systematically tested by extending the experiments to nuclei with one or two holes or particles outside closed shells.   
Considering the availability of suitable targets, this can be done in particular around  $^{40}$Ca and $^{90}$Zr.

It is clear that the anticipated new studies of the E1 response in a wide range of nuclei will also offer new insights into the much debated question of the structure underlying the formation of the PDR.   
The experiments provide the full low-energy IV E1 response independent of the problems of decay experiments discussed in sec.~\ref{subsec42}.
These new results will also permit to establish the systematic features of the IV spin-M1 resonance in medium-mass and heavy nuclei allowing, e.g., critical tests of new global parameterizations \cite{gor19}.
This is not only a genuine nuclear-structure problem \cite{hey10} but also impacts the physics of core-collapse supernovae. 
The IVSM1 strength represents the most important response in neutral-current neutrino-nucleus scattering \cite{lan04,sul18}, which serves as an important dissipative mechanism in the shock wave \cite{lan08}. 

With knowledge of the E1 and spin-M1 strength distribution the GSF can be constructed from the (p,p$^\prime$) experiments (except contributions from the orbital M1 scissors mode at low energies in deformed nuclei \cite{hey10,gut12}).  
These data contribute to the present discussion on the validity of the generalized BA hypothesis underlying all applications of GSFs in statistical model reaction calculations.
While the BA hypothesis seems to hold in the energy region of the IVGDR and in quasi-continuum $\gamma$ decay it has been questioned for the special case of g.s.\ absorption experiments at excitation energies around and below the neutron threshold.
Recent work trying to map the $\gamma$ decay in NRF experiments \cite{loe16,isa19} indicate non-statistical contributions, and it is speculated that these are related to the states forming the PDR.
Again systematics are needed, in particular on the role of g.s.\ deformation, which determines the relative importance of the IVGDR at low excitation energies.
A coordinated program of comparing GSFs deduced from the (p,p$^\prime$) experiments with those from the Oslo method in the same nuclei using the methods described in sec.~\ref{sec7} is foreseen.    

%%%% --- Figure ---%%%%
\begin{figure}
\begin{center}
\resizebox{0.40\textwidth}{!}{\includegraphics{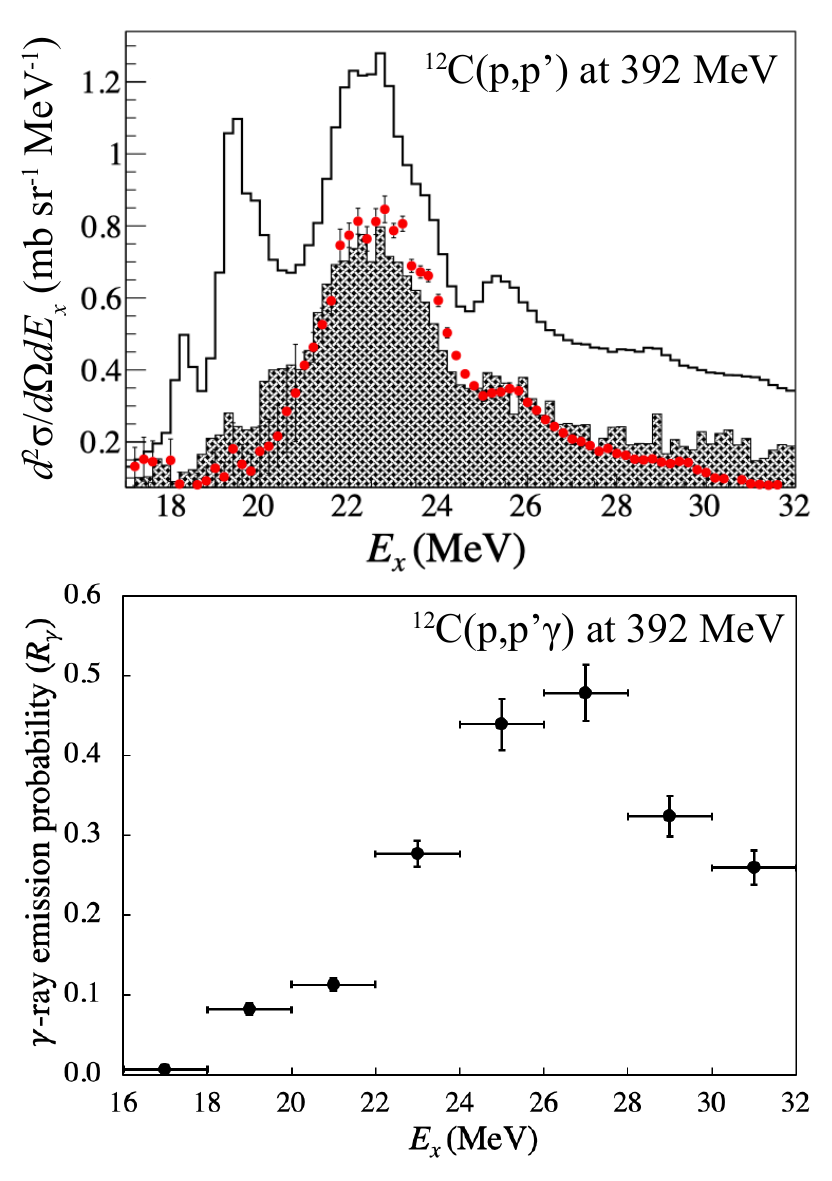}}
\caption{
Upper panel: the double differential cross section (solid line) of the $^{12}{\rm C}(p,p')$ reaction at 392 MeV and at the scattering angle of 0-3.5$^\circ$, non-spin-flip cross section (hatched area), and Coulomb-excitation cross section calculated from the photo-absorption cross section measured by ($\gamma$,total)~\cite{ahr75,ful85} (solid circles).
Lower panel: the gamma-emission probability from the excited $^{12}{\rm C}$.
The figures were taken from \cite{ree19}.} 
\label{fig811}  
\end{center} 
\end{figure}
%%%%%%%%%%%%%
One recent extension of the present studies are $\gamma$ coincidence measurements.
Since the beam is stopped far from the target position with efficient radiation shielding in the zero-degree setup, $\gamma$ detectors sensitive to the surrounding background radiation can be placed close to the target.
A first physics measurement has been carried out on $^{12}{\rm C}$ and $^{16}{\rm O}$ nuclei by placing twenty-five NaI detectors at a distance of 10~cm from the target~\cite{ree19}.
The experiment aims at a measurement of the gamma-emission probability after particle decay of the giant resonances for calibration of neutral-current neutrino detection~\cite{lan96} by organic liquid scintillators (KAMLAND)~\cite{KAMLAND} or  by water \v{C}erenkov detectors (Super Kamiokande)~\cite{SuperK}.
The double differential cross section for $^{12}{\rm C}$ at $E_0=392$ MeV is plotted by the solid curve in the upper panel of Fig.~\ref{fig811}.
The hatched region shows the non-spin-flip cross section determined using the spin-flip probability measured in the same kinematics~\cite{tam99}.
It is almost fully described by the cross section calculated from ($\gamma$,total) photoabsorption cross sections \cite{ahr75,ful85} with the eikonal approximation (see sec.~\ref{subsec33}). 
The $\gamma$-emission probability shown in the lower panel of fig.~\ref{fig811} is found to increase with excitation energy  up to 27 MeV with a maximum $\gamma$-emission probability of about 50\% and then to decrease again.

Furthermore, $\gamma$-coincidence measurements were performed with the CAGRA array (Clover Array Gamma-ray spectrometer at RCNP/RIBF for Advanced research) \cite{cagra,sul18} in a larger campaign at  RCNP combining it with the Grand Raiden spectrometer.
One of the motivations of the campaign was to study the IS electric dipole strength distribution and the $\gamma$ decay of the PDR in nickel, zirconium, tin and lead isotopes.
Recently, a pilot experiment testing the feasibility of measurements of direct $\gamma$ decay from the IVGDR to the ground or low-lying states with large-volume LaBr$_3$:Ce detectors~\cite{gia13}  has been performed on $^{90}$Zr \cite{scylla}.
Also, setups  for coincidence measurements of particle (CAKE)~\cite{cake} and $\gamma$ (BAGEL~\cite{bagel}, ALBA~\cite{alba}) decay at the zero-degree setup have been developed at iThemba LABS, and first experiments were successfully conducted.

Finally, the capability to perform these experiments with high energy resolution provides unique information.
In light nuclei, it allows an analysis on a state-by-state basis up to high excitation energies \cite{mat15,bir16,mat17}.
The fine structure observed in the energy region of the IVGDR carries information on the role of different damping mechanisms contributing to the width of the resonance.  
A quantitative description of giant resonance widths remains a challenge to state-of-the-art nuclear structure calculations.
Furthermore, the magnitude of the fine-structure fluctuations depends on the LD, which can be extracted in an energy region where a single excitation mode dominates.   
Unlike many other experimental techniques, where the LD is inferred from the comparison of experimental observables to statistical-model predictions, the fluctuation analysis provides LDs for a given spin and parity.
This allows a direct comparison with LD models.

Since LDs are an essential ingredient of astrophysical network calculations, such data provide an important test of microscopic model predictions \cite{gor18,zel19} needed to describe reaction paths in exotic nuclei like the $r$ and $p$ processes.    
Experimentally, $fp$-shell nuclei, for which a shell-model approach has been recently developed \cite{sen16}, are of interest.
In heavy nuclei, collective enhancement factors  in EDF-based models  \cite{gor08} and shell-model Monte Carlo calculations \cite{alh15} can be experimentally tested.
Finally, systematic studies of the $J = 1$ LDs can be combined with results for $J = 0$ and 2 from analogous high-resolution studies of the ISGMR \cite{vnc19a} and ISGQR \cite{she09} to experimentally constrain the LD spin distribution.

In conclusion, the present review demonstrates that the experimental techniques developed at RCNP and iThemba LABS to enable the measurement of inelastic proton scattering at very forward angles including $0^\circ$ are extremely fruitful providing contributions to a variety of current problems in nuclear structure, many of them with relevance to nuclear astrophysics.
The ongoing and future programs promise a deeper insight into the physics related to E1 and M1 strengths in nuclei.     

\begin{acknowledgement}

The development and realization of the zero-degree programs at RCNP and iThemba LABS would not have been possible without  the contributions of many colleagues, simply too numerous to name them all. 
Instead, we acknowledge the former and present graduate students working on the (p,p$^\prime$) projects as representatives: Sergej Bassauer, Jonny Birkhan, Lindsay Donaldson, Chihiro Iwamoto, Maxwell Jingo, Andreas Krugmann, Anna Maria Krumbholz, Muftahou Latif, Dirk Martin, Michael Mathy, Hiroaki Matsubara, Iryna Poltoratska and Gerhart Steinhilber. 
Their enthusiasm, ingenuity and commitment has been a true inspiration.
We thank M.N.~Harakeh for a careful reading of the manuscript and many constructive comments.
This work was funded by the Deutsche Forschungsgemeinschaft (DFG, German
Research Foundation) -- Projektnummer 279384907 -- SFB 1245, by JSPS KAKENHI, Grant No.\ JP14740154 and by MEXT KAKENHI, Grant No.\ JP25105509. 
      
\end{acknowledgement}


\begin{thebibliography}{abc99x}

\bibitem{kne06}
U. Kneissl, N. Pietralla, A. Zilges, J. Phys. G {\bf 32}, R217 (2006).

\bibitem{lov81}
W.G. Love, M.A. Franey, Phys. Rev. C {\bf 24}, 1073 (1981).

\bibitem{fuj11}
Y. Fujita, B. Rubio, W. Gelletly, Prog. Part. Nucl. Phys. {\bf 66}, 549 (2011).

\bibitem{mcc84}
J.B. McClelland, T.A. Carey, S. Seestrom-Morris, LAMPF Progress Report 1984, pp.~61.

\bibitem{mcc85}
J.B. McClelland, S.K. Nanda, T.A. Carey, K.W. Jones, M. Murray, M. Plum, T.N. Taddeucci, R.W. Fergerson, D.E. Ciskowski, L. Bimbot, B. Bonin, LAMPF  Progress Report 1985, pp.~58; {\em ibid.} 1987, pp.~37.

\bibitem{IUCF}
G.P.A. Berg, L.C. Bland, B.M. Cox, D. DuPlantis, D.W. Miller, K. Murphy, P. Schwandt, K.A. Solberg, E. J. Stephenson, B. Flanders, IUCF Sci. Tech. Rep. 1986 -- 1987 (unpublished), p.~152.

\bibitem{tam09}
A. Tamii, Y. Fujita, H. Matsubara, T. Adachi, J. Carter, M. Dozono, H. Fujita, K. Fujita, H. Hashimoto, K. Hatanaka, T. Itahashi, M. Itoh, T. Kawabata, K. Nakanishi, S. Ninomiya, A. Perez-Cerdan, L. Popescu, B. Rubio, T. Saito, H. Sakaguchi, Y. Sakemi, Y. Sasamoto, Y. Shimbara, Y. Shimizu, F. Smit, Y. Tameshige, M. Yosoi, J. Zenihiro, Nucl. Instrum. Methods A  {\bf 605}, 326 (2009).

\bibitem{nev11} 
R. Neveling, H. Fujita, F.D. Smit, T. Adachi, G.P.A. Berg, E.Z. Buthelezi, J. Carter, J.L. Conradie, M. Couder, R.W. Fearick, S.V. F\"{o}rtsch, D.T. Fourie, Y. Fujita, J. G\"{o}rres, K. Hatanaka, M. Jingo, A.M. Krumbholz, C.O. Kureba, J P. Mira, S.H.T. Murray, P. von Neumann-Cosel, S. O$'$Brien, P. Papka, I. Poltoratska, A. Richter, E. Sideras-Haddad, J.A. Swartz, A. Tamii, I. Usman, J.J. van Zyl, Nucl. Instrum. Methods A \textbf{654}, 29 (2011); erratum \textbf{662}, 101 (2012).

\bibitem{gar18}
U. Garg, G. Col\`o, Prog. Part. Nucl. Phys. {\bf 101}, 55 (2018).

\bibitem{fre18}
D. Frekers, M. Alanssari, Eur. Phys. J. A {\bf 54}, 177 (2018).

\bibitem{lon18}
A.M. Long, T. Adachi, M. Beard, G.P.A. Berg, M. Couder, R.J. de Boer, M. Dozono, J. G\"{o}rres, H. Fujita, Y. Fujita, K. Hatanaka, D. Ishikawa, T. Kubo, H. Matsubara, Y. Namiki, S. O'Brien, Y. Ohkuma, H. Okamura, H.J. Ong, D. Patel, Y. Sakemi, Y. Shimbara, S. Suzuki, R. Talwar, A. Tamii, A. Volya, T. Wakasa, R. Watanabe, M. Wiescher, R. Yamada, J. Zenihiro,  Phys. Rev. C {\bf 97}, 054613 (2018).

\bibitem{ads17}
P. Adsley, J.W. Br\"ummer, K.C.W. Li, D.J. Marin-Lambarri, N.Y. Kheswa, L.M. Donaldson, R. Neveling, P. Papka, L. Pellegri, V. Pesudo, L.C. Pool, F.D. Smit, J.J. van Zyl,  Phys. Rev. C {\bf  96}, 055802 (2017).

\bibitem{mou11}
B. Mouginot, E. Khan, R. Neveling, F. Azaiez, E.Z. Buthelezi, S.V. F\"ortsch, S. Franchoo, H. Fujita, J. Mabiala, J.P. Mira, P. Papka, A. Ramus, J.A. Scarpaci, F.D. Smit, I. Stefan, J.A. Swartz, I. Usman,  Phys. Rev. C {\bf 83}, 037302 (2011).

\bibitem{sav13}
D. Savran, T. Aumann, A. Zilges, Prog. Part. Nucl. Phys. {\bf 70}, 210 (2013).

\bibitem{adr05}
P.~Adrich, A.~Klimkiewicz, M.~Fallot, K.~Boretzky, T.~Aumann, D.~Cortina-Gil, U.~Datta Pramanik, Th.W.~Elze, H.~Emling, H.~Geissel, M.~Hellstr{\"o}m, K.L. Jones, J.V.~Kratz, R.~Kulessa, Y.~Leifels, C.~Nociforo, R.~Palit, H.~Simon, G.~Surowka, K.~S{\"u}mmerer, W.~Walus, and the LAND-FRS~Collaboration, Phys. Rev. Lett {\bf 95}, 132501 (2005).

\bibitem{kli07}
A.~Klimkiewicz, N.~Paar, P.~Adrich, M.~Fallot, K.~Boretzky, T.~Aumann, D.~Cortina-Gil, U.~Datta Pramanik, Th.W. Elze, H.~Emling, H.Geissel, M.~Hellstr{\"o}m, K.L. Jones, J.V. Kratz, R.~Kulessa, C.~Nociforo, R.~Palit, H.~Simon, G.~Surowka, K.~S{\"u}mmerer, D.~Vretenar, W.~Walus for~the LAND-FRS~Collaboration, Phys. Rev. C {\bf 76}, 051603(R) (2007).

\bibitem{wie09}
O. Wieland {\it et al.},
%O. Wieland, A. Bracco, F. Camera, G. Benzoni, N. Blasi, S. Brambilla, F. C. L. Crespi, S. Leoni, B. Million, R. Nicolini, A. Maj, P. Bednarczyk, J. Grebosz, M. Kmiecik, W. Meczynski, J. Styczen, T. Aumann, A. Banu, T. Beck, F. Becker, L. Caceres, P. Doornenbal, H. Emling, J. Gerl, H. Geissel, M. Gorska, O. Kavatsyuk, M. Kavatsyuk, I. Kojouharov, N. Kurz, R. Lozeva, N. Saito, T. Saito, H. Schaffner, H. J. Wollersheim, J. Jolie, P. Reiter, N. Warr, G. deAngelis, A. Gadea, D. Napoli, S. Lenzi, S. Lunardi,, D. Balabanski, G. LoBianco, C. Petrache, A. Saltarelli, M. Castoldi, A. Zucchiatti, J. Walker, A. B\"urger,
Phys. Rev. Lett. {\bf 102}, 092502 (2009).

\bibitem{ros13}
D.M. Rossi {\it et al.},
%D.M. Rossi, P.~Adrich, F.~Aksouh, H.~Alvarez-Pol, T.~Aumann, J.~Benlliure, M.~B\"ohmer, K.~Boretzky, E.~Casarejos, M.~Chartier, A.~Chatillon, D.~Cortina-Gil, U.~Datta~Pramanik, H.~Emling, O.~Ershova, B.~Fernandez-Dominguez, H.~Geissel, M.~Gorska, M.~Heil, H.T. Johansson, A.~Junghans, A.~Kelic-Heil, O.~Kiselev, A.~Klimkiewicz, J.V. Kratz, R.~Kr\"ucken, N.~Kurz, M.~Labiche, T.~Le~Bleis, R.~Lemmon, Y.~A. Litvinov, K.~Mahata, P.~Maierbeck, A.~Movsesyan, T.~Nilsson, C.~Nociforo, R.~Palit, S.~Paschalis, R.~Plag, R.~Reifarth, D.~Savran, H.~Scheit, H.~Simon, K.~S\"ummerer, A.~Wagner, W.~Walu\ifmmode~\acute{s}\else \'{s}\fi{}, H.~Weick, M.~Winkler,  
Phys. Rev. Lett. {\bf 111}, 242503 (2013).

\bibitem{pie06}
J. Piekarewicz, Phys. Rev. C {\bf 73}, 044325 (2006).

\bibitem{tso08}
N. Tsoneva, H. Lenske, Phys. Rev. C {\bf 77}, 024321 (2008).

\bibitem{pie11}
J. Piekarewicz, Phys. Rev. C {\bf 83}, 034319 (2011).

\bibitem{ina11}
T. Inakura, T. Nakatsukasa, K. Yabana, Phys. Rev. C {\bf 84}, 021302(R) (2011).

\bibitem{car10}
A. Carbone, G. Col\`{o}, A. Bracco, L.-G. Cao, P. F. Bortignon, F.  Camera, O. Wieland, Phys. Rev. C {\bf 81}, 041301 (2010).

\bibitem{fat12}
F.J. Fattoyev, J. Piekarewicz, Phys. Rev. C {\bf 86}, 015802  (2012).
 
\bibitem{tsa12}
M.B. Tsang, J.R. Stone, F. Camera, P. Danielewicz, S. Gandolfi, K. Hebeler, C.J. Horowitz, J. Lee, W.G. Lynch, Z. Kohley, R. Lemmon, P. M\"oller, T. Murakami, S. Riordan, X. Roca-Maza, F. Sammarruca, A.W. Steiner, I. Vida\~na,  S.J. Yennello, Phys. Rev. C {\bf 86}, 015803 (2012). 

\bibitem{rei10}
P.-G. Reinhard, W. Nazarewicz, Phys. Rev. C {\bf 81}, 051303(R) (2010).

\bibitem{rei13}
P.-G. Reinhard, W. Nazarewicz, Phys. Rev. C {\bf 87}, 014324 (2013).

\bibitem{rei14}
P.-G. Reinhard, V.O. Nesterenko, A. Repko, J. Kvasil,  Phys. Rev. C {\bf 89}, 024321 (2014).

\bibitem{gor04}
S. Goriely, E. Khan, M. Samyn, Nucl. Phys. A {\bf 739}, 331 (2004).

\bibitem{lit09}
E. Litvinova, H. P. Loens, K. Langanke, G. Mart\'{i}nez-Pinedo, T. Rauscher, P. Ring, F.-K. Thielemann, V. Tselyaev, Nucl. Phys. A {\bf 823,} 26 (2009).

\bibitem{dao12}
I. Daoutidis, S. Goriely,  Phys. Rev. C {\bf 86}, 034328 (2012).

\bibitem{ben03}
M. Bender, P.-H. Heenen, P.-G. Reinhard, Rev. Mod. Phys. {\bf 75}, 121 (2003).

\bibitem{paa07}
N. Paar, D. Vretenar, E. Khan, G. Col\`{o}, Rep. Prog. Phys. {\bf 70}, 691 (2007).

\bibitem{pap14}
P. Papakonstantinou, H. Hergert, V.Yu. Ponomarev, R. Roth, Phys. Rev. C {\bf 89}, 034306 (2014).

\bibitem{rye02} 
N. Ryezayeva, T. Hartmann, Y. Kalmykov, H. Lenske, P. von Neumann-Cosel, V.Yu. Ponomarev, A. Richter, A. Shevchenko, S. Volz, J. Wambach, Phys. Rev. Lett. \textbf{89}, 272502 (2002).

\bibitem{ton10}
A.P. Tonchev, S.L. Hammond, J.H. Kelley, E. Kwan, H. Lenske, G. Rusev, W. Tornow, N. Tsoneva, Phys. Rev. Lett. {\bf 104}, 072501 (2010).

\bibitem{lit10}
E. Litvinova, P. Ring, V. Tselyaev, Phys. Rev. Lett. {\bf 105}, 022502  (2010).

\bibitem{pol92}
T.D. Poelhekken, S.K.B. Hesmondhalgh, H.J. Hofmann, A. van der Woude, M.N. Harakeh,
Phys. Lett.  B {\bf 278}, 423 (1992).

\bibitem{sav06}
D. Savran, M. Babilon, A.M. van den Berg, M.N. Harakeh, J. Hasper, A. Matic, H.J. W\"ortche, A. Zilges,  Phys. Rev. Lett. {\bf 97}, 172502 (2006).
 
 \bibitem{end10}
J. Endres, E. Litvinova, D. Savran, P.A. Butler, M.N. Harakeh, S. Harissopulos, R.-D. Herzberg, R. Kr\"{u}cken, A. Lagoyannis, N. Pietralla, V.Yu. Ponomarev, L. Popescu, P. Ring, M. Scheck, K. Sonnabend, V.I. Stoica, H.J. W\"{o}rtche, A. Zilges, Phys. Rev. Lett. {\bf 105}, 212503 (2010).

\bibitem{bra15}
A. Bracco, F.C.L. Crespi, E.G. Lanza,  Eur. Phys. J. A {\bf 51}, 99 (2015).

\bibitem{bra19}
A. Bracco, E. Lanza, A. Tamii, Prog. Part. Nucl. Phys. (2019); https://doi.org/10.1016/j.ppnp.2019.02.001.

\bibitem{rus08}
G. Rusev, R. Schwengner, F. D\"{o}nau, M. Erhard, E. Grosse, A.R. Junghans, K. Kosev, K.D. Schilling, A. Wagner, F. Be\c{c}var, M. Krticka, Phys. Rev. C {\bf 77}, 064321 (2008).

\bibitem{rom15}
C. Romig, D. Savran, J. Beller, J. Birkhan, A. Endres, M. Fritzsche, J. Glorius, J. Isaak, N. Pietralla, M. Scheck, L. Schnorrenberger, K. Sonnabend, M. Zweidinger, Phys. Lett. B {\bf 744}, 369 (2015).

\bibitem{loe16}
B. L\"{o}her, D. Savran, T. Aumann, J. Beller, M. Bhike, N. Cooper, V. Derya, M. Duch\^{e}ne, J. Endres, A. Hennig, P. Humby, J. Isaak, J.H. Kelley, M.Kn\"{o}rzer, N. Pietralla, V.Yu. Ponomarev, C. Romig, M. Scheck, H. Scheit, J. Silva, A.P. Tonchev, W. Torn, F. Wamers, H. Weller, V. Werner, A. Zilges, Phys. Lett. B {\bf 756}, 72 (2016).

\bibitem{isa19}
J. Isaak, D. Savran, B. L\"{o}her, T. Beck, M. Bhike, U. Gayer, Krishichayan, N. Pietralla, M. Scheck, W. Tornow, V. Werner, A. Zilges, M. Zweidinger, Phys. Lett. B {\bf 788}, 225 (2019).

\bibitem{lat12}
J.M. Lattimer, Annu. Rev. Nucl. Part. Sci. {\bf 62}, 485 (2012).

\bibitem{heb15}
K. Hebeler, J. Holt, J. Men\'endez, A. Schwenk, Annu. Rev. Nucl. Part. Sci. {\bf 65}, 457 (2015).

\bibitem{wen09} 
De-Hua Wen, Bao-An Li, Lie-Wen Chen, Phys. Rev. Lett. {\bf 103}, 211102 (2009).

\bibitem{pol99} 
S.J. Pollock, M.C. Welliver, Phys. Lett. B {\bf 464}, 177 (1999).

\bibitem{dan02} 
P. Danielewicz,  R. Lacey, W.G. Lynch, Science {\bf 298}, 1592 (2002).

\bibitem{Sto06}
J.~Stone, P.-G. Reinhard, Prog. Part. Nucl. Phys. \textbf{58}, 587 (2006).

\bibitem{lat04} 
J.M. Lattimer, M. Prakash, Science {\bf 304}, 536 (2004).

\bibitem{lat14}
J.M. Lattimer, Nucl. Phys. {\bf A928}, 276 (2014).

\bibitem{heb10}
K. Hebeler, J.M. Lattimer, C.J. Pethick, A. Schwenk, Phys. Rev. Lett. {\bf 105}, 161102 (2010).

\bibitem{mol95ra}
P. M\"oller, J.R. Nix, W.D. Myers, W.J. Swiatecki,  At. Data Nucl. Data Tables \textbf{59}, 185 (1995).

\bibitem{roc18}
X. Roca-Maza, N. Paar, Prog. Part. Nucl. Phys. {\bf 101}, 96 (2018).

\bibitem{epj50}
{\it Topical Issue on Nuclear Symmetry Energy}, Eds.\ Bao-An Li, A. Ramos, G. Verde, I. Vida\~{n}a, Eur. Phys. J. A {\bf 50(2)}, (2014).

\bibitem{abb17}
B.P. Abbott {\it et al.}, Phys. Rev. Lett.  {\bf 119}, 161101 (2017).

\bibitem{fat18}
F.J. Fattoyev, J. Piekarewicz, C.J. Horowitz, Phys. Rev. Lett.  {\bf 120}, 172702 (2018).
 
\bibitem{mos18}
E.R. Most, L.R. Weih, L. Rezzolla, J. Schaffner-Bielic, Phys. Rev. Lett. {\bf 120}, 261103 (2018).

\bibitem{klu09}
P. Kl\"{u}pfel, P.-G. Reinhard, T.J. B\"{u}rvenich, J.A. Maruhn, Phys. Rev. C {\bf 79}, 034310 (2009).

\bibitem{erl13}
J. Erler, C J. Horowitz, W. Nazarewicz, M. Rafalski, P.-G. Reinhard, Phys. Rev. C  {\bf 87}, 044320 (2013).

\bibitem{naz14}
W. Nazarewicz,  P.-G. Reinhard, W. Satu{\l}a, D. Vretenar, Eur. Phys. J. A {\bf 50}, 20 (2104).

\bibitem{boh81}
O. Bohigas, N. Van Giai, D. Vautherin, Phys. Lett. B {\bf 102}, 105 (1981).

\bibitem{ber75}
B.L. Berman, S.C. Fultz, Rev. Mod. Phys. {\bf 47}, 713 (1975).

\bibitem{hey10}
K. Heyde, P. von Neumann-Cosel, A. Richter, Rev. Mod. Phys. {\bf 82}, 2365 (2010).

\bibitem{lan04}
K. Langanke, G. Mart\'inez-Pinedo, P. von Neumann-Cosel, A. Richter, Phys. Rev. Lett. {\bf 93}, 202501 (2004). 

\bibitem{lan08}
K. Langanke, G. Mart\'{i}nez-Pinedo, B. M\"uller, H.-Th. Janka, A. Marek, W.R. Hix, A. Juodagalvis, J.M. Sampaio, Phys. Rev. Lett. {\bf 100}, 011101 (2008).

\bibitem{cha11}
M.B. Chadwick {\it et al.}, Nucl. Data Sheets {\bf 112}, 2887 (2011).

\bibitem{loe12}
H.P. Loens, K. Langanke, G. Mart\'{i}nez-Pinedo, K. Sieja, Eur. Phys. J. A {\bf 48}, 34 (2012).

\bibitem{ots05}
T. Otsuka, T. Suzuki, R. Fujimoto, H. Grawe, Y. Akaishi, Phys. Rev. Lett. {\bf 95}, 232502 (2005). 

\bibitem{ots10}
T. Otsuka, T. Suzuki, M. Honma, Y. Utsuno, N. Tsunoda, K. Tsukiyama,  M. Hjorth-Jensen,
Phys. Rev. Lett. {\bf 104}, 012501 (2010). 

\bibitem{ost92}
F. Osterfeld, Rev. Mod. Phys. {\bf 64}, 491 (1992).

\bibitem{ver12}
J.D. Vergados, H. Ejiri, F. Simkovic, Rep. Prog. Phys. {\bf 75}, 103601 (2012).

\bibitem{ich06}
M. Ichimura, H. Sakai, T. Wakasa, Prog. Part. Nucl. Phys. {\bf 56}, 446 (2006).

\bibitem{ric90}
A. Richter, A. Weiss, B.A. Brown, O. H\"{a}usser, Phys. Rev. Lett. {\bf 65}, 2515 (1990).

\bibitem{lue96}
C. L\"uttge, P. von Neumann-Cosel, F. Neumeyer, C. Rangacharyulu, A. Richter, G. Schrieder, E. Spamer, D.I. Sober, S.K. Matthews, B.A. Brown, Phys. Rev. C {\bf 53}, 127 (1996).

\bibitem{vnc97}
P. von Neumann-Cosel, A. Richter, Y. Fujita, B.D. Anderson, Phys. Rev. C {\bf 55}, 532 (1997).

\bibitem{hof02}
F. Hofmann, P. von Neumann-Cosel, F. Neumeyer, C. Rangacharyulu, B. Reitz, A. Richter, G. Schrieder, D.I. Sober, L.W. Fagg, B.A. Brown, Phys. Rev. C {\bf 65}, 024311 (2002).

\bibitem{mar96}
G. Mart\'{i}nez-Pinedo, A. Poves, E. Caurier, A.P. Zuker,  Phys. Rev. C {\bf 53}, 2602(R) (1996).

\bibitem{vnc98}
P. von Neumann-Cosel, A. Poves, J. Retamosa, A. Richter,  Phys. Lett. B {\bf 443}, 1 (1998).

\bibitem{las86}
R.M. Laszewski, P. Rullhusen, S.D. Hoblit, S.F. LeBrun, Phys. Rev. C {\bf 34}, 2013(R) (1986).

\bibitem{las87}
R.M. Laszewski, R. Alarcon, S.D. Hoblit,  Phys. Rev. Lett. {\bf 59}, 431 (1987).

\bibitem{las88}
R.M. Laszewski, R. Alarcon, D.S. Dale, S.D. Hoblit, Phys. Rev. Lett. {\bf 61}, 1710 (1988).

\bibitem{ala89}
R. Alarcon, R.M. Laszewski,  D.S. Dale,  Phys. Rev. C {\bf 40}, 1097(R) (1989).

\bibitem{rus13}
G. Rusev,  N. Tsoneva, F. D\"{o}nau, S. Frauendorf, R. Schwengner, A.P. Tonchev, A.S. Adekola, S.L. Hammond, J.H. Kelley, E. Kwan, H. Lenske, W. Tornow, A. Wagner, Phys. Rev. Lett. {\bf 110}, 022503 (2013).

\bibitem{dja82}
C. Djalali, N. Marty, M. Morlet, A. Willis, J. C. Jourdain, N. Anantaraman, G.M. Crawley, A. Galonsky, P. Kitching, Nucl. Phys. A {\bf 388}, 1 (1982).

\bibitem{fre90}
D. Frekers, H.J. W\"ortche, A. Richter, R. Abegg, R.E. Azuma, A. Celler, C. Chan, T.E. Drake, R. Helmer, K.P.  Jackson, J.D.  King, C.A. Miller, R. Schubank, M.C. Vetterli, S. Yen, Phys. Lett.  B {\bf 244}, 178 (1990).

\bibitem{tad87}
T.N. Taddeucci, C.A. Goulding, T.A. Carey, R.C. Byrd, C.D. Goodman, C. Gaarde, J. Larsen, D. Horen, J. Rapaport, E. Sugarbaker, Nucl. Phys.  A {\bf 469}, 125 (1987).

\bibitem{zeg07}
R.G.T. Zegers, T. Adachi, H. Akimune, S.M. Austin, A.M. van den Berg, B.A. Brown, Y. Fujita, M. Fujiwara, S. Gales, C.J. Guess, M.N. Harakeh, H. Hashimoto, K. Hatanaka, R. Hayami, G.W. Hitt, M.E. Howard, M. Itoh, T. Kawabata, K. Kawase, M. Kinoshita, M. Matsubara, K. Nakanishi, S. Nakayama, S. Okumura, T. Ohta, Y. Sakemi, Y. Shimbara, Y. Shimizu, C. Scholl, C. Simenel, Y. Tameshige, A. Tamii, M. Uchida, T. Yamagata, M. Yosoi, Phys. Rev. Lett. {\bf 99}, 202501 (2007).

\bibitem{bir16}
J. Birkhan, H. Matsubara, P. von Neumann-Cosel, N. Pietralla, V.Yu. Ponomarev, A. Richter,
A. Tamii, J. Wambach, Phys. Rev. C {\bf 93}, 041302(R) (2016).

\bibitem{bro87}
B.A.~Brown, B.H.~Wildenthal,  Nucl. Phys. {\bf A474}, 290 (1987). 

\bibitem{tow87}
I. Towner, Phys. Rep. {\bf 155}, 263 (1987).

\bibitem{ari87} 
A.~Arima, K.~Shimizu, W.~Bentz, H.~Hyuga,  Adv. Nucl. Phys. {\bf 18}, 1 (1987). 

\bibitem{ana84} 
N.~Anantaraman, B.A.~Brown, G.M.~Crawley, A.~Galonsky, B.H.~Wildenthal, C.~Djalali, N.~Marty, M.~Morlet, A.~Willis, J.C.~Jourdain,  Phys. Rev. Lett. {\bf 31}, 1409 (1984). 

\bibitem{cra89} 
G.M.~Crawley, C.~Djalali, N.~Marty, M.~Morlet, A.~Willis, N.~Anantaraman, B.A.~Brown, A.~Galonsky, Phys. Rev. C {\bf 39}, 311 (1989). 

\bibitem{bar72}
 G.A. Bartholomew, E.D. Earle, A.J. Ferguson, J.W. Knowles, M.A. Lone, 
Adv. Nucl. Phys., Vol. 7, Ed. M. Baranger and E. Vogt, Plenum Press, New York-London (1973) p. 229


\bibitem{arn07}
M. Arnould, S. Goriely, K. Takahashi, Phys. Rep. {\bf 450}, 97 (2007).

\bibitem{sal11}
M. Salvatore, G. Palmiotti, Prog. Part. Nucl. Phys. {\bf 66}, 144 (2011).

\bibitem{wie12}
M. Wiescher, F. K\" {a}ppeler, K. Langanke, Annu. Rev. Astron. Astrophys. {\bf 50}, 165 (2012).

\bibitem{bri55}
D.M. Brink, PhD thesis, Oxford University (1955).

\bibitem{axe62}
P. Axel, Phys. Rev. {\bf 126}, 671 (1962).

\bibitem{bbb98}
P.F. Bortignon, A. Bracco, R.A. Broglia, {\it Giant Resonances: Nuclear Structure at Finite Temperature} (Harwood Academic, Amsterdam, 1998).

\bibitem{joh15}
C.W. Johnson, Phys. Lett. B {\bf 750}, 72 (2015).

\bibitem{hun17}
N. Quang Hung, N. Dinh Dang, L.T. Quynh Huong, Phys. Rev. Lett. {\bf 118}, 022502 (2017).

\bibitem{bas16}
S. Bassauer, P. von Neumann-Cosel, A. Tamii, Phys. Rev. C {\bf 94}, 054313 (2016).

\bibitem{lar17}
A.C. Larsen, M. Guttormsen, N. Blasi, A. Bracco, F. Camera, L. Crespo Campo, T.K. Eriksen, A. G\"orgen, T.W. Hagen, V.W. Ingeberg, B.V. Kheswa, S. Leoni, J.E. Midtbo, B. Million, H.T. Nyhus, T. Renstr{\o}m, S.J. Rose, I.E. Ruud, S. Siem, T.G. Tornyi, G.M. Tveten, A.V. Voinov, M. Wiedeking, F. Zeiser, J. Phys. G {\bf 44}, 064005 (2017).

\bibitem{mar17}
D. Martin, P. von Neumann-Cosel, A. Tamii, N. Aoi, S. Bassauer, C.A. Bertulani, J. Carter, L.M. Donaldson, H. Fujita, Y. Fujita, T. Hashimoto, K. Hatanaka, T. Ito, A. Krugmann, B. Liu, Y. Maeda, K. Miki, R. Neveling, N. Pietralla, I. Poltoratska, V.Yu. Ponomarev, A. Richter, T. Shima, T. Yamamoto, M. Zweidinger, Phys. Rev. Lett. {\bf 119}, 182503 (2017).

\bibitem{vnc19a}
P. von Neumann-Cosel, Acta Phys. Pol. B {\bf 50}, 439 (2019).

\bibitem{aib99}
H. Aiba, M. Matsuo, Phys. Rev. C {\bf 60}, 034307 (1999).

\bibitem{aib11}
%H. Aiba {\it et al.},
H. Aiba, M. Matsuo, S. Nishizaki, T. Suzuki, Phys. Rev. C {\bf 83}, 024314 (2011).

\bibitem{lac99}
D. Lacroix, P. Chomaz, Phys. Rev. C {\bf 60}, 064307 (1999).

\bibitem{lac00}
D. Lacroix, A. Mai, P. von Neumann-Cosel, A. Richter, J. Wambach, Phys. Lett. B {\bf 479}, 15 (2000).

\bibitem{she08} 
A. Shevchenko,  J. Carter, G.R.J. Cooper, R.W. Fearick, S.V. F\"{o}rtsch, Y. Kalmykov, P. von Neumann-Cosel, V.Yu. Ponomarev, A. Richter, I. Usman, J. Wambach, Phys. Rev. C \textbf{77}, 024302 (2008).

\bibitem{hei10}
W.D. Heiss, R.G. Nazmitdinov, F.D. Smit, Phys. Rev. C {\bf 81}, 034604 (2010).

\bibitem{she04} 
A. Shevchenko, J. Carter, R.W. Fearick, S.V. F\"{o}rtsch, H. Fujita, Y. Fujita, Y. Kalmykov, D. Lacroix, J.J. Lawrie, P. von Neumann-Cosel, R. Neveling, V.Yu. Ponomarev, A. Richter, E. Sideras-Haddad, F.D. Smit, J. Wambach, Phys. Rev. Lett \textbf{93}, 122501 (2004).
 
\bibitem{she09} 
A. Shevchenko, O. Burda, J. Carter, G.R.J. Cooper, R.W. Fearick, S.V. F\"{o}rtsch, H. Fujita, Y. Fujita, Y. Kalmykov, D. Lacroix, J.J. Lawrie, P. von Neumann-Cosel, R. Neveling, V.Yu. Ponomarev, A. Richter, E. Sideras-Haddad, F.D. Smit, J. Wambach, Phys. Rev. C \textbf{79}, 044305 (2009). 

\bibitem{usm11} 
I. Usman, Z. Buthelezi, J. Carter, G.R.J. Cooper, R.W. Fearick, S.V. F\"{o}rtsch, H. Fujita, Y. Fujita, Y. Kalmykov, P. von Neumann-Cosel, R. Neveling, P. Papakonstantinou, A. Richter, R. Roth, A. Shevchenko, E. Sideras-Haddad, F.D. Smit, Phys. Lett. B {\bf 698}, 191 (2011).

\bibitem{han90}
P.G. Hansen, B. Jonson, A. Richter, Nucl. Phys. A {\bf 518}, 13 (1990).

\bibitem{alh15}
Y. Alhassid,  Eur. Phys. J. A {\bf 51}, 171 (2015).

\bibitem{kal07}
Y. Kalmykov, C. \"Ozen, K. Langanke, G. Mart\'{i}nez-Pinedo, P. von Neumann-Cosel,  A. Richter, Phys. Rev. Lett. {\bf 99}, 202502 (2007).

\bibitem{sch00}
A. Schiller, L. Bergholt, M. Guttormsen, E. Melby, J. Rekstad, and S. Siem, Nucl. Instrum. Methods A {\bf 447}, 498 (2000).

\bibitem{berg93}
G.P.A. Berg, C.C. Foster, E.J. Stephenson, B.F. Davis, Indiana University Cyclotron Facility Scientific and Technical Report, May 1993 - April 1994, pp.~106.

\bibitem{mer94}
D.J. Mercer, G.M. Crawley, S. Danczyk, A. Galonsky, J. Wang, A. Bacher, G.P.A. Berg, A.C. Betker, W. Schmidt, E.J. Stephenson, Indiana University cyclotron Facility Scientific and Technical Report, May 1994 - April 1995, pp.~20

\bibitem{sak95}
Y. Sakemi, H. Sakaguchi, M. Yosoi, H. Akimune, T. Takahashi, A. Yamagoshi, A. Tamii, M. Fujiwara, K. Hatanaka, K. Hosono, T. Noro, H. Togawa, I. Daito, Y. Fujita, T. Inomata, Phys. Rev. C {\bf 51}, 3162 (1995).

\bibitem{tam99}
A. Tamii, H. Akimune, I. Daito, Y. Fujita, M. Fujiwara, K. Hatanaka, K. Hosono, F. Ihara, T. Inomata, T. Ishikawa, M. Itoh, M. Kawabata, T. Kawabata, M. Nakamura, T. Noro, E. Obayashi, H. Sakaguchi, H. Takeda, T. Taki, H. Toyokawa, H.P. Yoshida, M. Yoshimura, M. Yosoi, Phys. Lett. B {\bf 459}, 61 (1999).

\bibitem{kaw02}
T. Kawabata, T. Ishikawa, M. Itoh, M. Nakamura, H. Sakaguchi, H. Takeda, T. Taki, M. Uchida, Y. Yasuda, M. Yosoi, H. Akimune, K. Yamasaki, G.P.A. Berg, H. Fujimura, K. Hara, K. Hatanaka, J. Kamiya, T. Noro, E. Obayashi, T. Wakasa, H.P. Yoshida, B.A. Brown, H. Fujita, Y. Fujita, Y. Shimbara, H. Ueno, M. Fujiwara, K. Hosono, A. Tamii, H. Toyokawa, Phys. Rev. C {\bf 65}, 064316 (2002).

\bibitem{fuj07}
H. Fujita,  Y. Fujita, T. Adachi, A.D. Bacher, G.P.A. Berg, T. Black, E. Caurier, C.C. Foster, H. Fujimura, K. Hara, K. Harada, K. Hatanaka, J. J\"{a}necke, J. Kamiya, Y. Kanzaki, K. Katori, T. Kawabata, K. Langanke, G. Mart\'inez-Pinedo, T. Noro, D.A. Roberts, H. Sakaguchi, Y. Shimbara, T. Shinada, E.J. Stephenson, H. Ueno, T. Yamanaka, M. Yoshifuku, M. Yosoi, Phys. Rev. C {\bf 75}, 034310 (2007).

\bibitem{ree19}
M.S. Reen, I. Ou, T. Sudo, Y. Yamada, T. Shirahige, D. Fukuda, T. Mori, A. Ali, Y. Koshio, M. Sakuda, A. Tamii, N. Aoi, M. Yosoi, E. Ideguchi, T. Suzuki, T. Yamamoto, C. Iwamoto, T. Kawabata, S. Adachi, M. Tsumura, M. Murata, T. Furuno, H. Akimune, T. Yano, T. Suzuki, R. Dhir, submitted to Phys. Rev. C (2019).

\bibitem{cagra}
E. Ideguchi {\it et al.}, 
CAGRA+GR Campaign Experiments, RCNP Annual Report (2016); http://www.rcnp.osaka-u.ac.jp/Divisions/np1-a/CAGRA/index.html.

\bibitem{scylla}
A. Bracco, P. von Neumann-Cosel, A. Tamii, RCNP proposal E498; S. Nakamura {\it et al.}, to be published.

\bibitem{bagel}
Proposal PR251, iThemba LABS. 

\bibitem{alba}
 African LaBr3:Ce array (ALBA), iThemba LABS.

\bibitem{wak02} 
T. Wakasa, K.~Hatanaka, Y.~Fujita, G.P.A.~Berg, H.~Fujimura, H.~Fujita, M.~Itoh, J.~Kamiya, T.~Kawabata, K.~Nagayama, T.~Noro, H.~Sakaguchi, Y.~Shimbara, H.~Takeda, K.~Tamura, H.~Ueno, M.~Uchida, M.~Uraki, M.~Yosoi, Nucl. Instrum. Methods A {\bf 482}, 79 (2002). 

\bibitem{fuj99}
M. Fujiwara, H. Akimune, I. Daito, H. Fujimura, Y. Fujita, K. Hatanaka, H. Ikegami, I. Katayama, K. Nagayama, N. Matsuoka, S. Morinobu, T. Noro, M. Yoshimura, H. Sakaguchi, Y. Sakemi, A. Tamii,  and M. Yosoi, Nucl. Instrum. Methods A {\bf 422}, 488 (1999).

\bibitem{fuj97} 
Y. Fujita, H. Akimune, I. Daito, M. Fujiwara, M.N. Harakeh, T. Inomata, J. J\"anecke, K. Katori, C. L\"uttge, S. Nakayama, P. von Neumann-Cosel, A. Richter, A. Tamii, M. Tanaka, H. Toyakawa, H. Ueno, M. Yosoi, Phys. Rev. C {\bf 55}, 1137 (1997).

\bibitem{fuj02} 
H.~Fujita, Y.~Fujita, G.P.A.~Berg, A.D.~Bacher, C.C.~Foster, K.~Hara, K.~Hatanaka, T.~Kawabata, T.~Noro, H.~Sakaguchi, Y.~Shimbara, T.~Shinada, E.J.~Stephenson, H.~Ueno, M.~Yosoi, Nucl. Instrum. Methods A {\bf 484}, 17 (2002). 

\bibitem{tam96}
A. Tamii, H. Sakaguchi, H. Takeda, M. Yosoi, H. Akimune, M. Fujiwara, H. Ogata, M. Tanaka, H. Togawa, IEEE Trans. on Nucl Sci. NS-{\bf 43}, 2488 (1996).

%\bibitem{LeCroyFERA}
%LeCroy Corporation, model 4300B 16 input fast encoding and readout charge ADC (FERA), model 4301 fast encoding and readout driver module, and model 4303 16 input fast encoding time-to-charge converter.

%\bibitem{LeCroy1190}
%LeCroy Corporation, model 1190 dual port memory.

%\bibitem{REPICRPA220}
%Hayashi-REPIC Corporation, model RPA-220 preamplifier and discriminator card.

%\bibitem{CAENV1190A}
%CAEN, Model V1190A, 128 channel multievent time-to-digital converter.

%\bibitem{LeCroyPCOSIII}
%LeCroy Corporation, PCOS-III proportional chamber operating system.

%\bibitem{CAENV792NC}
%CAEN, model V792NC, 32 channel charge-to-digital converter.

%\bibitem{TechP-TM005}
%Technoland Corporation, model P-TM 005, 16 channel preamplifier and discriminator card.

\bibitem{MIDAS}
MIDAS (Maximum Integration Data Acquisition System), Paul Scherrer Institute, 
https://midas.psi.ch.

\bibitem{ohl72}
G.G. Ohlsen, Rep. Prog. Phys. {\bf 35}, 717 (1972).

\bibitem{apr83}
E. Aprile-Giboni, R. Hausammann, E. Heer, R. Hess, C. Lechanoine-Le Luc, W. Leo, S. Morenzoni, Y. Onel,  D. Rapin, Nucl. Instrum. Methods {\bf 215}, 147 (1983).

\bibitem{mcn85}
M.W. McNaughton, B.E. Bonner, H. Ohnuma, O.B. van Dijk, S. Tsu-Hsun, C.L. Hollas, D.J. Cremans, K.H. McNaughton, P.J. Riley, R.F. Rodebaugh S.-W. Xu, S.E. Turpin, B. Aas, G.S. Weston, Nucl. Instrum. Methods A {\bf 241}, 435 (1985).

\bibitem{tam11}
 A. Tamii, I. Poltoratska, P. von Neumann-Cosel, Y. Fujita, T. Adachi, C.A. Bertulani, J. Carter, M. Dozono, H. Fujita, K. Fujita, K. Hatanaka, D. Ishikawa, M. Itoh, T. Kawabata, Y. Kalmykov, A.M. Krumbholz, E. Litvinova, H. Matsubara, K. Nakanishi, R. Neveling, H. Okamura, H. J. Ong, B. \"{O}zel-Tashenov, V.Yu. Ponomarev, A. Richter, B. Rubio, H. Sakaguchi, Y. Sakemi, Y. Sasamoto, Y. Shimbara, Y. Shimizu, F.D. Smit, T. Suzuki, Y. Tameshige, J. Wambach, R. Yamada, M. Yosoi, J. Zenihiro,  Phys. Rev. Lett. \textbf{107}, 062502 (2011).

\bibitem{bes79}
D. Besset, B. Favier, L.G. Greeniaus, R. Hess, C. Lechanoine, D. Rapin,  D.W. Werren, Nucl. Instrum. Methods {\bf 166}, 515 (1979). 

\bibitem{has15}
T. Hashimoto, A.M. Krumbholz, P.-G. Reinhard, A. Tamii, P. von Neumann-Cosel, T. Adachi, N. Aoi, C.A. Bertulani, H. Fujita, Y. Fujita, E. Ganio\v{g}lu, K. Hatanaka, C. Iwamoto T. Kawabata, N.T. Khai, A. Krugmann, D. Martin, H. Matsubara, K.~Miki, R. Neveling, H. Okamura, H.J. Ong, I. Poltoratska, V.Yu. Ponomarev, A.~Richter, H. Sakaguchi, Y. Shimbara, Y. Shimizu, J. Simonis, F.D. Smit, G.~S\"usoy, J.H. Thies, T. Suzuki, M. Yosoi, J. Zenihiro, Phys. Rev. C {\bf 92}, 031305(R) (2015).

\bibitem{kru15}
A.M. Krumbholz, P. von Neumann-Cosel, T. Hashimoto, A. Tamii, T. Adachi, C.A. Bertulani, H. Fujita, Y. Fujita, E. Ganio\v{g}lu, K. Hatanaka, C. Iwamoto, T. Kawabata, N.T. Khai, A. Krugmann, D. Martin, H. Matsubara, R. Neveling, H. Okamura, H.J. Ong, I. Poltoratska, V.Yu. Ponomarev, A. Richter, H. Sakaguchi, Y. Shimbara, Y. Shimizu, J. Simonis,  F.D. Smit, G. S\"usoy, J.H. Thies, T. Suzuki, M. Yosoi, J. Zenihiro, Phys. Lett. B {\bf 744}, 7 (2015).

\bibitem{mat15}
H. Matsubara, A. Tamii, H. Nakada, T. Adachi, J. Carter, M. Dozono, H. Fujita, K.~Fujita, Y. Fujita, K. Hatanaka, W. Horiuchi, M. Itoh, T. Kawabata, S.~Kuroita, Y. Maeda, P. Navrat\'{i}l, P. von Neumann-Cosel, R. Neveling, H. Okamura, L. Popescu, I. Poltoratska, A. Richter, B. Rubio, H. Sakaguchi, S. Sakaguchi, Y. Sakemi, Y. Sasamoto, Y. Shimbara, Y. Shimizu, F.D. Smit, K. Suda, Y.~Tameshige, H. Tokieda, Y. Yamada, M. Yosoi, J. Zenihiro, Phys. Rev. Lett. {\bf 115}, 102501 (2015).

\bibitem{ber88} 
C.A. Bertulani, G. Baur,  Phys. Rep. \textbf{163}, 299 (1988).

\bibitem{fea18}
R.W. Fearick, B. Erler, H. Matsubara, P. von Neumann-Cosel, A. Richter, R. Roth, A. Tamii, Phys. Rev. C {\bf 97}, 044325 (2018).

\bibitem{era86} 
R.A. Eramzhyan, B.S. Ishkhanov, I.M. Kapitonov, V.G. Neudatchin, Phys. Rep. \textbf{136}, 229 (1986).

\bibitem{wak97}
T. Wakasa, H. Sakai, H. Okamura, H. Otsu, S. Fujita, S. Ishida, N. Sakamoto, T. Uesaka, Y. Satou, M.B. Greenfield, K. Hatanaka Phys. Rev. {\bf C 55}, 2909 (1997).

\bibitem{li09}
T. Li, U. Garg, Y. Liu, R. Marks, B.K. Nayak, P.V. Madhusudhana Rao, M. Fujiwara, H. Hashimoto, K. Nakanishi, S. Okumura, M. Yosoi, M. Ichikawa, M. Itoh, R. Matsuo, T. Terazono, M. Uchida, Y. Iwao, T. Kawabata, T. Murakami, H. Sakaguchi, S. Terashima, Y. Yasuda, J. Zenihiro, H. Akimune, K. Kawase, M.N. Harakeh, Phys. Rev. C {\bf 81}, 034309 (2010).

\bibitem{str00} 
S. Strauch, P. von Neumann-Cosel, C. Rangacharyulu, A. Richter, G. Schrieder, K. Schweda, J. Wambach, Phys. Rev. Lett. \textbf{85}, 2913 (2000).

\bibitem{vnc99} 
P. von Neumann-Cosel, F. Neumeyer, S. Nishizaki, V.Yu. Ponomarev, C. Rangacharyulu, B. Reitz, A. Richter, G. Schrieder, D.I. Sober, T. Waindzoch, J. Wambach, Phys. Rev. Lett. \textbf{82}, 1105 (1999).
  
\bibitem{dwba07}
J. Raynal, program DWBA07, NEA Data Service NEA1209/08.

\bibitem{fra85}
M.A. Franey, W.G. Love, Phys. Rev. C {\bf 31}, 488 (1985).

\bibitem{pol11}
I. Poltoratska, Doctoral thesis D17, TU Darmstadt (2011); 
http://tuprints.ulb.tu-darmstadt.de/2671.

\bibitem{hof07}
F. Hofmann, C. B\"aumer, A.M. van den Berg, D. Frekers, V.M. Hannen, M.N. Harakeh, M. de Huu, Y. Kalmykov, P. von Neumann-Cosel, V.Yu. Ponomarev, S. Rakers, B. Reitz, A. Richter, K. Schweda, A. Shevchenko, J. Wambach, H.J. W\"ortche, Phys. Rev. C {\bf 76}, 014314 (2007).

\bibitem{pol12} 
I. Poltoratska, P. von Neumann-Cosel, A. Tamii, T. Adachi, C.A. Bertulani, J. Carter, M. Dozono, H. Fujita, K. Fujita, Y. Fujita, K. Hatanaka, M. Itoh, T. Kawabata, Y. Kalmykov, A.M. Krumbholz, E. Litvinova, H. Matsubara, K. Nakanishi, R. Neveling, H. Okamura, H.J. Ong, B. \"{O}zel-Tashenov, V.Yu. Ponomarev, A. Richter, B. Rubio, H. Sakaguchi, Y. Sakemi, Y. Sasamoto, Y. Shimbara, Y. Shimizu, F.D. Smit, T. Suzuki, Y. Tameshige, J. Wambach, M. Yosoi, J. Zenihiro, Phys. Rev. C \textbf{85},  041304(R) (2012).

\bibitem{iwa12}
C. Iwamoto, H. Utsunomiya, A. Tamii, H. Akimune, H. Nakada, T. Shima, T. Yamagata, T. Kawabata, Y. Fujita, H. Matsubara, Y. Shimbara, M. Nagashima, T. Suzuki, H. Fujita, M. Sakuda, T. Mori, T. Izumi, A. Okamoto, T. Kondo, B. Bilgier, H.C. Kozer, Y.-W. Lui, K. Hatanaka, Phys. Rev. Lett. {\bf 108}, 262501(2012).

\bibitem{bro14}
B.A. Brown, W.D.M. Rae, Nucl. Data Sheets {\bf 120}, 115 (2014).

\bibitem{bro06}  
B.A.~Brown, W.A.~Richter, Phys. Rev. C {\bf 74}, 034315 (2006). 

\bibitem{ric08}
W.A. Richter, S. Mkhize, B.A. Brown, Phys. Rev. C {\bf 78}, 064302 (2008).

\bibitem{suz00}
T.~Suzuki, Prog. Theor. Phys. {\bf 321}, 859 (2000).

\bibitem{ensdf}
Evaluated Nuclear Structure Data File, https://www.nndc.bnl.gov/ensdf/.

\bibitem{bak97}
F.T. Baker, L. Bimbot, C. Djalali, C. Glashausser, H. Lenske, W.G. Love, M. Morlet, E. Tomasi-Gustafsson, J. Van de Wiele, J. Wambach, A. Willis, Phys. Rep. {\bf 289}, 235 (1997).

%\bibitem{ber85}
%C.A.~Bertulani, G.~Baur, Nucl.~Phys.~A {\bf 442}, 739 (1985).

\bibitem{end03}
J. Enders, P. von Brentano, J. Eberth, A. Fitzler, C. Fransen, R.-D. Herzberg, H. Kaiser, L. K\"aubler, P. von Neumann-Cosel, N. Pietralla, V.Yu. Ponomarev, A. Richter, R. Schwengner, I. Wiedenh\"over, Nucl. Phys. A {\bf 724}, 243 (2003).

\bibitem{shi08}
T. Shizuma, T. Hayakawa, H. Ohgaki, H. Toyokawa, T. Komatsubara, N. Kikuzawa, A.~Tamii, H. Nakada, Phys. Rev. C {\bf 78}, 061303(R) (2008).

\bibitem{sch10}
R. Schwengner, R. Massarczyk, B.A. Brown, R. Beyer, F. D\"{o}nau, M. Erhard, E. Grosse, A.R. Junghans, K. Kosev, C. Nair, G. Rusev, K.D. Schilling, A. Wagner, Phys. Rev. C {\bf 81}, 054315 (2010).

\bibitem{koh87}
R. K\"ohler, J.A. Wartena, H. Weigmann, L. Mewissen, F. Poortmans, J.P. Theobald, S. Raman, Phys. Rev. C {\bf 35}, 1646 (1987).

\bibitem{vey70} 
A. Veyssi\`{e}re, H. Beil, R. Berg\`{e}re, P. Carlos, A. Lepr\^{e}tre,  Nucl. Phys. A {\bf 159}, 561 (1970). 

\bibitem{sch88} 
K.P. Schelhaas, J.M. Henneberg, M. Sanzone-Arenh\"ovel, N. Wieloch-Laufenberg, U. Zurm\"uhl, B. Ziegler, M. Schumacher, F. Wolf,  Nucl. Phys. A {\bf 489}, 189 (1988).

\bibitem{bir17}
J. Birkhan, M. Miorelli, S. Bacca, S. Bassauer, C.A. Bertulani, G. Hagen, H. Matsubara, P. von Neumann-Cosel, T. Papenbrock, N. Pietralla, V.Yu. Ponomarev, A. Richter, A. Schwenk, A. Tamii, Phys. Rev.  Lett. {\bf 118}, 252501 (2017).

\bibitem{ber93} 
C.A. Bertulani, A.M. Nathan, Nucl. Phys. A \textbf{554}, 158 (1993).

\bibitem{bas14}
S. Bassauer, MSc thesis, Technische Universit\"at Darmstadt (2014), unpublished.

\bibitem{sas09} 
 M.~Sasano, H.~Sakai, K.~Yako, T.~Wakasa, S.~Asaji, K.~Fujita, Y.~Fujita, M.B.~Greenfield, Y.~Hagihara, K.~Hatanaka, T.~Kawabata, H.~Kuboki, Y.~Maeda, H.~Okamura, T.~Saito, Y.~Sakemi, K.~Sekiguchi, Y.~Shimizu, Y.~Takahashi, Y.~Tameshige, A.~Tamii, Phys. Rev. C {\bf 79}, 024602 (2009). 

%\bibitem{doz09}
%M. Dozono, T. Wakasa, E. Ihara, S. Asaji, K. Fujita, K. Hatanaka, M. Ichimura, T. Ishida, T. Kaneda, H. Matsubara, Y. Nagasue, T. Noro, Y. Sakemi, Y. Shimizu, H. Takeda, Y. Tameshige, A. Tamii, Y. Yamada, Phys. Rev. C {\bf 80}, 024319 (2009).

\bibitem{zeg06}
R.G.T. Zegers, H. Akimune, S.M. Austin, D. Bazin, A.M. van den Berg, G.P.A. Berg, B.A. Brown, J. Brown, A.L. Cole, I. Daito, Y. Fujita, M. Fujiwara, S. Gal\`es, M.N. Harakeh, H. Hashimoto, R. Hayami, G.W. Hitt, M.E. Howard, M. Itoh, J. J\"anecke, T. Kawabata, K. Kawase, M. Kinoshita, T. Nakamura, K. Nakanishi, S. Nakayama, S. Okumura, W.A. Richter, D.A. Roberts, B.M. Sherrill, Y. Shimbara, M.  Steiner, M. Uchida, H. Ueno, T. Yamagata, M. Yosoi, Phys. Rev. C  {\bf 74}, 024309 (2006).

\bibitem{end05}
J. Enders, P. von Neumann-Cosel, C. Rangacharyulu, A. Richter, Phys. Rev.  C {\bf 71}, 014306 (2005).

\bibitem{fay97}
M.S. Fayache, P. von Neumann-Cosel, A. Richter, Y.Y. Sharon, L. Zamick, Nucl. Phys. A {\bf 627}, 14 (1997).

\bibitem{ric85}
A. Richter, Prog. Part. Nucl. Phys. {\bf 13}, 1 (1985).

\bibitem{tok80}
H. Toki, W. Weise, Phys. Lett. B {\bf 97}, 12 (1980).

\bibitem{kaw04} 
T.~Kawabata, H.~Akimune, H.~Fujimura, H.~Fujita, Y.~Fujita, M.~Fujiwara, K.~Hara, K.Y.~Hara, K.~Hatanaka, T.~Ishikawa, M.~Itoh, J.~Kamiya, S.~Kishi, M.~Nakmura, K.~Nakanishi, T.~Noro, H.~Sakaguchi, Y.~Shimbara, H.~Takeda, A.~Tamii, S.~Terashima, H.~Toyokawa, M.~Uchida, H.~Ueno, T.~Wakasa, Y.~Yasuda, H.P.~Yoshida, M.~Yosoi, Phys. Rev. C {\bf 70}, 034318 (2004). 

\bibitem{mat10}
H. Matsubara, Doctoral thesis, Osaka University (2010). 

\bibitem{pie12}
J. Piekarewicz, B.K. Agrawal, G. Col\`o, W. Nazarewicz, N. Paar, P.-G.  Reinhard, X. Roca-Maza, D. Vretenar, Phys. Rev. C {\bf 85}, 041302(R) (2012).

\bibitem{ang13}
I. Angeli, K.P. Marinova, At. Data. Nucl. Data Tables {\bf 99}, 69 (2013).

\bibitem{tar14} 
%
C.M. Tarbert {\it et al.}, Phys. Rev. Lett. {\bf 112}, 242502 (2014).

\bibitem{klo07} 
%
B. K{\l}os, A. Trzci\'{n}ska, J. Jastr\c{e}bski, T. Czosnyka, M. Kisieli\'{n}ski, P. Lubi\'{n}ski, P. Napiorkowski, L. Pie\'{n}kowski, F.J. Hartmann, B. Ketzer, P. Ring, R. Schmidt, T. von Egidy, R. Smola\'{n}czuk, S. Wycech, K. Gulda, W. Kurcewicz, E. Widmann, B.A. Brown, Phys. Rev. C {\bf 76}, 014311 (2007).

\bibitem{bro07} 
B.A. Brown, G. Shen, G.C. Hillhouse, J. Meng, A. Trzci\ifmmode \acute{n}\else \'{n}\fi{}ska, Phys. Rev. C {\bf 76}, 034305 (2007).

\bibitem{sak17}
H. Sakaguchi, J. Zenihiro,  Prog. Part. Nucl. Phys. {\bf 97}, 1 (2017).

\bibitem{hor01}
C.J.~Horowitz, J. Piekarewicz, Phys. Rev. Lett. {\bf 86}, 5647 (2001).

\bibitem{abr12} 
S. Abrahamyan {\it et al.}, Phys. Rev. Lett. {\bf 108}, 112502 (2012).

\bibitem{lev51}
J.S. Levinger, Phys. Rev. {\bf 84}, 43 (1951).

\bibitem{roc15}
X. Roca-Maza, X. Vi\~{n}as, M. Centelles, B.K. Agrawal, G. Col\`o, N. Paar, J. Piekarewicz, D. Vretenar, Phys. Rev. C {\bf 92}, 064304 (2015).

\bibitem{lat07}
J.M.~Lattimer, M. Prakash, Phys. Rep. {\bf 442}, 109 (2007).

\bibitem{gan12}
S. Gandolfi, J. Carlson, S. Reddy, Phys. Rev. C {\bf 85}, 032801 (2012).

\bibitem{lat16}
J.M. Lattimer, M. Prakash, Phys. Rep. {\bf 621}, 127 (2016). 

\bibitem{roc13}
X. Roca-Maza, M. Brenna, G. Col\`{o}, M. Centelles, X. Vi\~nas, B.K. Agrawal, N. Paar, D. Vretenar, J. Piekarewicz, Phys. Rev. C {\bf 88}, 024316 (2013).

\bibitem{hag16a}
G. Hagen, A. Ekstr\"om, C. Forss\'en, G R. Jansen, W. Nazarewicz, T. Papenbrock, K.A. Wendt, S. Bacca, N. Barnea, B. Carlsson, C. Drischler, K. Hebeler, M. Hjorth-Jensen, M. Miorelli, G. Orlandini, A. Schwenk, J. Simonis, Nat. Phys. {\bf 12}, 186 (2016).

\bibitem{mio16}
M. Miorelli, S. Bacca, N. Barnea, G. Hagen, G.R. Jansen, G. Orlandini, T. Papenbrock, Phys. Rev. C {\bf 94}, 034317 (2016).

\bibitem{heb11}
K. Hebeler, S.K. Bogner, R.J. Furnstahl, A. Nogga, A. Schwenk, Phys. Rev. C {\bf 83}, 031301 (2011).

\bibitem{eks15}
A. Ekstr\"om, G.R. Jansen, K.A. Wendt, G. Hagen, T. Papenbrock, B. Carlsson, C. Forss\'en, M. Hjorth-Jensen, P. Navr\'atil, W. Nazarewicz, Phys. Rev. C {\bf 91}, 051301 (2015).

\bibitem{oke87}
G.J. O'Keefe, M.N. Thompson, Y.I. Assari, R.E. Pywell, K. Shoda, Nucl. Phys. A {\bf 469}, 239 (1987).

\bibitem{ahr75} 
J. Ahrens, H. Borchert, K.H. Czock, H.B. Eppler, H. Gimm, H. Gundrum, M. Kroning, P. Riehn, G. Sita Ram, A. Zieger, B. Ziegler, Nucl. Phys. A \textbf{251}, 479 (1975).

\bibitem{har02}
T. Hartmann, J. Enders, P. Mohr, K. Vogt, S. Volz, A. Zilges, Phys. Rev. C {\bf 65}, 034301 (2002).

\bibitem{nazpc}
W. Nazarewicz, private communication.

\bibitem{sav18}
D. Savran, V. Derya, S. Bagchi, J. Endres, M.N. Harakeh, J.  Isaak, N. Kalantar-Nayestanaki, E.G. Lanza, B. L\"oher, A. Najafi, S. Pascu, S.G. Pickstone, N. Pietralla, V.Yu. Ponomarev, C. Rigollet, C. Romig, M. Spieker, A. Vitturi, A. Zilges,  Phys. Lett. B {\bf 786}, 16 (2018).

\bibitem{rus09} 
G. Rusev, R. Schwengner, R. Beyer, M. Erhard, E. Grosse, A.R. Junghans, K. Kosev, C. Nair, K.D. Schilling, A. Wagner, F. D\"{o}nau,  S. Frauendorf, Phys. Rev. C \textbf{79}, 061302(R) (2009).

\bibitem{oze14}
B. \"{O}zel-Tashenov, J. Enders, H. Lenske, A.M. Krumbholz, E. Litvinova, P. von Neumann-Cosel, I. Poltoratska, A. Richter, G. Rusev, D. Savran, N. Tsoneva, Phys. Rev. C {\bf 90}, 024304 (2014).

\bibitem{lit07}
E. Litvinova, P. Ring, V.I. Tselyaev, Phys. Rev. C {\bf 75}, 064308 (2007).

\bibitem{lit15}
E. Litvinova, Phys. Rev. C {\bf 91}, 034332 (2015).

\bibitem{rau97}
T. Rauscher, F.-K. Thielemann, K.-L. Kratz, Phys. Rev. C {\bf 56}, 1613 (1997).

\bibitem{mas14}
R.~Massarczyk, R.~Schwengner, F.~D\"onau, S.~Frauendorf, M.~Anders, D.~Bemmerer, R.~Beyer, C.~Bhatia, E.~Birgersson, M.~Butterling, Z.~Elekes, A.~Ferrari, E.~Gooden, M.R.~Hannaske, R.~Junghans, A.M.~Kempe, H.~Kelley, J.T.~K\"ogler, A.~Matic, L.~Menzel, M.S.~M\"uller, P.~Reinhardt, T.M.~R\"oder, G.~Rusev, D.~Schilling, K.K.~Schmidt, G.~Schramm, P.~Tonchev, W.~Tornow, A.~Wagner,  Phys. Rev. Lett. {\bf 112}, 072501 (2014).

\bibitem{sch13}
R. Schwengner, R. Massarczyk, G. Rusev, N. Tsoneva, D. Bemmerer, R. Beyer, R. Hannaske, A. R. Junghans, J.H. Kelley, E. Kwan, H. Lenske, M. Marta, R. Raut, K.D. Schilling, A. Tonchev, W. Tornow, A. Wagner,  Phys. Rev. C {\bf 87}, 024306  (2013).

\bibitem{pel14}
L. Pellegri {\it et al.},
%
Phys. Lett. B 738, 519 (2014).

\bibitem{bas18}
S.~Bassauer, P.~von~Neumann-Cosel, A.~Tamii, EPJ Web of Conferences {\bf 178}, 03008 (2018); and to be published.

\bibitem{oro98}
A.M. Oros, K. Heyde, C. De Coster, B. Decroix, Phys. Rev. C {\bf 57}, 990 (1998).

\bibitem{car71}
P. Carlos, H. Beil, R. Berg\`{e}re, A. Lepr\^{e}tre, A. Veyssi\`{e}re, Nucl. Phys. A {\bf 172}, 437 (1971).

\bibitem{car74}
P. Carlos, H. Beil, R. Berg\`{e}re, A. Lepr\^{e}tre, A. De Miniac, A. Veyssi\`{e}re, Nucl. Phys. A {\bf 225}, 171 (1974).

\bibitem{boh75}
A. Bohr and B. R. Mottelson, {\it Nuclear Structure, Vol. II} (Benjamin, Reading, 1975).

\bibitem{don18} 
L.M. Donaldson, C.A. Bertulani, J. Carter, V.O. Nesterenko, P. von Neumann-Cosel, R. Neveling, V.Yu. Ponomarev, P.-G. Reinhard, I.T. Usman, P. Adsley, J.W. Br\"ummer, E.Z. Buthelezi, G.R.J. Cooper, R.W. Fearick, S.V. F\"{o}rtsch, H. Fujita, Y. Fujita, M. Jingo, W. Kleinig, C.O. Kureba, J. Kvasil, M. Latif, K.C.W. Li, J.P. Mira, F. Nemulodi, P. Papka, L. Pellegri, N. Pietralla, A. Richter, E. Sideras-Haddad, F.D. Smit, G.F. Steyn, J.A. Swartz, A. Tamii, Phys. Lett. B \textbf{776}, 133 (2018).

\bibitem{ito03}
M. Itoh, H. Sakaguchi, M. Uchida, T. Ishikawa, T. Kawabata, T. Murakami, H. Takeda, T. Taki, S. Terashima, N. Tsukahara, Y. Yasuda, M. Yosoi, U. Garg, M. Hedden, B. Kharraja, M. Koss, B.K. Nayak, S. Zhu, H. Fujimura, M. Fujiwara, K. Hara, H.P. Yoshida, H. Akimune, M.N. Harakeh, M. Volkerts, Phys. Rev. C{\bf  68}, 064602 (2003).

\bibitem{nes06}
V.O. Nesterenko, W. Kleinig, J. Kvasil, P. Vesely, P.-G. Reinhard, D.S. Dolci, Phys. Rev. C {\bf 74}, (2006).

 \bibitem{cha98}
E. Chabanat, P. Bonche, P. Haensel, J. Meyer, R. Schaeffer, Nucl. Phys. A {\bf 635}, 231 (1998).

\bibitem{kle08}
W. Kleinig, V.O. Nesterenko, J. Kvasil, P.-G. Reinhard, P. Vesely, Phys. Rev. C {\bf 78}, 044313 (2008).


\bibitem{cas01}
R.F. Casten, N.V. Zamfir, Phys. Rev. Lett. {\bf 87}, 052503  (2001).

\bibitem{iac01}
F. Iachello, Phys. Rev. Lett. {\bf 87}, 052502  (2001).

\bibitem{kru14}
A. Krugmann, D. Martin, P. von Neumann-Cosel, N. Pietralla, A. Tamii, EPJ Web of Conferences {\bf 66}, 02060 (2014); and to be published. 

\bibitem{fil14}
D.M. Filipescu, I. Gheorghe, H. Utsunomiya, S. Goriely, T. Renstr\o{}m, H.-T. Nyhus, O. Tesileanu, T. Glodariu, T. Shima, K. Takahisa, S. Miyamoto, Y.-W. Lui, S. Hilaire, S. P\'eru, M. Martini, A.J. Koning, Phys. Rev. C {\bf 90}, 064616 (2014).

\bibitem{var14}
V.V. Varlamov, B.S. Ishkhanov, V.N. Orlin, K.A. Stopani, Eur. Phys. J. A {\bf 50}, 114 (2014).

\bibitem{utspc}
H. Utsunomiya, private communication. 
The data were taken as a part of the PHOENIX collaboration for IAEA-CRP F41032 and submitted to IAEA.

\bibitem{woe94}
H.J. W\"ortche, Doctoral thesis D17, Technische Universit\"at Darmstadt (1994).


\bibitem{ton17}
A.P. Tonchev, N. Tsoneva, C. Bhatia, C.W. Arnold, S. Goriely, S.L. Hammond, J.H. Kelley, E. Kwan, H. Lenske, J. Piekarewicz, R. Raut, G. Rusev, T. Shizuma, W. Tornow, Phys. Lett. B {\bf 773}, 20 (2017).

\bibitem{ste83}
W. Steffen, H.-D. Gr\"af, A. Richter, A. H\"arting, W. Weise, U. Deutschmann, G. Lahm, R. Neuhausen, Nucl. Phys. A {\bf 404}, 413 (1983).

\bibitem{tak88}
K. Takayanagi, K. Shimizu, A. Arima, Nucl. Phys. A {\bf 481}, 313 (1988).

\bibitem{tom11}
J.R. Tompkins, C.W. Arnold, H.J. Karwowski, G.C. Rich, L.G. Sobotka, C.R. Howell, Phys. Rev. C {\bf 84}, 044331 (2011).

\bibitem{lan03}
K. Langanke, G. Mart\'{i}nez-Pinedo, Rev. Mod. Phys. {\bf 75}, 819 (2003). 

\bibitem{deh84}
D. Dehnhard, D.H. Gay, C L. Blilie, S J. Seestrom-Morris, M.A. Franey, C.L. Morris, R.L. Boudrie, T.S. Bhatia, C. Fred Moore, L.C. Bland, H. Ohnuma, Phys. Rev. C {\bf 30}, 242 (1984).

\bibitem{lip84}
E. Lipparini, A. Richter, Phys. Lett. B {\bf 144}, 13 (1984).

\bibitem{cra83}
G.M. Crawley, N. Anantaraman, A. Galonsky, C. Djalali, N. Marty, M. Morlet, A. Willis, and J.-C. Jourdain, Phys. Lett. B {\bf 127}, 322 (1983).

\bibitem{yak09}
K. Yako, M. Sasano, K. Miki, H. Sakai, M. Dozono, D. Frekers, M.B. Greenfield, K. Hatanaka, E. Ihara, M. Kato, T. Kawabata, H. Kuboki, Y. Maeda, H. Matsubara, K. Muto, S. Noji, H. Okamura, T.H. Okabe, S. Sakaguchi, Y. Sakemi, Y. Sasamoto, K. Sekiguchi, Y. Shimizu, K. Suda, Y. Tameshige, A. Tamii, T. Uesaka, T. Wakasa, H. Zheng, Phys. Rev. Lett. {\bf 103}, 012503 (2009).

\bibitem{gre07}
E.-W. Grewe, D. Frekers, S. Rakers, T. Adachi, C. B\"aumer, N.T. Botha, H. Dohmann, H. Fujita, Y. Fujita, K. Hatanaka, K. Nakanishi, A. Negret, R. Neveling, L. Popescu, Y. Sakemi, Y. Shimbara, Y. Shimizu, F.D. Smit, Y. Tameshige, A. Tamii, J. Thies, P. von Brentano, M. Yosoi, R.G.T. Zegers, Phys. Rev. C {\bf 76}, 054307 (2007).

\bibitem{mat17}
M. Mathy, J. Birkhan, H. Matsubara, P. von Neumann-Cosel, N. Pietralla, V.Yu. Ponomarev, A. Richter, A. Tamii, Phys. Rev. C {\bf 95}, 054316 (2017).

\bibitem{and87} 
B.D.~Anderson, T. Chittrakarn, A.R. Baldwin, C. Lebo, R. Madey, P.C. Tandy, J.W. Watson, C.C. Foster, B.A. Brown, B.H. Wildenthal, Phys. Rev. C {\bf 36}, 2195 (1987). 

\bibitem{and91} 
B.D.~Anderson, N. Tamimi, A.R. Baldwin, M. Elaasar, R. Madey, D.M. Manley, M. Mostajabodda’vati, J.W. Watson, W.M. Zhang, C.C. Foster, Phys. Rev. C {\bf 43}, 50 (1991). 

\bibitem{bro83}  
B.A.~Brown, B.H.~Wildenthal, Phys. Rev. C {\bf 28}, 2397 (1983). 

\bibitem{suz08} 
Y.~Suzuki, W.~Horiuchi, M.~Orabi, K.~Arai,  Few-Body Syst. {\bf 42}, 33 (2008).

\bibitem{bar13} 
B.R.~Barrett, P. Navr\'atil, J.P. Vary, Prog. Part. Nucl. Phys. {\bf 69}, 131 (2013). 

\bibitem{pud97} 
B S.~Pudliner, V.R. Pandharipande, J. Carlson, S.C. Pieper, R.B. Wiringa, Phys. Rev. C {\bf 56}, 1720 (1997).

\bibitem{tam68}
R. Tamagaki, Prog. Theor. Phys. {\bf 39}, 91 (1968).

\bibitem{ent03} 
D.R.~Entem, R.~Machleidt, Phys. Rev. C {\bf 68}, 041001 (2003).

\bibitem{tho77} 
R.~Thompson, M. LeMere, Y.C. Tang, Nucl. Phys. A {\bf 286}, 53 (1977).

\bibitem{vnc00} 
P.~von~Neumann-Cosel, H.-D. Gr\"af, U. Kr\"amer, A. Richter, E. Spamer, Nucl. Phys. A {\bf 669}, 3 (2000). 

\bibitem{suz03}
T.~Suzuki, R.~Fujimoto, T.~Otsuka,  Phys. Rev. C {\bf 67}, 044302 (2003). 

\bibitem{rothpc}
R. Roth (private communication).

\bibitem{sag16}
H. Sagawa, T. Suzuki,  M. Sasano, Phys. Rev. C {\bf 94}, 041303(R) (2016).

\bibitem{sag18}
H. Sagawa, T. Suzuki, Phys. Rev. C {\bf 97}, 054333 (2018).

\bibitem{pol14} 
I. Poltoratska, R.W. Fearick, A.M. Krumbholz, E. Litvinova, H. Matsubara, P. von Neumann-Cosel, V.Yu. Ponomarev, A. Richter, A. Tamii, Phys. Rev. C \textbf{89}, 054322 (2014).

\bibitem{kal06} 
Y. Kalmykov, T. Adachi, G.P.A. Berg, H. Fujita, K. Fujita, Y. Fujita, K. Hatanaka, J. Kamiya, A.K. Nakanishi, P. von Neumann-Cosel, V.Yu. Ponomarev, A. Richter, N. Sakamoto, Y. Sakemi, A. Shevchenko, Y. Shimbara, Y. Shimizu, F.D. Smit, T. Wakasa, J. Wambach, M. Yosoi, Phys. Rev. Lett. \textbf{96}, 012502 (2006).

\bibitem{jin18} 
M. Jingo, E.Z. Buthelezi, J. Carter, G.R.J. Cooper, R.W. Fearick, S.V. F\"ortsch, C.O. Kureba A.M. Krumbholz P. von Neumann-Cosel, R. Neveling, P. Papka, I. Poltoratska, V.Yu. Ponomarev, A. Richter, E. Sideras-Haddad, F.D. Smit, J.A. Swartz, A. Tamii, I.T. Usman, Eur. Phys. J A {\bf 54}, 254 (2018).

\bibitem{usm16}
I.T. Usman, Z. Buthelezi, J. Carter, G.R.J. Cooper, R.W. Fearick, S.V. F\"ortsch, H. Fujita, Y. Fujita, P. von Neumann-Cosel, R. Neveling, P. Papakonstantinou, I.~Pysmenetska, A. Richter, R. Roth, E. Sideras-Haddad, F.D. Smit, Phys. Rev. C {\bf 94}, 024308 (2016).

\bibitem{kam97}
S. Kamerdzhiev, J. Lisantti,  P. von Neumann-Cosel,  A. Richter,  G. Tertychny, J. Wambach, Phys. Rev. C {\bf 55}, 2010 (1997).
   
\bibitem{sin65}
P.P. Singh, R.E. Segel,  L. Meyer-Sch\"utzmeister, S.S. Hanna, R.G. Allas,  Nucl. Phys. \textbf{65}, 577 (1965).

\bibitem{law65}
G.P. Lawrence, A.R. Quinton, Nucl. Phys. {\bf 65}, 275 (1965).

\bibitem{put68}
L.W. Put, J.D.A. Roeders, A. Van der Woude, Nucl. Phys. A {\bf 112}, 561 (1968).

\bibitem{dau92}
I. Daubechies, \emph{Ten lectures on wavelets}, SIAM, 1992.

\bibitem{byr92}
J.S.~Byrnes, \emph{Wavelets and their Applications}, (Kluwer Academic, Dordrecht, 1992).

\bibitem{tor98}
C. Torrence,  G. Compo, Bull. Amer. Meteor. Soc. \textbf{79}, 61 (1998).

\bibitem{ire99}
I. Ireland, R. Walsh, R. Harrison, E. Priest, Astron. Astrophys. \textbf{347}, 355 (1999).

\bibitem{fed98}
A. Fedorova, M. Zeitlin, Z. Parsa, Proc. Part. Acc. Conf. 97, Editors M. Comyn, M. K. Craddock, M.  Reiser, J. Thomson, Vol. 2 (1998) pp. 1502, 1505, 1508.

\bibitem{pet10}
I. Petermann, K. Langanke, G. Mart\'{i}nez-Pinedo, P. von~Neumann-Cosel, F. Nowacki, A. Richter,  Phys. Rev. C {\bf 81}, 014308 (2010).

\bibitem{kur18}
C.O. Kureba,  Z. Buthelezi, J. Carter, G.R.J. Cooper, R.W. Fearick, S.V. F\"ortsch, M. Jingo, W. Kleinig, A. Krugmann, A.M. Krumbholz, J. Kvasil, J. Mabiala, J.P. Mira, V.O. Nesterenko, P. von Neumann-Cosel,  R. Neveling, P. Papka,  P.-G. Reinhard, A. Richter, E. Sideras-Haddad, F.D. Smit, G.F. Steyn, J.A. Swartz, A. Tamii, I.T. Usman, Phys. Lett. B {\bf 779}, 269 (2018).

\bibitem{liu07}
Yonggang Liu, X. San Liang, R.H. Weisberg, J. Atmos. Oceanic Technol. {\bf 24}, 2093 (2007).

\bibitem{mal98}
S. Mallat, \textit{A Wavelet Tour of Signal Processing} (Academic Press,
San Diego, 1998).

\bibitem{har01} 
M.N. Harakeh, A. van der Woude,  \textit{Giant Resonances: Fundamental High-Frequency Modes of Nuclear Excitation}, (Oxford University, Oxford, 2001).

\bibitem{bol88}
G.O. Bolme, L.S. Cardman, R. Doerfler, L.J. Koester Jr., B.L. Miller, C.N. Papanicolas, H. Rothhaas, S.E. Williamson, Phys. Rev. Lett. {\bf 61}, 1081 (1988).

\bibitem{die94}
H. Diesener, U. Helm, G. Herbert, V. Huck, P. von Neumann-Cosel, C. Rangacharyulu, A. Richter, G. Schrieder, A. Stascheck, A. Stiller, J. Ryckebusch, J. Carter, Phys. Rev. Lett. {\bf 72}, 1994 (1994).

\bibitem{hun03}
M. Hunyadi, A.M. van den Berg, N. Blasi, C. B\"aumer, M. Csatlos, L. Csige, B. Davids, U. Garg, J. Gulyas, M.N. Harakeh, M.A. de Huu, B.C. Junk, A. Krasznahorkay, S. Rakers, D. Sohler, H.J. W\"ortche, Phys. Lett. B {\bf 576}, 253 (2003).

\bibitem{ber83}
G.F. Bertsch, P.F. Bortignon, R.A. Broglia, Rev. Mod. Phys. {\bf 55}, 287 (1983).

\bibitem{sol92}
 V.G. Soloviev, \textit{Theory of Atomic Nuclei: Quasiparticles and Phonons} (IOP Publishing, Bristol, 1992).

\bibitem{vnc19b}
P. von Neumann-Cosel, V.Yu. Ponomarev, A. Richter, J. Wambach, submitted to Eur. Phys. J. A; arXiv:1905.02579.

\bibitem{don16}
L.M. Donaldson, PhD thesis, University of the Witwatersrand (2016); and to be published.

\bibitem{jon76}
B. Jonson, E. Hagberg, P.G. Hansen, P. Hornsh{\o}j, P. Tidemand-Petersson, Proc. 3rd Int. Conf. on Nuclei far from Stability,  CERN report {\bf 76--13}, 277 (1976).

\bibitem{mue83}
S. M\"{u}ller, F. Beck, D. Meuer,  A. Richter, Phys. Lett. B {\bf 113}, 362 (1982).

\bibitem{kil87}
G. Kilgus, G. K\"uhner, S. M\"uller, A. Richter, W. Kn\"upfer,  Z. Phys. A {\bf 326}, 41 (1987).

\bibitem{end97}
J.~Enders, N.~Huxel, P.~von Neumann-Cosel, A.~Richter, Phys. Rev. Lett. {\bf 79}, 2010 (1997).

\bibitem{eri60}
T. Ericson, Phys. Rev. Lett. {\bf 5}, 430 (1960).

\bibitem{wei09}
H.A. Weidenm\"uller, G.E. Mitchell, Rev. Mod. Phys. {\bf 81}, 539 (2009).

\bibitem{mit10}
G.E. Mitchell, A. Richter, H.A. Weidenm\"uller, Rev. Mod. Phys. {\bf 82}, 2845 (2010).  

\bibitem{haq82}
R.U. Haq, A. Pandey, O. Bohigas, Phys. Rev. Lett. {\bf 48}, 1086 (1982).

\bibitem{egi05}
T. von Egidy, D. Bucurescu, Phys. Rev. C \textbf{72}, 044311 (2005).

\bibitem{egi09}
T. von Egidy, D. Bucurescu,  Phys. Rev. C {\bf 80}, 054310 (2009).

\bibitem{gor08}
S. Goriely, S. Hilaire,  A.J. Koning, Phys. Rev. C \textbf{78}, 064307 (2008).

\bibitem{dem01}
P. Demetriou, S. Goriely, Nucl. Phys. A \textbf{695}, 95 (2001).

\bibitem{nak97}
H. Nakada, Y. Alhassid, Phys. Rev. Lett. {\bf 79}, 2939 (1997).

\bibitem{hil06}
S. Hilaire, S. Goriely, Nucl. Phys. A {\bf 779}, 63 (2006).

\bibitem{hou09}
K. Van Houcke, S.M.A. Rombouts, K. Heyde, Y. Alhassid, Phys. Rev. C {\bf 79}, 024302 (2009).

\bibitem{cap09}
R. Capote, M. Herman, P. Oblo\v{z}insk\'{y}, P.G. Young, S. Goriely, T. Belgya, A.V. Ignatyuk, A.J. Koning, S. Hilaire, V.A. Plujko, M. Avrigeanu, O. Bersillon, M.B. Chadwick, T. Fukahori, Zh. Ge, Y. Han, S. Kailas, J. Kopecky, V.M. Maslov, G. Reffo, M. Sin, E.Sh. Soukhovitskii, P. Talou, Nucl. Data Sheets {\bf 110}, 3107 (2009).

\bibitem{kop93}
J. Kopecky, M. Uhl, R.E. Chrien, Phys. Rev. C {\bf 47}, 312 (1993).

\bibitem{plu99}
V.A. Plujko, Nucl. Phys. A {\bf 649}, 209c (1999).

\bibitem{kop87}
J. Kopecky, R.E. Chrien, Nucl. Phys. A {\bf 468}, 285 (1987). 

\bibitem{you04}
D.H. Youngblood, Y.-W. Lui, H.L. Clark, B. John, Y. Tokimoto, X. Chen, Phys.~Rev.~C {\bf 69}, 034315 (2004).

\bibitem{gut16}
M. Guttormsen, A.C. Larsen, A. G\" {o}rgen, T. Renstr{\o}m, S. Siem, T.G. Tornyi, G.M. Tveten, Phys. Rev. Lett.  {\bf 116}, 012502 (2016).

\bibitem{cre18}
L. Crespo Campo, M. Guttormsen, F.L. Bello Garrote, T.K. Eriksen, F. Giacoppo, A. G\" {o}rgen, K. Hadynska-Klek, M. Klintefjord, A.C. Larsen, T. Renstr{\o}m, E. Sahin, S. Siem, A. Springer, T.G. Tornyi,  G.M. Tveten, Phys. Rev. C {\bf 98}, 054303 (2018).

\bibitem{boh84}
D. Bohle, A. Richter, W. Steffen, A.E.L. Dieperink, N. Lo Iudice, F. Palumbo, O. Scholten,
Phys. Lett. B {\bf 137}, 27 (1984).

\bibitem{gut12}
M. Guttormsen,  L.A. Bernstein, A. B\"urger, A. G\"orgen, F. Gunsing, T.W. Hagen, A.C. Larsen, T. Renstr{\o}m, S. Siem, M. Wiedeking, J.N. Wilson, Phys. Rev. Lett. {\bf 109}, 162503 (2012).

\bibitem{ang16}
C.T. Angell, R. Hajima, T. Shizuma, B. Ludewigt, B.J. Quiter, Phys. Rev. Lett. {\bf 117}, 142501 (2016).

\bibitem{voi04}
A. Voinov, E. Algin, U. Agvaanluvsan, T. Belgya, R. Chankova, M. Guttormsen, G.E. Mitchell, J. Rekstad, A. Schiller,  S. Siem, Phys. Rev. Lett. {\bf 93}, 142504 (2004).

\bibitem{ang12}
C.T. Angell, S L. Hammond, H.J. Karwowski, J.H. Kelley, M. Krti\v{c}ka, E. Kwan, A. Makinaga, G. Rusev, Phys. Rev. C {\bf 86}, 051302(R) (2012).

\bibitem{ang15}
Erratum to ref.~\cite{ang12}, Phys. Rev. C {\bf 91}, 039901(E) (2015).

\bibitem{sye09}
N.U.H. Syed, M. Guttormsen, F. Ingebretsen, A.C. Larsen, T. L\"onnroth, J. Rekstad, A. Schiller, S. Siem, A. Voinov, Phys.~Rev.~C {\bf 79}, 024316 (2009).

\bibitem{heu16}
A. Heusler, R.V. Jolos, T. Faestermann, R. Hertenberger, H.-F. Wirth, P. von Brentano, 
Phys. Rev. C {\bf 93}, 054321 (2016).

\bibitem{gut05}
M. Guttormsen, R. Chankova, U. Agvaanluvsan, E. Algin, L.A. Bernstein, F. Ingebretsen, T. L\" {o}nnroth, S. Messelt, G.E. Mitchell, J. Rekstad, Phys. Rev. C {\bf 71}, 044307 (2005).

\bibitem{lar10}
A.C. Larsen, S. Goriely, Phys. Rev. C {\bf 82}, 014318 (2010).

\bibitem{bei74}
H. Beil, R. Berg\`ere, P. Carlos, A. Lepr\^etre, A. De Miniac, A. Veyssiere, Nucl. Phys. A {\bf 227}, 427 (1974).

\bibitem{uts13}
H. Utsunomiya,  S. Goriely, T. Kondo, C. Iwamoto, H. Akimune, T. Yamagata, H. Toyokawa, H. Harada, F. Kitatani, Y.-W. Lui, A.C. Larsen, M. Guttormsen, P.E. Koehler, S. Hilaire, S. P\'eru, M. Martini, A.J. Koning, Phys. Rev. C {\bf 88}, 015805 (2013).

\bibitem{kal90}
C. Kalbach, Phys. Rev. C {\bf 41}, 1656 (1990).

\bibitem{hag14}
G. Hagen, T. Papenbrock, M. Hjorth-Jensen, D. Dean, Rep. Prog. Phys. {\bf 77}, 096302 (2014).

\bibitem{hag16b}
G. Hagen, G.R. Jansen, T. Papenbrock, Phys. Rev. Lett. {\bf 117}, 172501 (2016).

\bibitem{gys19}
P. Gysbers, G. Hagen, J.D. Holt, G.R. Jansen, T D. Morris, P. Navr\'atil, T. Papenbrock, S. Quaglioni, A. Schwenk, S.R. Stroberg K.A. Wendt,  Nat. Phys. (2019); https://doi.org/10.1038/s41567-019-0450-7.

\bibitem{mio18}
M. Miorelli, S. Bacca, G. Hagen, T. Papenbrock, Phys. Rev. C {\bf 98}, 014324 (2018).

\bibitem{gor19}
S. Goriely, V. Plujko, Phys. Rev. C {\bf 99}, 014303 (2019).

\bibitem{sul18}
C. Sullivan, R.G.T. Zegers, S. Noji, S.M. Austin, J. Schmitt, N. Aoi, D. Bazin, M. Carpenter, J.J. Carroll, H. Fujita, U. Garg, G. Gey, C.J. Guess, T.H. Hoang, M.N. Harakeh, E. Hudson, N. Ichige, E. Ideguchi, A. Inoue, J. Isaak, C. Iwamoto, C. Kacir, T. Koike, N. Kobayashi, S. Lipschutz, M. Liu, P. von Neumann-Cosel, H.J. Ong, J. Pereira, M. Kumar Raju, A. Tamii, R. Titus, V. Werner, Y. Yamamoto, Y.D. Fang, J.C. Zamora, S. Zhu, X. Zhou, Phys. Rev. C {\bf 98}, 015804 (2018).

\bibitem{lan96}
K. Langanke, P. Vogel, E. Kolbe, Phys. Rev. Lett. {\bf 76}, 2629 (1996).

\bibitem{KAMLAND}
A. Suzuki, Nucl. Phys. B (Proc. Suppl.) {\bf 77}, 171 (1999).

\bibitem{SuperK}
Y. Totsuka, Rep. Prog. Phys. {\bf 55}, 377 (1992).

\bibitem{ful85}
E.G. Fuller, Phys. Rep. {\bf 127}, 185 (1985).

\bibitem{gia13}
A. Giaz, L. Pellegri, S. Riboldi, F. Camera, N. Blasi, C. Boiano, A. Bracco, S. Brambilla, S. Ceruti, S. Coelli, F.C.L. Crespi, M. Csatl\`{o}s, S. Frega, J. Guly\`{a}s, A. Krasznahorkay, S. Lodetti, B. Million, A.  Owens, F. Quarati, L. Stuhl, O. Wieland, Nucl. Instrum. Methods A {\bf 729}, 910 (2013).

\bibitem{cake}
P. Adsley, R. Neveling, P. Papka, Z. Dyers, J.W.  Br\"ummer, C.Aa. Diget, N.J. Hubbard, K.C.W. Li, A. Long, D.J. Marin-Lambarri, L. Pellegri, V. Pesudo, L.C. Pool, F.D. Smit, S. Triambak,
JINST {\bf 12}, T02004 (2017).

\bibitem{gor18}
S. Goriely, S. Hilaire, S. P\'eru, K. Sieja,  Phys. Rev. C {\bf 98}, 014327 (2018).

\bibitem{zel19}
V. Zelevinsky, M. Horoi, Prog. Part. Nucl. Phys. {\bf 105}, 180 (2019).

\bibitem{sen16}
R. Sen'kov, V. Zelevinsky,  Phys. Rev. C {\bf 93}, 064304 (2016).



\end{thebibliography}
\end{document}